\documentclass[fleqn,twoside]{article}
\usepackage{amsmath,amssymb,espcrc2,color,cite}
\usepackage[usenames,dvipsnames]{xcolor}

\usepackage{graphicx,soul}
\usepackage[figuresright]{rotating}

\newcommand{\url}[1]{{\tt #1}}
\newcommand{\lsim}
{\;\raisebox{-.3em}{$\stackrel{\displaystyle <}{\sim}$}\;}
\newcommand{\gsim}
{\;\raisebox{-.3em}{$\stackrel{\displaystyle >}{\sim}$}\;}

\newcommand{\gmt}{\ensuremath{(g-2)_\mu}}
\newcommand{\br}{{\rm BR}}
\newcommand{\bsg}{BR($b \to s \gamma$)}
\newcommand{\cls}{\ensuremath{{\rm CL}_s}}
\newcommand{\btn}{BR($B_u \to \tau \nu_\tau$)}

\newcommand{\bmm}{\ensuremath{\br(B_s \to \mu^+\mu^-)}}
\newcommand{\bsmm}{\ensuremath{\br(B_s \to \mu^+\mu^-)}}

\newcommand{\bsdmm}{\ensuremath{\br(B_{s, d} \to \mu^+\mu^-)}}

\newcommand{\ssi}{\ensuremath{\sigma^{\rm SI}_p}}
\newcommand{\ssin}{\ensuremath{\sigma^{\rm SI}_n}}

\newcommand{\MW}{\ensuremath{M_W}}
\newcommand{\MZ}{\ensuremath{M_Z}}
\newcommand{\Mh}{\ensuremath{M_h}}

\newcommand{\MA}{\ensuremath{M_A}}

\newcommand{\mt}{m_t}

\newcommand{\msusy}{M_{\rm SUSY}}
\newcommand{\mgl}{\ensuremath{m_{\tilde g}}}
\newcommand{\msq}{\ensuremath{m_{\tilde q}}}
\newcommand{\msqone}{\ensuremath{m_{\tilde q_1}}}
\newcommand{\msqtwo}{\ensuremath{m_{\tilde q_2}}}
\newcommand{\sto}[1]{\ensuremath{\tilde t_{#1}}}

\newcommand{\stau}[1]{\ensuremath{\tilde \tau_{#1}}}
\newcommand{\stopone}{\ensuremath{\tilde t_{1}}}
\newcommand{\mstop}[1]{\ensuremath{m_{\tilde t_{#1}}}}
\newcommand{\msbot}[1]{\ensuremath{m_{\tilde b_{#1}}}}
\newcommand{\msqt}{\ensuremath{m_{\tilde q_3}}}
\newcommand{\msl}{\ensuremath{m_{\tilde l}}}

\newcommand{\cha}[1]{\tilde \chi^\pm_{#1}}
\newcommand{\champ}[1]{\tilde \chi^\mp_{#1}}
\newcommand{\mcha}[1]{\ensuremath{m_{\tilde \chi^\pm_{#1}}}}
\newcommand{\neu}[1]{\tilde \chi^0_{#1}}
\newcommand{\mneu}[1]{\ensuremath{m_{\tilde \chi^0_{#1}}}}
\newcommand{\mst}[1]{m_{\tilde t_{#1}}}
\newcommand{\msb}[1]{m_{\tilde b_{#1}}}
\newcommand{\mstau}[1]{\ensuremath{m_{\tilde \tau_{#1}}}}
\newcommand{\msmu}[1]{\ensuremath{m_{\tilde \mu_{#1}}}}

\newcommand{\mslep}{\ensuremath{m_{\tilde \ell}}}

\newcommand{\smu}[1]{\tilde \mu_{#1}}

\newcommand{\slep}[1]{\tilde \ell_{#1}}
\newcommand{\tb}{\ensuremath{\tan\beta}}

\newcommand{\tev}{\ensuremath{\,\, \mathrm{TeV}}}
\newcommand{\gev}{\ensuremath{\,\, \mathrm{GeV}}}

\newcommand{\ifb}{\ensuremath{{\rm fb}^{-1}}}
\def\order#1{\ensuremath{{\cal O}(#1)}}
\def\refeq#1{\mbox{Eq.~(\ref{#1})}}

\def\reffi#1{\mbox{Fig.~\ref{#1}}}

\def\refta#1{\mbox{Table~\ref{#1}}}

\def\refse#1{\mbox{Sect.~\ref{#1}}}

\def\citere#1{\mbox{Ref.~\cite{#1}}}

\newcommand{\lhccol}{\ensuremath{{\rm LHC8}_{\rm col}}}
\newcommand{\lhcewk}{\ensuremath{{\rm LHC8}_{\rm EWK}}}
\newcommand{\lhcstop}{\ensuremath{{\rm LHC8}_{\rm stop}}}
\newcommand{\atom}{{\tt Atom}}
\newcommand{\scorpion}{{\tt Scorpion}}
\newcommand{\chisqlhccol}{{\ensuremath{\chi^2(\lhccol)}}}
\newcommand{\chisqlhcewk}{{\ensuremath{\chi^2(\lhcewk)}}}
\newcommand{\chisqlhcstop}{{\ensuremath{\chi^2(\lhcstop)}}}
\newcommand{\chisqatom}{{\ensuremath{\chi^2(\atom)}}}
\newcommand{\chisqscorpion}{{\ensuremath{\chi^2(\scorpion)}}}
\newcommand{\chisqatomscorpion}{{\ensuremath{\chi^2(\atom{\rm~and~}\scorpion)}}}

\definecolor{orange}{rgb}{1,0.5,0}

\definecolor{Gray}{named}{Gray}
\newcommand{\gray}[1]{\color{Gray}#1 \color{Black}}

\newcommand{\ETslash}{\ensuremath{/ \hspace{-.7em} E_T}}

\graphicspath{{figs/}}

\title{\vspace{-4.5cm}
\bf \LARGE The pMSSM10 after LHC Run 1 \\ \vspace{0.5em}}

\author{
{\bf K.J.~de~Vries}\address[Imperial]
   {High\,Energy\,Physics\,Group,\,Blackett\,Laboratory,\,Imperial\,College,\,Prince\,Consort\,Road,\,London\,SW7\,2AZ,\,UK},
{\bf E.A.~Bagnaschi}\address[DESY]
   {DESY, Notkestra{\ss}e 85, D--22607 Hamburg, Germany},
{\bf O.~Buchmueller}{\addressmark[Imperial],
\bf R.~Cavanaugh}\address[FNAL]
   {Fermi National Accelerator Laboratory, P.O. Box 500, 
    Batavia, Illinois 60510, USA}\hbox{$^{\rm ,}$}\address[UIC]
   {Physics Department, University of Illinois at Chicago, Chicago, 
    Illinois 60607-7059, USA},
{\bf M.~Citron}\addressmark[Imperial],
{\bf A.~De~Roeck}\address[CERN]
   {Physics Department, CERN, CH--1211 Geneva 23, Switzerland}\hbox{$^{\rm ,}$}\address[Antwerpen]
   {Antwerp University, B--2610 Wilrijk, Belgium},
 {\bf M.J.~Dolan}\address[SLAC]
{Theory Group, SLAC National Accelerator Laboratory,
2575 Sand Hill Road, Menlo Park, \\ CA 94025-7090, USA \&
ARC Centre of Excellence for Particle Physics at the Terascale, School of Physics, University of Melbourne, 3010, Australia},
{\bf J.R.~Ellis}\address[KCL]{Theoretical Particle Physics
  and Cosmology Group, Department of Physics, King's College London, London~WC2R~2LS, UK}\hbox{$^{\rm ,}$}\addressmark[CERN], 
{\bf H.~Fl\"acher}\address[Bristol]
   {H.H.~Wills Physics Laboratory, University of Bristol, Tyndall Avenue, Bristol BS8 1TL, UK},
{\bf S.~Heinemeyer}\address[Santander]
   {Instituto de F\'{\i}sica de Cantabria (CSIC-UC), 
    E--39005 Santander, Spain},
{\bf G.~Isidori}\address[Zurich]
{Physik-Institut, Universit\"at Z\"urich, CH-8057 Z\"urich, Switzerland},
{\bf S.~Malik}\addressmark[Imperial],
{\bf J.~Marrouche}\addressmark[CERN],
{\bf D.~Mart\'inez~Santos}\address[NIKHEF]{Nikhef National Institute for Subatomic Physics and VU University Amsterdam, Amsterdam, The Netherlands \& Universidade de Santiago de Compostela, 
E-15706 Santiago de Compostela, Spain},
{\bf K.A.~Olive}\address[Minnesota] 
{William I.\ Fine Theoretical Physics Institute, School of Physics and
 Astronomy, University of Minnesota, Minneapolis, Minnesota 55455, USA}, 
{\bf K.~Sakurai}\addressmark[KCL],
{\bf G.~Weiglein}\addressmark[DESY]
}

\begin{document}
\begin{abstract}
\vspace{-0.5em}
We present a frequentist analysis of the parameter space of the pMSSM10, in
which the following 10 soft SUSY-breaking parameters are specified
independently at the mean scalar top mass
scale $\msusy \equiv \sqrt{\mst1 \mst2}$: the
gaugino masses $M_{1,2,3}$,  the first-and second-generation squark
masses $\msqone = \msqtwo$, the third-generation squark mass $\msqt$, a
common slepton mass $\mslep$ and a common trilinear mixing parameter
$A$, as well as the Higgs mixing parameter $\mu$, the pseudoscalar Higgs
mass $\MA$ and $\tb$, the ratio of the two Higgs vacuum expectation values. We use the {\tt MultiNest} sampling algorithm with
$\sim 1.2 \times 10^9$ points to sample the pMSSM10 parameter space. A dedicated study shows that
the sensitivities to strongly-interacting  sparticle masses of ATLAS and
CMS searches for jets, leptons + $\ETslash$ signals depend only weakly on
many of the other pMSSM10 parameters. With the aid of the
{\tt Atom} and {\tt Scorpion} codes, we also implement the LHC searches for
electroweakly-interacting sparticles and light stops, so as to confront the pMSSM10 parameter space with all relevant SUSY searches. 
In addition, our analysis
includes Higgs mass and rate
measurements {using the {\tt HiggsSignals} code}, 
SUSY Higgs exclusion bounds, the measurements of \bmm\ by LHCb 
and CMS, other $B$-physics observables, electroweak precision
observables, the cold dark matter density and the XENON100 and LUX
searches for spin-independent dark matter scattering, assuming
that the cold dark matter is mainly provided by the lightest
neutralino $\neu1$. We show that the pMSSM10 is able to provide a supersymmetric interpretation of \gmt,
unlike the CMSSM, NUHM1 and NUHM2.
As a result, we find (omitting Higgs rates) 
that the minimum {$\chi^2 = 20.5$ with 18 
degrees of freedom (d.o.f.) in the pMSSM10, corresponding to a
$\chi^2$ probability of $30.8$\%, to be compared with 
$\chi^2/{\rm d.o.f.} = 32.8/24 \ (31.1/23) \ (30.3/22)$} in the CMSSM
(NUHM1) (NUHM2). 
We display the one-dimensional likelihood functions for
sparticle masses, and show that they may be significantly lighter in the
pMSSM10 than in the other models, e.g., the gluino may be as light
as $\sim 1250 \gev$ at the 68\% CL, and squarks, stops, electroweak gauginos and
sleptons may be much lighter than in the CMSSM, NUHM1 and NUHM2.
{We discuss the discovery potential of future LHC runs,
 $e^+e^-$ colliders and direct detection experiments.}
\vspace{-0.5em}
\begin{center}
{\tt KCL-PH-TH/2015-15, LCTS/2015-07, CERN-PH-TH/2015-066, \\
DESY 15-046, FTPI-MINN-15/13, UMN-TH-3427/15, SLAC-PUB-16245, FERMILAB-PUB-15-100-CMS}
\end{center}

\end{abstract}

\newpage

\maketitle

\tableofcontents

\section{Introduction}
\label{sec:intro}

The quest for supersymmetry (SUSY) has been among the principal
objectives of the ATLAS and CMS experiments during Run~1 of the Large
Hadron Collider (LHC). However, despite searches in many production and
decay channels, no significant signals have been 
observed~\cite{ATLAS20,CMS20}. 
These negative results impose strong
constraints on $R$-conserving SUSY models,
in particular, which are also constrained by measurements of the mass
and other properties of the Higgs boson~\cite{lhch}, by
precision measurements of rare decays such as  
$B_s \to \mu^+ \mu^-$~\cite{LHCbBsmm,CMSBsmm,BsmmComb,CMSLHCbBsmm}
{and other measurements}. 
Overall, these constraints tend to reduce the capacity of SUSY models to 
alleviate the hierarchy problem. {However},
their impact on a possible resolution of the 
discrepancy between the experimental measurement of \gmt\ 
and theoretical calculations in the Standard Model (SM) depends on
further assumptions as will be discussed below.

There have been many analyses that combine these constraints in global
statistical fits within specific SUSY models based on the minimal
supersymmetric extension of the Standard Model (MSSM)~\cite{HK}. Many of
these analyses assume that the low-energy soft SUSY-breaking parameters
of the MSSM may be extrapolated using the renormalization-group
equations (RGEs) up to some grand unified theory (GUT) scale, where they
are postulated to satisfy some universality conditions. Examples of such
models include the constrained MSSM
(CMSSM)~\cite{funnel,cmssm,AbdusSalam:2011fc}, in which the soft
SUSY-breaking mass parameters $m_0$ and $m_{1/2}$ are assumed to be
universal at the GUT scale, as are the trilinear parameters $A_0$. Other
examples include models that relax the universality assumptions for the
soft SUSY-breaking contributions to the Higgs masses, the NUHM1 \cite{nuhm1} and
NUHM2 \cite{nuhm2} (see also, e.g., \citere{AbdusSalam:2011fc}), 
but retain universality for the slepton, squark and gaugino
masses. Such models are particularly severely constrained by the LHC
searches for colored sparticles, the squarks and gluino,
which also place indirect limits on the masses of sleptons and electroweak
gauginos and higgsinos via the GUT scale constraints, while the
  direct search limits on these particles have much less impact.

An alternative approach is to make no assumption about the RGE extrapolation to
very high energies, but take a purely phenomenological approach in which the
 soft SUSY-breaking parameters are specified at low energies, and are not required to be universal at any
input scale, a class of models referred to as the phenomenological MSSM
with $n$ free parameters (pMSSM$n$) \cite{pMSSM}.
This is the framework explored in this paper.
Favoured mass patterns in a pMSSM$n$ analysis might then give hints
for (alternative) GUT-scale scenarios.

In the absence of any assumptions, the pMSSM has so many parameters that
a thorough analysis of its multi-dimensional parameter space is
computationally prohibitive. Here we restrict our attention to a
ten-dimensional version, the pMSSM10, in which the following assumptions
are made. Motivated by the absence of significant flavor-changing
neutral interactions (FCNI) beyond those in the Standard Model (SM), we
assume that the soft SUSY-breaking contributions to the masses of the
squarks of the first two generations are equal, {which we also assume for} the
three generations of sleptons. The FCNI argument does not motivate any
relation between the soft SUSY-breaking contributions to the masses of
left- and right-handed sfermions, but here we assume for simplicity that
they are equal. As a result, we consider the following 10 parameters in 
our analysis (where ``mass'' is here used as a synonym for a soft
SUSY-breaking parameter{, and the gaugino masses and trilinear couplings are
taken to be real}):
\begin{align}
{\rm 3~gaugino~masses}: & \; M_{1,2,3} \, ,  \nonumber \\
{\rm 2~squark~masses}: & \; m_{\tilde q_1} \, = \, m_{\tilde q_2} \, \ne \, m_{\tilde q_{3}}, \nonumber \\
{\rm 1~slepton~mass}: & \; \mslep \, , \nonumber \\
\label{pMSSM10}
{\rm 1~trilinear~coupling}: & \; A \, ,  \\
{\rm Higgs~mixing~parameter}: & \; \mu \, ,  \nonumber \\
{\rm Pseudoscalar~Higgs~mass}: & \; \MA \, ,  \nonumber \\
{\rm Ratio~of~vevs}: & \; \tb \, .   \nonumber
\end{align}
All of these parameters are specified at a low renormalisation scale,
the mean scalar top mass scale,
$\msusy \equiv \sqrt{\mst1 \mst2}$, close to that of electroweak symmetry
breaking. 

In any pMSSM scenario such as this, the disconnect between the different
gaugino masses allows, for example, the U(1) and SU(2) gauginos to be
much lighter than is possible in GUT-universal models, where their
masses are related to the gluino mass and hence constrained by gluino
searches at the LHC. Likewise, the disconnect between the different
squark masses opens up more possibilities for light stops, and the
disconnect between squark and slepton masses largely frees the latter
from LHC constraints. 

{An important feature of our global analysis is} that the possibilities
for light electroweak gauginos and sleptons reopen an opportunity for an
{significant} SUSY contribution to \gmt\ in the pMSSM, a possibility that is
precluded in simple GUT-universal models 
such as the CMSSM, NUHM1 and NUHM2 by the LHC searches for
strongly-interacting sparticles. As we discuss in detail in this paper,
the pMSSM10's flexibility removes the tension between LHC constraints and the
measured value of \gmt \cite{newBNL}, with the result that the best fit in the pMSSM10
has a global $\chi^2$ probability that is considerably better than in the
CMSSM, NUHM1, NUHM2 or SM. 

The main challenges for a global fit of the pMSSM10 are the efficient sampling of the ten-dimensional parameter space 
and the accurate implementation of the various SUSY searches by ATLAS and CMS.
As in~\cite{mc10}, here we use the sampling algorithm 
{\tt MultiNest}~\cite{multinest} to   
scan efficiently the pMSSM10 parameter space. To achieve sufficient coverage of the relevant parameter space,
approximately  $1.2 \times 10^9$ pMSSM10 points were sampled. 
However, confronting all these sample points individually with all relevant collider searches is computationally impossible.
In order to overcome this problem and still to 
apply the SUSY searches in a consistent and precise manner, we split the LHC searches into three categories. 
In the first category we consider inclusive 
SUSY searches that mainly constrain the production of coloured sparticles, namely the gluino and squarks. 
To apply these searches to the pMSSM10 
parameter space, we follow closely an approach proposed in~\cite{Buchmueller:2013exa},
which uses a variety of inclusive SUSY searches covering different final 
states to establish a simple but accurate look-up table that depends only on the gluino, squark and LSP masses.
Then, in order to implement the other two categories of LHC constraints on the SUSY electroweak
sector and compressed stop spectra,
we treat the LHC searches for electroweakly-interacting sparticles via trileptons and dileptons, and for light stops,
separately using dedicated algorithms validated using the 
{\tt Atom}~\cite{Atom} and {\tt Scorpion}~\cite{Scorpion} codes. 
In all cases we consider the latest SUSY searches from ATLAS and CMS that are based on the full Run~1 data set,
as detailed later in the paper. 
We perform extensive validations of the applications of these searches to the pMSSM10, 
so as to ensure that we make an accurate and comprehensive set of implementations 
of the experimental constraints on the model.   

More information about the scan of the pMSSM10 parameter space using the {\tt MultiNest} technique,
as well as details about our implementations of the LHC searches, are provided in Section~\ref{sec:method}.
Section~\ref{sec:results} discusses the results of the pMSSM10 analysis, {including the best-fit
point and other benchmark points with low sparticle masses that could serve to focus analyses at Run~2 of the LHC.}
Section~\ref{sec:tachyons}
discusses the extent to which the preferred ranges of pMSSM10 parameters permit
renormalization-group extrapolation to GUT scales.
Section~\ref{sec:run2prospects} 
analyses the prospects for  
discovering SUSY in future runs of the LHC{,
  Section~\ref{sec:e+e-prospects} {analyses} the  
prospects for discovering SUSY at possible future $e^+ e^-$ colliders,
and} our conclusions are summarised in Section~\ref{sec:summary}.


\section{Method}
\label{sec:method}

We describe in this Section how we perform a global fit of the pMSSM10 taking into account constraints from 
direct searches for SUSY particles, {the Higgs boson mass and rate measurements, SUSY Higgs 
  exclusion bounds, precision electroweak} observables, $B$-physics observables, and
astrophysical and cosmological constraints on cold dark matter. 
We describe the scanned parameters and their ranges, 
the framework that we use to calculate the observables, 
and the treatment of the various constraints.


\subsection{Parameter Ranges}

As described above we consider a ten-dimension subset (pMSSM10) of the full 
pMSSM parameter space. 
The selected SUSY parameters were listed in \refeq{pMSSM10}, and
the ranges of these parameters that we sample are shown in
Table~\ref{tab:ranges}. We also indicate in the right column of this Table
how we divide the ranges of most of these parameters into segments, as we did previously
for our analyses of the CMSSM, NUHM1 and NUHM2~\cite{mc9,mc10}.

The combinations of these segments constitute boxes, in which we
sample the parameter space using the {\tt MultiNest} package~\cite{multinest}. 
For each box, we choose a prior for which 80\% of the sample has a flat distribution
within the nominal range, and 20\% of the sample is outside the box in normally-distributed tails in
each variable. In this way, our total sample exhibits a smooth overlap between boxes,
eliminating features associated with box boundaries. 
An initial scan over all mass parameters with absolute values 
$ \le 4000\gev$ showed that non-trivial behaviour of the global likelihood function
was restricted to $|M_1|\lesssim500\gev$ and $\msl\lesssim1000\gev$. 
In order to achieve high resolution efficiently, we restricted the ranges of these parameters to
$|M_1|<1000\gev$ and $0<\msl<2000\gev$ in the full scan. 

\begin{table*}[htb!]
\begin{center}
\begin{tabular}{|c|c|c|} \hline
Parameter   &  \; \, Range      & Number of  \\ 
            &             & segments   \\ 
\hline         
$M_1$       &  (-1 ,  1 )\tev  & 2 \\
$M_2$       &  ( 0 ,  4 )\tev  & 2 \\
$M_3$       &  (-4 ,  4 )\tev  & 4 \\
\msq        &  ( 0 ,  4 )\tev  & 2 \\
\msqt       &  ( 0 ,  4 )\tev  & 2 \\
\msl        &  ( 0 ,  2 )\tev  & 1 \\
\MA         &  ( 0 ,  4 )\tev  & 2 \\
$A$         &  (-5  , 5 )\tev  & 1 \\
$\mu$        &  (-5  , 5 )\tev  & 1 \\
\tb         &  ( 1  , 60)      & 1 \\
\hline \hline
Total number of boxes &   & 128     \\
\hline
\end{tabular}
\caption{\it Ranges of the pMSSM10 parameters sampled, together with the numbers of
segments into which each range was divided, and the corresponding number of sample boxes.} 
\label{tab:ranges}
\end{center}
\end{table*}


\subsection{MasterCode Framework}

We calculate the observables that go into the likelihood using the 
{\tt MasterCode} framework~\cite{mc7,mc8,mc8.5,mc9,mc10,mcweb},
which interfaces various public and private codes: {\tt SoftSusy~3.3.9}~\cite{Allanach:2001kg} 
for the spectrum, {\tt FeynWZ}~\cite{Svenetal} for the electroweak precision observables, 
{\tt FeynHiggs~2.10.0}~\cite{FeynHiggs,Mh-logresum} 
for the Higgs sector and \gmt, {\tt SuFla}~\cite{SuFla},
{\tt SuperIso}~\cite{SuperIso}
for the $B$-physics observables, {\tt Micromegas~3.2}~\cite{MicroMegas} for the dark matter
relic density, {\tt SSARD}~\cite{SSARD} for the spin-independent cross-section
\ssi, {\tt SDECAY~1.3b}~\cite{Muhlleitner:2003vg} for calculating sparticle branching ratios, and {\tt HiggsSignals~1.3.0}~\cite{HiggsSignals} 
and {\tt HiggsBounds~4.2.0}~\cite{HiggsBounds} for calculating constraints on the Higgs sector. 
The codes are linked using the SUSY Les Houches Accord (SLHA)~\cite{SLHA}.


\subsection{Electroweak, Flavour, Cosmological and Dark Matter Constraints}

For many of these constraints, we follow very closely our previous implementations, which
were summarized recently in Table~1 in~\cite{mc10}. 
Specifically, we treat all electroweak precision observables, all $B$-physics
observables (except for \bsdmm), \gmt, and the relic density as
Gaussian constraints. 
The $\chi^2$ contribution from \bsdmm, combined here in the quantity
  $R_{\mu\mu}$~\cite{mc9},  is calculated using the combination of 
CMS~\cite{CMSBsmm} and LHCb~\cite{LHCbBsmm} results described
in~\cite{CMSLHCbBsmm}. {We incorporate the current world average of
the branching ratio for \bsg\ from~\cite{HFAG14} combined with the theoretical estimate in the SM from~\cite{Misiak15},
and the recent measurement of the branching ratio for \btn\ by the Belle Collaboration~\cite{Bellebtn}
combined with the SM estimate from~\cite{UTfit15}.} We use the upper limit on the
spin-independent cross section as a function of the lightest neutralino
mass $\mneu1$ from LUX~\cite{lux}, which is slightly stronger than
that from XENON100~\cite{XENON100}, 
taking into account the theoretical uncertainty on \ssi\ as
described in~\cite{mc9}.


\subsection{Higgs Constraints}

We use the recent combination of ATLAS and CMS
measurements of the mass of the Higgs boson: 
$\Mh = 125.09 \pm 0.24 \gev$~\cite{Aad:2015zhl}, which we combine with a {one-$\sigma$}
uncertainty of $1.5 \gev$ in the {\tt FeynHiggs} calculation of $\Mh$ in the MSSM.  

In addition, we refine substantially our treatment of
the Higgs boson 
constraints, as compared with previous analyses in the {\tt MasterCode} framework.
In order to include the observed Higgs signal rates we have
incorporated {\tt HiggsSignals}~\cite{HiggsSignals}, which evaluates the
$\chi^2$ contribution of 77 channels from the Higgs
boson searches at the LHC and the Tevatron (see \citere{HiggsSignals}
for a complete list of references).
A discussion of the effective number of contributing channels is given
in \refse{sec:best-fit} below.

We also take into account the relevant searches for heavy neutral MSSM Higgs bosons via
the $H/A \to \tau^+\tau^-$ channels~\cite{CMSHA,ATLASHA}. We evaluate the corresponding $\chi^2$
contribution using the code 
{\tt HiggsBounds}~\cite{HiggsBounds}, which includes the latest CMS
results~\cite{CMSHA} based on $\sim 25~\ifb$ of data~\footnote{The corresponding ATLAS results~\cite{ATLASHA} have
similar sensitivity, but are documented less completely.}. These results include a combination of the two
possible production modes, $gg \to H/A$ and $b \bar b \to b \bar b H/A$,
which is consistently evaluated depending on the MSSM parameters. Their
implementation in {\tt HiggsBounds} has been tested against the
published CMS data, and very good qualitative and quantitative agreement
had been found~\cite{HBtautau}. Other Higgs boson searches are not
taken into account, as they turn out to be weaker in the
pMSSM10 that we study.


\subsection{LHC Constraints on Sparticle Masses}

A comprehensive and accurate application of the SUSY searches with the full
Run~1 data of the LHC to the pMSSM10 parameter space 
is a central part of this paper. 
As most of these searches have been interpreted by ATLAS and CMS
only in simplified model frameworks, we have introduced supplementary procedures
in order to apply these searches to the complicated sparticle spectrum
content of a full SUSY model such as the pMSSM10.  For this we consider
three separate categories of particle  mass constraints that arise from
the LHC searches: a) generic constraints on coloured sparticles (gluinos
and squarks), b) dedicated constraints on
electroweakly-interacting gauginos, Higgsinos and sleptons, c) dedicated
constraints on stop production in scenarios with compressed spectra.  We
refer to the combination of all these constraints from direct SUSY
searches as the LHC8 constraint, with sectors labelled as \lhccol,
\lhcewk, and \lhcstop, respectively. In the following subsections we
provide further details about 
our implementations of these individual constraints,
discussing in detail the validations of our procedures and the corresponding uncertainties.

We use two dedicated software frameworks for recasting the LHC analyses used in this paper.
Both frameworks implement the full list of cuts of a given experimental
search to obtain yields in the 
respective signal regions of the search. These signal yields are then confronted with the SM 
background yields and observations in data, as reported by the experimental searches. 
Based on these comparisons we construct the standard statistical
estimator \cls~\cite{ATLAS:2011tau}, 
which is also used by the experiments to determine the compatibility of their data with a given signal hypothesis.
In this way it is possible to interpret the various LHC searches in any given SUSY model,
such as those explored in our pMSSM10 scans. 

To recast the ATLAS searches considered in this paper we use 
{\tt Atom}~\cite{Atom}, which is a {\tt Rivet}~\cite{rivet} based
framework.  
{\tt Atom} models the resolutions of LHC detectors by
mapping from the truth-level particles found for example in {\tt PYTHIA~6}~\cite{PYTHIA6} event samples to the reconstructed objects, 
such as $b$-jets and isolated leptons, according to the reported detector performances. 
In particular, the efficiencies of object reconstruction and the parameters associated with the 
momentum smearing are implemented in the form of analytical functions or numerical grids.  
The program has already been used in several studies \cite{atom_application}, and the validation of the code can be found in \cite{atom_validation}.

For the CMS searches we use a private code called {\tt Scorpion}~\cite{Scorpion}
that was already used in~\cite{Buchmueller:2013exa}.
{\tt Scorpion} obtains signal yields for a number of CMS searches 
based on events generated with {\tt PYTHIA~6}~\cite{PYTHIA6} 
that are passed through the 
{\tt DELPHES~3}~\cite{DELPHES3} detector simulation package
using an appropriate data card to emulate the response of the CMS detector.
A significant effort was made to validate the modelling of these analyses by comparing the results 
obtained with the published results of the experimental collaboration. 
For further information on the validation of the CMS searches see~\cite{Buchmueller:2013exa}.

The signal yields from {\tt Atom} and {\tt Scorpion} are confronted with the background
yields and observations obtained from the individual ATLAS and CMS searches, and the corresponding \cls\ is calculated using the 
{\tt LandS} package~\cite{LandS}.
We convert the calculated \cls\ value for a generic spectrum in the MSSM 
into a $\chi^2$ contribution by interpreting it as a p-value for the
signal hypothesis assuming one degree of freedom.


\subsubsection{LHC constraints on coloured sparticles}
\label{LHC8colour} 

In the cases of the CMSSM, NUHM1 and NUHM2, we showed in~\cite{mc9,mc10}
that it was sufficient to extrapolate to other parameter values the exclusion contour in the CMSSM
$(m_0,m_{1/2})$ plane from the ATLAS search for jets+\ETslash~\cite{ATLAS20}
that was given for specified values of $\tan \beta$ and $A_0$. 
We showed that the ATLAS exclusion is, to good approximation,
independent of $\tb$ and $A_0$~\cite{mc8,mc10,cms0l-aT} and, for the
applications to the NUHM1 and the NUHM2, we checked that these limits in the
$(m_0,m_{1/2})$ plane were independent of the degrees of non-universality of the
soft SUSY-breaking contributions to the Higgs masses, within the intrinsic
sampling uncertainties.

In the case of the pMSSM10, however, the implementation of the direct searches for coloured sparticles is less straightforward.
It is computationally impossible to apply all the LHC search constraints
individually to each of the $\sim 1.2 \times 10^9$
parameter choices in our sample. For example, {\tt PYTHIA~6} and {\tt DELPHES~3} take several minutes for the generation
of 10,000 events followed by detector simulation, which is required to determine the signal acceptance and \cls\ of each point sampled in the parameter space.
Instead, we follow an approach outlined in~\cite{Buchmueller:2013exa},
which constructs universal mass 
limits on coloured sparticles by combining an inclusive set of jets + $X$ + 
\ETslash\ searches, as we now describe.  

As was shown in~\cite{Buchmueller:2013exa}, it is possible to establish lower limits on the gluino mass, $\mgl$, and the
third-generation squark mass, $\msqt$, that are independent of the details of the underlying spectrum,
within the intrinsic sampling uncertainties, by combining a suitable set of inclusive SUSY searches. 
In this approach the limits only depend on $\mgl$, $\msqt$ and the mass of 
the lightest sparticle $\mneu1$. The essence of the idea is that strongly-interacting
sparticles decay through a variety of different cascade channels, whose relative probabilities
depend on other model parameters. However, if one combines a sufficiently complete set of
channels of the form jets + $X$ + \ETslash, one will capture essentially all the relevant decay channels.

In order to apply this idea to the pMSSM10 parameter space, we have to 
extend this  approach to include also the generic first- and second- 
generation squark mass, $\msq$, as a free parameter. 
We then construct a `universal' $\chi^2$ function that depends only on
$\mneu1$, \mgl, \msq, and \msqt, as detailed below. 
This function defines our implementation of this \lhccol\ constraint. 
There are two caveats to this approach. One is that the region
of parameter space where $\mst1 - \mneu1$ is small, which is
the object of dedicated searches, requires special attention. The other is
that searches for electroweakly-produced sparticles (sleptons, neutralinos
and charginos) fall outside the scope of the \lhccol\
constraint. We have developed dedicated approaches to establish
accurate LHC limits for the special cases of electroweakly-produced sparticles
and the compressed-stop scenario with
$\mst1 - \mneu1 < \mt$, as described in Sections~\ref{LHC8EWK} and~\ref{LHC8stop}, respectively.

\begin{figure*}[htb!]
\begin{center}
\hspace{-1.3cm}
\resizebox{10cm}{!}{\includegraphics{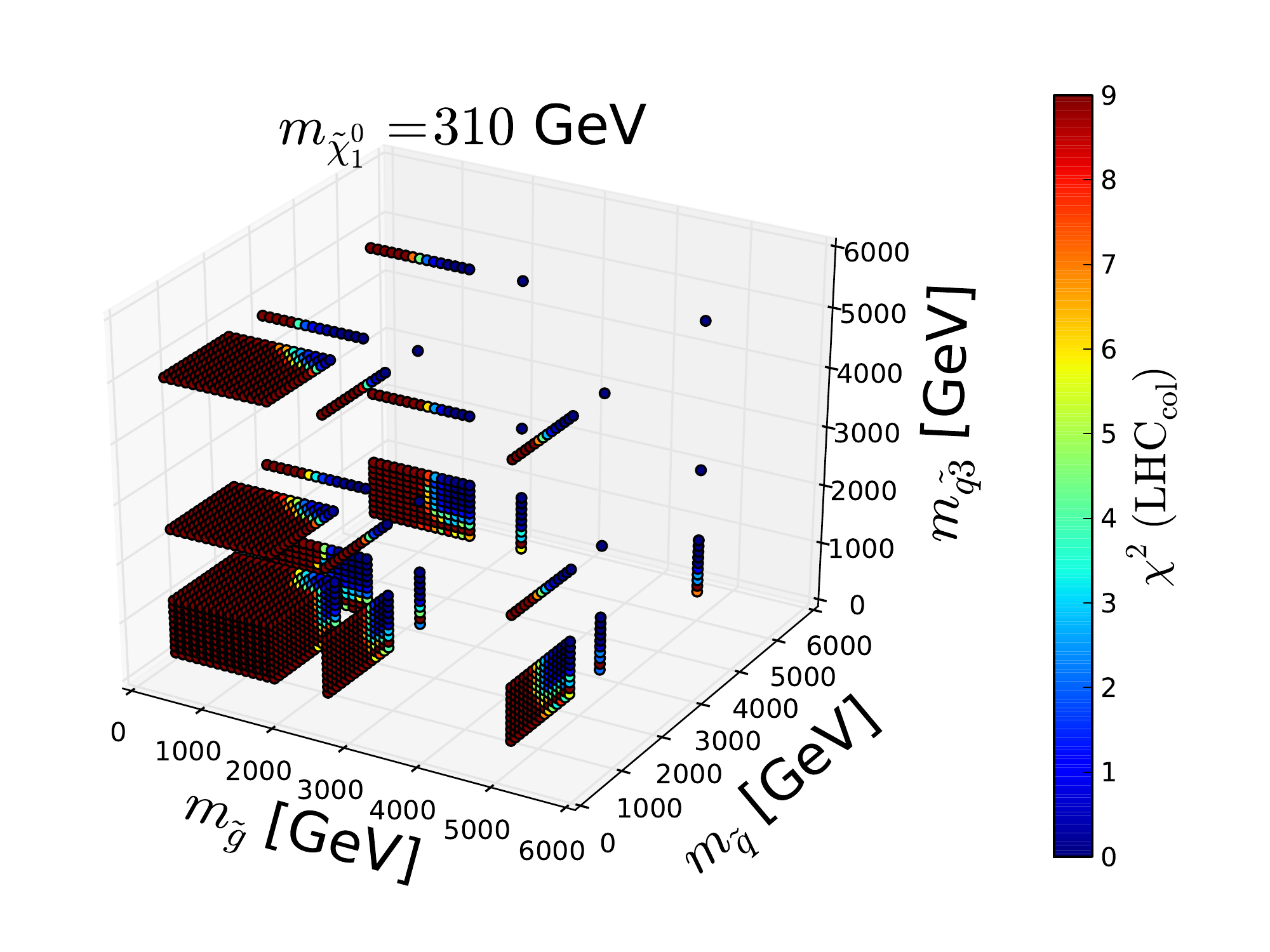}} \\
\hspace{-0.7cm}
\resizebox{7cm}{!}{\includegraphics{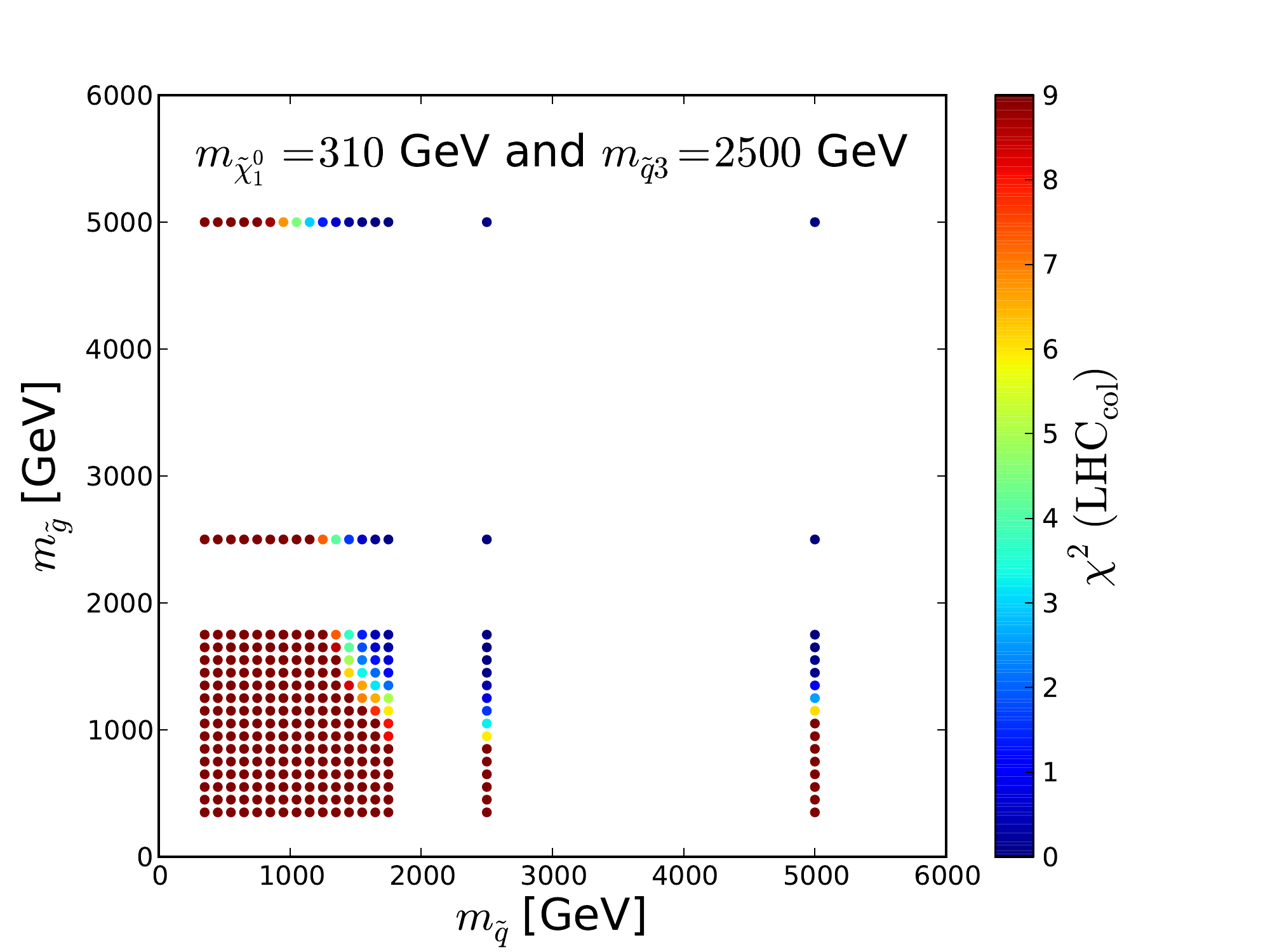}}
\resizebox{7cm}{!}{\includegraphics{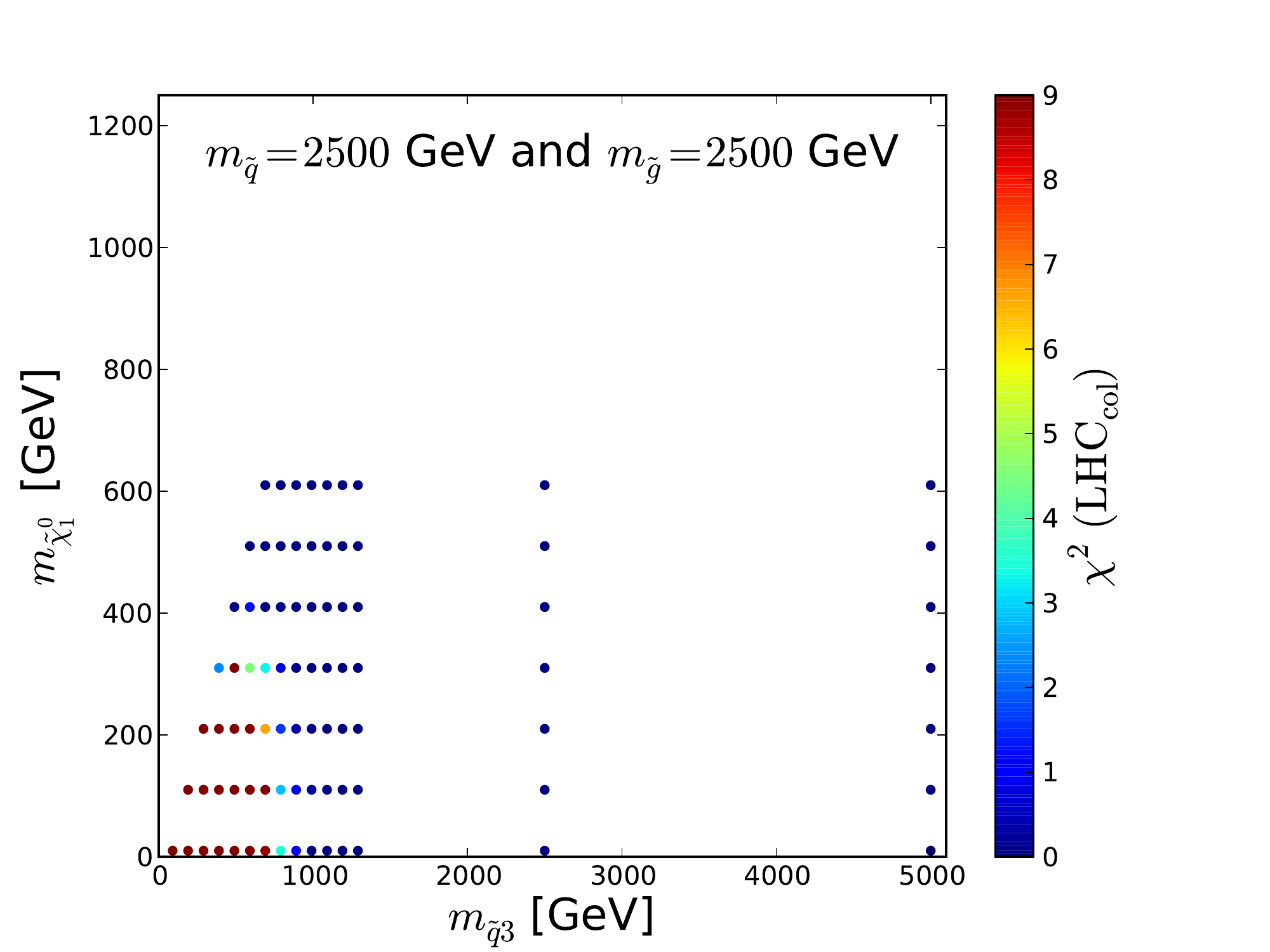}}
\end{center}
\vspace{-1cm}
\caption{\it Illustration of the grid in \mneu{1}, \mgl, \msq, and \msqt\ 
on which $\chi^2(\scorpion)$ is evaluated in order to construct \lhccol.
The upper  panel shows  the three-dimensional grid for $\mneu1=310\gev$, 
the lower left panel shows a two-dimensional slice through the grid, and the lower right panel
is another two-dimensional slice that illustrates the dependence on
$\mneu1$, see the text.
}
\label{fig:chi2-lookup}
\end{figure*}

In order to construct $\chi^2$ as a function of $\mneu1$, \mgl,
\msq, and \msqt, we first generate a sample of points on a 
$1+3$~dimensional grid, which we use for linear
interpolation. 
We construct this grid starting from values of $\mneu1 =
\{10,~110,~\ldots,~610\}~\gev$. 
For each of these values of $\mneu1$, we select the following values of $\mgl$ and $\msq$:
$\{\mneu1+40,~\mneu1+140,~\ldots,~1750,~2500,~5000\}~\gev$, 
whereas $\msqt$ takes values 
$\{\mneu1+80,~\mneu1+180,~\ldots,~1290,~2500,~5000\}~\gev$, where the
dots indicate steps of 100\gev, so that the total number of points in the
grid is 25,564. 
The choice for this grid is motivated by the need for a fine granularity at low
masses, while also capturing the parameter behaviours at higher masses. 

We associate a SUSY spectrum to each point on the grid, by setting the
first- and second-generation squark masses equal to $\msq$, and
the third-generation squark masses equal to $\msqt$.
For each SUSY spectrum we generate {coloured sparticle production}
events using {\tt PYTHIA~6}~\cite{PYTHIA6} and pass 
them through the {\tt DELPHES~3}~\cite{DELPHES3} detector simulation code using a detector card that 
emulates the CMS detector response. 
We then pass the resulting events through {\tt Scorpion}~\cite{Scorpion}, which emulates the 
monojet, MT2, single-lepton, same- and opposite-sign dilepton (SS and OS) 
and 3-lepton CMS searches~\cite{cms_mt2,CMSsearches},
to estimate the numbers of signal events in each of the signal regions.
After this we calculate the \cls\ using the {\tt LandS} package~\cite{LandS}, by 
combining all signal regions from these searches. 
{If searches have overlapping signal regions, we take the strongest
expected limit, as is the case for the CMS monojet and single-lepton searches.}

In \reffi{fig:chi2-lookup} we show a three-dimensional overview and
a pair of two-dimensional slices through this grid. 
The top panel shows the full three-dimensional grid for $\mneu1=310\gev$ and 
illustrates the fine and coarse granularity of the grid at  low
and high values of \mgl\, \msq, and \msqt, respectively.
The lower left panel shows the two-dimensional slice for the same
neutralino mass and $\msqt=2500\gev$, highlighting that there is
only a small, though non-negligible, dependence of the $\chi^2$ function on \msq\
for values of $\mgl\gtrsim2500\gev$. 
The lower right panel shows the $\chi^2$ function as a function of \msqt\ and \mneu1,
for fixed $\msq=2500\gev$ and $\mgl=2500\gev$, illustrating that for
different values of \mneu1 different grids are defined in \mgl, \msq, and \msqt.

In order to apply the \lhccol\ constraint to a generic pMSSM10 spectrum, 
we calculate $\msq$ ($\msqt$) as the cross-section-weighted
average of the first- and second- (third-)generation squark masses, 
to ensure that the \lhccol\ constraint reflects the actual 
production cross-sections. 
This is especially relevant for the third-generation squark masses, as they 
generally have large splittings. 
The $\chi^2$ contribution for \lhccol\ is obtained by linear interpolation of
the $\chi^2$ values on the $1+3$-dimensional grid.
There is one special case when $\mst1-\mneu1 < \mt$:
here the standard searches listed above are less sensitive, and the universality of the limits is
expected to break down. In this case, we calculate \msqt\ assuming zero cross-section for the lighter
stop, and consider separately the impacts of dedicated stop searches in this region, 
as described in Section~\ref{LHC8stop}.

\begin{figure*}[htb!]
\resizebox{8cm}{!}{\includegraphics{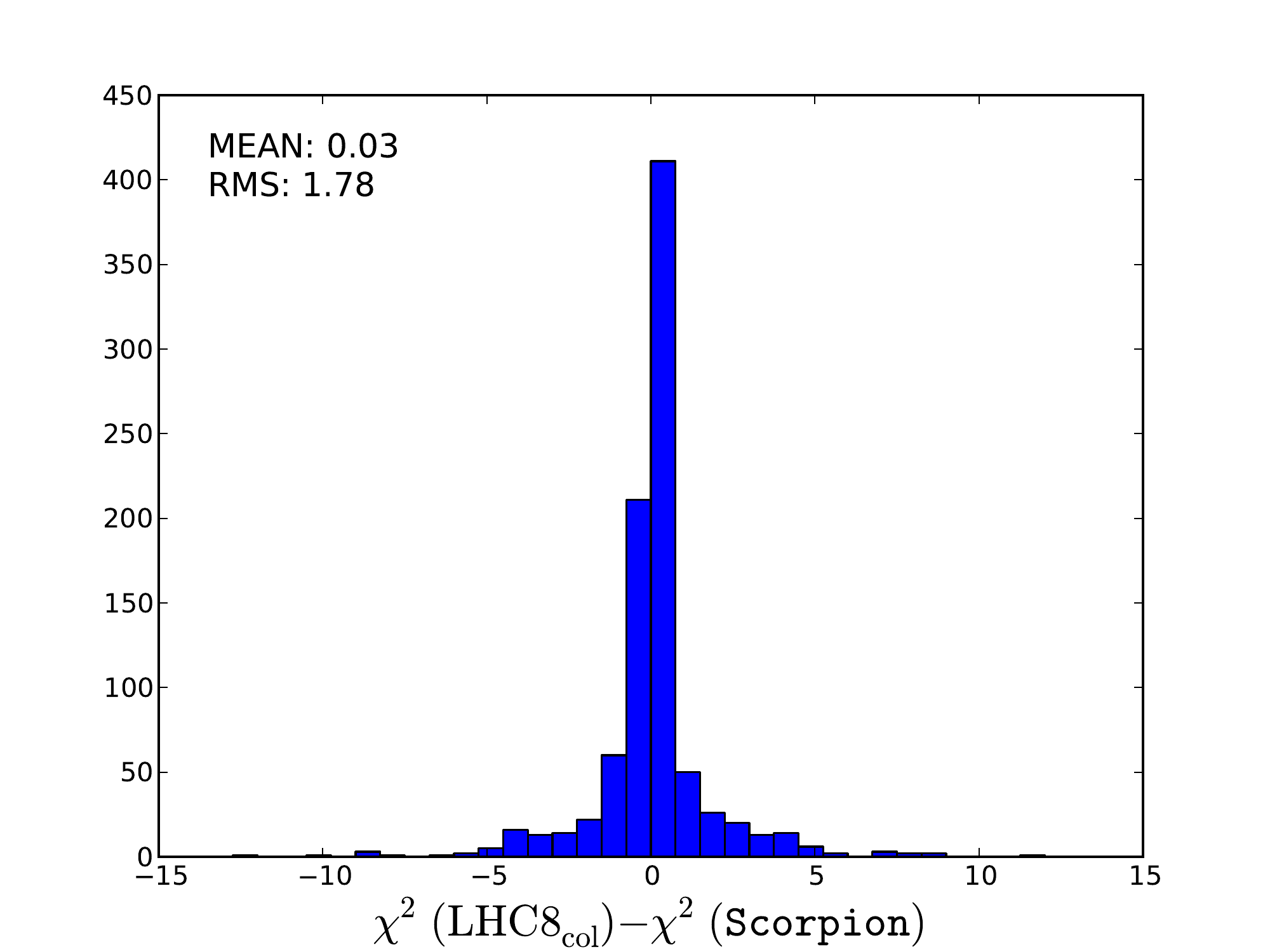}}
\resizebox{8cm}{!}{\includegraphics{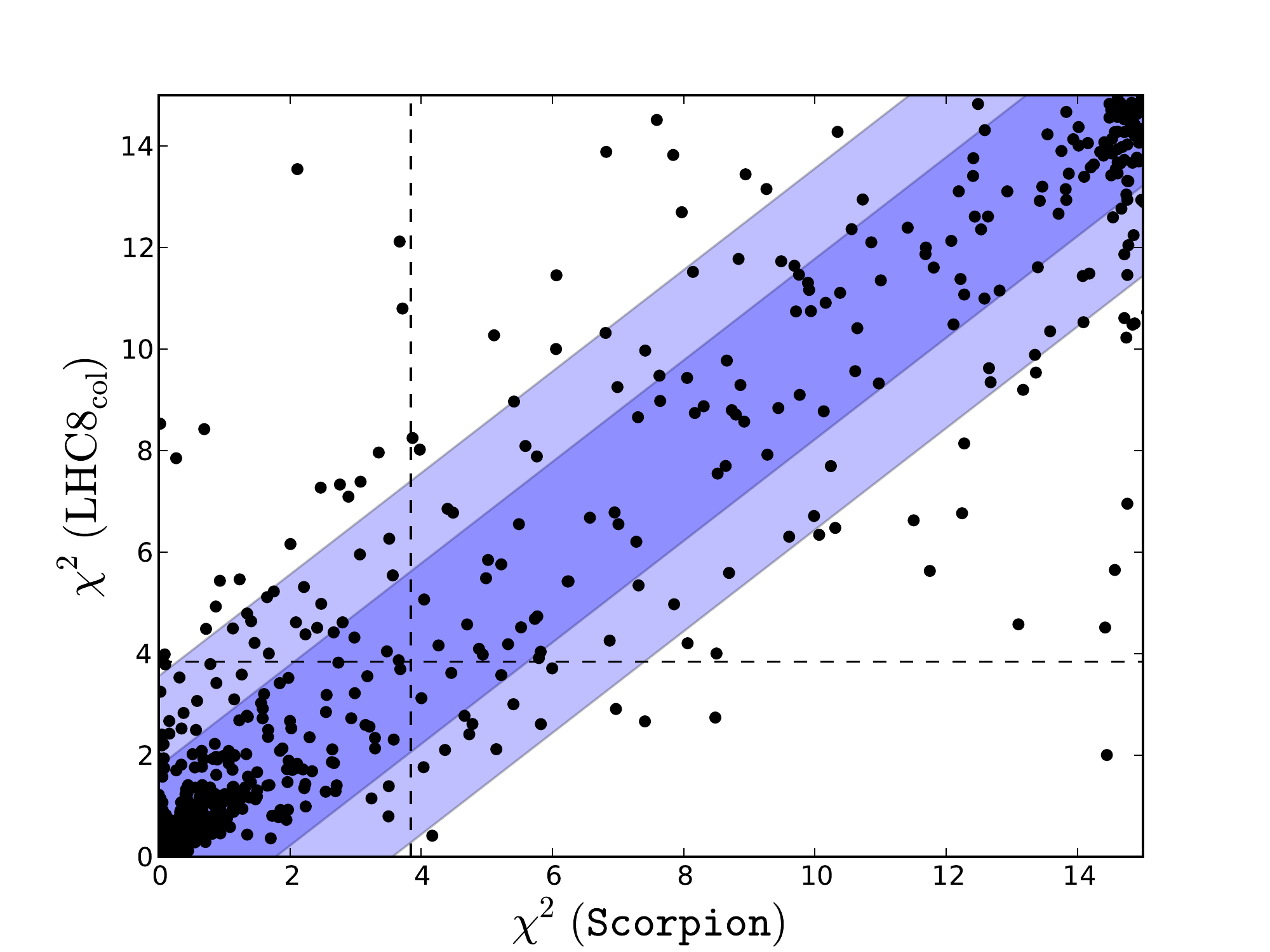}}
\caption{\it Left panel: Histogram of the differences between the values of
the likelihood function $\chisqscorpion$ evaluated using individual \lhccol\ searches for 1000 randomly-selected points
and the estimate $\chisqlhccol$ obtained by interpolation from a look-up table as described in the text.
Right panel: Scatter plot in the $(\chisqscorpion, \chisqlhccol)$ plane of the
$\chi^2$ values obtained from the two approaches; the vertical and horizontal dashed lines in this plot correspond to the
95\% \cls\ in each approach.}
\label{fig:1000random}
\end{figure*}

In order to validate the \lhccol\ constraint and to gauge
quantitatively its uncertainty, we have performed a number of studies and tests. 
First, we randomly selected 1000 model points from our sample where
at least one of the sparticle masses is low enough to have been 
within the reach of LHC Run 1 
($\mneu1<600 \gev$ and either $\mgl<1500 \gev$, $\msq<1600 \gev$ or $\msqt<900 \gev$)  
and $\Delta \chi^2 < 10$ relative to the global minimum.
For these points we compare the $\chi^2$ values interpolated from the look-up table ($\chisqlhccol$) with
the $\chi^2$ obtained by running the full chain of event generation, detector simulation and
analyses ($\chisqscorpion$). 
The left panel of \reffi{fig:1000random} shows a histogram of the
differences for the 1000 randomly-selected points. 
As indicated in the legend of this figure, the standard deviation on this distribution is
$\sigma_{\chi^2}=1.8$.

The right panel of \reffi{fig:1000random} shows a scatter plot in the $(\chisqscorpion, \chisqlhccol)$ plane of the
$\chi^2$ values obtained from the two approaches. 
They would agree perfectly along the diagonal where $\chisqscorpion = \chisqlhccol$,
and the lighter- and darker-shaded blue strips are the 
$\pm 1 \sigma_{\chi^2}$ and $\pm 2 \sigma_{\chi^2}$ bands around this diagonal.
The vertical and horizontal dashed lines in this plot correspond to the
95\% $\cls$ in each approach.
For the majority of points, the interpolation and the full analysis agree
whether the point is excluded at the 95\% $\cls$, or not, and
most of the remaining points lie within $\pm 2 \sigma_{\chi^2}$.

\begin{figure*}[htb!]
\begin{center}
\resizebox{7.5cm}{!}{\includegraphics{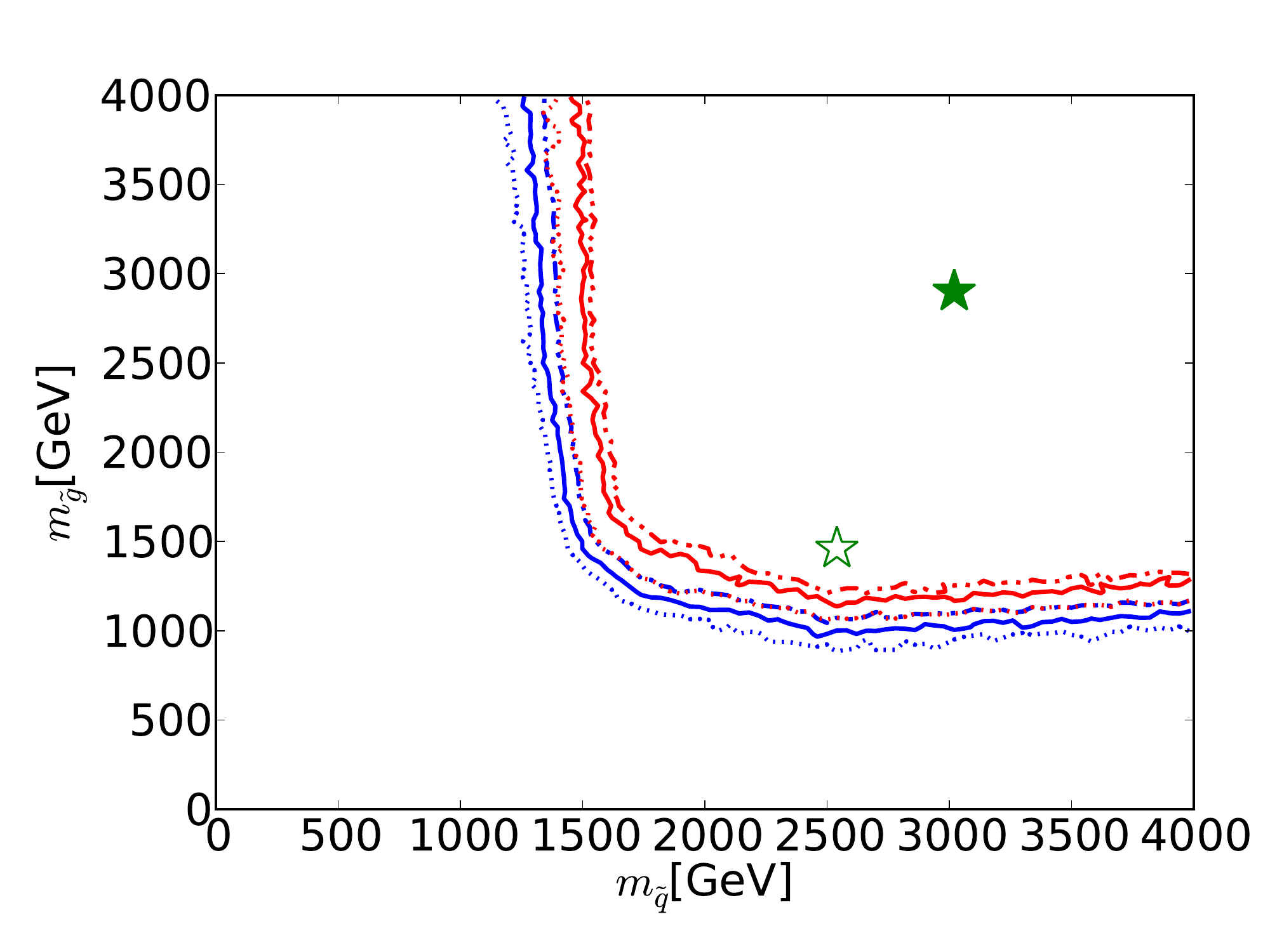}}
\resizebox{7.5cm}{!}{\includegraphics{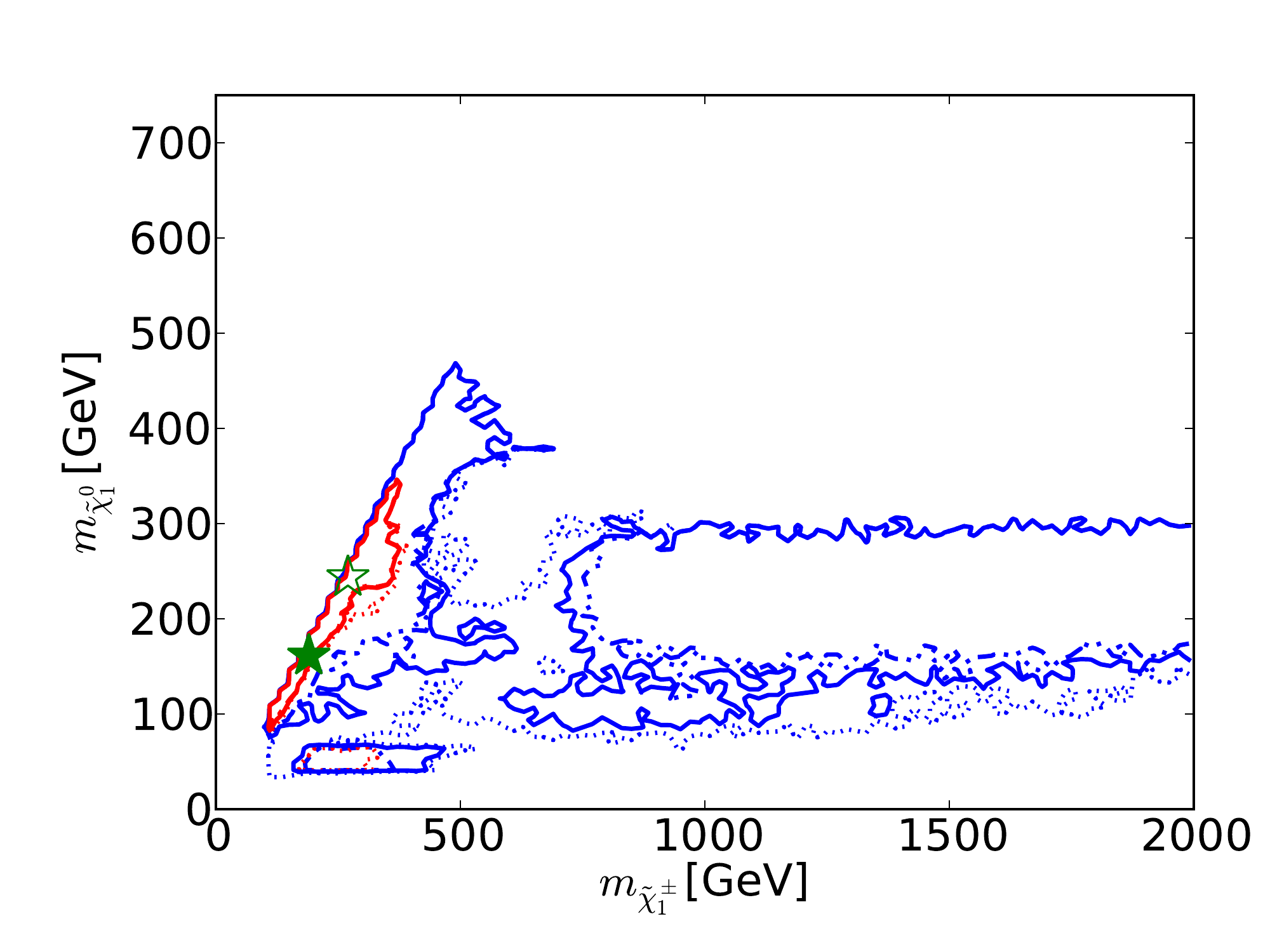}}\\
\resizebox{7.5cm}{!}{\includegraphics{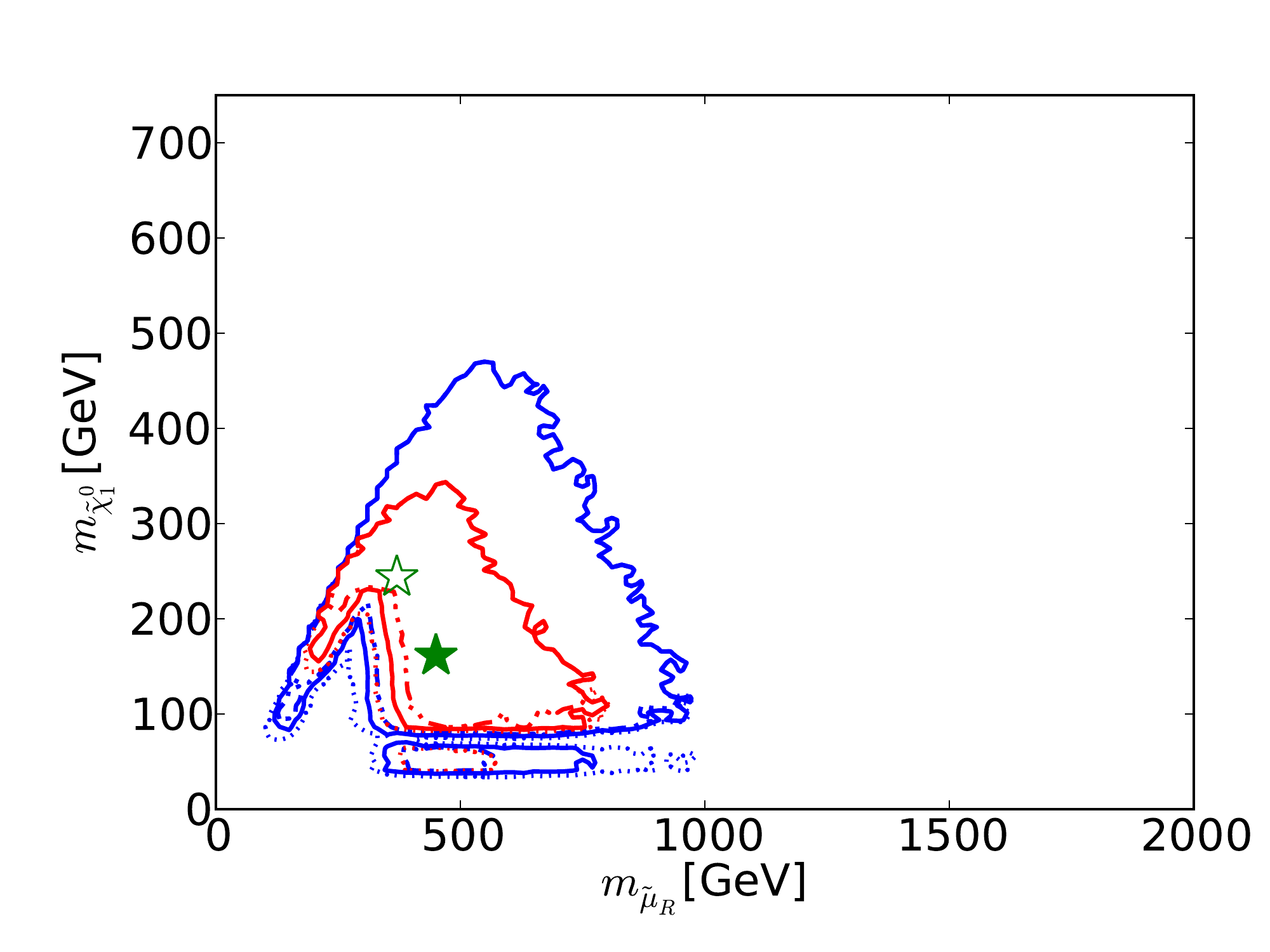}}
\resizebox{7.5cm}{!}{\includegraphics{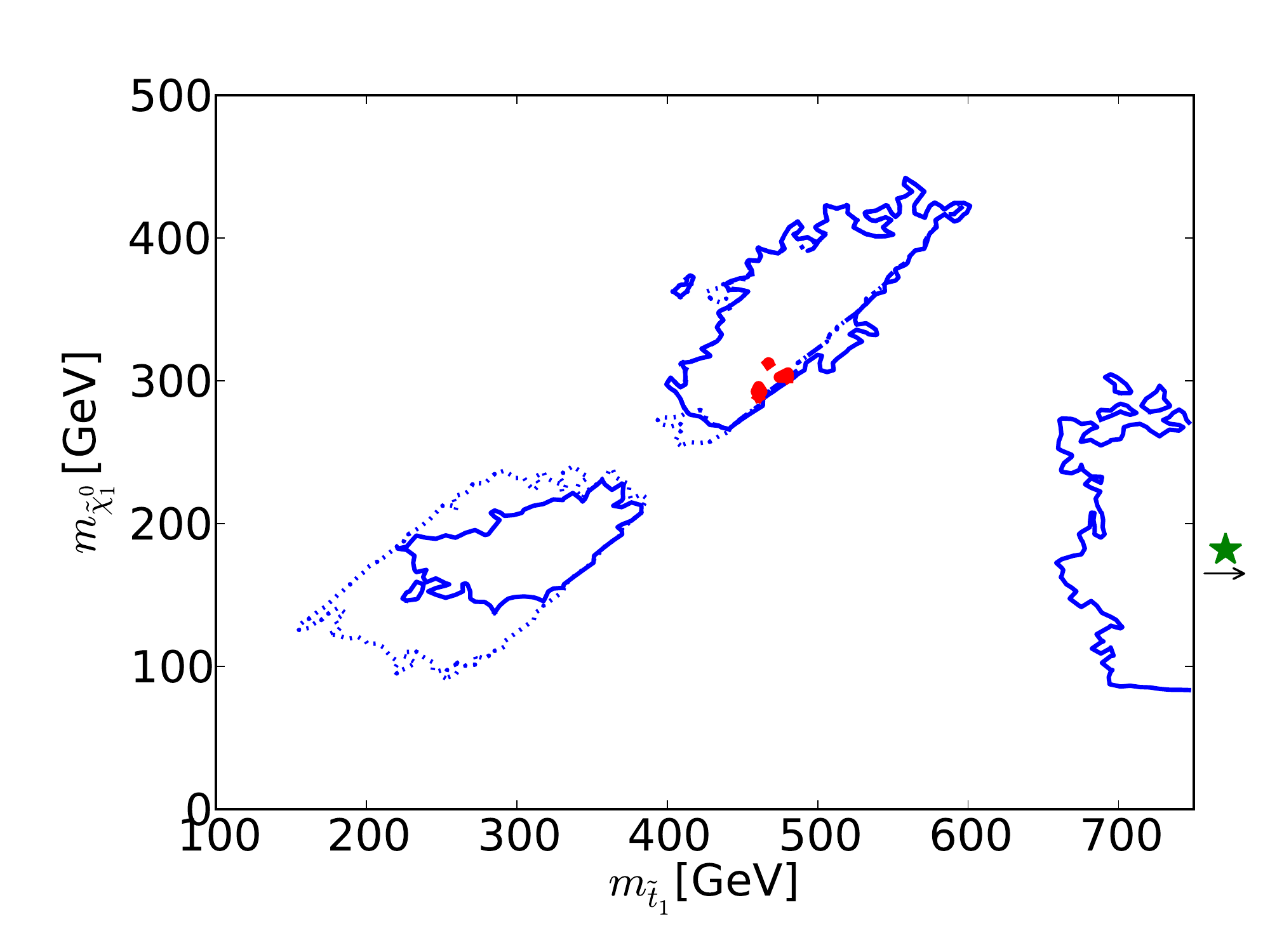}}
\end{center}
\vspace{-1cm}
\caption{\it Impacts of the $\pm$~1~$\sigma$ uncertainties in our implementations of the \lhccol,
\lhcewk\ and \lhcstop\ constraints on the 68 and 95\% CL regions
(indicated by the red and blue contours) in the corresponding relevant mass planes: $(\msq, \mgl)$
(upper left panel),  $(\mcha1, \mneu1)$ (upper right panel), $(\smu{R}, \mneu1)$
(lower left panel), and $(\mstop1, \mneu1)$ (lower right panel).
In each case, the dot-dashed and dashed contours are obtained by
shifting the respective $\chi^2$
penalty up and down by one standard deviation $\sigma_{\chi^2}$,
as discussed in the text. 
The filled green stars correspond to the nominal best-fit point and the
open stars (shown if not overlapping) to those which were obtained from 
shifting the $\chi^2$ up or down with $\sigma_{\chi^2}$.
{We note that in the lower right panel the best-fit points lie outside the
displayed parameter range.}
}
\label{fig:up-down}
\end{figure*}

We then assess how the uncertainty
$\sigma_{\chi^2}$ in our implementation of the  
\lhccol\ constraint translates into uncertainties in sparticle mass limits:
see the upper left panel of \reffi{fig:up-down}~\footnote{The other panels of
\reffi{fig:up-down} show the corresponding uncertainties in our treatments of the \lhcewk\
and \lhcstop\ constraints, which are discussed later.}. 
For this estimate, we bin the 1000 points of the first test, and calculate the
standard deviation, $\sigma_{\chi^2}$, for points with $\chisqlhccol\leq1$, $1<\chisqlhccol\leq4$ and
$\chisqlhccol>4$.
We then apply the \lhccol\ constraint in three ways: with the nominal implementation, and 
shifting the  $\chisqlhccol$ penalty up and down according to these binned standard deviations. 
The results are shown in the upper left panel of \reffi{fig:up-down} as solid and
dotted red (blue) contours in the $(\mgl, \msq)$ plane corresponding to the nominal and up- and down-shifted
cases for the 68 (95)\% CL, respectively~\footnote{Here and in subsequent analogous parameter
planes, we treat the $\Delta \chi^2 = 2.30$ and 5.99 contours as proxies for the
68\% and 95\% CL contours.}.
A dedicated study of points within the 68\% and 95\% CL regions confirms that our implementation
of the \lhccol\ constraint is valid within these uncertainties, and our estimate of $\chi^2$ at the
best-fit point differs from the {\tt Scorpion} evaluation by less than one~\footnote{Our \lhcewk\ and \lhcstop\ analyses
described later also differ by less than one from the corresponding {\tt Scorpion/Atom} evaluations.
This is also true for the benchmark points introduced later.}.

We conclude that the uncertainty $\sigma_{\chi^2}$ in our estimate $\chisqscorpion$ {is generally reliable, and} translates
into an uncertainty of \order{50\gev} in the limits on the gluino and squark masses, which is fully sufficient for the purpose of our studies.

\subsubsection{LHC constraints on electroweak gauginos, Higgsinos and sleptons}
\label{LHC8EWK}

Unlike the searches for coloured sparticles, where we were able to construct
a computationally-efficient, approximately universal limit, 
the LHC constraints on electroweakly-produced sparticles vary strongly in
sensitivity, depending on the mass hierarchy of sparticles and their corresponding
decay modes and final states.
For example, searches in the three-lepton plus missing energy channel  
constrain the chargino and neutralino masses up to $\mcha1 = \mneu2 \lsim 700$\,GeV for $\mneu1 \lsim 300$\,GeV, 
if $\cha1$ and $\neu2$ decay exclusively into on-shell sleptons~\cite{atlas_3L, cms_EWK},
whereas a much weaker limit, $\mcha1 = \mneu2 \lsim 450$ GeV for $\mneu1 \lsim 100$\,GeV,
was found in an analysis of the two-lepton plus missing energy channel~\cite{cms_EWK, atlas_2L},
assuming that the $\cha1$ and $\neu2$ decay exclusively into the $\neu1$ in association with $W$ and $Z$, 
respectively, and not taking into account the decay
$\neu2 \to \neu1 h$~\cite{Bharucha:2013epa,Han:2014nba}.
The same two-lepton analyses constrain slepton pair production,
leading to the limits 
$m_{\tilde \ell_{L(R)}} \lsim 270$ (200)\,GeV for $\mneu1 \lsim 100$ (50)\,GeV~\cite{cms_EWK, atlas_2L}.
Therefore, the universal limit approach that we use to combine and characterise searches for coloured sparticles
is inapplicable to searches for electroweakly-produced sparticles, and we use an alternative method.

For model points where the production of electroweakly-produced sparticles provides a non-trivial constraint,
they must be much lighter than the coloured sparticles, since otherwise the much higher rates of
production of coloured sparticles would already exclude the model points.
Therefore, in the region of interest, there can be only a few particles lighter than
the electroweakly-produced sparticles, implying that one can use
a combination of a few simplified models (SMS)
to approximate the sensitivities of the
LHC searches for the production of these sparticles.
Depending on the decay mode and final state,  we select ATLAS and/or CMS limits
derived from relevant simplified models to calculate
the contributions of these searches to our global $\chi^2$ function. 
For the LHC searches that constrain electroweakly-produced gauginos, Higgsinos and sleptons,
to a good approximation all relevant $\chi^2$ contributions can be
extracted from  simplified chargino-neutralino and simplified
smuon and selectron models. 

For each simplified model limit we construct a function 
$\chi_{\rm SMS}^2$ that depends on the two relevant masses: 
($\mcha1\simeq\mneu2, \mneu1 $) for the simplified chargino-neutralino model
and ($\mslep, \mneu1$) for the simplified slepton $(\slep{} \equiv \tilde e,
\tilde \mu)$ model. 
We assume that $\chi_{\rm SMS}^2 = 15$ 
in the bulk of the region excluded in the simplified model,
and that this $\chi^2$ penalty vanishes exponentially when
crossing the boundary to the allowed region, with the general form
\begin{equation}
\chi_{\rm SMS}^2 \; = \; \min_{l,r} \left[15\cdot B
    \cdot\frac{1}{e^{(d_{l,r}-\mu_{l,r})/\sigma_{l,r}}+1} \right] \, ,
\label{exponentialform}
\end{equation}
where the subscripts $l,r$ indicate the simplified
model exclusion contour to the left and right (in the horizontal direction,
i.e., $\mcha1\simeq\mneu2$ or $\mslep$) of the point on the contour 
with the largest value of \mneu1,
$B$ is the branching ratio of the decay in question (as calculated with 
\texttt{SDECAY}~\cite{Muhlleitner:2003vg}),  
$d$ is the closest distance in \gev\ to the contour, 
and $\mu$ and $\sigma$ control the precise fall-off of the
$\chi^2$ function, so as to mimic the experimental uncertainty bands,
{and are functions of \mneu{1}.}
We note that if one sets $\mu=-\sigma$ then $\chi^2_{\rm SMS}(d=0)\approx4$, 
so that the exclusion on the contour corresponds approximately to the 95\% \cls.
Finally, to avoid an unphysically slow fall-off outside the 95\% \cls\ limit 
we set $\sigma=50\gev$ and
adjust $d$ accordingly if $\sigma>50\gev$ and $d-\mu>\sigma$ (and hence
$\chi^2_{\rm SMS}\lesssim4$). 

In order to illustrate \refeq{exponentialform}, we display in \reffi{fig:sms-chi2}
$\chi^2_{\rm SMS}/B$ for the $\cha1 \neu2$ decay via sleptons.
In the left panel $\chi^2_{\rm SMS}/B$ is shown for a fixed value of
$\mneu1=300\gev$ where the green (blue) line corresponds to $d_l, \mu_l,\sigma_l$,
($d_r, \mu_r, \sigma_r$), whereas 
vertical dashed lines indicate the position of the contour. 
The right panel shows the same $\chi^2_{\rm SMS}/B$ (in colour) as a function of
$\mcha1\simeq\mneu2$ and $\mneu1$, and the 95\% \cls\ exclusion contour found in Fig.~7(a)
of~\cite{atlas_3L} (blue line). 
Note that we apply no constraint for $\mneu{1} \gtrsim 380\gev$, the highest value on
the blue experimental contour.

\begin{figure*}[htb!]
\begin{center}
\resizebox{7.5cm}{!}{\includegraphics{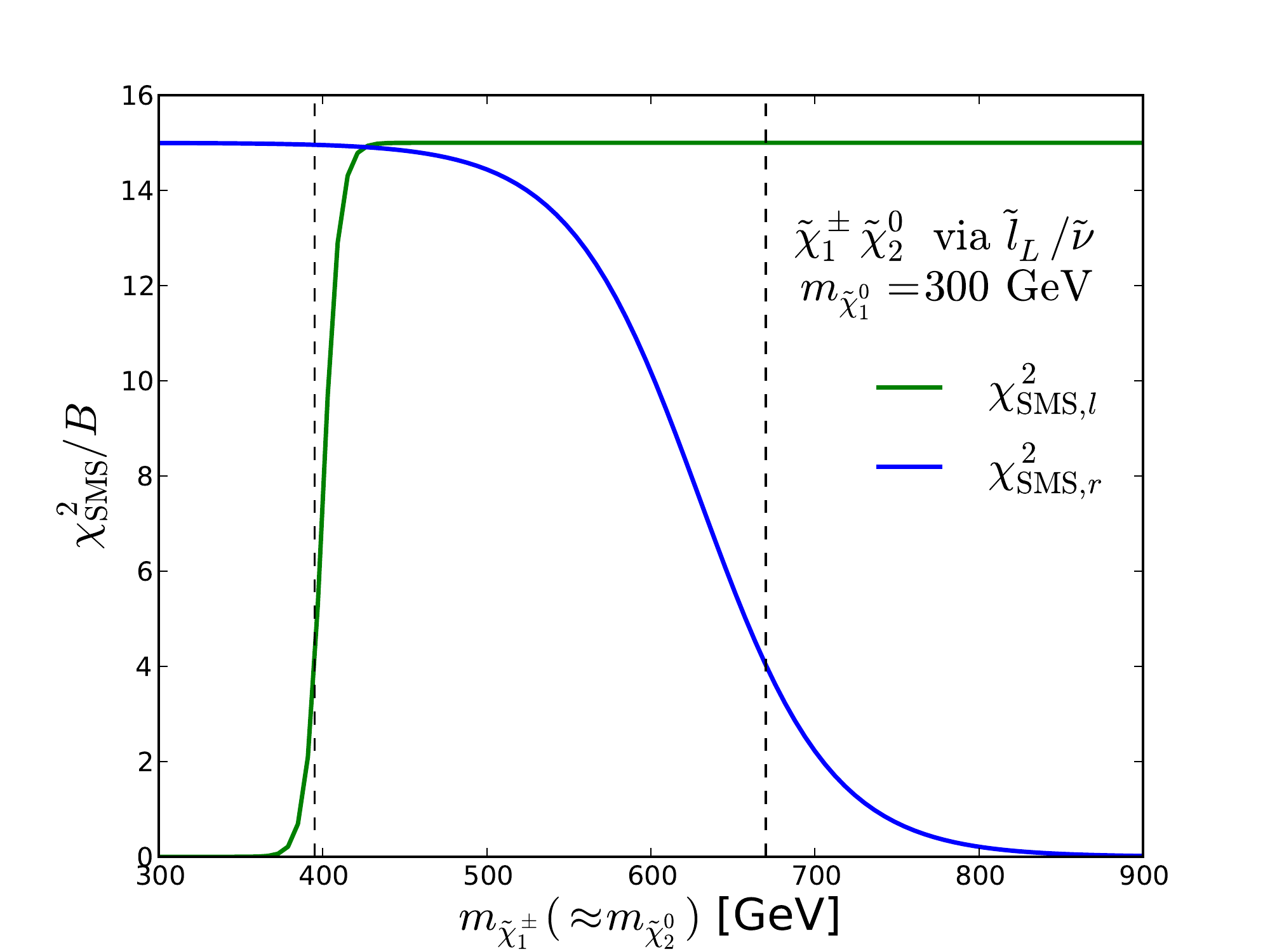}}
\resizebox{7.5cm}{!}{\includegraphics{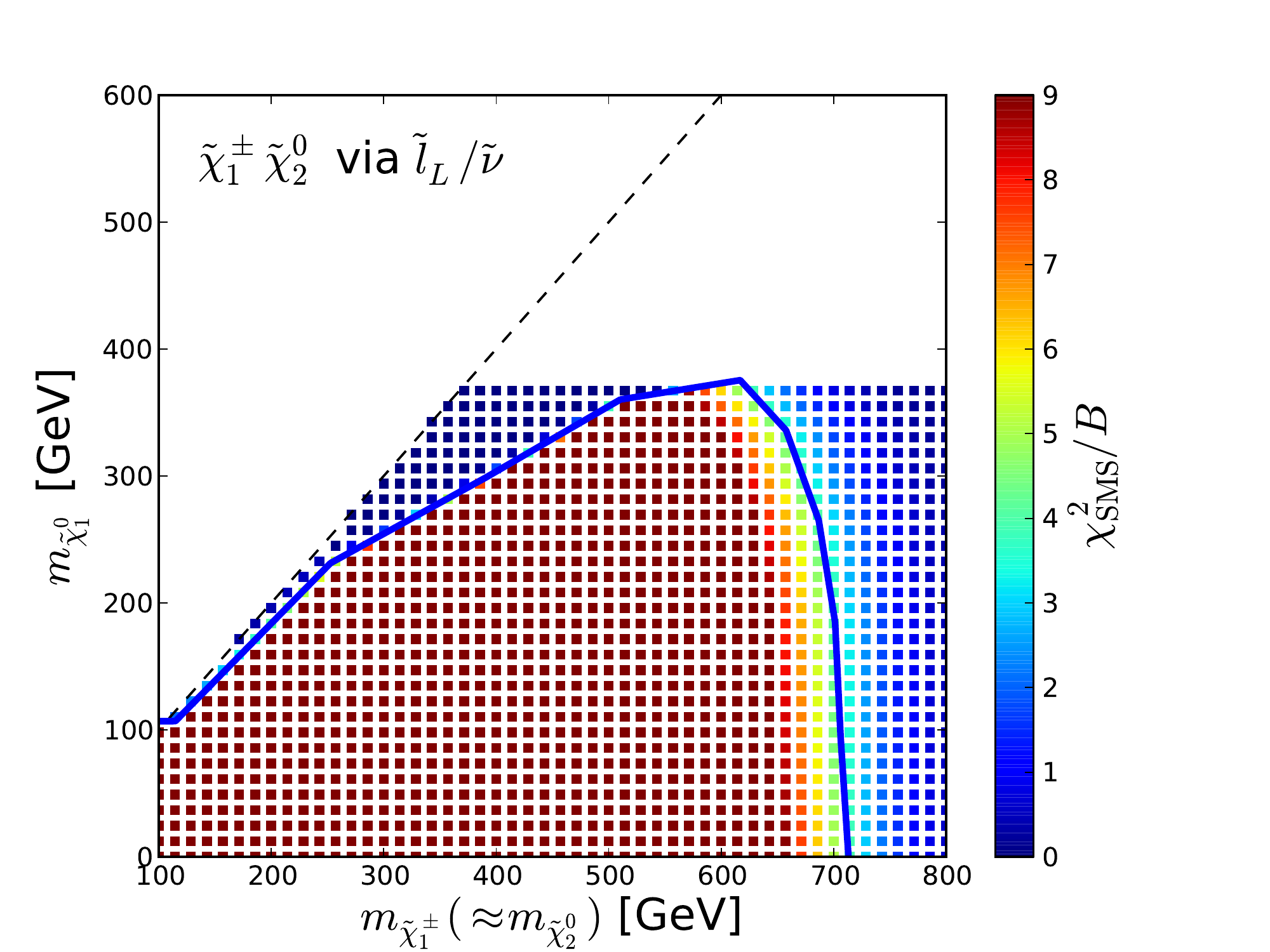}}\\
\end{center}
\vspace{-1cm}
\caption{\it 
Illustration of $\chi^2_{\rm SMS}/B$, as defined in
\refeq{exponentialform}, for $\cha1 \neu2$ production and decay via sleptons.
In the left panel $\chi^2_{\rm SMS}/B$ is shown for a fixed value of
$\mneu1=300\gev$, where the green (blue) line corresponds to $d_l, \mu_l,\sigma_l$,
($d_r, \mu_r, \sigma_r$) and 
vertical dashed lines indicate the position of the contour. 
The right panel shows the same $\chi^2_{\rm SMS}/B$ (in colour) as a function of
$\mcha1\simeq\mneu2$ and $\mneu1$, and the 95\% \cls\ exclusion contour found in Fig.~7(a)
of~\cite{atlas_3L} (blue line).
}
\label{fig:sms-chi2}
\end{figure*}

In order to establish \lhcewk\ we tuned the $\mu$ and $\sigma$ parameters 
for each simplified model to reproduce best the $\chi^2$
values that we obtained using \atom\ for a representative set of model points 
from our sample. 
Table~\ref{tab:SMS-EWkino} summarises the implementations of the simplified model exclusion limits
that contribute to \lhcewk. 
Note that, as described above, the large value of 
$\sigma_r=300\gev$ for the limit from $\cha1 \neu2$ production and decay via $WZ$
is replaced by setting $\sigma_r=50\gev$ and
adjusting $d_r$ accordingly when $d_r-\mu_r>\sigma_r$ 
(and hence $\chi^2_{\rm SMS}\lesssim4$).
Also, we had to produce our own contour for the direct production of right- and
left-handed sleptons (selectrons and smuons), corresponding to their production
cross-sections. 
Note that this simplified model contour is also applied when left-handed
sleptons decay via $\neu2$ and $\cha1$.

\begin{table*}[htb!]
\begin{center}
\renewcommand{\arraystretch}{1.1}
\begin{tabular}{|c|c|c|c|} \hline
Simplified Model   &  Limit & $(\mu_l, \sigma_l)$~[GeV] & $(\mu_r, \sigma_r)$~[GeV]\\ 
\hline         
$\cha1 \neu2$ via $\slep{}$   & Fig.~7(a) in~\cite{atlas_3L} & (-5, 5) & (-40, 40) \\
$\cha1 \neu2$ via $WZ$  & Fig.~7(b) in~\cite{atlas_3L} & (-20, 20) & (-300, 300) \\
$\slep{} \to \ell \neu{1,2},~\nu_\ell \cha1 $  &  {Gener}ated using \atom\ & (-20, 10) & (-40, 30) \\
\hline 
\end{tabular}
\caption{\it The simplified model limits used to constrain electroweak
  gauginos, Higgsinos and sleptons.
}  
\label{tab:SMS-EWkino}
\renewcommand{\arraystretch}{1.0}
\end{center}
\end{table*}

In order to validate our method and to determine quantitatively its
uncertainty, we compare the contributions to the global $\chi^2$
function calculated with this \lhcewk\ limit approach, 
\begin{align}
\chisqlhcewk &= \sum_{\rm SMS} \chi^2_{\rm SMS}~,
\end{align}
to results from a full recast of all the above-listed searches as implemented in {\tt Atom}.
In this recast the full analysis is simulated, so that it is possible to determine for any arbitrary SUSY spectrum
the \cls\ value (and hence the corresponding $\chi^2$) with which a given search penalizes the SUSY spectrum. 
We obtain a set of 1000 model points from our sample by binning the 
$(\mneu2\approx\mcha1, \mneu1)$ plane in $100\times100$ bins, selecting one
point randomly per bin, and then take a random subset of 1000 of these points. 
This procedure was employed to ensure a representative set of the decay modes
in our sample.

\reffi{fig:1000randomEWcolour}
displays scatter plots in the $(\mcha1, \mneu1)$ plane of the contributions to the global $\chi^2$
function for these 1000 model points
as calculated using the \lhcewk\ method ($\chisqlhcewk$) 
(left panel) and the {\tt Atom} code $\chisqatom$ (right
panel), with the indicated 
colour code in each plot. The immediate visual impression is that the colours in the two scatter plots are 
generally quite similar, indicating that the two procedures deliver similar $\chi^2$ contributions
overall. A closer inspection of the plots reveals similar bands of low-$\chi^2$ points with small $\mcha1 - \mneu1$ in
a chargino coannihilation strip region, while elsewhere we see similar disfavouring of points with
low $\mneu1 \lsim 150 \gev$ and larger $\mcha1$. However, even within
this band we see a sparse set of points with relatively low $\chi^2$
that appear similarly in both the \lhcewk\ analysis based on simplified
models and the {\tt Atom} implementation of the full searches. 
These are mainly due to the decay $\neu2 \to \neu1 h$, thus
weakening the stronger $\neu2 \to \neu1 Z$-based limit.

\begin{figure*}[htb!]
\resizebox{8.5cm}{!}{\includegraphics{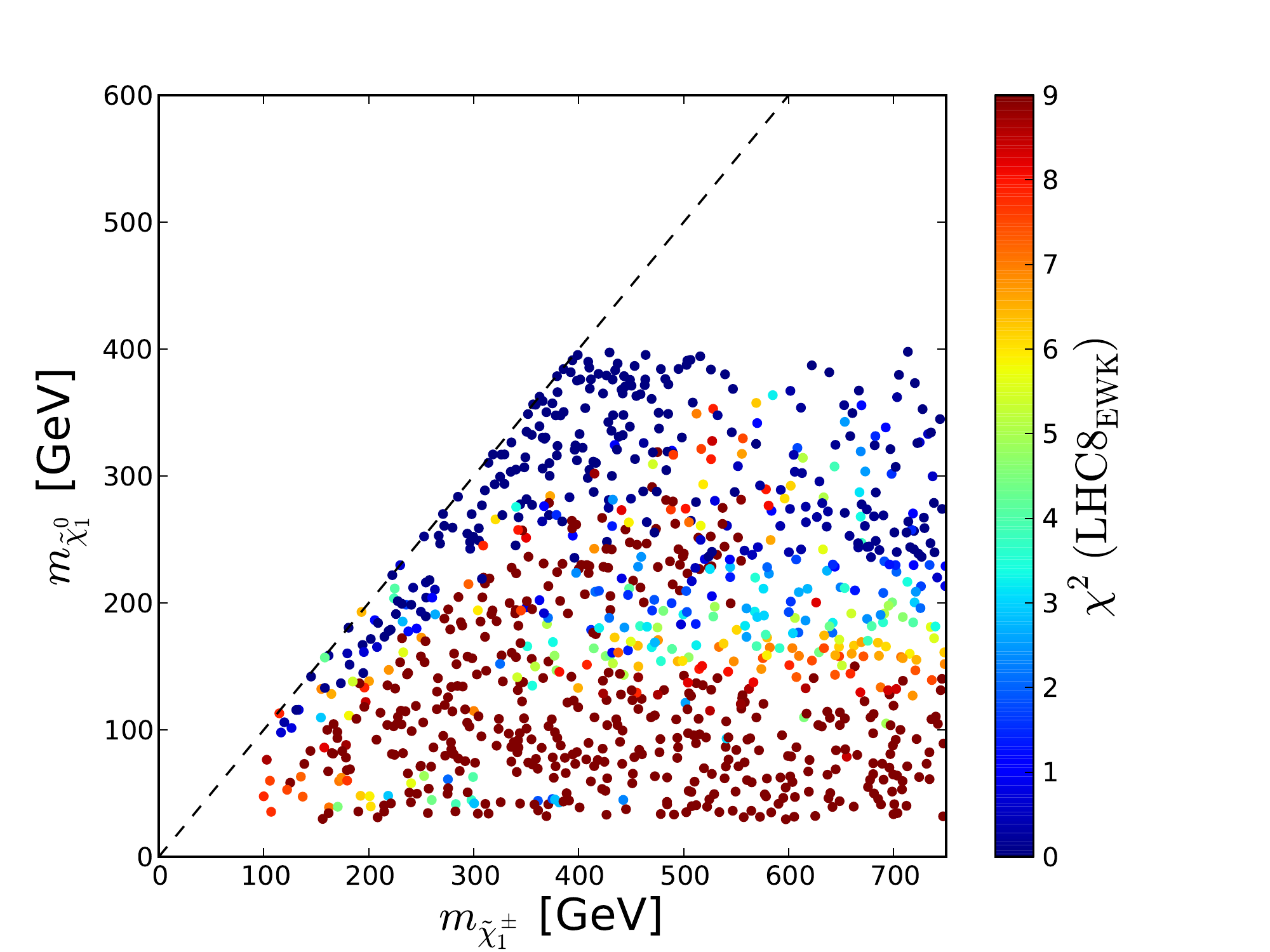}}
\resizebox{8.5cm}{!}{\includegraphics{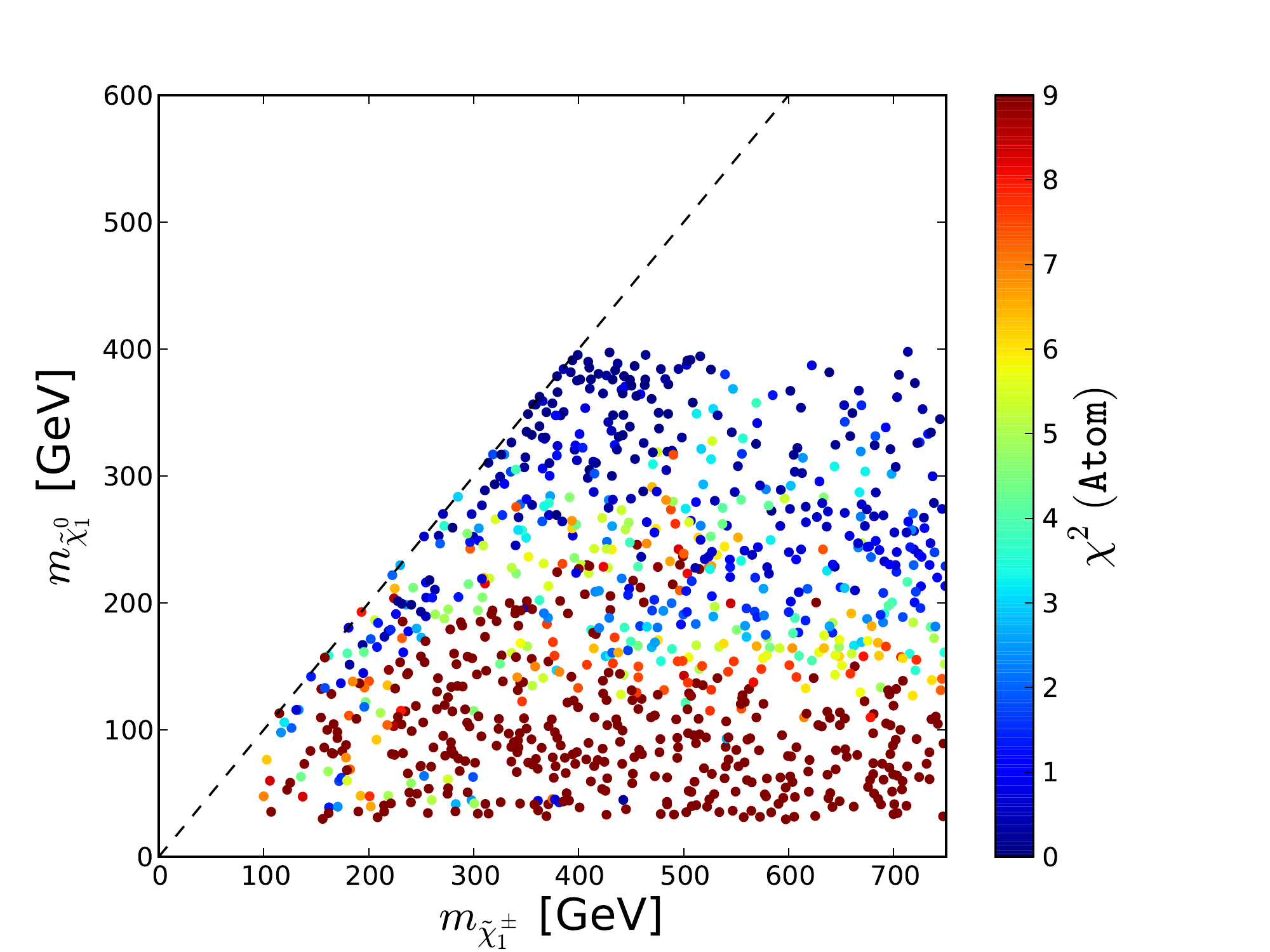}}
\caption{\it Scatter plots in the $(\mcha1, \mneu1)$ plane of the contributions to the global $\chi^2$
functions from the electroweakly-interacting sparticle constraints for 1000 randomly-selected points
accessible to LHC searches,
as calculated using the \lhcewk\ method based on simplified model searches ($\chisqlhcewk$, left panel)
and the {\tt Atom} code ($\chisqatom$, right panel).}
\label{fig:1000randomEWcolour}
\vspace{1em}
\end{figure*}

For a more quantitative comparison of our \lhcewk\ method and {\tt Atom} we
turn to \reffi{fig:1000randomEW}. 
We see in the left panel that the difference between $\chisqlhcewk$ and
$\chisqatom$ is relatively small, with an r.m.s.~difference $\sigma_{\chi^2}=2.31$. The correlation between
$\chisqlhcewk$ and $\chi^2({\rm Atom})$ is visible in the scatter plot in the right panel of \reffi{fig:1000randomEW}.
We see that most points are either excluded with $\Delta \chi^2 > 4$
in both analyses, or allowed with $\Delta \chi^2 < 4$ in both cases. Last but not least, there are
relatively few `off-diagonal' points with large $\Delta \chi^2$, which form
the small non-Gaussian tail of the
$\chisqlhcewk - \chisqatom$ distribution seen in the left panel of \reffi{fig:1000randomEW}.

To quantify the impact of this uncertainty on our analysis, 
we follow the same procedure as for our limits on coloured sparticles,
and translate the $\sigma_{\chi^2}$ (binned analogously) into 
a $\pm~1~\sigma$ band for our 68\% and 95\% CL contours in the important
$(\mcha1, \mneu1)$ and $(\msmu{R}, \mneu1)$ planes. As can be seen
in the upper right and lower left panels of \reffi{fig:up-down},
the uncertainty associated with \lhcewk\ is in general
small in the 68\% CL region of our fit, although it is larger at the 95\% CL
level in the $(\mcha1, \mneu1)$ plane. {The effects on the
best-fit point of these upward and downward shifts in the $\chi^2$ treatment are
shown in these panels as open green stars. The downward shift has very little effect,
and is essentially invisible in the $(\mcha1, \mneu1)$ plane. The upward shift
increases the best-fit values of $\mneu1$ and $\mcha1$ while reducing that of $\msmu{R}$,
though the variations are contained well within the 68\% CL region, clearly indicating that the
corresponding uncertainties do not impact the overall conclusions.}

\begin{figure*}[htb!]
\resizebox{8cm}{!}{\includegraphics{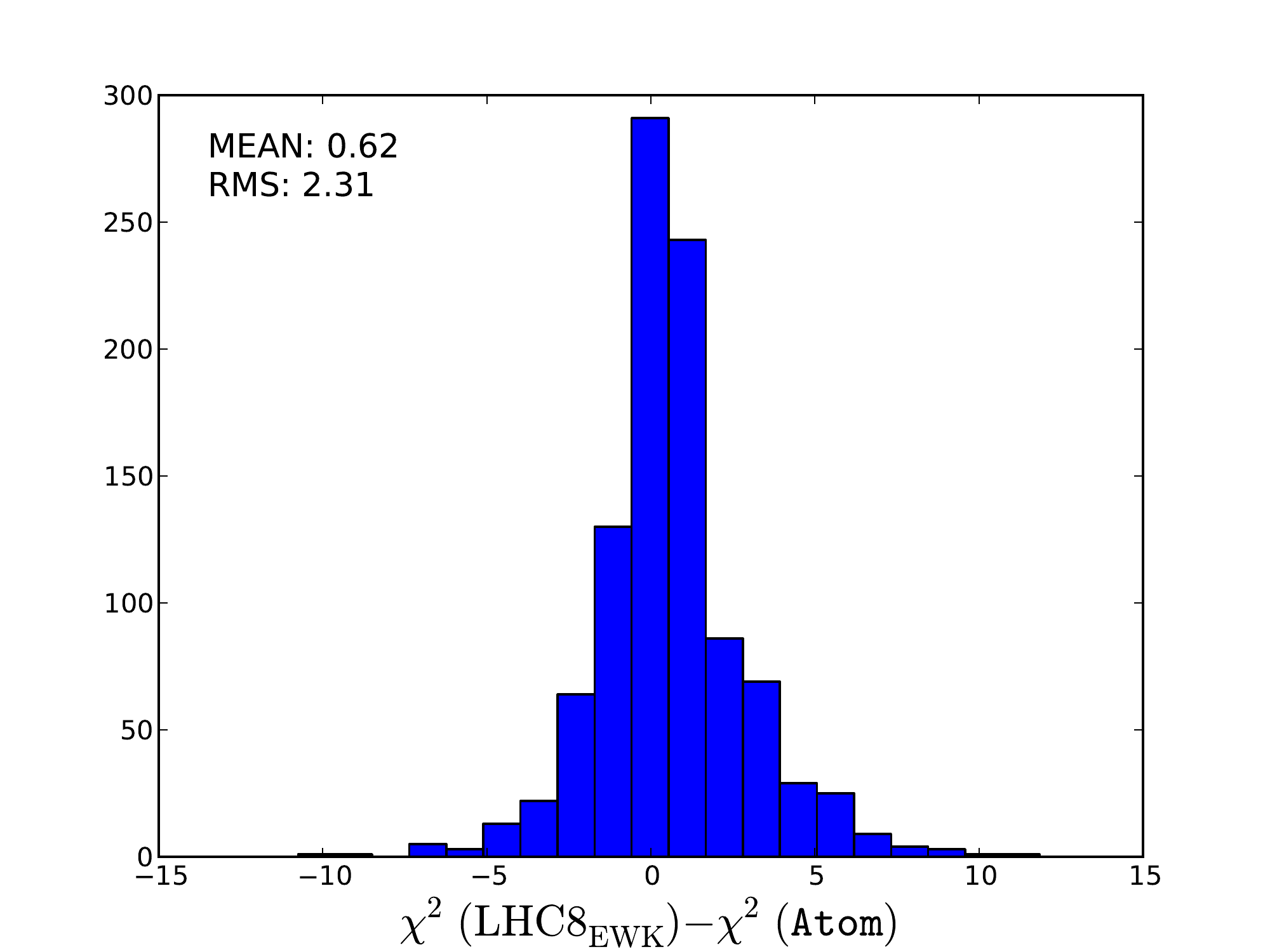}}
\resizebox{8cm}{!}{\includegraphics{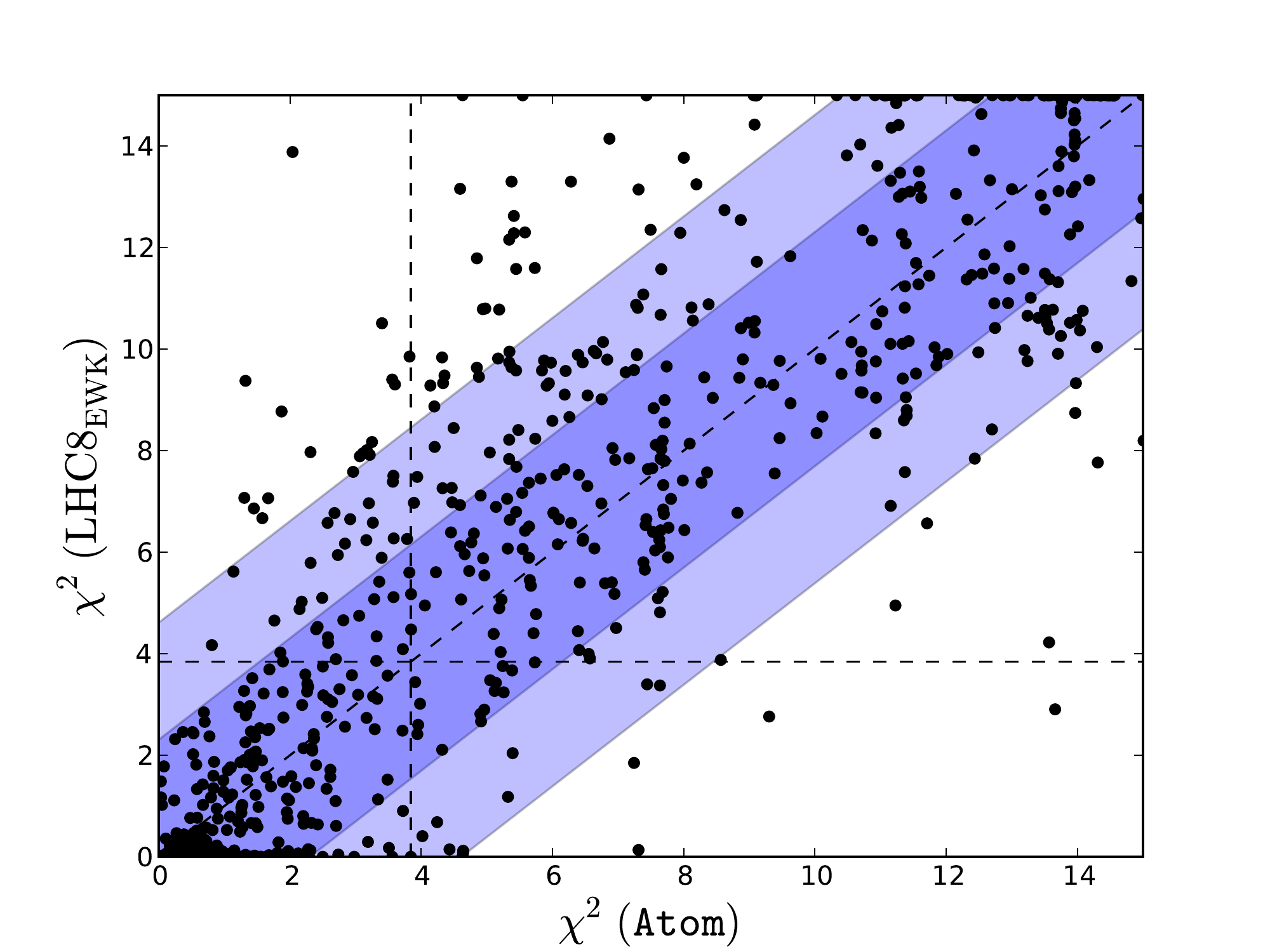}}
\caption{\it Left panel: Histogram of the differences between the values of the contributions of
the electroweakly-interacting sparticle constraints to
the global likelihood function $\chisqlhcewk$ evaluated using simplified model searches for the 1000 randomly-selected points
and the estimate $\chisqatom$ obtained using the {\tt Atom} code.
Right panel: Scatter plot in the $(\chisqatom, \chisqlhcewk)$ plane of the
$\chi^2$ values obtained from the two approaches; the vertical and horizontal dashed lines in this plot correspond to the
95\% $\cls$ in each approach.
}
\label{fig:1000randomEW}
\vspace{1em}
\end{figure*}

\subsubsection{LHC constraints on compressed stop spectra}
\label{LHC8stop}
In their searches for stop production, ATLAS and CMS have placed special
emphasis on compressed spectra, which pose particular challenges for LHC searches.
Whilst limits on stop production in the region  where $\mst1 - \mneu1 > m_t$ are fully included in 
the \lhccol\ limits described in Section~\ref{LHC8colour},
a dedicated treatment of the compressed-spectrum
region $\mst1 - \mneu1 < m_t$ is required in order to include properly all the relevant collider limits.
In this region we calculate the contribution of stop searches to the global $\chi^2$ 
in a similar way as for the for electroweakly~produced sparticles described in
Section~2.5.2. We refer to this dedicated limit-setting procedure as
\lhcstop .

\begin{figure*}[htb!]
\centering
\resizebox{10cm}{!}{\includegraphics{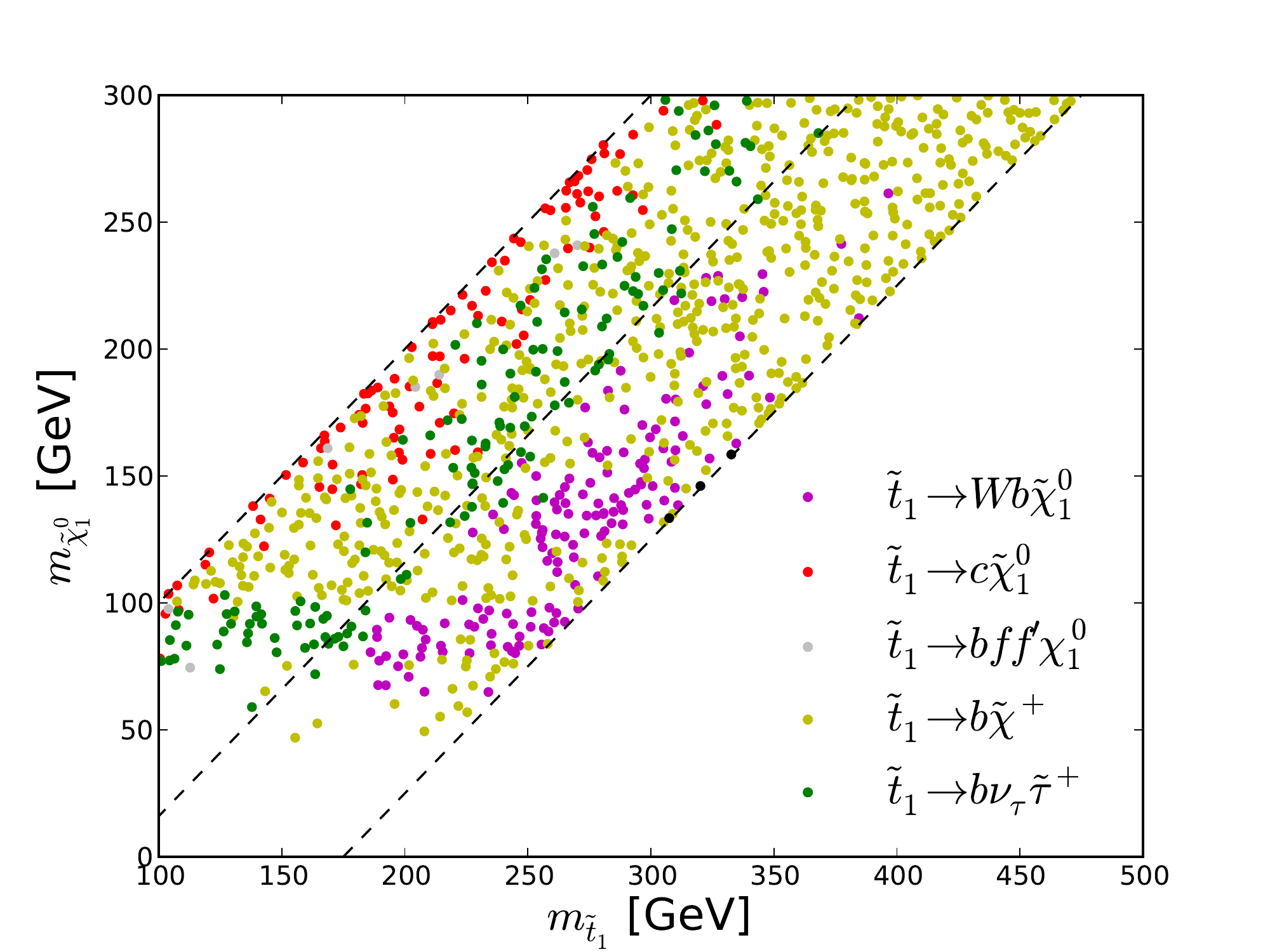}}
\caption{\it Scatter plot in the $(\mstop1, \mneu1)$ plane of the
  $\sto1$ decay modes with branching ratios $> 50$\% for 1000 randomly-selected points with
  $\mst1 - \mneu1 < \mt$.} 
\label{fig:stop_decay}
\end{figure*}

We show in \reffi{fig:stop_decay} a colour-coded scatter plot 
in the $(\mstop1, \mneu1)$ plane of the $\sto1$ decay modes with branching ratios $> 50$\% for
1000 randomly-selected pMSSM10 points in the region of interest.
We see that the $\sto1 \to b \cha1$ mode (shown in light green)
dominates for the majority of points, 
and that this decay can be important throughout the parameter region displayed.
We also find that, when this is the dominant stop decay mode,
in most cases the $\cha1$ and $\neu1$ are almost mass degenerate. 
To constrain the final states with this decay mode we implement 
the simplified model limit presented in Fig.~6 of the ATLAS di-bottom analysis \cite{arXiv:1308.2631},
where $\mcha1 - \mneu1 = 5$ GeV is assumed,
applying this for the model points with $\mcha1 - \mneu1 < 30$ GeV.

If $\mst1 - \mneu1 > \MW + m_b$, the 3-body $\sto1 \to b W \neu1$ mode can dominate stop decay.
The points for which this mode is dominant are shown by purple dots in \reffi{fig:stop_decay}.
For this decay mode we implement the simplified model limit presented for $\MW + m_b < \mst1 - \mneu1 < m_t$ in Fig.~15 
of the ATLAS single-lepton analysis~\cite{arXiv:1407.0583}.      

In the $\mstop1 - \mneu1 < \MW + m_b$ region, the decays
$\sto1 \to c \neu1$ (red dots in \reffi{fig:stop_decay}) and $\sto1 \to b f f' \neu1$ (grey dots)
can be the dominant stop decay modes.
The $\sto1 \to b \nu_\tau \stau1$ mode (green dots) may also
dominate stop decay in this region, as well as in the 
$\mstop1 - \mneu1 \gsim \MW + m_b$ region, as can also be seen in
\reffi{fig:stop_decay}.  

Due to the variety of different stop decay modes that are relevant in this compressed region,
we cannot use only the limits from simplified models provided by the experiments,
as they do not cover all relevant decay chains {and assume branching ratios of 100\%}.  However,
these missing, in part rather complex, decay chains can effectively be constrained by hadronic inclusive searches
such as those we have already used for our \lhccol\ limits.
In particular, the CMS hadronic $m_{T2}$ search~\cite{cms_mt2} has rather high sensitivity for these decay chains,
as the kinematic phase space covered by the search makes no special assumptions on the final state,
other than it having a purely hadronic signature.

Based on these inclusive searches, we derive limits for simplified models for $\sto1 \to c \neu1$
and $\sto1 \to b \nu_\tau \stau1$ decays.
For the $\sto1 \to b \nu_\tau \stau1$ simplified model
we assume $m_{\stau1} - m_{\neu1} \lsim 40$ GeV when creating the limit in the ($m_{\sto1}$, $m_{\neu1}$) plane.
We do not implement a simplified model limit for $\sto1 \to b f f' \neu1$ because
this decay mode has negligible impact on our study, as can be seen in \reffi{fig:stop_decay}.
Using these simplified model limits, we constrain the
stop decay modes following a procedure very similar to what we used for
\lhcewk,
using an interpolating function of the form (\ref{exponentialform}) to mimic 
the uncertainty (yellow) band in, e.g., Fig.~6c in \cite{arXiv:1308.2631}.
We summarise our implementation of the simplified model limits in \refta{tab:SMS-stop}.
When establishing these limits we use values of the parameters $\mu_{l,r}$ and $\sigma_{l,r}$ that
depend on \mneu1. 
Whenever multiple values of these parameters are given for different values of $\mneu1$, 
the parameters for intermediate values of $\mneu1$ are obtained by 
linear interpolation, and taken as constants elsewhere.

\begin{table*}[htb!]
\begin{center}
\renewcommand{\arraystretch}{1.1}
\begin{tabular}{|c|c|c|c|c|c|} \hline
Decay              &  Limit                                     & \mneu1~[GeV] & $(\mu_l, \sigma_l)$~[GeV] & $(\mu_r, \sigma_r)$~[GeV]  & Condition/Remark     \\ 
\hline                                                            
$\sto{} \to b \cha1$    & Fig.~6(c) in~\cite{arXiv:1308.2631}   & 210 & (10, 20) & (-50, 50) &$\mcha1 - \mneu1 < 30 \gev$  \\
                        &                                       & 300 & (-250, 200) & (-200, 200) &   \\
\hline
$\sto{} \to b W \neu1$  & Fig.~15 in~\cite{arXiv:1407.0583}     & 100 & (-20,50) & (-70, 50) & $\MW < \mst1 - \mneu1 < m_t$  \\
                        &                                       & 150 & (-50, 50)& (-100,50) & \\
\hline
$\sto{} \to b \nu \tilde \tau_1$  & {Gener}ated using               & -   & (-50, 50)& (-20, 50) & Based on~\cite{cms_mt2}, assuming \\
                                  & \scorpion                   & & & & $\mstau1 - \mneu1 \lsim 40 \gev$ \\
\hline
$\sto{} \to c \neu1$       &         {Gener}ated using              &-   & (-20, 20) & (-20, 20)&  Based on~\cite{cms_mt2} \\
                           &        \scorpion    & & & & \\
\hline
\end{tabular}
\caption{\it The simplified model limits used to constrain scenarios with
  compressed stop spectra. 
When establishing these limits we use values of $\mu_{l,r}$ and $\sigma_{l,r}$
in \protect\refeq{exponentialform} that
in some cases depend on \mneu1. 
Whenever multiple values of these parameters are specified for different values of $\mneu1$, 
the parameters for intermediate values of $\mneu1$ are obtained by 
linear interpolation, and taken as constants elsewhere.}
\label{tab:SMS-stop}
\renewcommand{\arraystretch}{1.0}
\end{center}
\end{table*}

As for our \lhccol\ and \lhcewk\ limit implementations,
it is also important to determine accurately the uncertainty in 
the dedicated limit procedure for the compressed stop region.  
Note that in the compressed region not only the constraints from \lhcstop\ but
also those from \lhcewk\ play a role. 
Therefore we first assess the qualitative agreement between \chisqlhcstop\ and
the ``true'' \chisqatomscorpion\ as calculated using the {\tt Scorpion} 
and {\tt Atom} codes, for points with $\chisqlhcewk<2$. 
\reffi{fig:1000randomstopcolour} compares scatter plots in the
$(\mst1, \mneu1)$ plane of $\chisqlhcewk+\chisqlhcstop$ (left panel) 
and \chisqatomscorpion\ (right panel). 
The colour code used is indicated on the right-hand sides of the panels, 
and we see that the patterns of 
colours in the two scatter plots are qualitatively similar. 
This is remarkable, given the interplay of so many different decay
chains.

\begin{figure*}[htb!]
\resizebox{8cm}{!}{\includegraphics{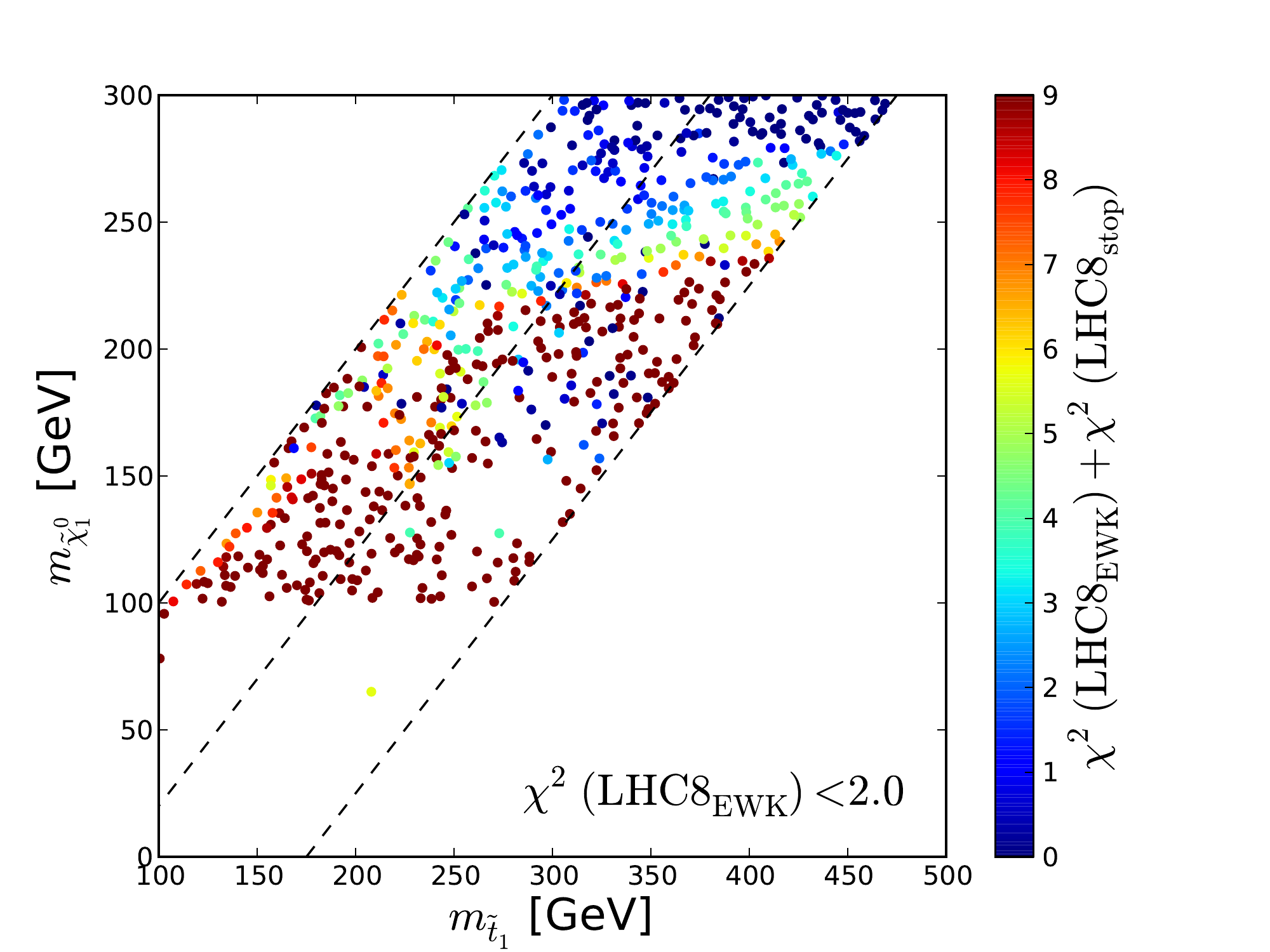}}
\resizebox{8cm}{!}{\includegraphics{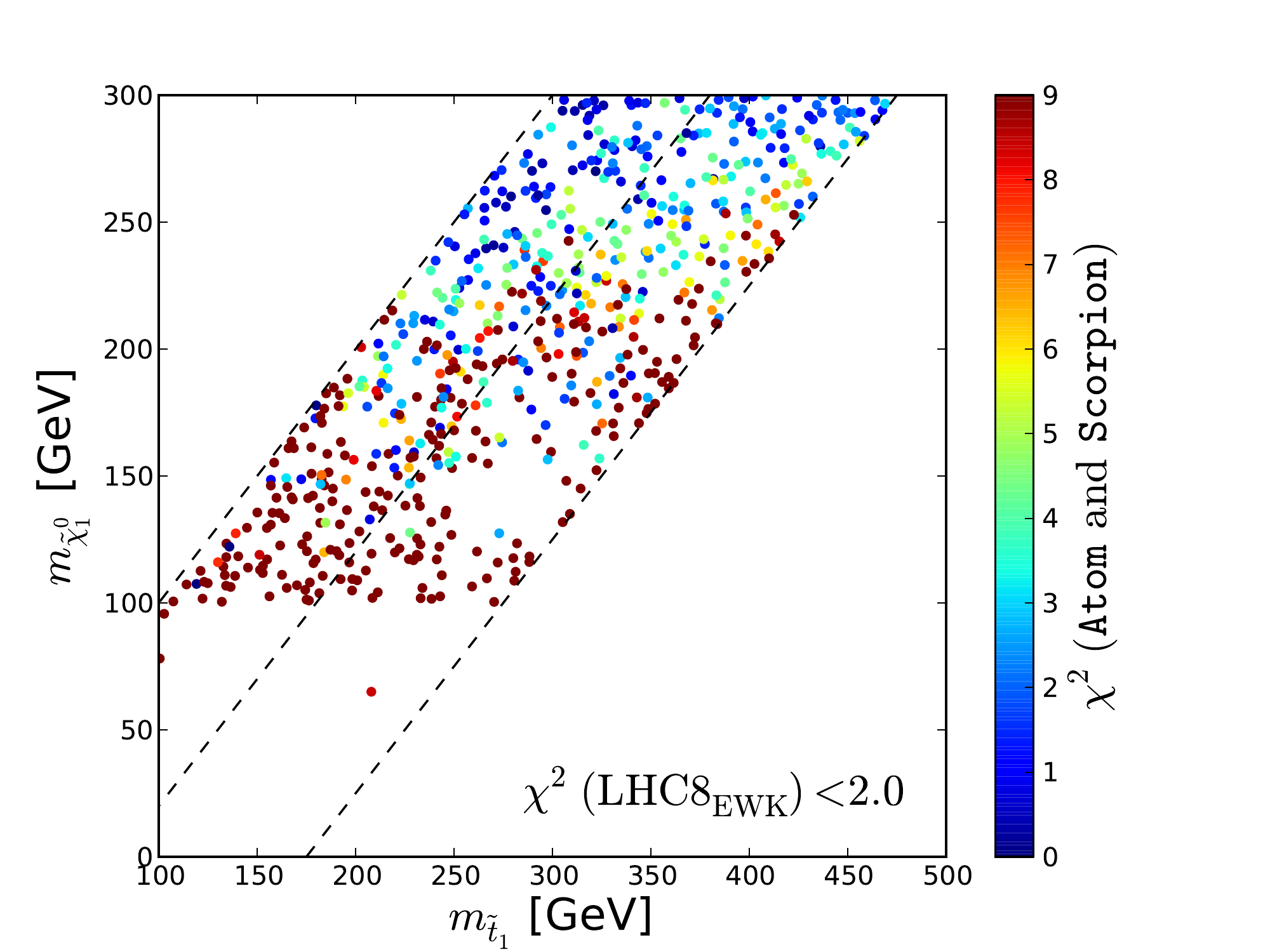}}
\caption{\it Scatter plots in the $(\mst1, \mneu1)$ plane of the contributions to the global $\chi^2$
functions from the ATLAS mono-jet~\protect\cite{atlas_stop-monojet}
and single-lepton~\cite{atlas_stop-1L} searches for 1000 randomly-selected points in the regions of interest.
The left panel shows calculations using simplified model searches ($\chisqlhcewk$)
and the right panel shows results from the {\tt Scorpion} and {\tt Atom}
codes ($\chi^2({\rm true})$).}
\label{fig:1000randomstopcolour}
\end{figure*}

More quantitative comparisons of the contributions to the global $\chi^2$ function calculated on the basis of 
the simplified model searches for stops and electroweakly produced sparticles 
($\chisqlhcewk+\chisqlhcstop$) with results from {\tt Scorpion} and {\tt Atom} for these
1000 randomly-selected pMSSM10 points ($\chi^2({\rm true})$) are shown in \reffi{fig:1000randomstop}.
The left panel shows a histogram of the difference between $\chisqlhcewk+\chisqlhcstop$ and
$\chi^2({\rm true})$, showing that it is
relatively small, with an r.m.s.\ difference $\sigma_{\chi^2}=3.15$.
The right panel of \reffi{fig:1000randomstop} displays a scatter plot in the
$(\chisqatomscorpion, \chisqlhcewk+\chisqlhcstop)$
plane. We see that points that are (dis)favoured at the 95\% $\cls$ level in the simplified approach
are, in general,  also (dis)favoured at the 95\% $\cls$ level in the more sophisticated approach
based on {\tt Scorpion} and {\tt Atom}. 

To determine quantitatively the effect of the uncertainty in the \lhcstop\ procedure,
we translate the impact of the above-mentioned $\sigma_{\chi^2}=3.15$ uncertainty into the
$(\mst1, \mneu1)$ plane in the lower right panel of \reffi{fig:up-down}.
This shows the impacts of $\pm~1~\sigma_{\chi^2}$ variations on our 68\% and 95\% contours in this plane,
which is rather small except for small values of $\mst1$ and $\mneu{1}$.

Based on this study, we conclude that the computationally-manageable simplified approach \lhcstop\ is sufficiently
reliable for our physics purposes. Specifically, we note that there are
points with low $\mst1$ that
survive the full LHC constraints with relatively low $\chi^2$. 

\begin{figure*}[htb!]
\resizebox{8cm}{!}{\includegraphics{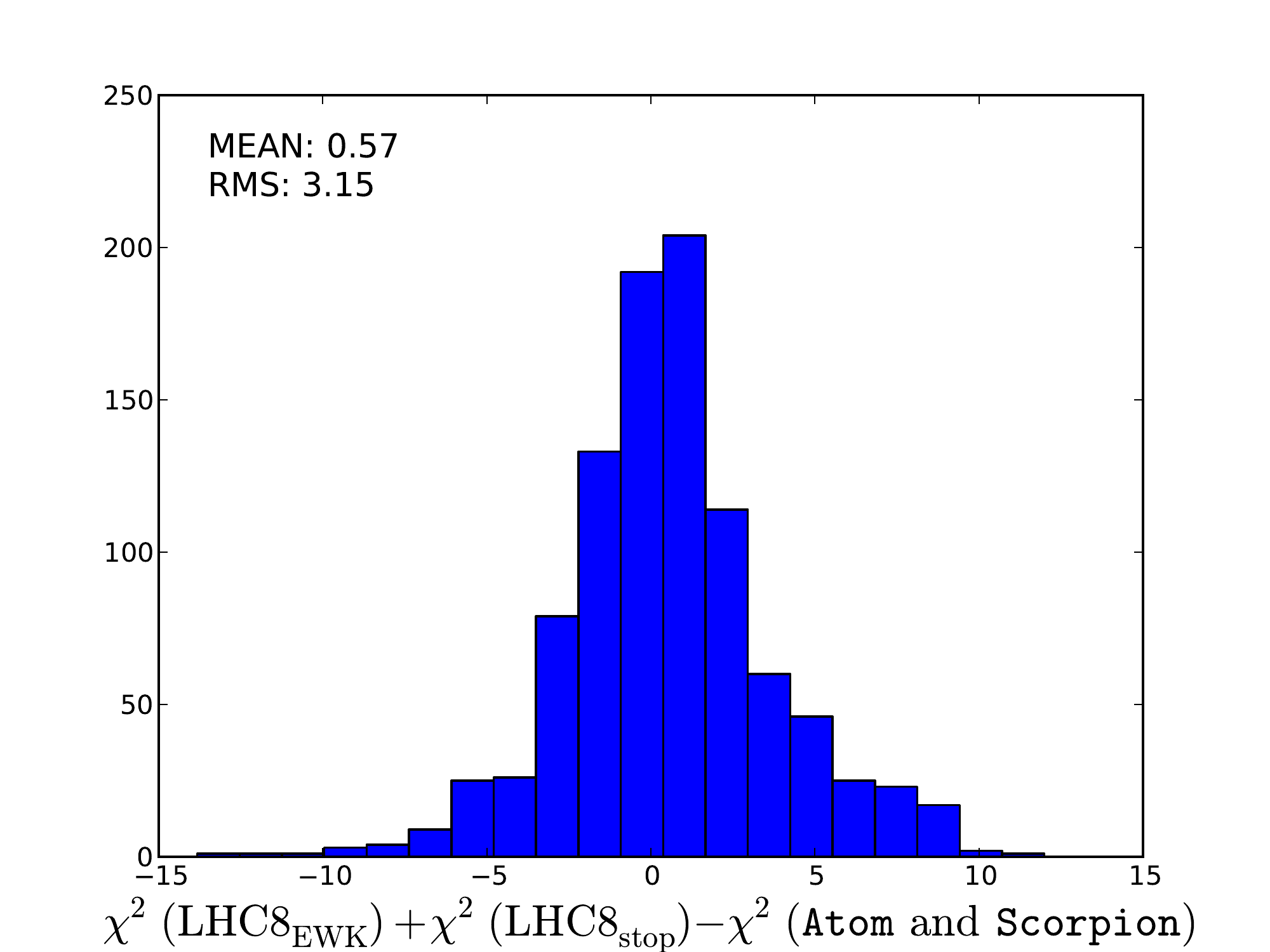}}
\resizebox{8cm}{!}{\includegraphics{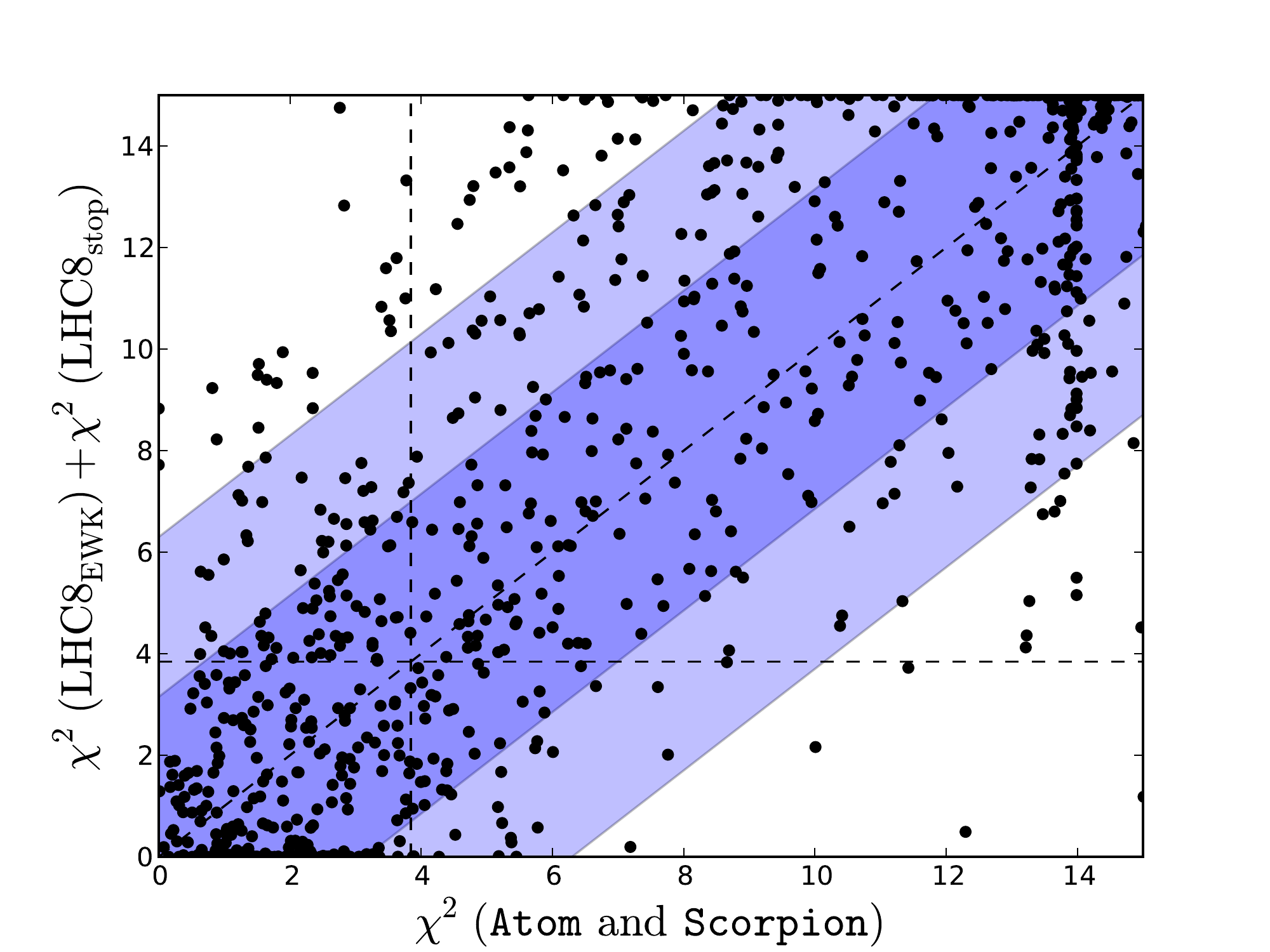}} \\
\vspace{-1cm}
\caption{\it Left panel: Histogram of the difference between the values of the contributions of
the stop constraints to the global likelihood function $\chisqlhcewk+\chisqlhcstop$ evaluated using
simplified model searches for 1000 randomly-selected points
and the estimates of $\chi^2$ found using {\tt Scorpion} and {\tt Atom}.
Right panel: Scatter plot in the $(\chi^2({\rm true}), \chisqlhcewk+\chisqlhcstop)$
plane of the values obtained from the two approaches; the vertical and horizontal dashed lines in these
plots correspond to the 95\% $\cls$ in each approach.}
\label{fig:1000randomstop}
\end{figure*}

\section{Results}
\label{sec:results}

\subsection{Mass Planes}

\begin{figure*}[htb!]
\vspace{-1cm}
\hspace {0.5cm}
\resizebox{7cm}{!}{\includegraphics{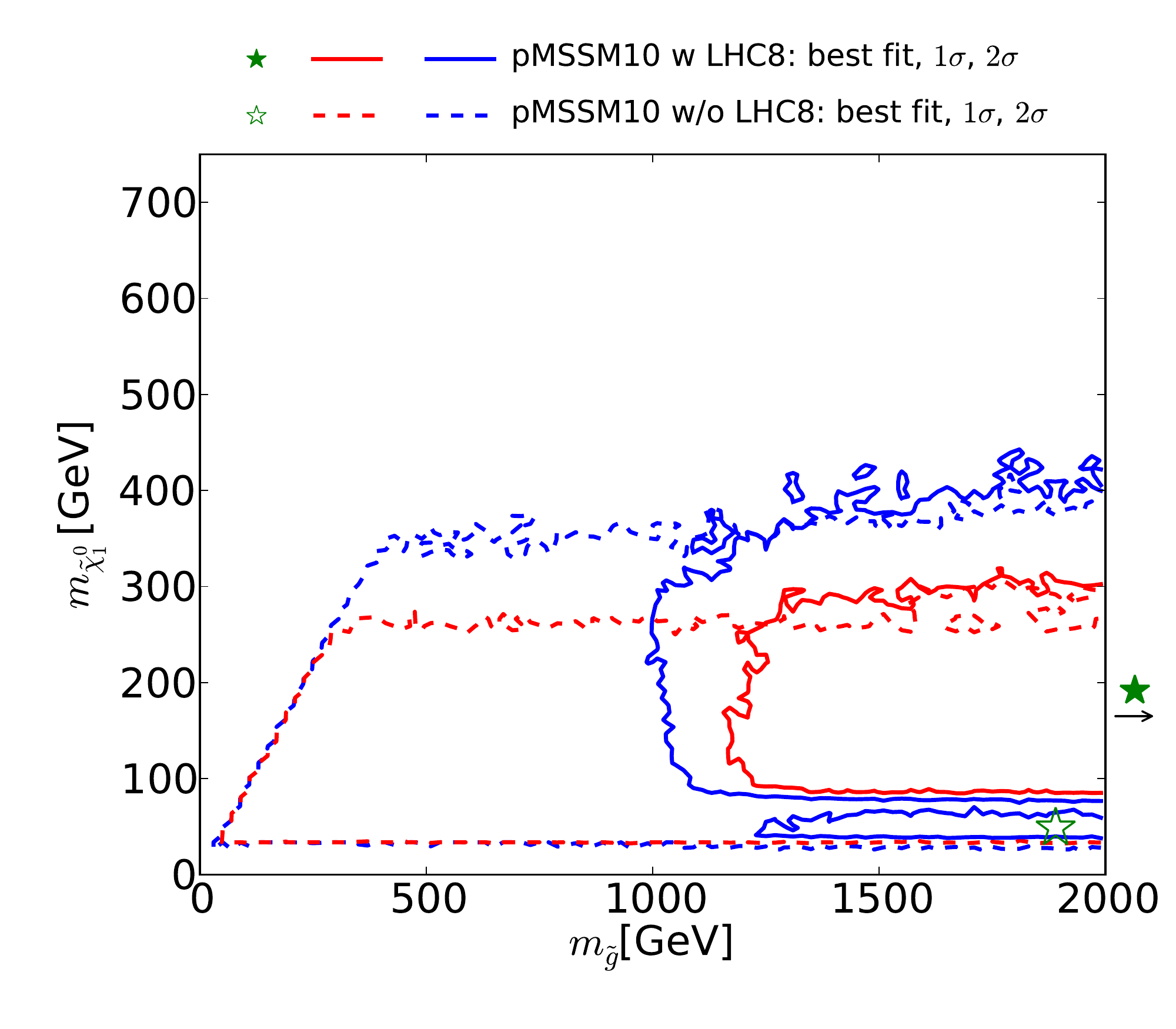}}
\resizebox{7cm}{!}{\includegraphics{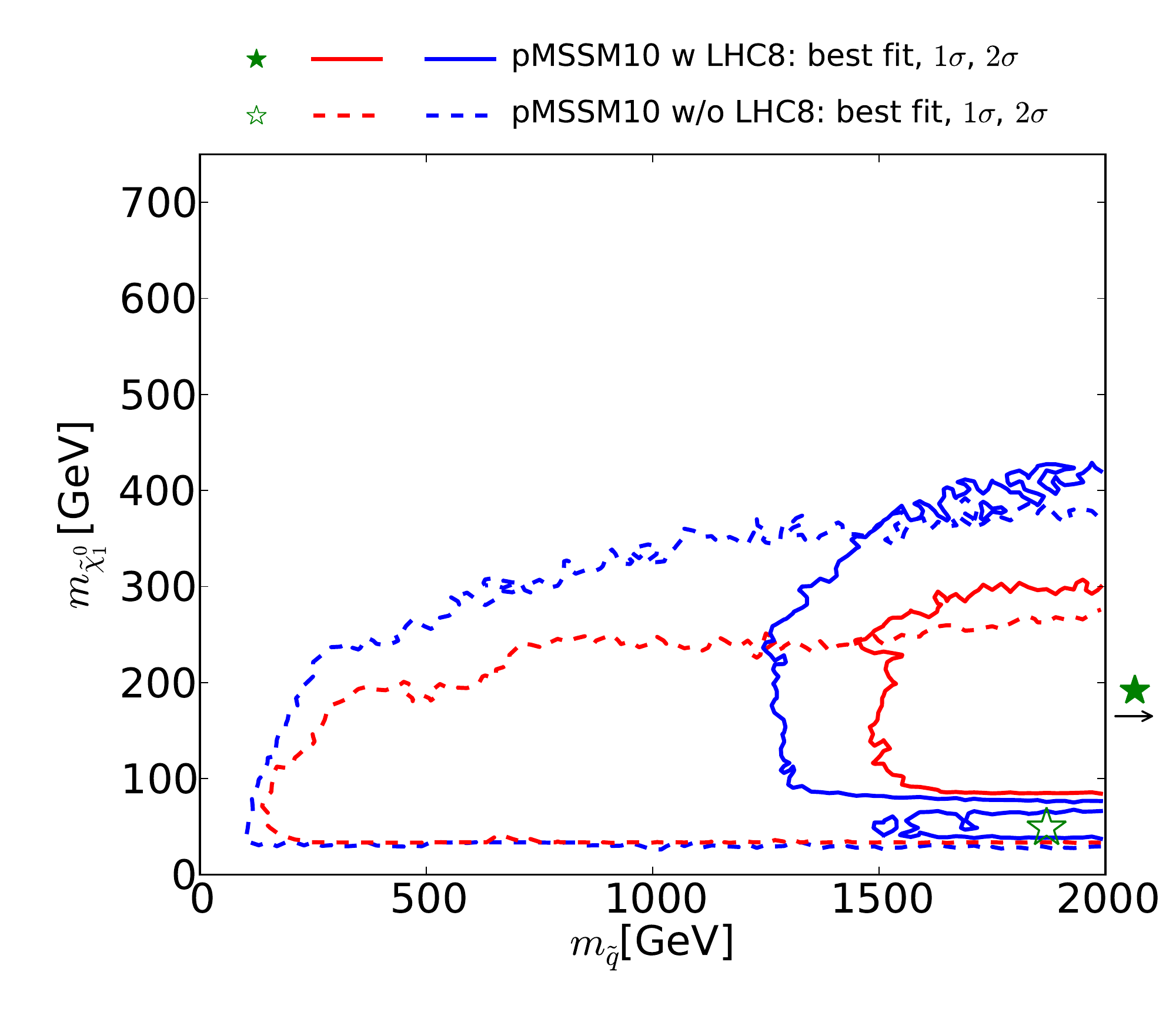}}\\
\vspace{-0.5cm}
\hspace {0.5cm}
\resizebox{7cm}{!}{\includegraphics{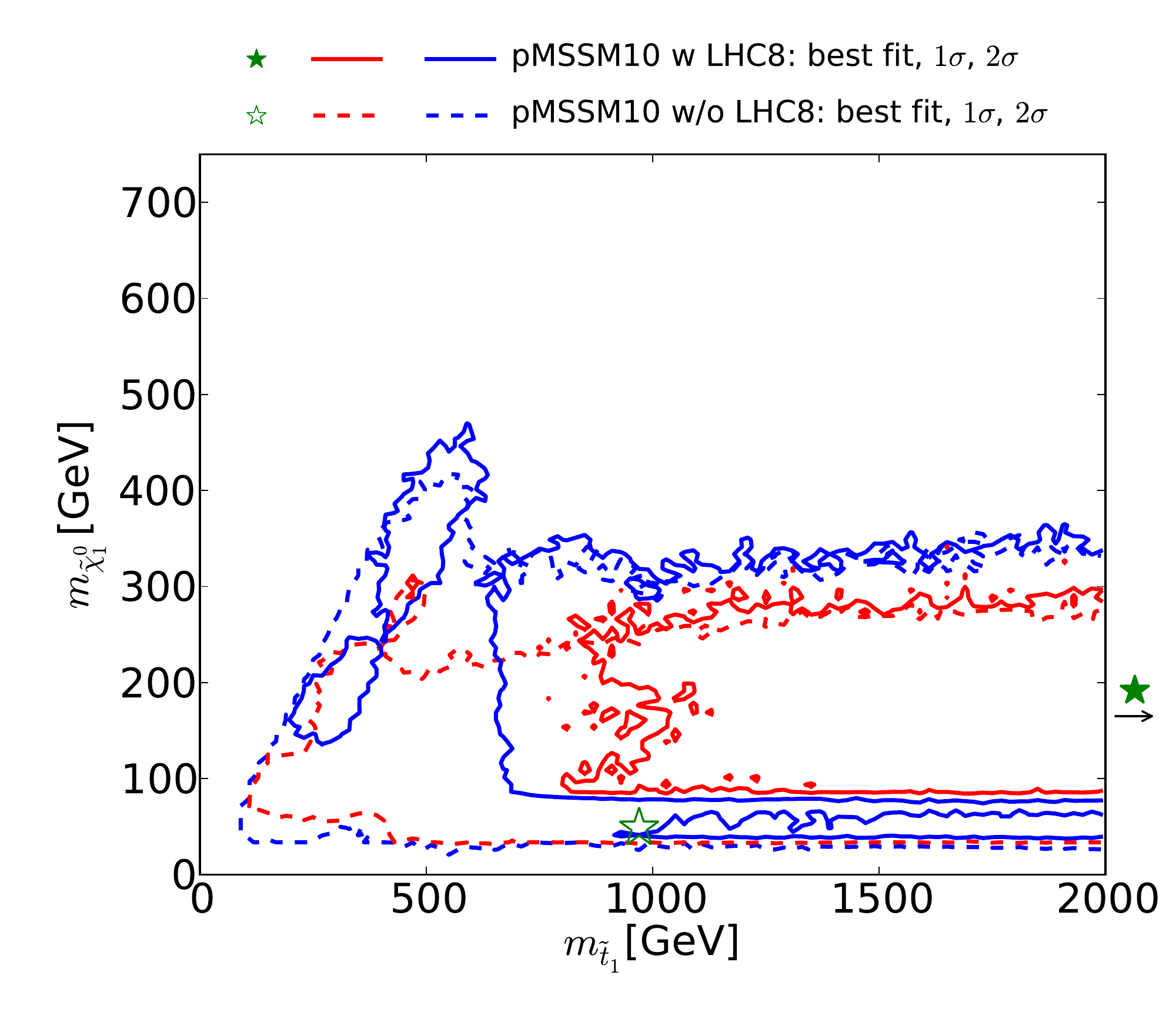}}
\resizebox{7cm}{!}{\includegraphics{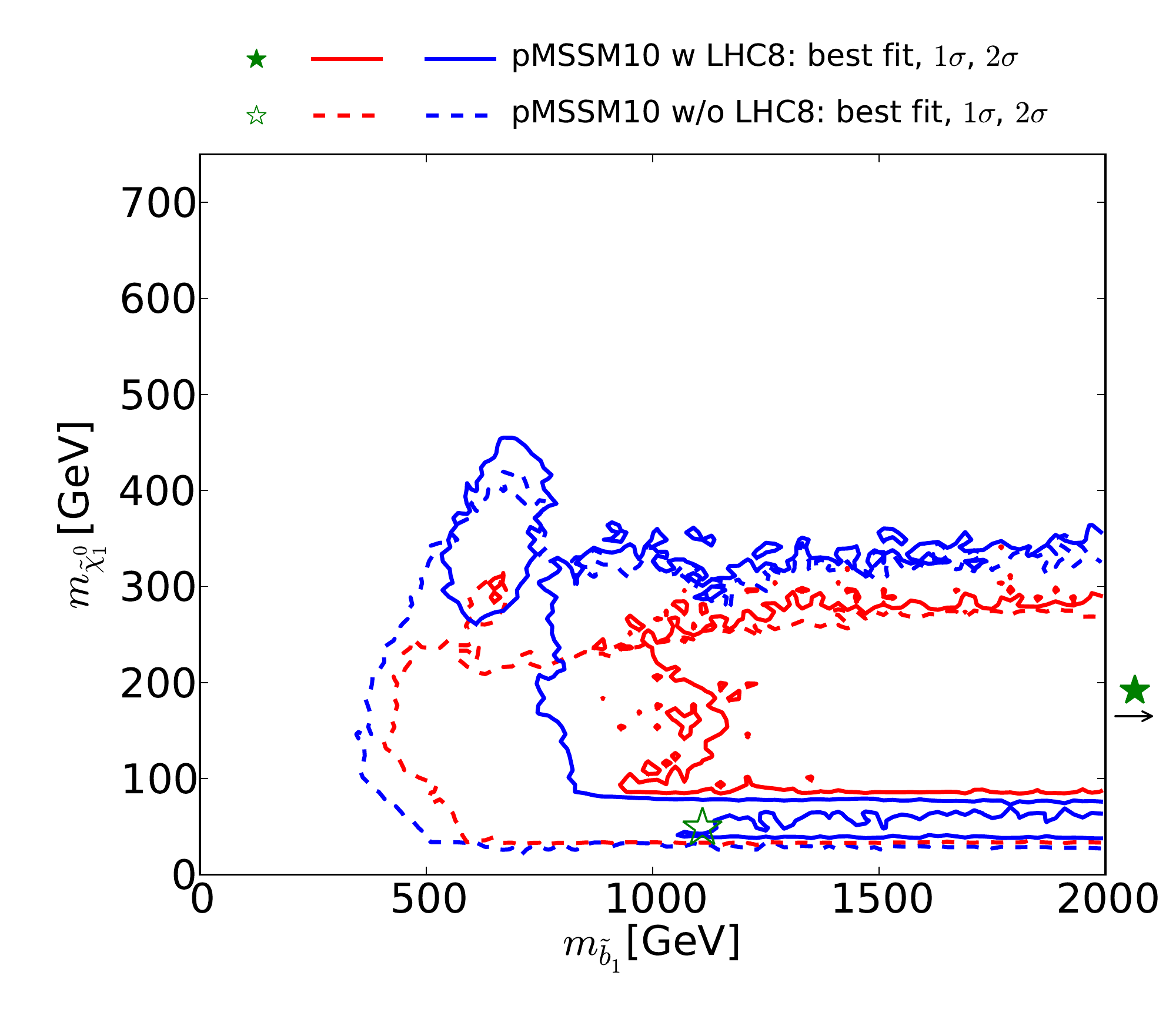}}\\
\vspace{-0.5cm}
\hspace {0.5cm}
\resizebox{7cm}{!}{\includegraphics{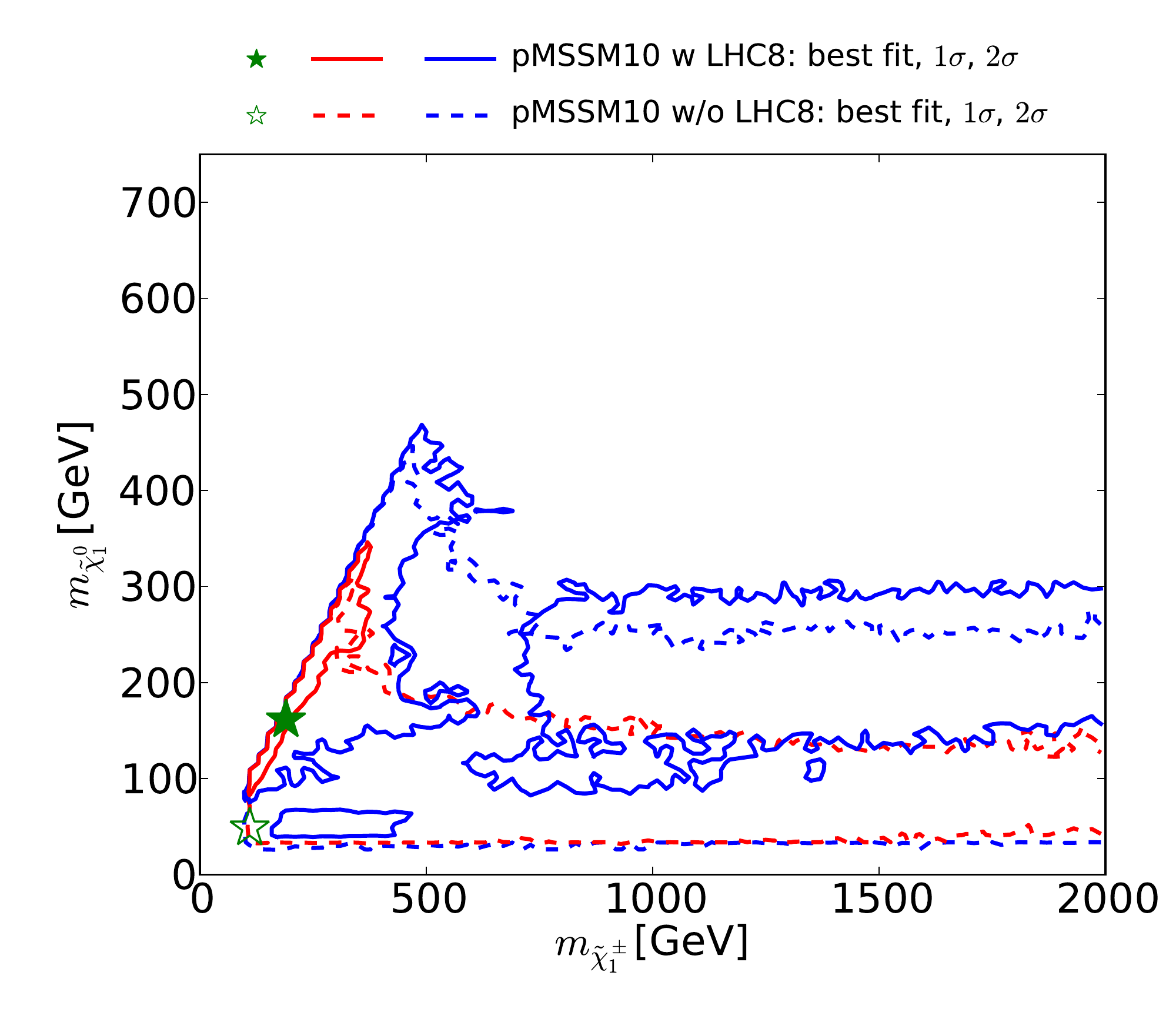}}
\resizebox{7cm}{!}{\includegraphics{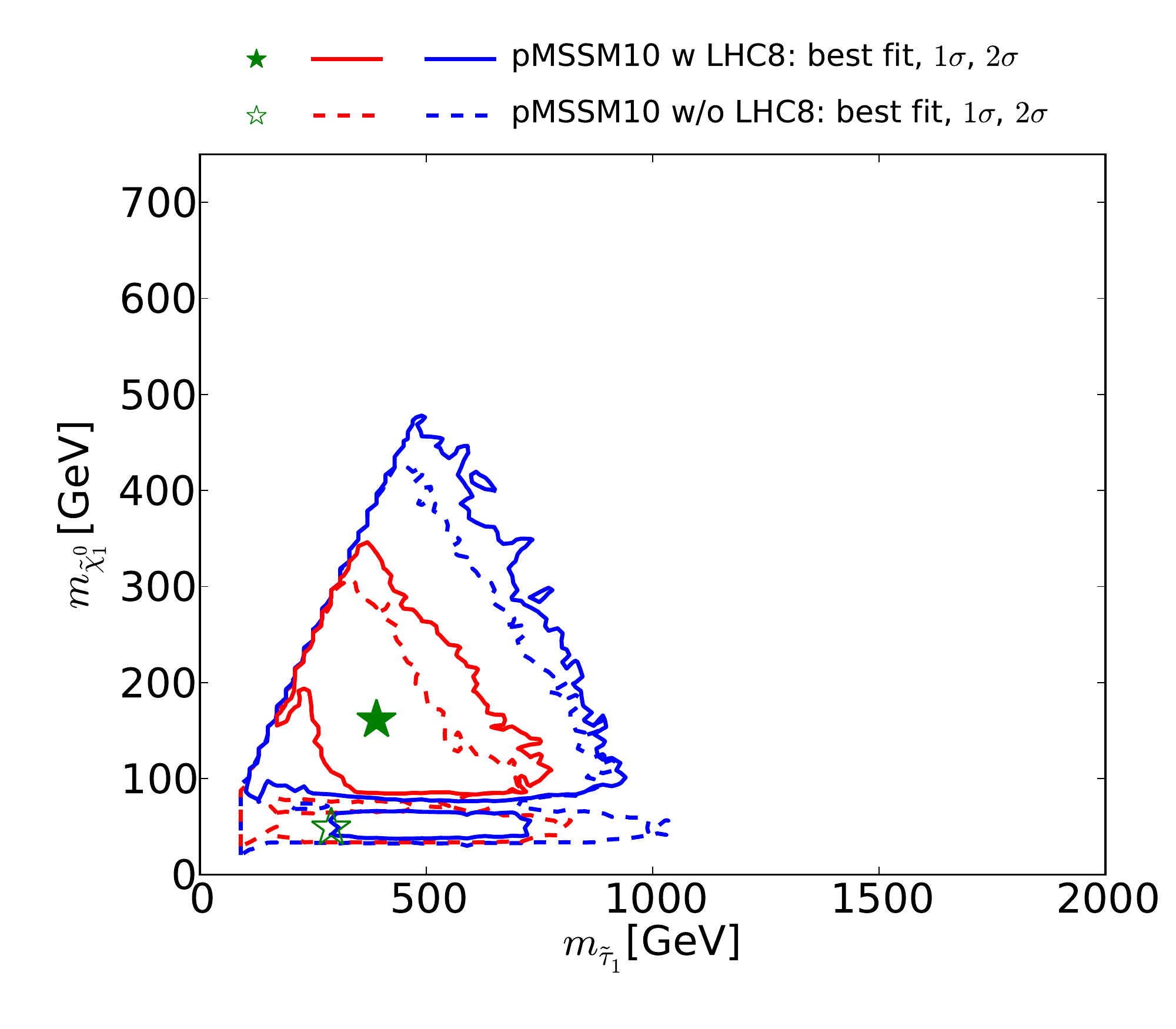}}\\
\vspace{-0.5em}
\caption{\it The two-dimensional profile likelihood functions for (top left to bottom right)
the masses of the gluino, the first- and second-generation squarks, the lighter stop and sbottom
squarks, the lighter chargino and the lighter stau, each
versus the lightest neutralino mass $\mneu1$. 
In each panel the solid (dashed) red/blue contours denote the 
$\Delta \chi^2 = 2.30/5.99$ level contours for the case where we
do (not) apply the LHC8 constraints, respectively.
The green filled and empty stars indicate the corresponding best-fit points.
}
\label{fig:planes}
\end{figure*}

\reffi{fig:planes} displays the two-dimensional profile likelihood functions {in planes of}
(from top left to bottom right) the masses of 
the gluino, the first- and second-generation squarks, the lighter stop and sbottom
squarks, the lighter chargino and the lighter stau, each
versus the lightest neutralino mass $\mneu1$. 
In each panel the solid (dashed) red/blue contours denote the 
68\%/95\% CL contours for the case where we
do (not) apply any LHC constraints, respectively~\footnote{{However, 
the LEP SUSY constraints~\cite{LEPSUSY} are applied.}}.
The green filled and empty stars indicate the corresponding best-fit points. In the cases of the gluino and squarks, the
filled stars lie beyond the displayed parts of the corresponding planes, and their locations
are indicated by arrows. In these cases the likelihood function varies little as a function of the coloured sparticle mass.

On the other hand, we find that in general $\mneu1 \lesssim300\gev$ at the $\sim68\%$ CL,
increasing to $\sim 500 \gev$ at the $\sim95\%$~CL.
This and the preference for low stau masses ($\lesssim700\gev$ at the $\sim68\%$ CL,
$\lesssim 1000 \gev$ at the $\sim95\%$ CL) are reflections of the fulfilment of the
\gmt\ constraint in the pMSSM10, cf., \reffi{fig:gmt} below,
and (in the latter case) the restriction to a common slepton mass for all three generations.

We can distinguish two ranges of $\mneu1$ that are allowed at the 95\% CL:
a narrow band where $\mneu1 \lesssim 80\gev$ and a broader region at larger $\mneu1$
that also includes regions favoured at the 68\% CL. In the low-$\mneu1$ region,
before applying the LHC8 constraints the smuon, selectron and stau could have been
relatively light, and $t$-channel sfermion exchange could bring the relic density into the range allowed
by cosmology. However, after applying the LHC8 constraints only the 
$Z$- and $h$-funnels are allowed in this region. 
In the region where $\mneu1\gtrsim80\gev$, before implementing the LHC8 constraints stau coannihilation and
$t$-channel sfermion exchange were both possible. However, after applying the LHC8 constraints
the dominant processes controlling the dark matter density are $\neu1 - \neu2 - \cha1$
coannihilations, with the LSP having mainly a Bino composition.

The two top panels of \reffi{fig:planes} display clearly the direct
impacts of the LHC8 constraints, which are visible in the displacements to larger
masses of the $68\%$ and $95\%$ CL contours, as can be seen from
the comparison of the solid and dashed lines.  
On the other hand, the pictures in the two middle panels are more complex.
There are intermediate values of $\mst1$ that are disfavoured by
the LHC8 constraints, 
but there are regions with low values of $\mst1$ that are allowed by the
LHC8 constraints at the 95\% CL, and even some points with $\mst1$ and $\msb1$
that are favoured at the 68\% CL, though these are not prominent.
In the case of the lighter sbottom, the LHC8 constraints disfavour
{the region where both
$\msb1$ and $\mneu1$ have small values.
However, a small value of $\msb1$ is} {still allowed 
at the $\sim 95\%$~CL if $\mneu1 \gtrsim 300 \gev$ to $450 \gev$, where
some points are favoured at the 68\% CL.

Finally, the bottom two panels of \reffi{fig:planes} show the impacts of the
LHC8 constraints on the chargino and stau masses.
The main impact on the chargino mass is to disfavour most values except
some where $\mcha1 - \mneu1$ is small.
{This is an indirect effect of the LHC8 constraints, with the}
coannihilation of the dark matter particle with the lighter chargino
{playing an important role in bringing}
the dark matter density into the allowed range.
This compression of the spectrum can be attributed to the
\lhcewk\ limits on direct production of light  
sleptons, and to a lesser extent on charginos decaying via sleptons. 
These constraints on light sleptons disfavour the {$t$-channel sfermion
exchange} and 
stau coannihilation regions. The latter is a consequence of our choice
of a single mass parameter for the masses of all the
scalar leptons (see also \refse{sec:summary}).
In the case of the lighter stau, we see in the bottom right panel of \reffi{fig:planes} a triangular region that is
favoured at the $\sim 68\%$~CL, which is somewhat reduced {and
shifted towards higher mass values} by the LHC8
constraints.


\subsection{The Best-Fit Point}
\label{sec:best-fit}

\begin{table*}[htb!]
\renewcommand{\arraystretch}{1.1}
\begin{center}
\begin{tabular}{|c|r||r|r|r|r|} \hline
Parameter   &  Best-Fit & Low \mstop1  & Low \msq   & Low \mgl    & Low all   \\ 
\hline         
$M_1$       &  170   \gev & 300  \gev  & 210  \gev  & 190  \gev   & -120 \gev \\
$M_2$       &  170   \gev & 310  \gev  & 220  \gev  & 200  \gev   & 160  \gev \\
$M_3$       &  2600  \gev & 1660 \gev  & 3730 \gev  & - 1070 \gev & 1700 \gev \\
\msq        &  2880  \gev & 3700 \gev  & 1530 \gev  & 2430 \gev   & 1790 \gev \\
\msqt       &  4360  \gev & 720  \gev  & 1840 \gev  & 3780  \gev  & 1300 \gev \\
\msl        &  440   \gev & 390  \gev  & 430  \gev  & 410  \gev   & 740  \gev \\
\MA         &  2070  \gev & 3540 \gev  & 2810 \gev  & 2990 \gev   & 1350 \gev \\
$A$         &  790   \gev & 1790 \gev  & 2510 \gev  & 3000 \gev   & 1863 \gev \\
$\mu$       &  550   \gev & 1350 \gev  & 640  \gev  & 530 \gev    &  190 \gev \\
\tb         &  37.6~~~~~  & 37.3~~~~~  & 40.8~~~~~  & 33.9~~~~~   & 35.4~~~~~ \\
\hline
\end{tabular}
\caption{\it {Parameters of the pMSSM10 best-fit point and other comparison 
benchmark points at low $\mstop1$, low $\msq$ and/or $\mgl$.}}
\label{tab:bf-point}
\end{center}
\renewcommand{\arraystretch}{1.0}
\vspace{1em}
\end{table*}

{We now discuss the characteristics of the best-fit point, whose 
parameters are listed in \refta{tab:bf-point}, together with the
parameters of several benchmark points that are discussed below.
The best-fit} spectrum
is shown in \reffi{fig:bestfitspectrum}, and its SLHA file~\cite{SLHA} can be downloaded from
the MasterCode website~\cite{mcweb}. We note first the near-degeneracy
between the $\neu1, \neu2$ and $\cha1$, which is a general feature of
our 68\% CL region that occurs in order to bring the cold dark matter density
into the range allowed by cosmology: see the bottom left panel of
\reffi{fig:planes}. {Correspondingly, we see in \refta{tab:bf-point} that
$M_1 \simeq M_2$, though $M_3$ is very different.}
The overall $\neu1/\neu2/\cha1$ mass 
scale is bounded from below by the LEP and
\lhcewk\ constraints, and from above by \gmt, especially at the 68\% CL.}
We display in \reffi{fig:mass-summary} 
the 95\% (68\%) CL
intervals in our fit for the masses of pMSSM10 particles as lighter (darker) peach
shaded bars, with the best-fit values being indicated with blue horizontal
lines~\footnote{The striations in these bars reflect the non-monotonic behaviours of the $\chi^2$
function visible in Fig.~\ref{fig:onedimensional}.}.
Turning back to \reffi{fig:bestfitspectrum},
we note the near-degeneracy between the slepton masses, which reflects our
assumption of a common input slepton mass at the input scale $M_{\rm SUSY}$
that would not hold in more general versions of the pMSSM. The overall slepton
mass scale is {below 1 \tev}, as seen in \reffi{fig:mass-summary}, being bounded
from above by \gmt\ and from below by \lhcewk\ constraint.
The latter also provides the strongest upper bound on the $\neu1/\neu2/\cha1$. 
We also see in \reffi{fig:mass-summary}  that the gluino, squark,
stop and bottom masses are all very poorly constrained in our pMSSM10
analysis, though the \lhccol\ constraint forbids low masses.

{Concerning the Higgs sector, we note that the best-fit value for $M_A$ lies in the
multi-TeV region (where its actual value is only
weakly constrained) and is therefore far in the decoupling region. Accordingly,
the properties of the light Higgs boson at about 125~GeV resemble very
closely those of the Higgs boson of the SM.}

\begin{figure*}[htb!]
\vspace{0.5cm}
\centering
\resizebox{10cm}{!}{\includegraphics{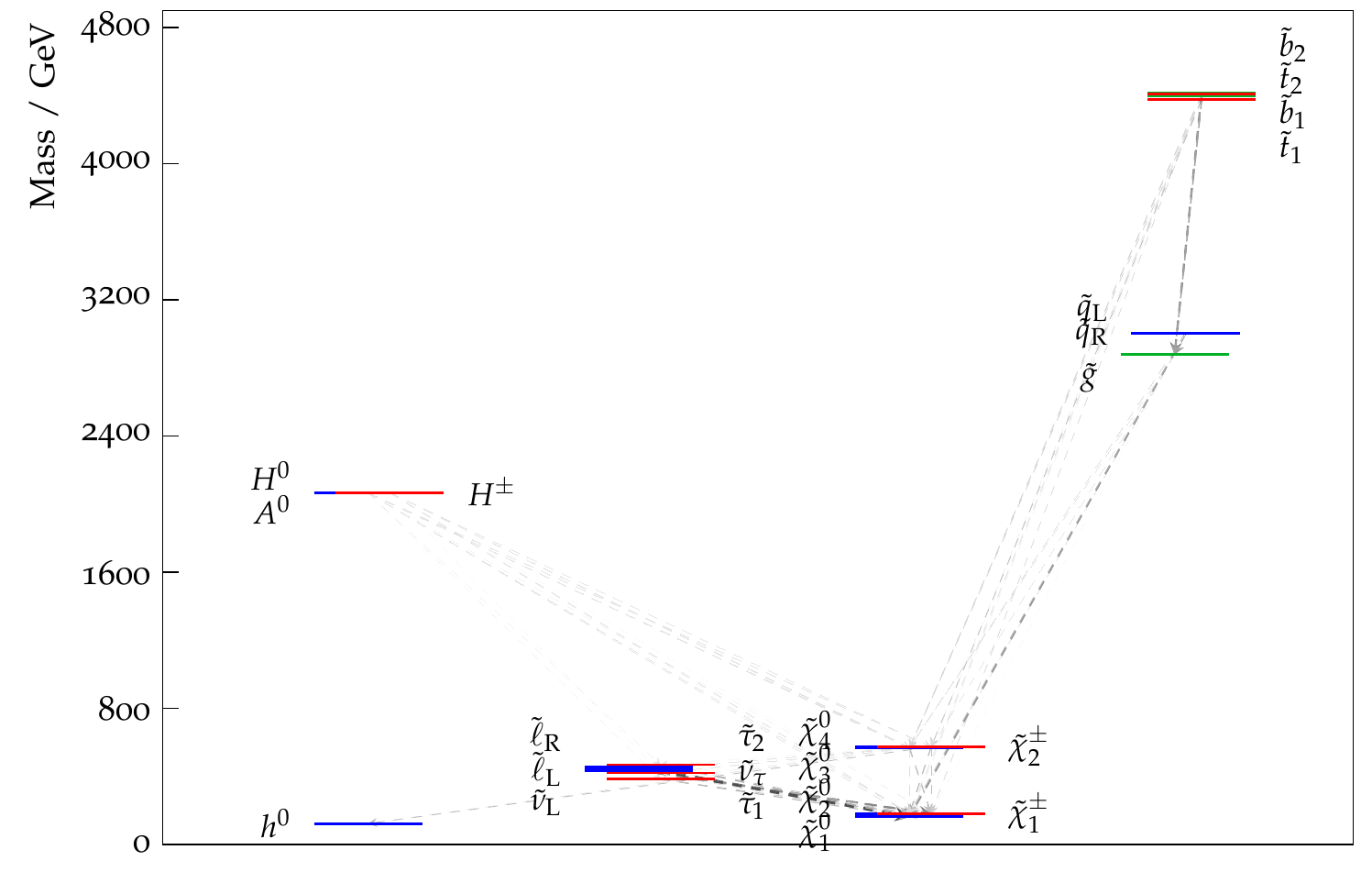}}
\caption{\it The particle spectrum and dominant decay branching ratios at our best-fit {pMSSM10} point. Note the
near-degeneracies between $\neu1, \neu2$ and $\cha1$, between the
sleptons, between $\neu3, \neu4$ and $\cha2$, between the ${\tilde q_L}$
and ${\tilde q_R}$, between the heavy Higgs bosons, and between the stops and
bottoms, which are general features of our 68\% CL region. On the other hand,
the overall sparticle mass scales, in particular of the coloured sparticles,
are poorly determined.
} 
\label{fig:bestfitspectrum}
\end{figure*}

\begin{figure*}[htb!]
\vspace{0.5cm}
\centering
\resizebox{18cm}{!}{\includegraphics{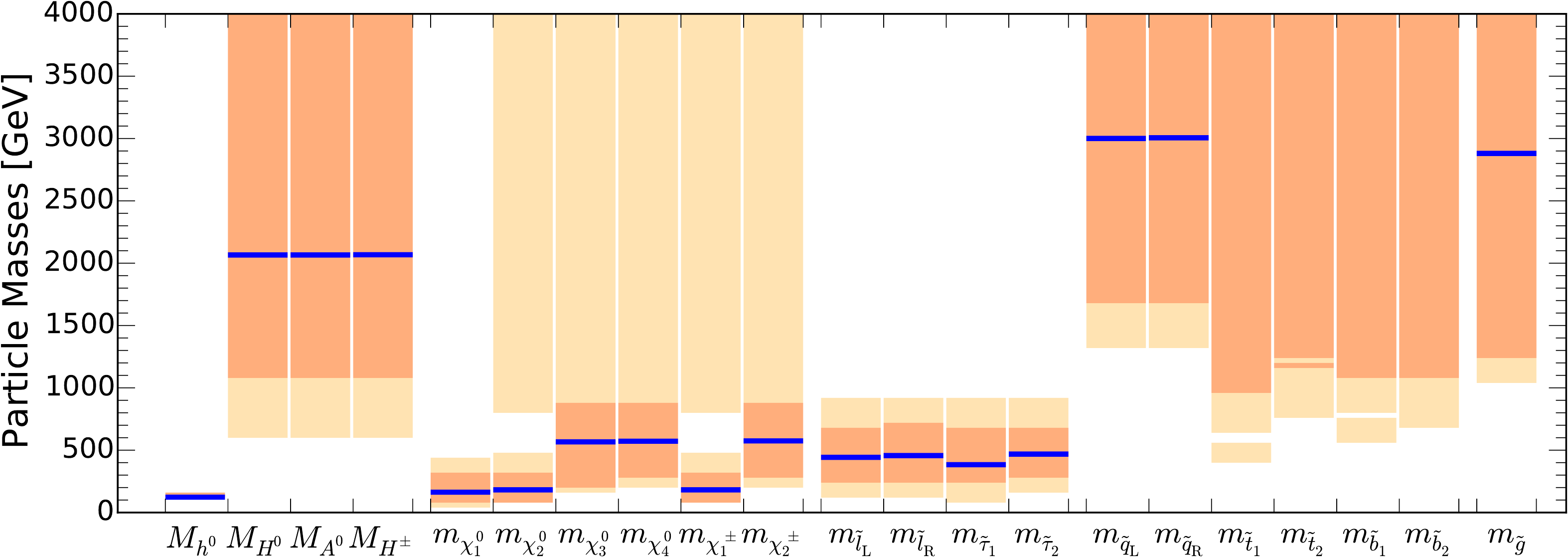}} 
\caption{\it Summary of mass ranges {predicted in the pMSSM10}. The light (darker) peach shaded bars
indicate the 95\% (68\%)~CL intervals, whereas the blue horizontal lines mark
the values of the masses at the best-fit point.
} 
\label{fig:mass-summary}
\end{figure*}

The first column of \refta{tab:breakdown} lists the most important contributions to the total $\chi^2$
function of different (groups of) constraints at the best-fit pMSSM10 point. 
The total $\chi^2$ value at the best-fit point is $\chi^2 = 83.3$, of which the largest part is due to the Higgs constraints evaluated using {\tt HiggsSignals}.

To convert the total $\chi^2$ of our fit into a $\chi^2$ probability estimate, we calculate the 
$\chi^2$ contribution and corresponding number of degrees of freedom (d.o.f.) 
by considering only constraints that have significant contributions to our 
global $\chi^2$ function in large regions of the relevant parameter space. 
We do not include in this procedure constraints from {\tt HiggsSignals}, 
which do not in general vary strongly in our preferred fit regions 
(see, e.g., $\chi^2(HS)$ in~\refta{tab:breakdown}). 
Therefore, to calculate the $\chi^2$ probability we consider in 
total 31 constraints, which translate into 18 d.o.f for the pMSSM10, 
24 d.o.f. for CMSSM, 23 d.o.f. for NUHM1, and 22 d.o.f. NUHM2.  Previous studies~\cite{mc1} showed that this definition
of the $\chi^2$ probability represents a good estimate of fit quality and enables a
comparison between different models on an equal footing. It also represents a reasonable 
approximation to the underlying absolute $p$-values of our fits.    

Comparing to the $\chi^2$ values for the CMSSM, NUHM1, and NUHM2
{shown in the last three columns of \refta{tab:breakdown}},
we see that the largest improvement is in the contribution from \gmt, though
there are also small improvements in \bsmm\ and the $Z$-pole observables. 
Overall, we see that the pMSSM10 has a $\chi^2$ probability of {30.8\% compared to 10.8\%, 12.1\% and 11.0\% for} the
CMSSM, NUHM1 and NUHM2, respectively, demonstrating that the pMSSM10 gives a
significantly better fit~\footnote{{The $\chi^2$ probabilities and values of $\chi^2$ differ from those
given in \cite{mc10}, as we have updated the CMSSM and NUHM analyses
with the most recent Higgs mass determination~\cite{Aad:2015zhl} and other new information~\cite{CMSLHCbBsmm,Bellebtn}.
We note, in particular, that the new \btn\ measurement~\cite{Bellebtn} improves the agreement with the SM and the SUSY models we study.}}.

We {stress}, however, that these $\chi^2$ probabilities are only approximate and assume an underlying
$\chi^2$-distribution with no correlations between the observables. 
A more proper treatment would be to smear the measurements around the best-fit
predictions, fit to these toy measurements and evaluate the fraction of
cases in which the resulting $\chi^2$ exceeds the observed $\chi^2$. 
We leave such an evaluation as a topic for future work.

\begin{table*}[htb!]
\renewcommand{\arraystretch}{1.1}
\begin{center}
{\small
\begin{tabular}{|c|c|c|c|c|c||c|c|c|} 
\hline
Constraint ~~~~{d.o.f.}     & \multicolumn{5}{|c||}{pMSSM10} & CMSSM  & NUHM1    & NUHM2   \\
& best fit & low $\mst1$ & low $\msq$ & low $\mgl$ & low all & \cite{mc9}  & \cite{mc9}  & \cite{mc10}  \\
\hline    \hline                                                                         
LHC8 ~~~~~~~~~~~~~~~1                      &   0.1    &   1.0    &   0.8    &   1.0   &  0.4    &  -       &  -       &  -      \\
Jets+\ETslash\ ~~~~~~~~~~~~1            &  -       &  -       &  -       &  -      & -       &   2.0    &   0.0    &   0.5   \\
$\Mh$ ~~~~~~~~~~~~~~~~~~\,1                     &   0.0    &   0.0    &   0.2    &   0.2   &  0.0    &   0.1    &   0.1    &   0.4   \\
$\MW$ ~~~~~~~~~~~~~~~~~\,1                     &   0.0    &   0.1    &   0.1    &   0.5   &  0.1    &   0.0    &   0.0    &   0.4   \\
$B_{s,d} \to \mu^+ \mu^-$ ~~~~~~1 &   0.2    &   0.0    &   0.0    &   0.2   &  0.0    &   0.5    &   0.3    &   0.4   \\
\bsg\  ~~~~~~~~1                    &   0.1    &   0.0    &   0.2    &   0.1   &  0.1    &   0.5    &   0.0    &   0.0   \\
\btn\  ~~~~1                    &   0.2    &   0.2    &   0.2    &   0.2   &  0.3    &   0.2    &   0.2    &   0.2   \\
{Other $B$ physics} ~~5   &   3.3    &   3.0    &   3.2    &   3.3   &  3.0    &   3.2    &   3.3    &   3.3   \\
$\Omega_{\neu1} h^2$ ~~~~~~~~~~~~~~~1      &   0.1    &   0.0    &   0.3    &   0.0   &  0.1    &   0.0    &   0.0    &   0.0   \\
\ssi\ ~~~~~~~~~~~~~~~~~~\,1                     &   0.0    &   0.1    &   0.0    &   0.1   &  0.5    &   0.0    &   0.0    &   0.0   \\
$A/H\to\tau^+\tau^-$ ~~~~\,1      &   0.0    &   0.0    &   0.0    &   0.0   &  0.0    &   0.0    &   0.0    &   0.0   \\
Nuisance  ~~~~~~~~~~~\,3                 &   0.0    &   0.1    &   0.0    &   0.1   &  0.8    &   0.1    &   0.0    &   0.1   \\
\gmt\ ~~~~~~~~~~~~\,1                     &   0.0    &   0.7    &   0.0    &   0.0   &  0.6    &   9.3    &  10.6    &   8.4   \\
$Z$ pole ~~~~~~~~~~~~~\,13                 &  16.3    &  17.0    &  17.1    &  16.8   & 16.4    &  16.8    &  16.5    &  16.7   \\
\hline                                                                                                             
\hline                                                                                                              
Parameters~~                   &  10 + 3  &  10 + 3  &  10 + 3  &  10 + 3 & 10 + 3  &   4 + 3  &   5 + 3  &   6 + 3 \\
$\chi^2/{\rm d.o.f.}$~~        &  20.5/18 &  22.2/18 &  22.0/18 &  22.3/18& 22.2/18 &  32.8/24 &  31.1/23 &  30.3/22\\
$\chi^2$ probability~~                    &  0.31    &  0.22    &  0.23    &  0.22   & 0.22    &  0.11    &  0.12    &  0.11   \\
\hline                                                                            
$\chi^2({\rm HS})$  ~~~~~~~~~~~77&  62.8    &  62.6    &  62.8    &  62.8   & 62.8    &  -       & -        & -       \\
\hline
\end{tabular}}
\caption{\it Table of the total $\chi^2$ breakdowns at the pMSSM10 {best-fit and
low-$\mst{1}$, low-$\msq$ and low-$\mgl$ points, and in the CMSSM, NUHM1, and NUHM2
(updated from \cite{mc9, mc10}, using in particular the current value of $\Mh$~\cite{Aad:2015zhl}). The \lhcstop, \lhcewk\
and \lhccol\ constraints were applied only to the pMSSM10, whereas a generic
jets + \ETslash\ constraint was applied to the CMSSM, NUHM1 and 
NUHM2~\protect\cite{mc9, mc10}. For each set of constraints, the (rounded) $\chi^2$ contribution and the number
of non-zero contributions is provided. The nuisance parameters are $\mt, \alpha_s(\MZ)$ and $\MZ$.
The bottom rows show the number of parameters (including the nuisance
parameters) and the total $\chi^2/{\rm d.o.f.}$ omitting Higgs signal
rates: the latter have been calculated only for the pMSSM10 points, and are given separately in the last line.
We also show an estimate of the corresponding $\chi^2$ probability, which is calculated as the $\chi^2$ probability
neglecting correlations between the observables.}
} 
\label{tab:breakdown}
\end{center}
\renewcommand{\arraystretch}{1.0}
\end{table*}

\subsection{Sparticle Masses}

\begin{figure*}[htb!]
\resizebox{8cm}{!}{\includegraphics{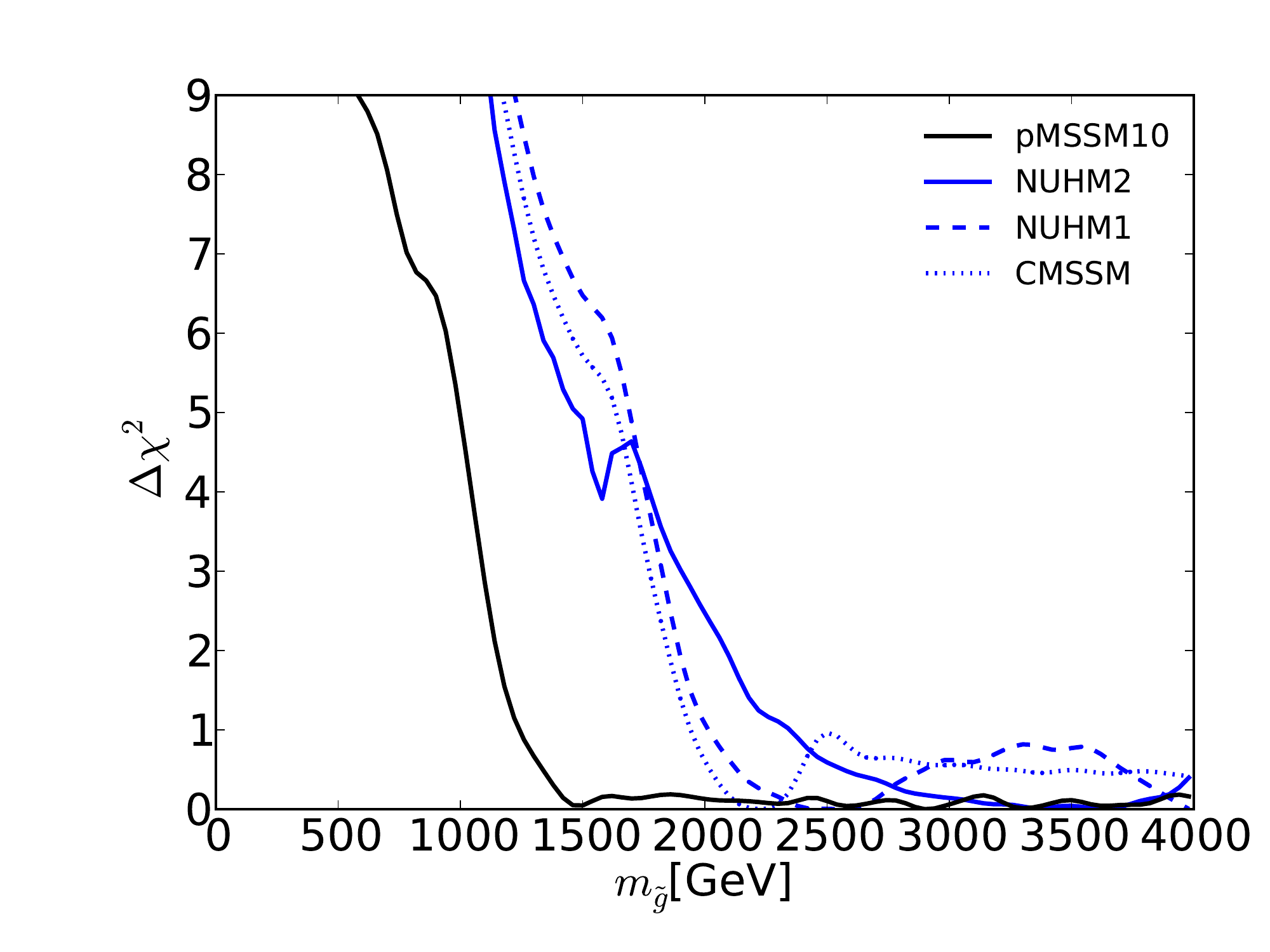}}
\resizebox{8cm}{!}{\includegraphics{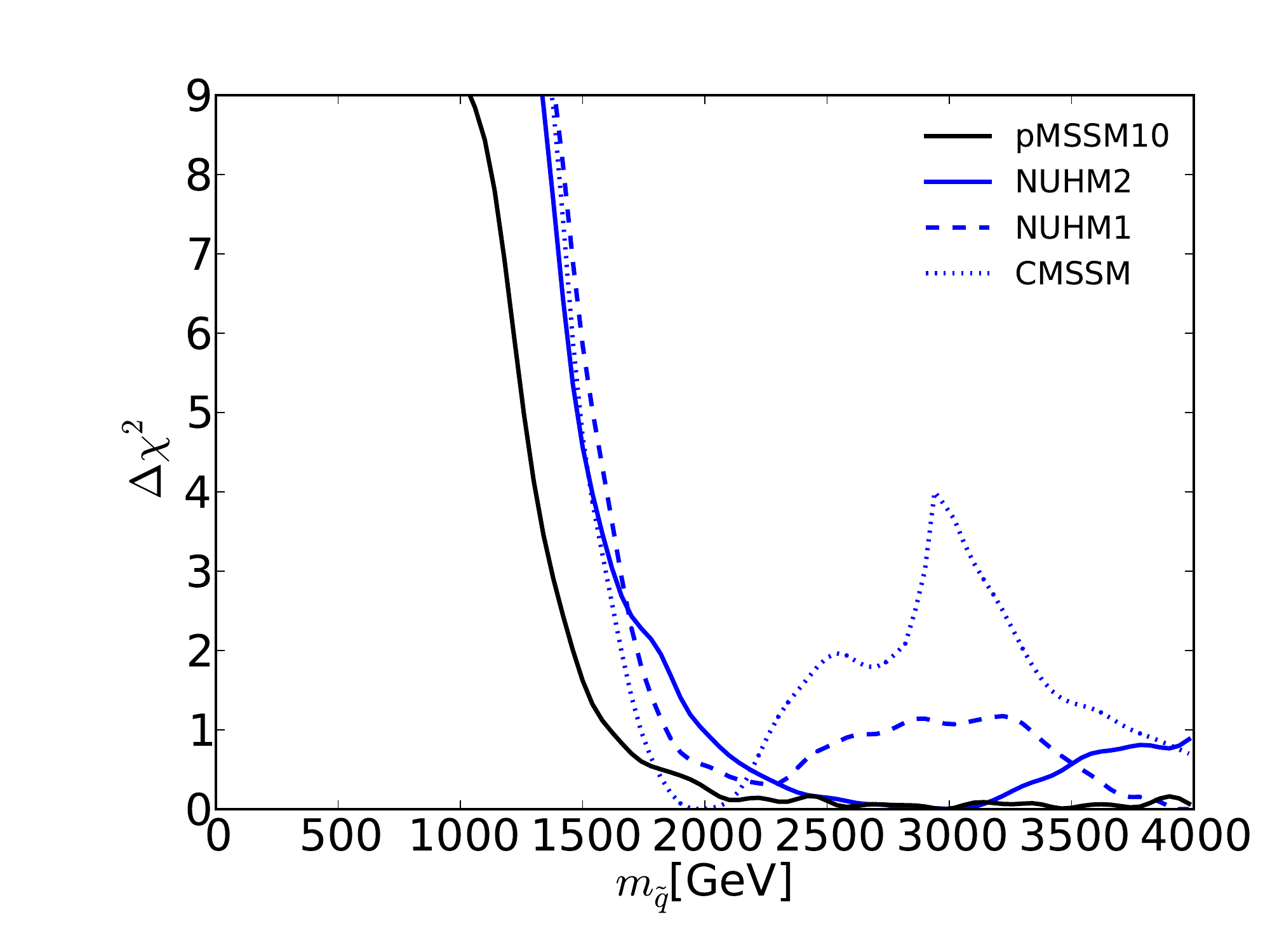}}\\
\hspace {0.5cm}
\resizebox{8cm}{!}{\includegraphics{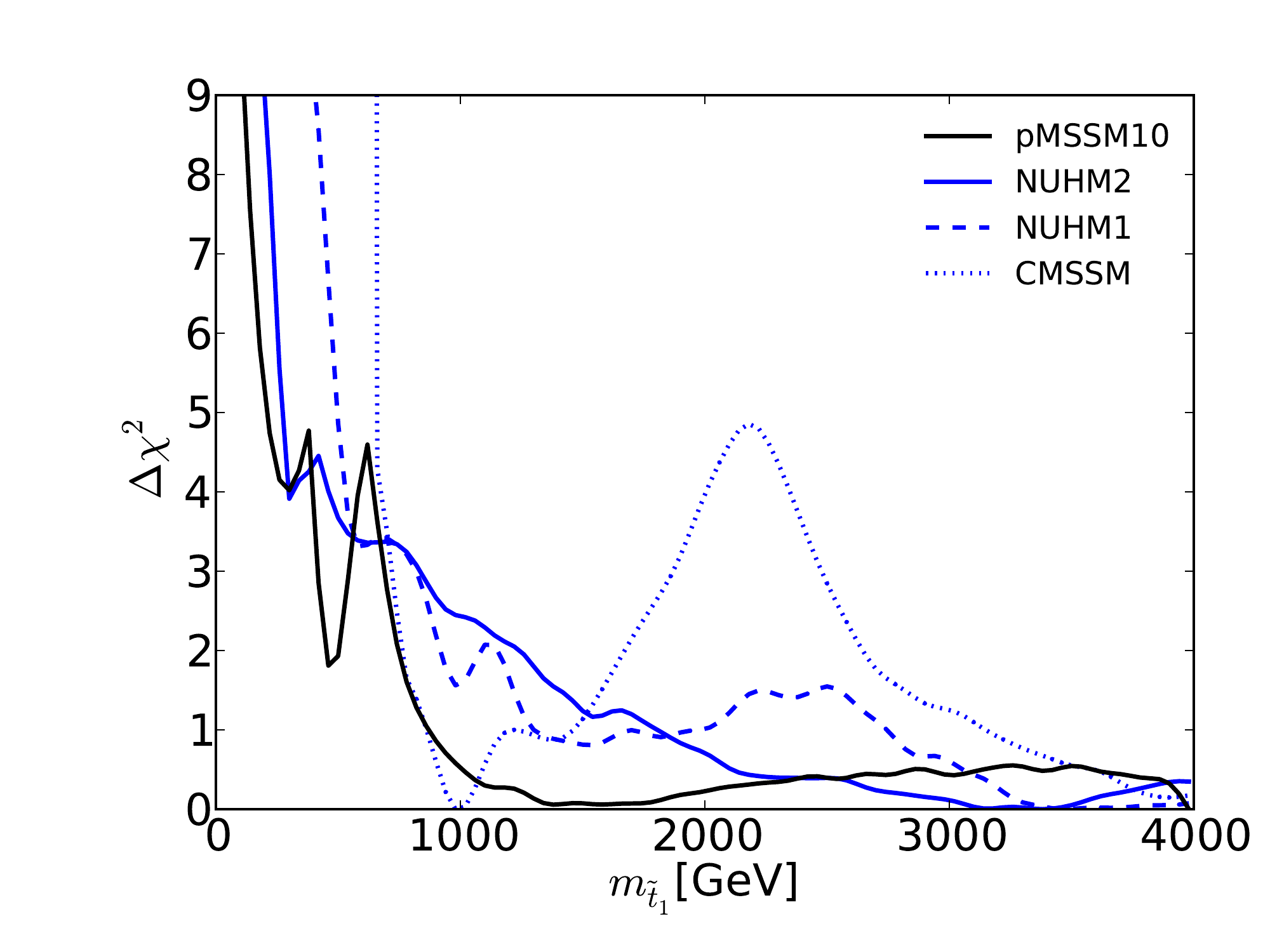}}
\resizebox{8cm}{!}{\includegraphics{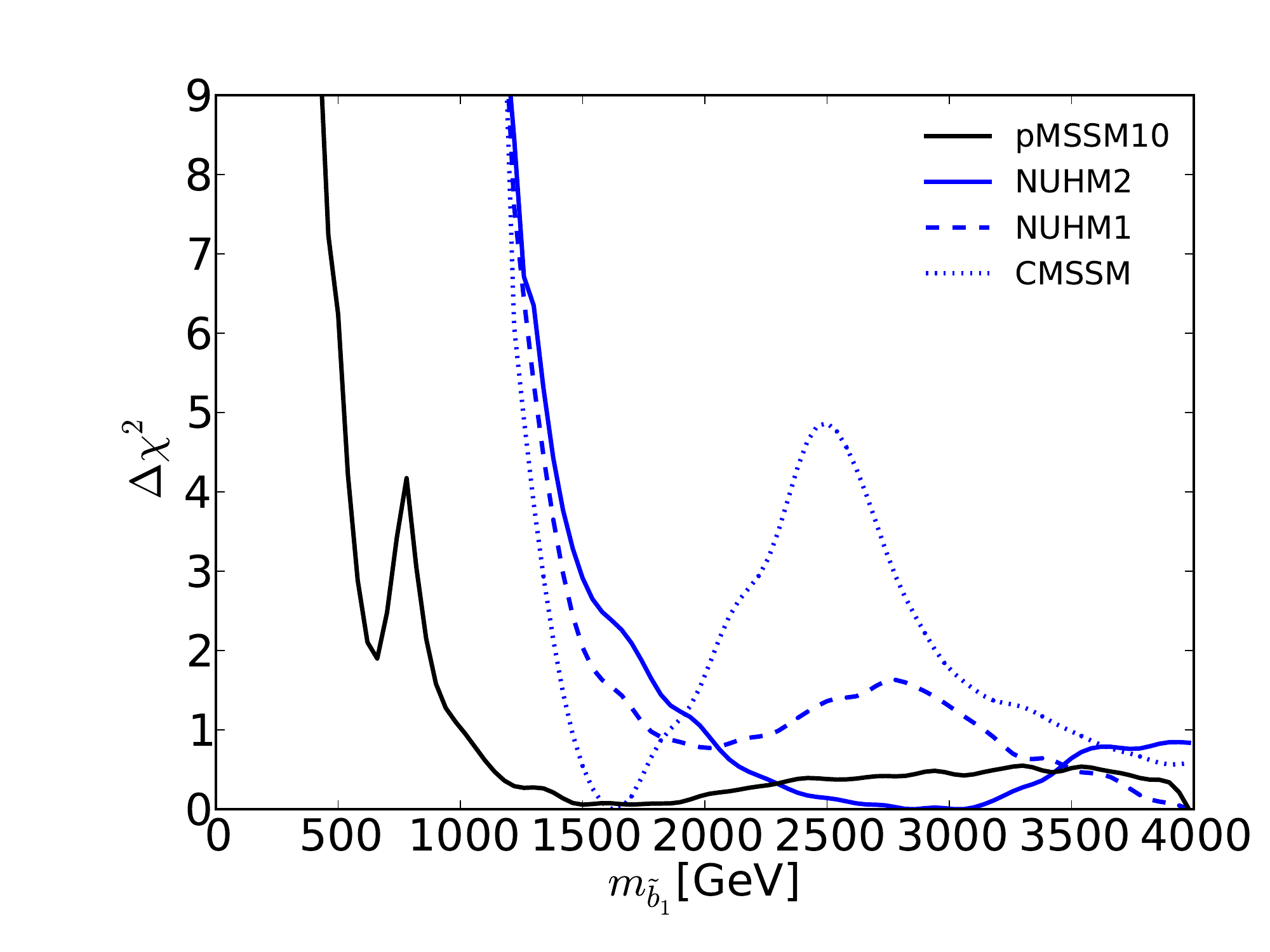}}\\
\hspace {0.5cm}
\resizebox{8cm}{!}{\includegraphics{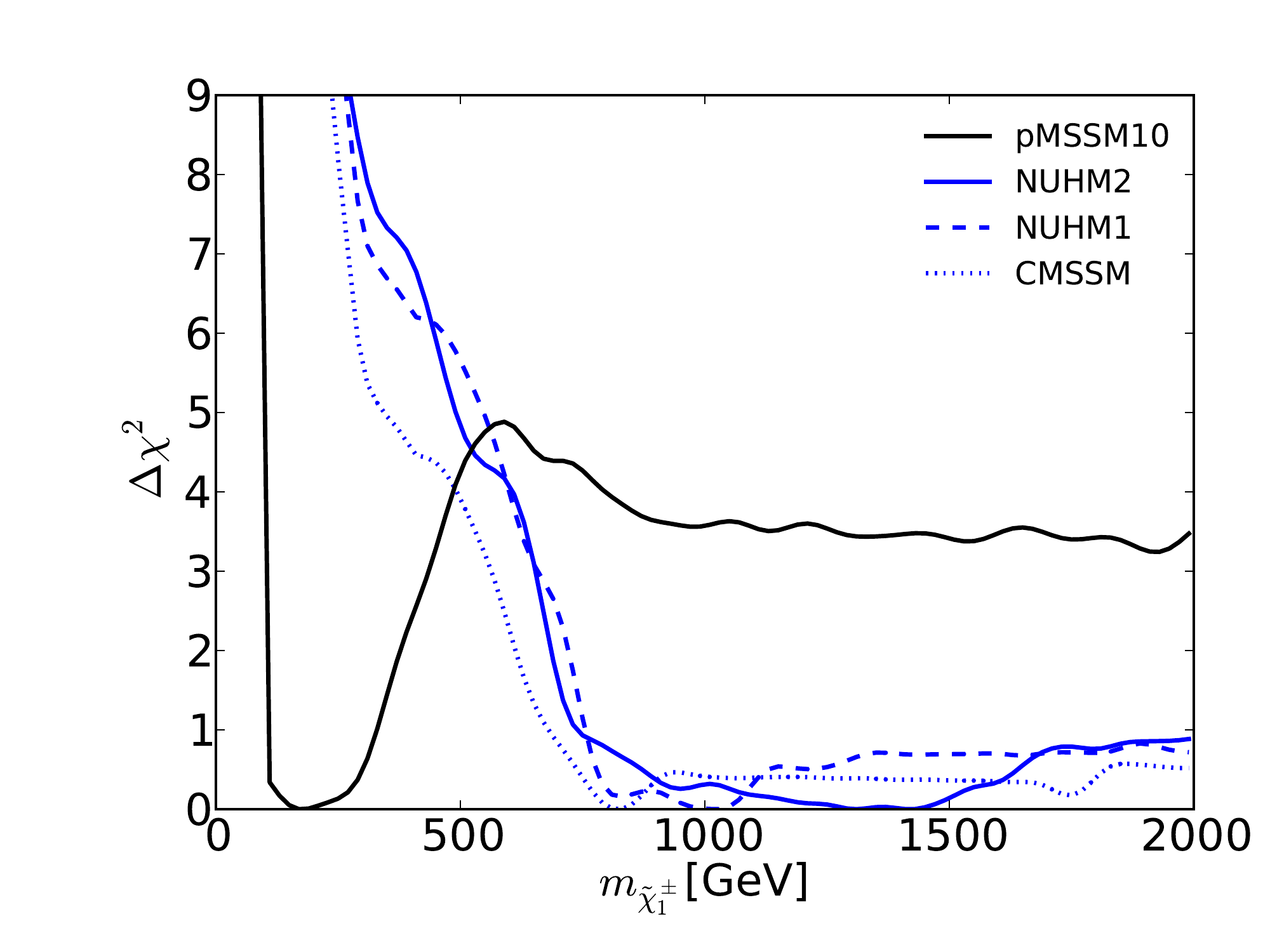}}
\resizebox{8cm}{!}{\includegraphics{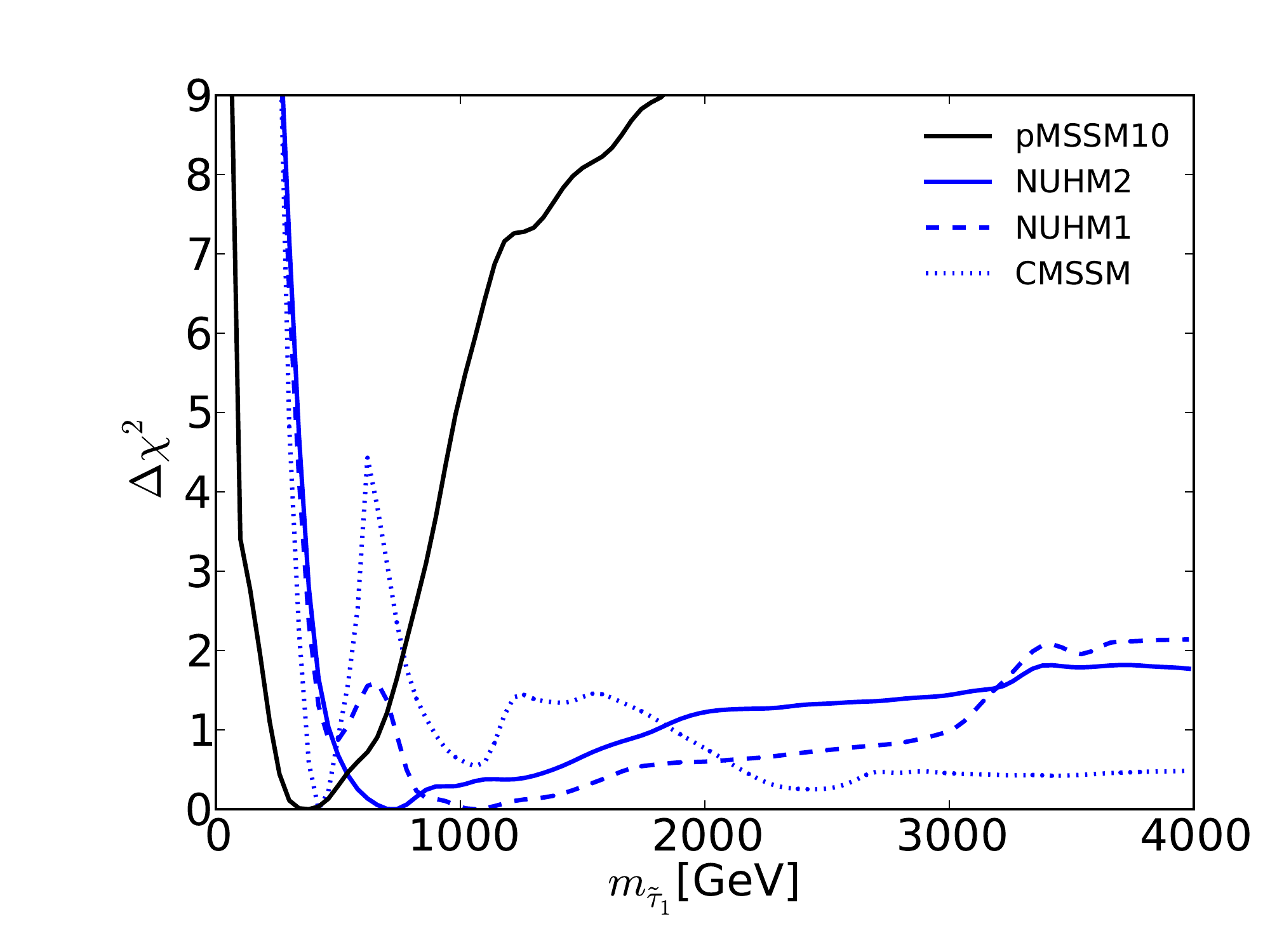}}\\
\vspace{-1cm}
\caption{\it The one-dimensional profile likelihood functions
for \mgl, \msq, \mstop1, \msbot1,
\mcha1 and \mstau1. 
In each panel the solid black line is for the pMSSM10, the solid blue line for the NUHM2,
the dashed blue line for the NUHM1 and the dotted blue line for the
CMSSM.
}
\label{fig:onedimensional}
\end{figure*}

\reffi{fig:onedimensional} displays (from top left to bottom right) the
one-dimensional profile likelihood functions for the masses of the
gluino, the first- and second-generation squarks, the lighter stop and
sbottom squarks, the lighter chargino and the lighter stau. In each
panel the solid black line is for the pMSSM10, the solid blue line for
the NUHM2, the dashed blue line for the NUHM1 and the dotted blue line
for the CMSSM (the latter three lines are {updated from \citere{mc10} to
include new constraints such as the LHC combined value of $\Mh$~\cite{Aad:2015zhl}}). In
the case of $\mgl$, we see that significantly lower masses are allowed
in the pMSSM10 than in the other models: $> 1250 \gev$ at the 68\%~CL
and  $\sim 1000 \gev$ at the 95\%~CL. We also see that there is a
similar, though smaller, reduction in the lower limit on $\msq$, to
$\sim 1500\gev$ at the 68\% CL and $\sim 1300\gev$ at the 95\% CL. 
The picture is more complicated for $\mstop1$, where we see structures in
the one-dimensional likelihood function for $\mstop1 < 1000 \gev$ that
reflect the low-mass islands in the corresponding panel of
\reffi{fig:planes} that are allowed at the 95\% CL. In the bottom row of
\reffi{fig:onedimensional}, the one-dimensional profile likelihood
functions for $\mcha1$ and $\mstau1$ in the pMSSM have minima
at the lower mass limits $\sim 100 \gev$ established at LEP, and there
is an upper limit $\mstau1 \lesssim 1000 \gev$ at the 95\% CL.
These effects are due to the \gmt\ constraint
and the choice of generation-independent slepton masses in the pMSSM10.
On the other hand, the light chargino (which is nearly degenerate in mass with the
second lightest neutralino), has an upper mass
limit below $500 \gev$ at the 95\%~CL. This would allow neutralino and
chargino pair production at an 1000~GeV $e^+e^-$~collider, as we discuss later.


\subsection{Benchmark pMSSM10 Models}

{In view of the variety of pMSSM10 parameters that are allowed at the 68\% CL,
we consider in this subsection various specific benchmark models that illustrate
the range of possibilities. Specifically, looking at the middle panels of Fig.~\ref{fig:planes}, we see that
a very low stop mass in the compressed-stop region is possible,
and the top panels of Fig.~\ref{fig:planes} show the possibilities for a gluino
{\it or} squark mass that is lower than at the best-fit point. Also, we see in the upper left panel of Fig.~\ref{fig:up-down}
that SUSY may well appear with {\it both} the squark and gluino masses having lower masses than at the best-fit point.
We investigate these possibilities with the benchmark points discussed below}, {whose SLHA files~\cite{SLHA}
can be downloaded from the MasterCode website~\cite{mcweb}.}

\boldmath
\subsubsection{{Low-$\mst1$ point}}
\unboldmath

{We display in the upper left panel of Fig.~\ref{fig:benchmarkspectra} the spectrum at the point that
minimizes $\chi^2$ locally within the low-$\mst{1}$ (and low-$\msb{1}$) 68\% CL region visible in
the middle planes of Fig.~\ref{fig:planes}. Like the pMSSM10 best-fit point shown in
Fig.~\ref{fig:bestfitspectrum}, this point also exhibits near-degeneracies 
between $\neu1, \neu2$ and $\cha1$, between the
sleptons, between $\neu3, \neu4$ and $\cha2$ (reflected also in the fact that $M_1 \simeq M_2$,
as seen in the second column of \refta{tab:bf-point}), and between the ${\tilde q_L}$
and ${\tilde q_R}$. However, all the stops and sbottoms are light
at this point. As in Fig.~\ref{fig:bestfitspectrum}, the dominant decay modes are illustrated
in fifty shades of grey~\cite{James}. The second column of \refta{tab:breakdown} lists the contributions to the total $\chi^2$
function of different (groups of) constraints at this low-$\mst1$ pMSSM10 point. Comparing with
the corresponding breakdown for the best-fit point shown in the first column of \refta{tab:breakdown},
we see larger contributions from the {LHC8 constraint
(principally from \lhccol)} and from \gmt,
which are largely responsible for the increase in the total $\chi^2$ to 22.2 (omitting the
{\tt HiggsSignals} contributions) {and the corresponding decrease
in the $\chi^2$ probability to {0.22}}. However, we emphasize that this point provides a perfectly acceptable fit to
all the constraints.}

\begin{figure*}[htb!]
\centering
\resizebox{7.5cm}{!}{\includegraphics{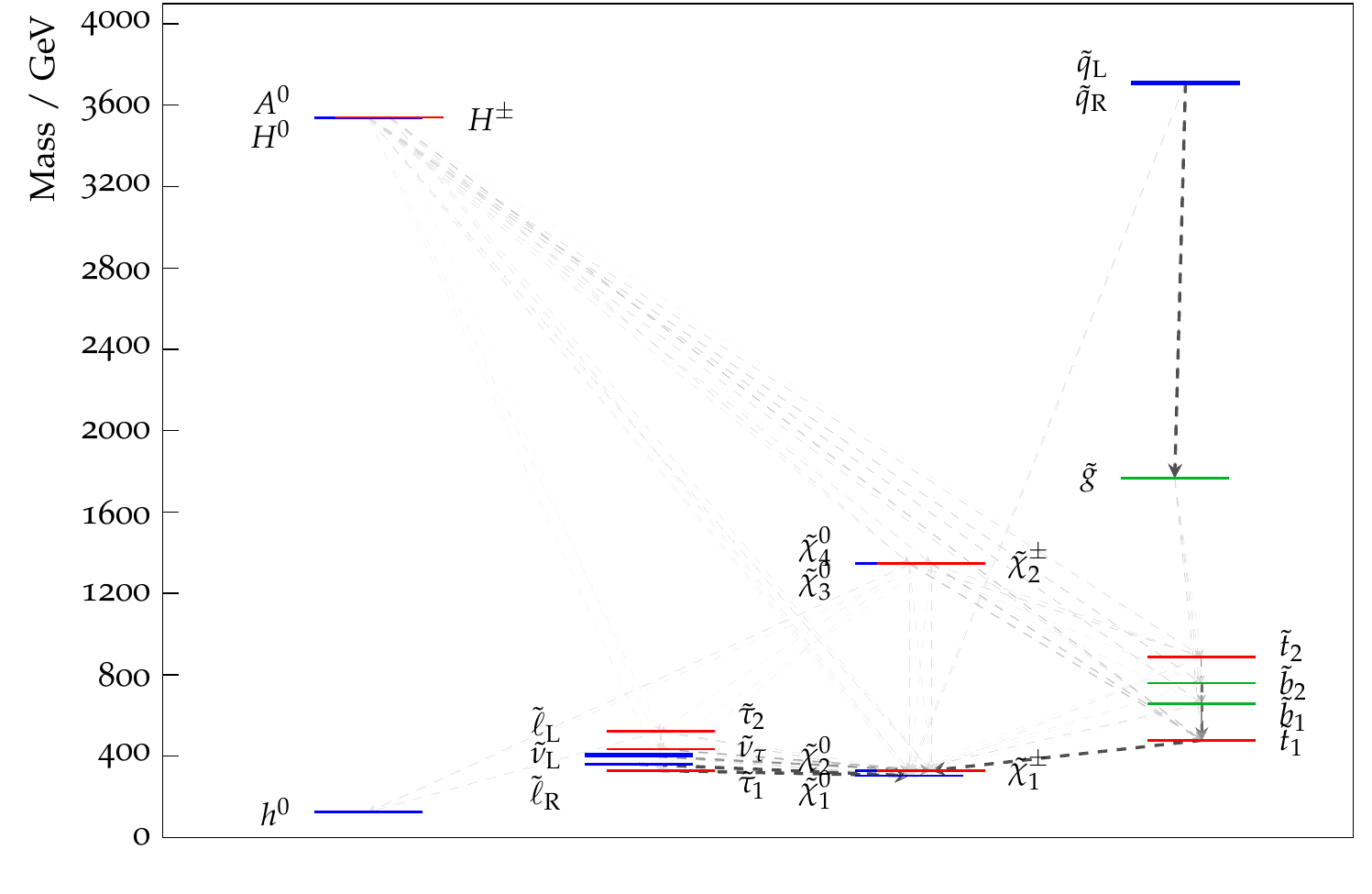}}
\resizebox{7.5cm}{!}{\includegraphics{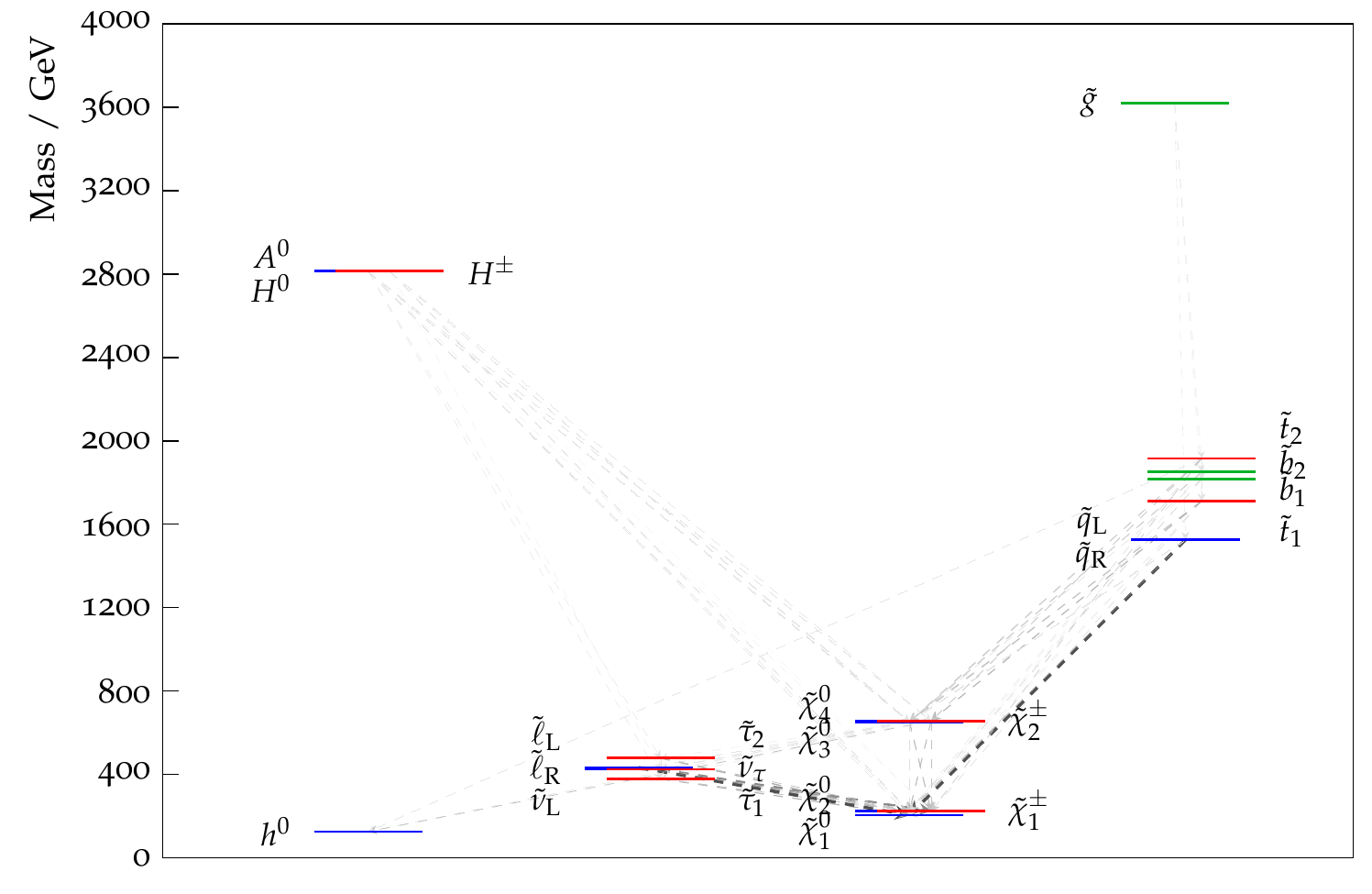}} \\
\resizebox{7.5cm}{!}{\includegraphics{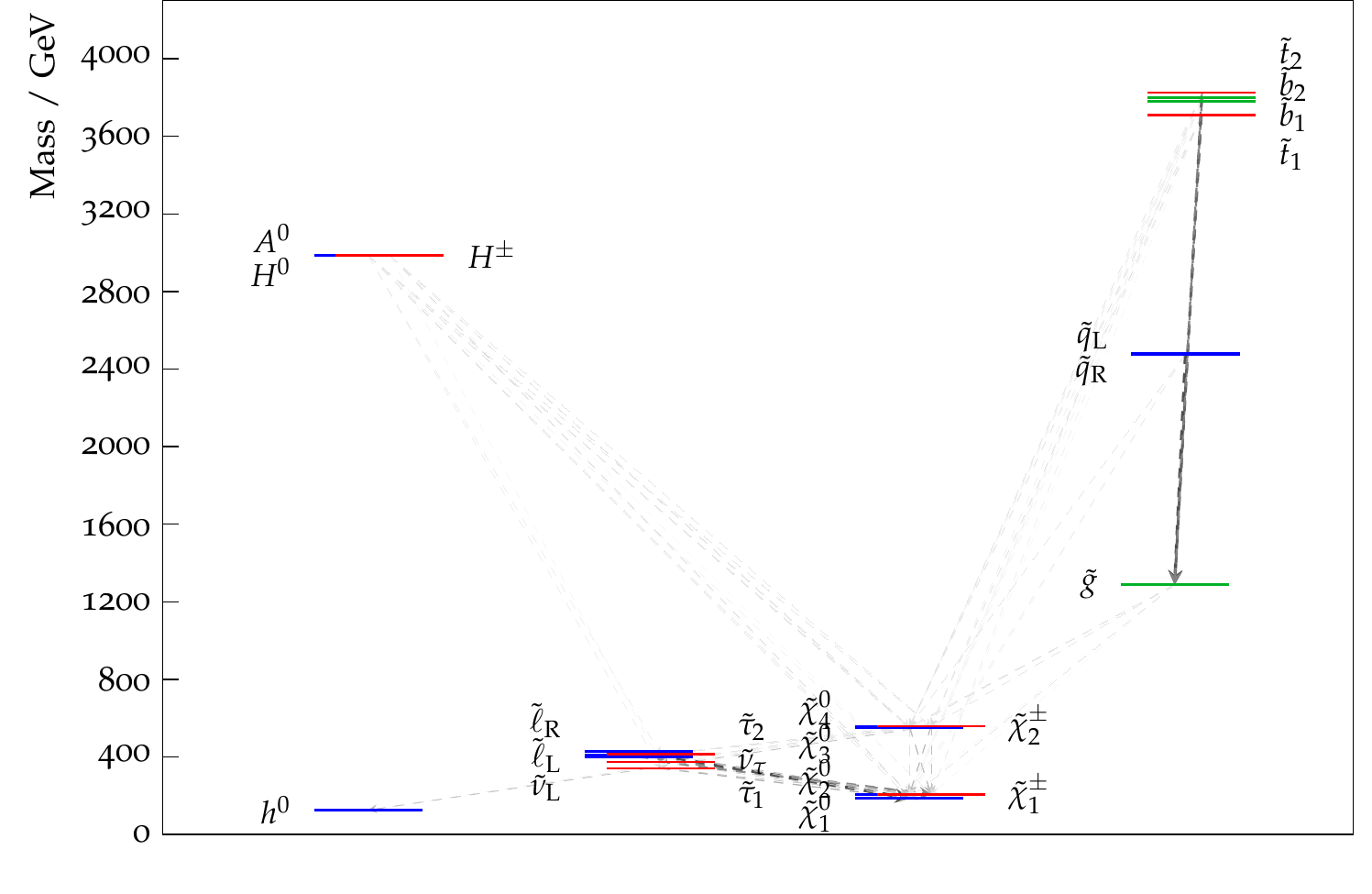}}
\resizebox{7.5cm}{!}{\includegraphics{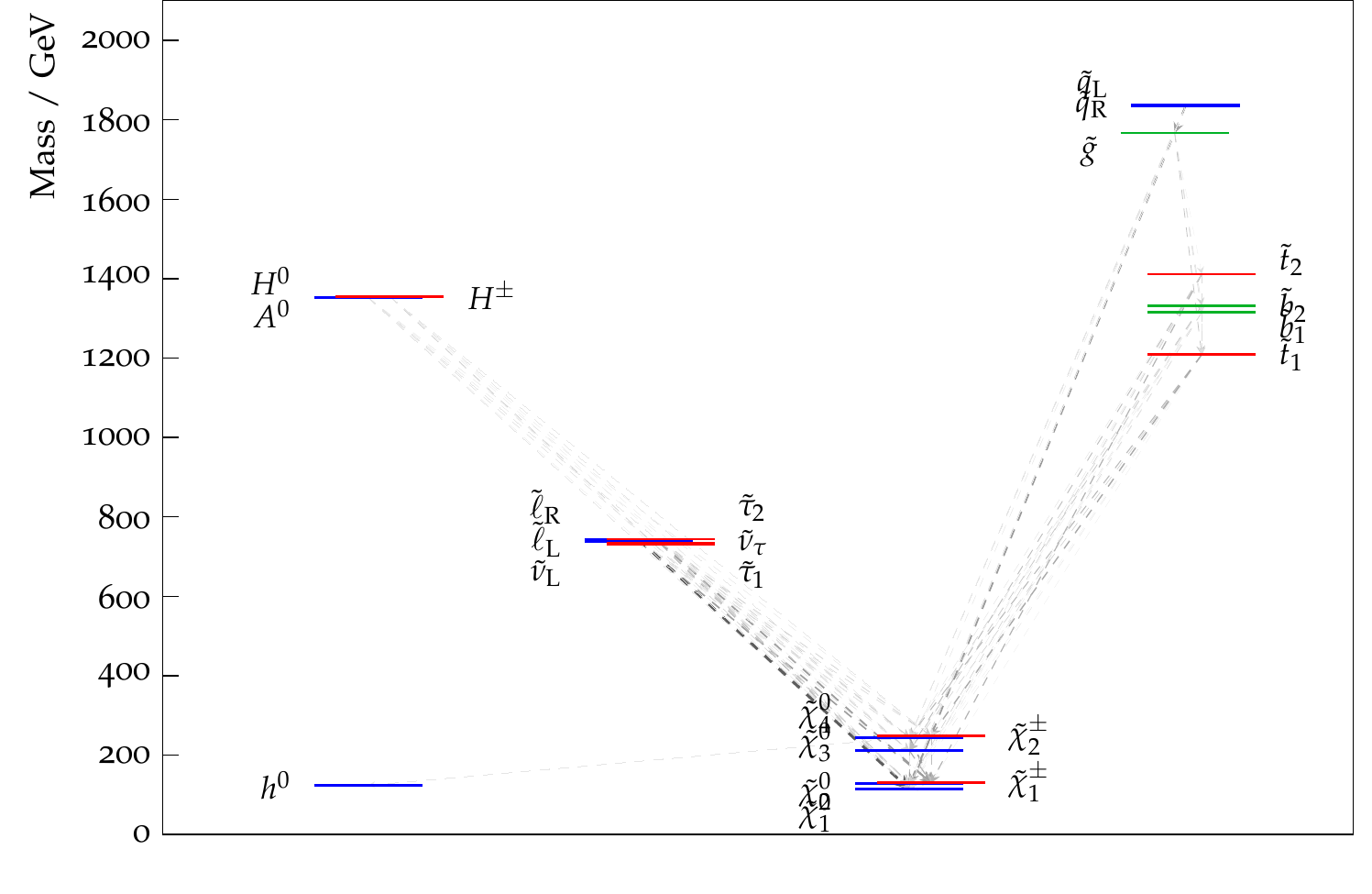}} \\
\caption{\it {The particle spectra and dominant decay branching ratios at the benchmark
points discussed in the text. Upper left panel: the low-$\mst1$
pMSSM10 point, where the stops and bottoms are relatively light. Upper right panel: similarly for the low-$\msq$
benchmark point, where all the squarks are relatively light.
Lower left panel: similarly for the  low-$\mgl$ benchmark point.
Lower right panel: similarly for the point where all squarks and the gluino masses are $< 2 \tev$.
Note in each case the near-degeneracies between $\neu1, \neu2$ and $\cha1$, between the
sleptons, between $\neu3, \neu4$ and $\cha2$, between the ${\tilde q_L}$
and ${\tilde q_R}$, and between the heavy Higgs bosons.}
} 
\label{fig:benchmarkspectra}
\end{figure*}

\boldmath
\subsubsection{Low-{$\msq$} point}
\unboldmath

{We consider next a benchmark point with relatively
low masses for the first- and second-generation squarks.
As can be seen in the top right panel of \reffi{fig:planes}, the lowest value of $\msq$ that is allowed
at the 68\% CL is $\simeq 1500 \gev$, and we have chosen
as benchmark a point that also has $\mneu1 \simeq 200 \gev$,
whose spectrum is shown in the upper right panel of \reffi{fig:benchmarkspectra}.
We see there that the near-degeneracies between $\neu1, \neu2$ and $\cha1$, between the
sleptons, between $\neu3, \neu4$ and $\cha2$, and between the heavy Higgs bosons
are very similar to those at the best-fit and low-$\mst1$ points. By choice, the masses of the
first- and second-generation squarks are much lighter than at either of these points,
and the third-generation squarks have masses intermediate between the
best-fit and low-$\mst1$ points. As seen in \refta{tab:breakdown}, the largest part of the
increase in $\chi^2$ to 22.0, compared to the best-fit point, and the corresponding
decrease in the $\chi^2$ probability to {0.22}, is again due to the LHC8 constraint.}

\boldmath
\subsubsection{Low-{$\mgl$} point}
\unboldmath

{We consider next a benchmark point with a relatively
low gluino mass.
As can be seen in the top left panel of \reffi{fig:planes}, our global fit requires
$\mgl \gtrsim 1250 \gev$ at the 68\% CL. We have chosen
as benchmark a point that has this value of $\mgl$ and also
$\mneu1 \simeq 200 \gev$,
whose spectrum is shown in the right panel of \reffi{fig:benchmarkspectra}.
We see again the near-degeneracies within groups of MSSM particles,
as for the benchmark points considered previously. We see a clear hierarchy of masses between
the groups of strongly-interacting sparticles, with the third-generation sparticles being
much heavier than those of the first and second generation, which are in turn much
heavier than the gluino. Again as seen in \refta{tab:breakdown}, the largest part of the
increase in $\chi^2 \to 22.3$ compared to the best-fit point is again due to the \lhccol\ constraint,
with increases also from \lhcewk, and $\MW$.}
{The total $\chi^2$ probability of 21.7\% is comparable to those of the low-$\mst1$ and -$\msq$ points.}


\boldmath
\subsubsection{Point with squark and gluino masses below $2 \tev$}
\unboldmath

{Finally, we display in the lower right panel of Fig.~\ref{fig:benchmarkspectra} the spectrum at a point
from near the turning-point in Fig.~\ref{fig:up-down}, which can be regarded as a `compromise'
between the two previous benchmarks where the gluino and all the squarks (including those
in the third generation) have masses $< 2\tev$, as do the heavy $A/H$ Higgs bosons.
Like the previous pMSSM10 benchmark points, this point also exhibits near-degeneracies 
between $\neu1, \neu2$ and $\cha1$, between the
sleptons, between $\neu3, \neu4$ and $\cha2$ , and between the ${\tilde q_L}$
and ${\tilde q_R}$. In addition, the sbottom squarks are also nearly degenerate,
whereas the stops exhibit a greater mass splitting, due to the $\mt$-dependence 
in the off-diagonal stop mass matrix elements.
The contributions to the total $\chi^2$
function of different (groups of) constraints at this low-mass pMSSM10 point
are shown in the fifth column of \refta{tab:breakdown}. Comparing with
the corresponding breakdown for the best-fit point, we see larger contributions
from \lhccol, \ssi\ and \gmt. All these contributions to $\chi^2$ are $< 1$, but they
do suggest possibilities for SUSY discovery in jets + \ETslash searches early
during LHC Run~2, and in upcoming direct dark matter detection experiments.
The total $\chi^2 = 22.2$ at this point, and the corresponding $\chi^2$ probability is 22.4\%.}

\boldmath
\subsection{The Anomalous Magnetic Dipole Moment of the Muon \gmt}
\unboldmath

It is well-known that there is a discrepancy of $\sim3.5\sigma$ between
the measured value of \gmt~\cite{newBNL} and the value predicted in the
SM~\cite{g-2,Jegerlehner}. Sizeable contributions to \gmt\ from
SUSY can occur when smuons, charginos and the 
lightest neutralino have masses of~\order{100 \gev}. 
It is known from previous analyses of the CMSSM, NUHM1 and
NUHM2~\cite{mc9,mc10} that in these models there is tension between SUSY
interpretations of the discrepancy in the anomalous magnetic dipole
moment of the muon \gmt\ (which favour lower electroweak sparticle masses)
and LHC constraints from direct searches for
sparticles and the measured value of the lightest Higgs boson
(which favour higher coloured sparticle masses).
This tension arises from the universality relations imposed in these
models at the GUT scale between the soft SUSY-breaking contributions to
the masses of the strongly- and electroweakly-interacting sparticles. 
In the pMSSM10 there are no such assumptions, and thus one might hope to
resolve this tension. 

This point is apparent in \reffi{fig:gmt}, where we display in the left
panel the 
contributions $\Delta \chi^2$ from \gmt\ to the global $\chi^2$
functions of our fits to the  CMSSM (blue dotted line), the NUHM1 (blue
dashed line), the NUHM2 (blue solid line) and the pMSSM10 (black solid
line), as well as the experimental likelihood function that we assume
(solid red line). We see that the pMSSM10 is able to fit \gmt\ perfectly
with $\Delta \chi^2 \simeq 0$ at the best-fit point, whereas the other
``universal" models exhibit contributions $\Delta \chi^2 \gsim 9$ from
\gmt.

\begin{figure*}[htb!]
\begin{center}
\resizebox{8cm}{!}{\includegraphics{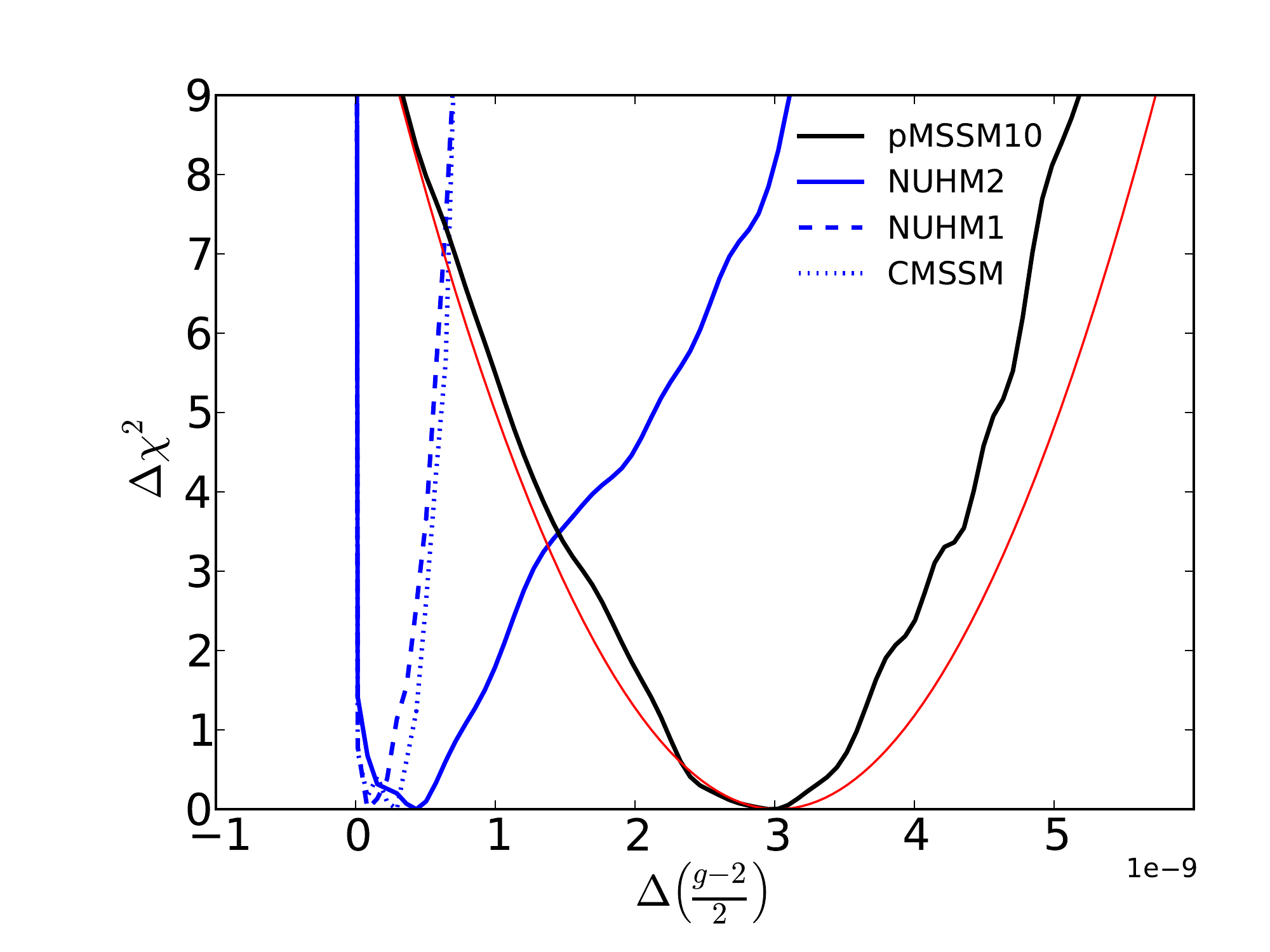}}
\hspace{-0.5cm}
\resizebox{8cm}{!}{\includegraphics{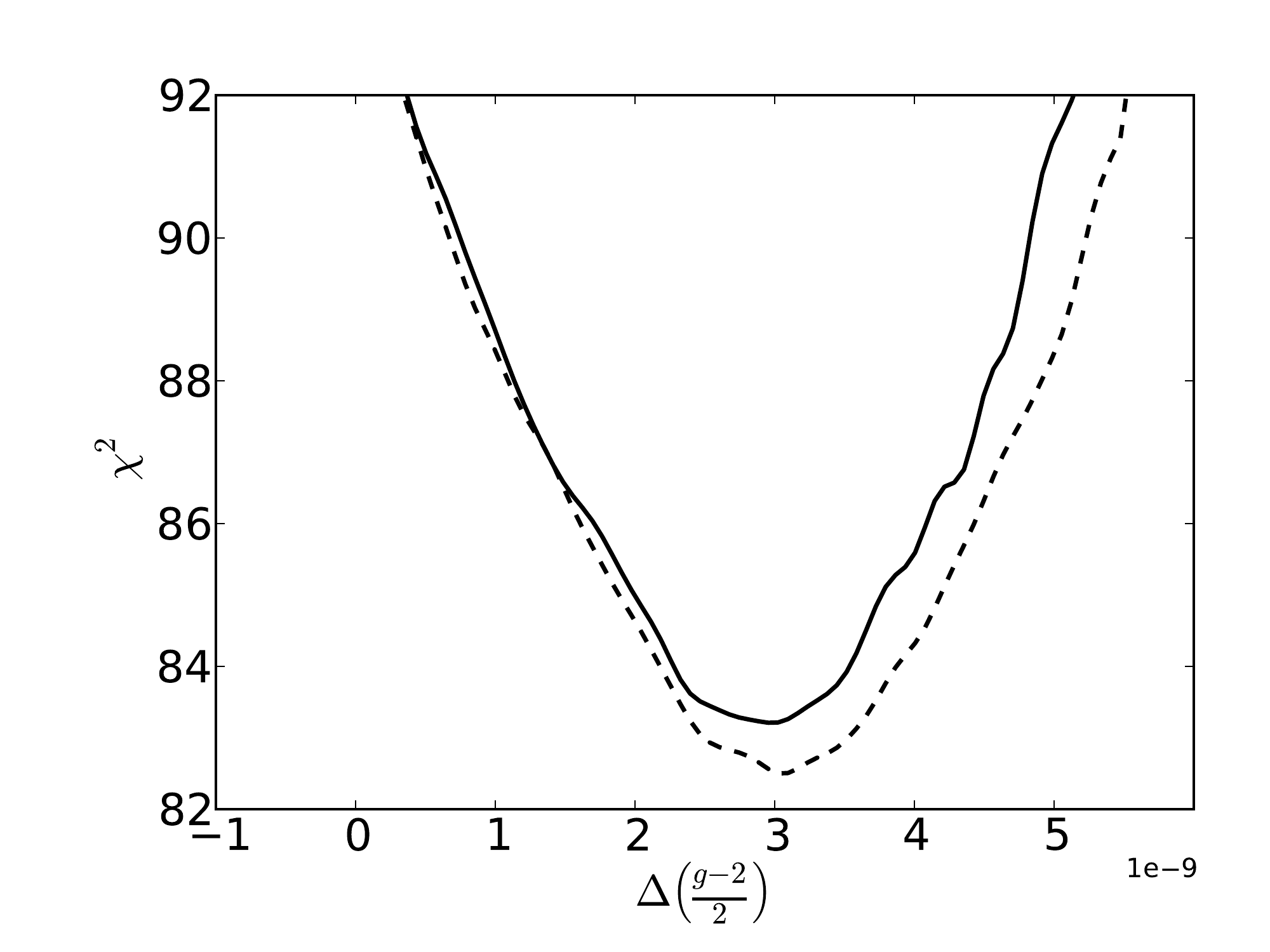}}
\end{center}
\vspace{-1cm}
\caption{\it Profile likelihoods for the SUSY contribution to \gmt.
The left panel shows the $\Delta \chi^2$ contributions from \gmt\ to the global likelihood functions
of our fits to the  CMSSM (blue dotted line), the NUHM1 (blue dashed line), the NUHM2 (blue solid line) and
the pMSSM10 (black solid line), as well as the experimental likelihood
function that we assume (solid red line). 
The right panel displays the global $\chi^2$ function calculated without (dashed line) and with (solid line) the contribution
of the electroweakly-interacting sparticle searches implemented via \lhcewk.
}
\label{fig:gmt}
\end{figure*}

\begin{figure*}[htb!]
\vspace{1cm}
\resizebox{7.5cm}{!}{\includegraphics{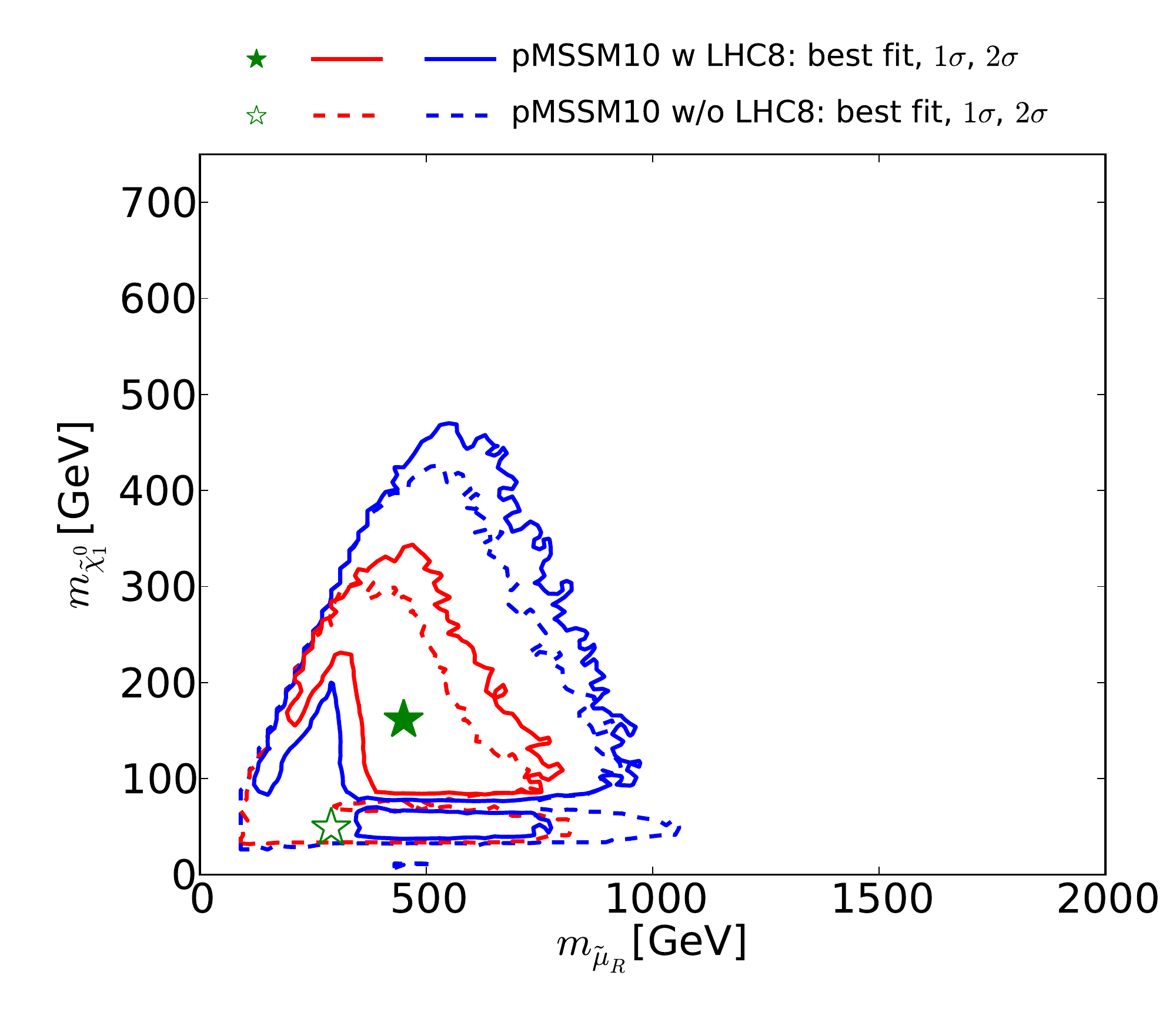}}
\resizebox{8.5cm}{!}{\includegraphics{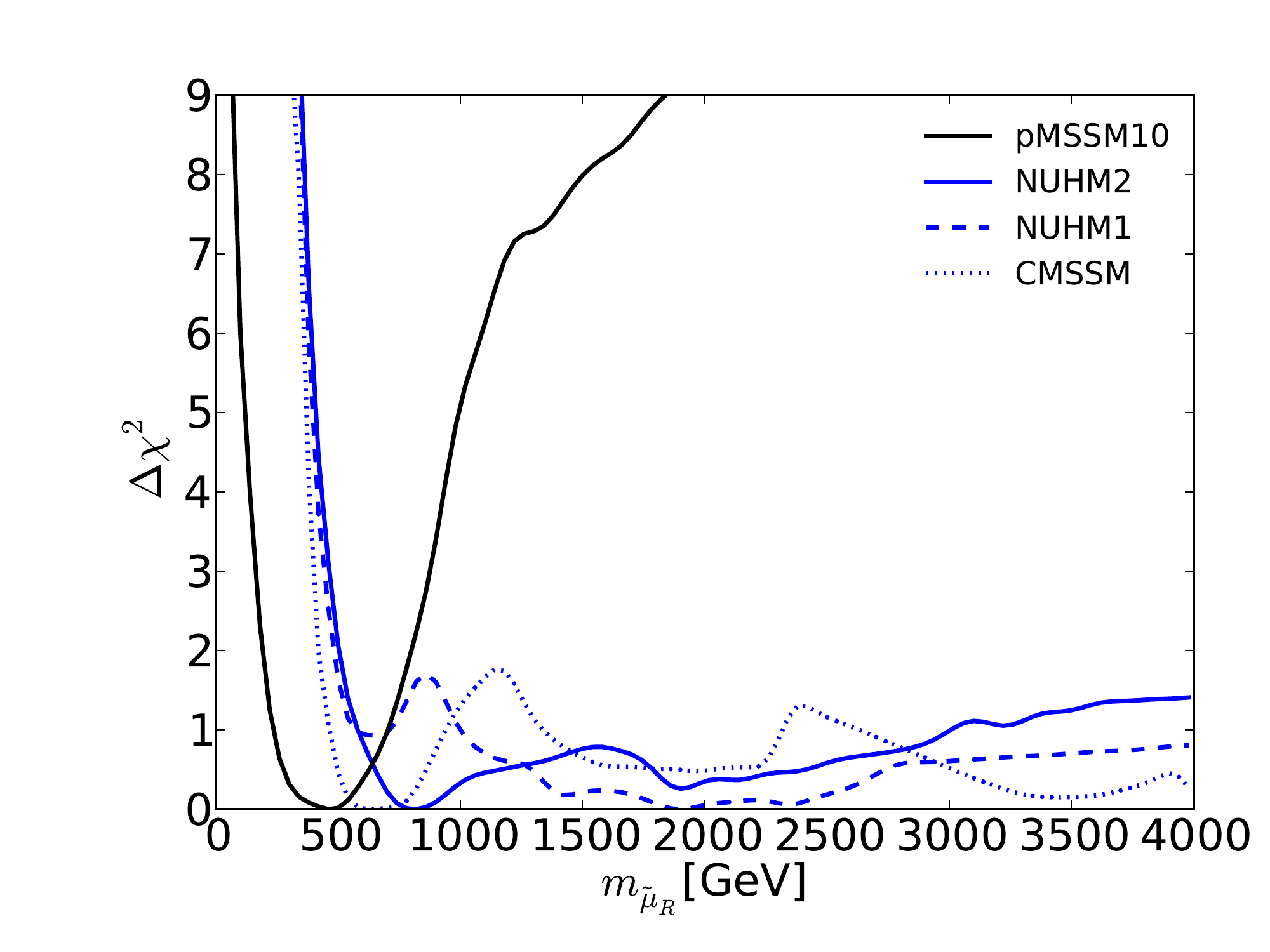}}\\
\vspace{-1cm}
\caption{\it Left panel: The two-dimensional profile likelihood function
for $\msmu{R}$ versus the lightest neutralino mass $\mneu1$.
The solid (dashed) red/blue contours denote the 
$\Delta \chi^2 = 2.30/5.99$ level contours for the case where we
do (not) apply the LHC8 constraints, respectively, and
the green filled and empty stars indicate the corresponding best-fit points.
Right panel: The one-dimensional profile likelihood function
for $\msmu{R}$ in the pMSSM10 (solid black line), the NUHM2, (solid blue line),
the NUHM1 (dashed blue line) and the CMSSM (dotted blue line).
}
\label{fig:smuon}
\end{figure*}

We display in the right panel of \reffi{fig:gmt}
the impact on the global $\chi^2$ as a function of \gmt\ of implementing
the LHC constraints on electroweakly-interacting
sparticles using the \lhcewk\ method described earlier (which, as we have shown, provides a reasonably accurate
as well as computationally economical representation of the LHC8 constraints on electroweakly-interacting sparticles).
The solid line is the global
$\chi^2$ function with the \lhcewk\ constraint included, and the dashed line when they are omitted. The minimum value of
$\chi^2$ increases from 82.6 to 83.3, and the value of \gmt\ at the minimum is essentially unchanged.
We conclude that the impacts of the LHC searches for electroweakly-interacting particles are limited, and the
pMSSM10 resolution of the \gmt\ puzzle survives the LHC electroweak constraints
with flying colours.

The left panel of \reffi{fig:smuon} displays the two-dimensional profile likelihood function
in the $(\msmu{R}, \mneu1)$ plane, with the solid (dashed) red/blue contours denoting the 
$\Delta \chi^2 = 2.30/5.99$ level contours for the case where we
do (not) apply the LHC8 constraints, respectively, and
the green filled and empty stars indicating the corresponding best-fit
points~%
\footnote{We do
not show the corresponding results for the $\smu{L}$, which are
very similar.}%
. Qualitatively, this plane is quite similar to the corresponding $(m_{\tilde \tau_1}, \mneu1)$
plane shown in the bottom right panel of \reffi{fig:planes}, though we note, e.g.,
that the best-fit value of $\msmu{R}$ is $\sim 100 \gev$ larger than the
best-fit value of $m_{\tilde \tau_1}$. This feature is apparent also when one compares
the right panel of \reffi{fig:smuon}, which displays the one-dimensional profile likelihood function
for $\msmu{R}$ with the corresponding plot for $m_{\tilde \tau_1}$ in the bottom right
panel of \reffi{fig:onedimensional}. In both cases, the one-dimensional profile likelihood function
in the pMSSM10 is shown as a solid black line, that in the NUHM2 as a solid blue line,
that in the NUHM1 as a dashed blue line, and that in the CMSSM as a dotted blue line.

\begin{figure*}[htb!]
\centering
\resizebox{10cm}{!}{\includegraphics{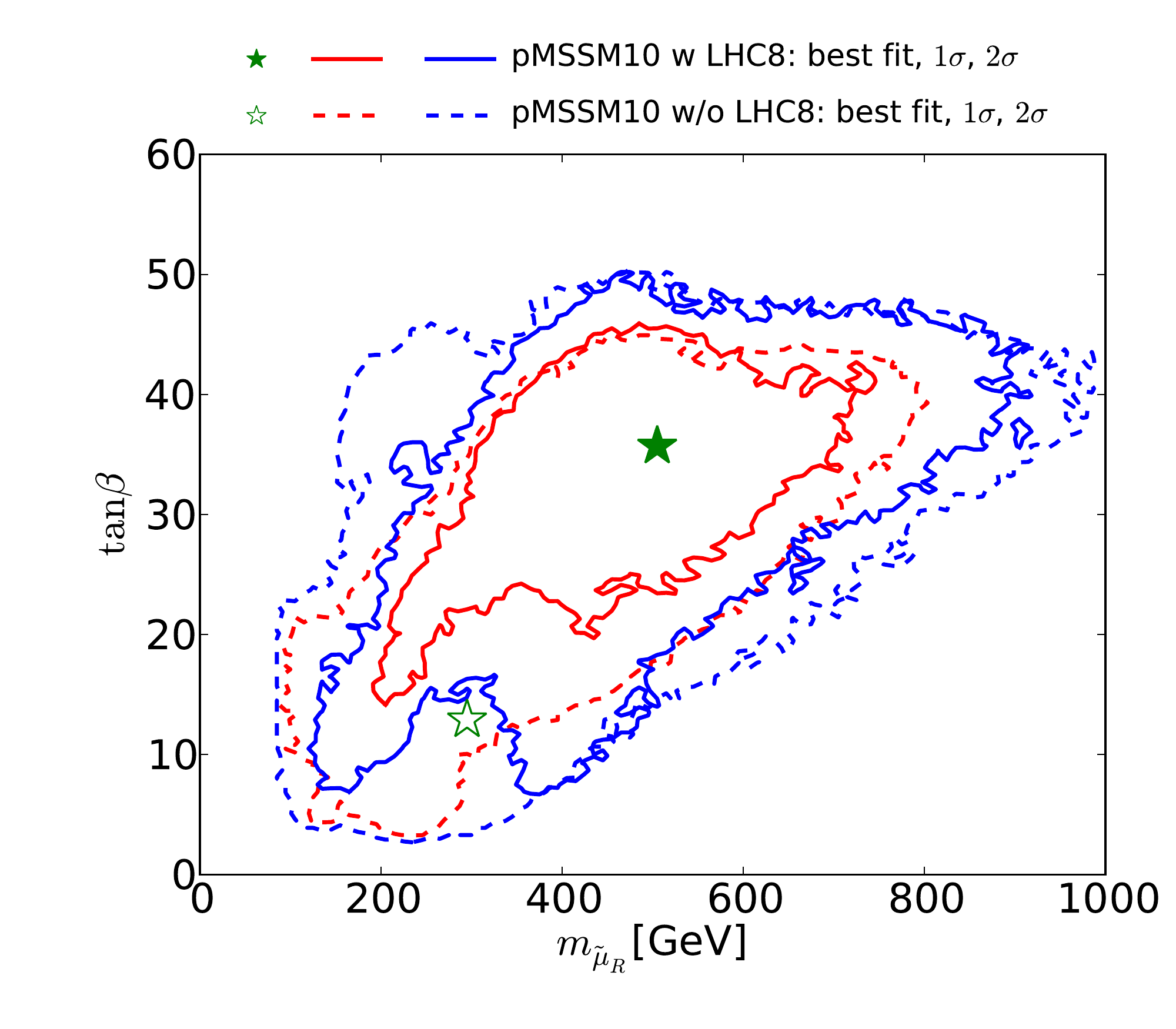}}
\vspace{-1cm}
\caption{\it The 68\% and 95\% CL regions in the
$(m_{\tilde \mu_R}, \tb)$ plane before (dashed lines) and after (solid lines) implementation of the LHC8 and other constraints.
}
\label{fig:smuon-tanb}
\end{figure*}


\boldmath
\subsection{Interplay of the \hbox{\lhcewk, \gmt} and Dark Matter Constraints}
\label{sec:gm2-DM}
\unboldmath

The 68\% and 95\% CL regions in the
$(m_{\tilde \mu_R}, \tb)$ plane before (dashed lines) and after (solid lines) implementation of the LHC8 and other constraints
are displayed in Fig.~\ref{fig:smuon-tanb}. We see that 
{the lowest values of $\tb$ receive a $\chi^2$ penalty, which is
due to a combination of different effects. In particular, 
the \lhcewk\ constraint disfavours lower values of $\msmu{R,L}$ which, in
  combination with \gmt, results in a $\chi^2$ penalty for} $\tb \lsim 10$. 
Because we impose slepton mass universality in the pMSSM10, stau masses are also pushed to
higher values. In this way
the \lhcewk\ constraints eliminate pMSSM10 models with a Bino-like LSP and
small $\ssi$, for which stau coannihilation and $t$-channel slepton
exchanges brought the relic LSP density into the allowed range. 
The remaining models with $\tb \lsim 30$ then
fall foul of the LUX upper limit \cite{lux} on $\ssi$, because the
LSP has a substantial Higgsino component, which enhances $\ssi$.
The overall combined effect of the \lhcewk, \gmt\ and dark matter
constraints {is to prefer values of $\tb$ between about 15
and 45 at the 68\% CL, though $\tb$ values below 10 are
still allowed at the 95\% CL. We note that this feature is an}
effect of the choice of a single slepton
mass scale, which could be avoided in more general versions of the pMSSM.


\subsection{Higgs Physics}

\begin{figure*}[htb!]
\begin{center}
\resizebox{10cm}{!}{\includegraphics{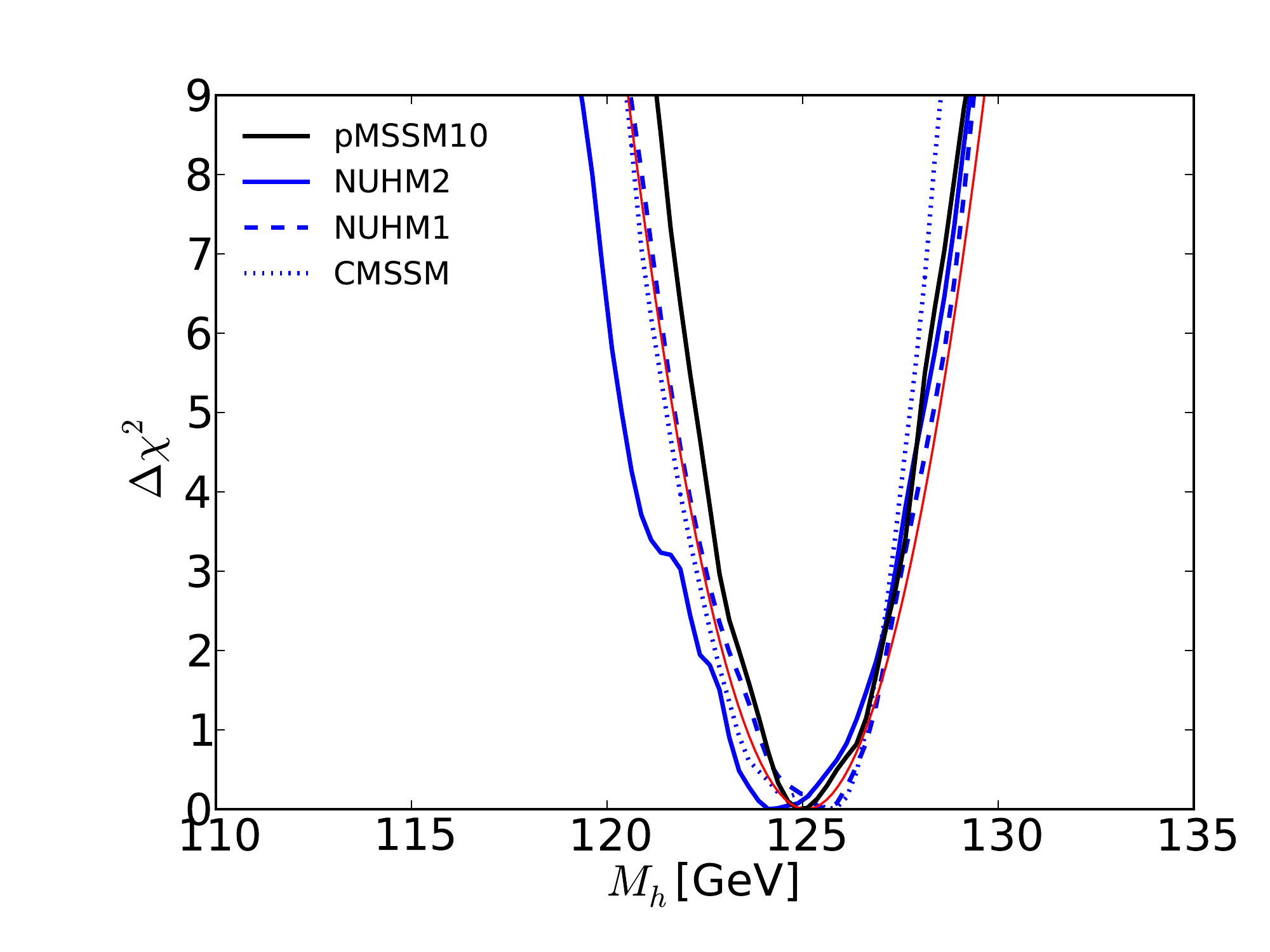}}
\end{center}
\vspace{-1cm}
\caption{\it The one-dimensional profile likelihood function
for $\Mh$: the solid black line is for the pMSSM10, the solid blue line
for the NUHM2, the dashed blue line for the NUHM1, the dotted blue line
for the CMSSM, and the red line is the $\chi^2$ penalty from the
experimental measurement of $\Mh$ with the assumed theoretical
uncertainty of $1.5 \gev$.}
\label{fig:higgs}
\end{figure*}

\reffi{fig:higgs} displays one-dimensional profile likelihood for $\Mh$
when the LHC constraints are applied. 
We see that the likelihood for $\Mh$ in the pMSSM10 (black line) is very
similar to the experimental value smeared by the theoretical uncertainty
in the {\tt FeynHiggs} calculation of $\Mh$ for specific values of the
MSSM input parameters. 

\begin{figure*}[htb!]
\resizebox{7.5cm}{!}{\includegraphics{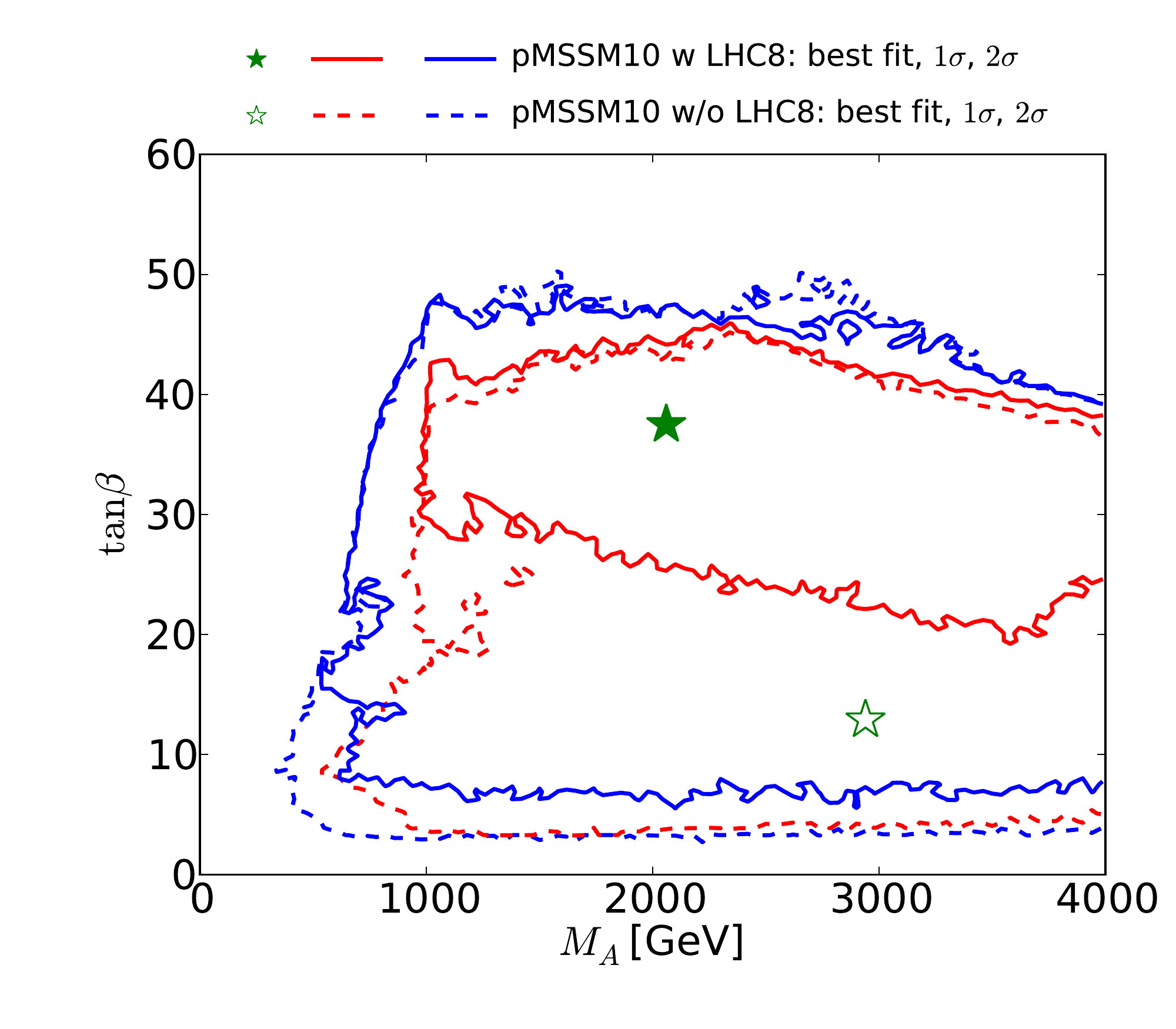}}
\resizebox{8.5cm}{!}{\includegraphics{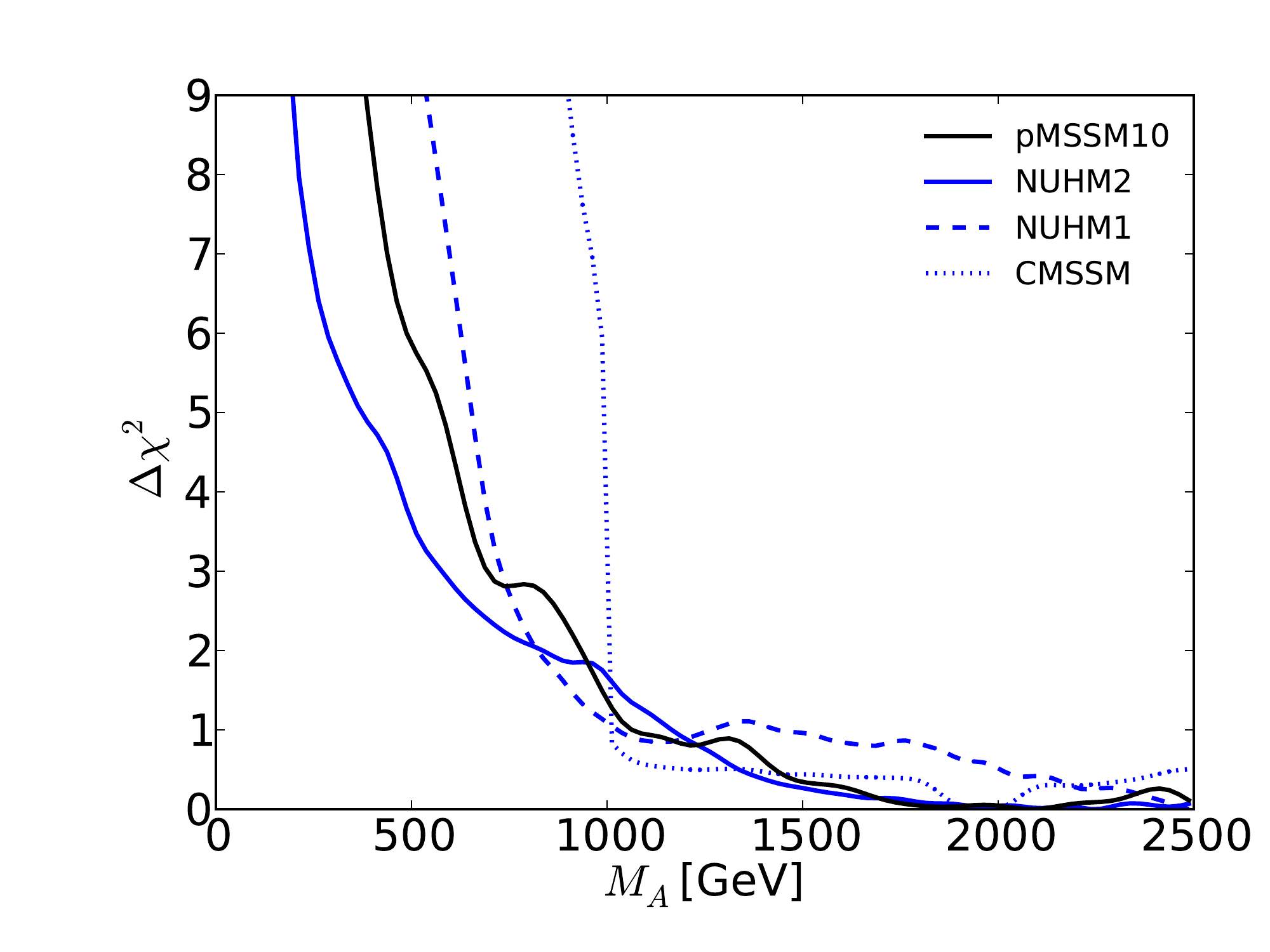}}
\caption{\it Left panel: The two-dimensional profile likelihood function
for $\MA$ versus $\tb$.
The solid (dashed) red/blue contours denote the 
$\Delta \chi^2 = 2.30/5.99$ level contours for the case where we
do (not) apply the LHC constraints, respectively, and
the green filled and empty stars indicate the corresponding best-fit points.
Right panel: The one-dimensional profile likelihood function
for $\MA$: the solid black line is for the pMSSM10, the solid blue line
for the NUHM2, the dashed blue line for the NUHM1 and the dotted blue
line for the CMSSM.
}
\label{fig:MA}
\end{figure*}

The left panel of \reffi{fig:MA} displays the two-dimensional profile
likelihood function in the $(\MA, \tb)$ plane. As before, the solid
(dashed) red/blue contours denote the $\Delta \chi^2 = 2.30/5.99$ level
contours for the case where we do (not) apply the LHC8 constraints,
respectively, and the green filled and empty stars indicate the
corresponding best-fit points. Comparing the dashed and solid 68\%
contours, we see that lower values of $\tb$ are disfavoured at the 68\%
CL by the combination of \lhcewk, \gmt\ and Dark Matter constraints,
as discussed in the previous subsection. 
{Those constraints, in combination with the choice of a single slepton 
mass scale for all three generations, lead to limits of
$\MA \gsim 1000 (500) \gev$ at the 68 (95)\%~CL, whereas otherwise
low CP-odd Higgs boson masses down to $\MA \sim 500 (350) \gev$ 
would be found in the 68 (95)\%~CL area.}

The right panel of \reffi{fig:MA} displays the corresponding
one-dimensional profile likelihood function for $\MA$: as before,
the solid black line is for the pMSSM10, the solid blue line for the
NUHM2, the dashed blue line for the NUHM1 and the dotted blue line for
the CMSSM. 
Lower $\MA$ values for $\tb \lsim 30$ are in particular disfavoured
by the LUX and other limits, as discussed in the previous subsection.


\boldmath
\subsection{{\bsmm} Decay}
\unboldmath

\begin{figure*}[htb!]
\begin{center}
\resizebox{10cm}{!}{\includegraphics{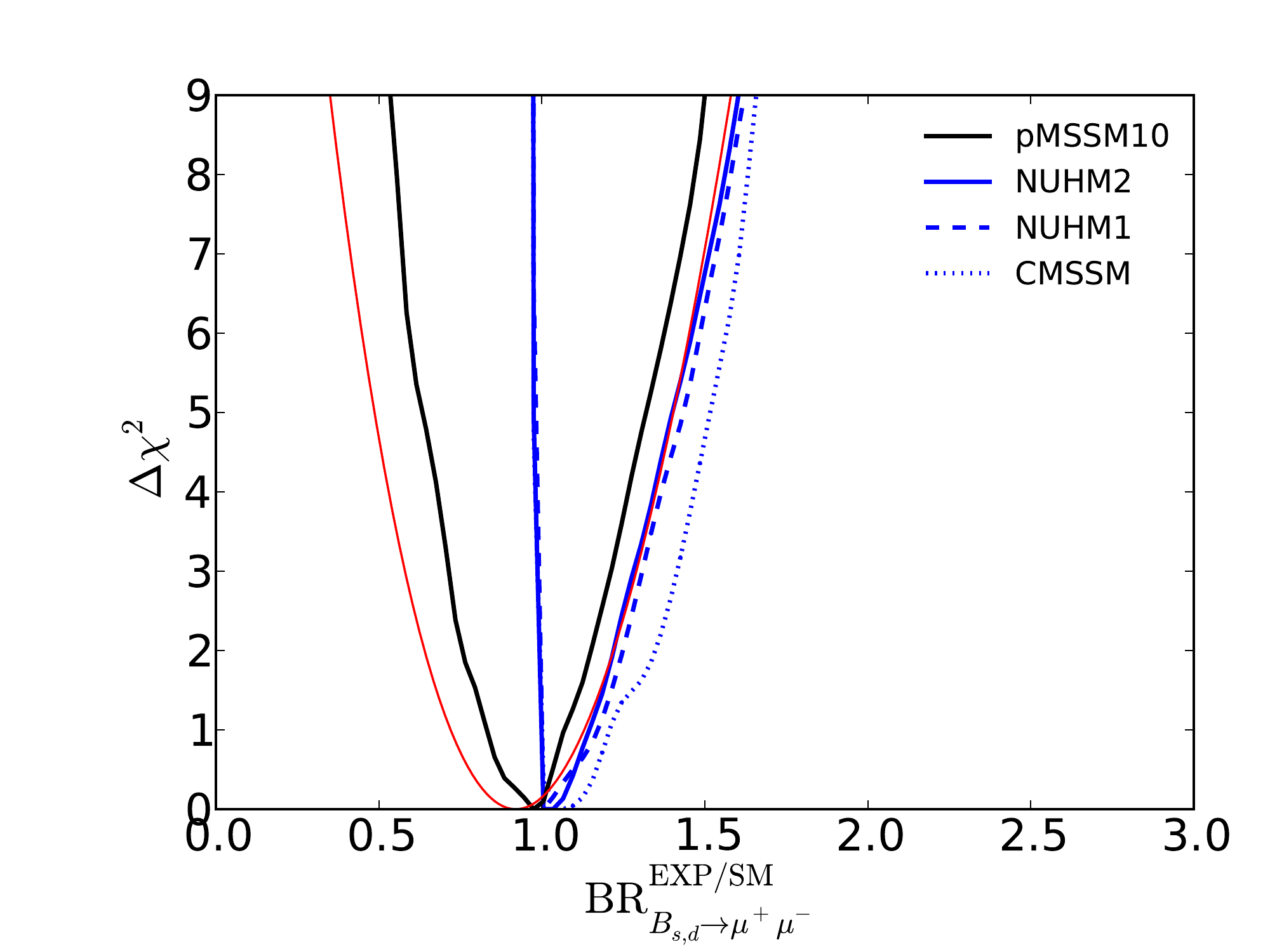}}
\end{center}
\vspace{-1cm}
\caption{\it The one-dimensional profile likelihood functions
for \bsmm\ relative to the SM value.
The solid black line is for the pMSSM10, the solid blue line for the NUHM2,
the dashed blue line for the NUHM1, the dotted blue line for the CMSSM, and the
red line is the $\chi^2$ penalty from the experimental constraint.}
\label{fig:bsmm}
\end{figure*}

We display as a black line in \reffi{fig:bsmm} the profile likelihood in the pMSSM10 for the ratio of \bsmm\ to
the SM value. This can be compared with the $\chi^2$ penalty from the experimental constraint
on \bsmm, which is shown as a red line. It is interesting to note
that in the pMSSM10 both enhancement and suppression are possible, as opposed
to the CMSSM, the NUHM1 and the NUHM2~\cite{mc9,mc10}, in which a suppression
was not possible and only an enhancement was allowed.
This comes about because the extra parameters in the pMSSM10 make possible some
negative interference between the SM and SUSY amplitudes, which is not possible in the other
models when the various other constraints are implemented.


\subsection{Direct Dark Matter Detection}

\begin{figure*}[htb!]
\begin{center}
\resizebox{7.5cm}{!}{\includegraphics{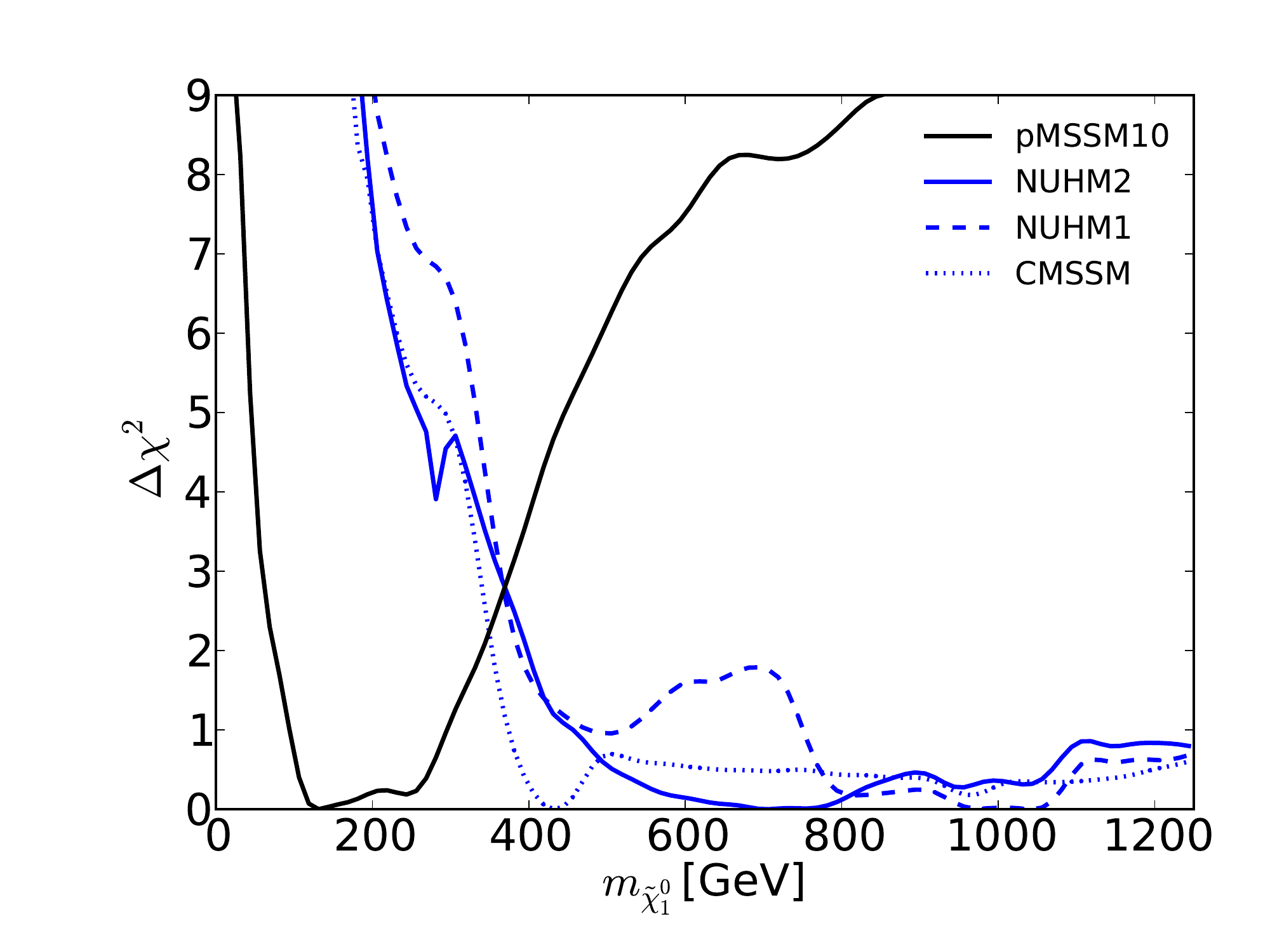}}
\resizebox{7cm}{!}{\includegraphics{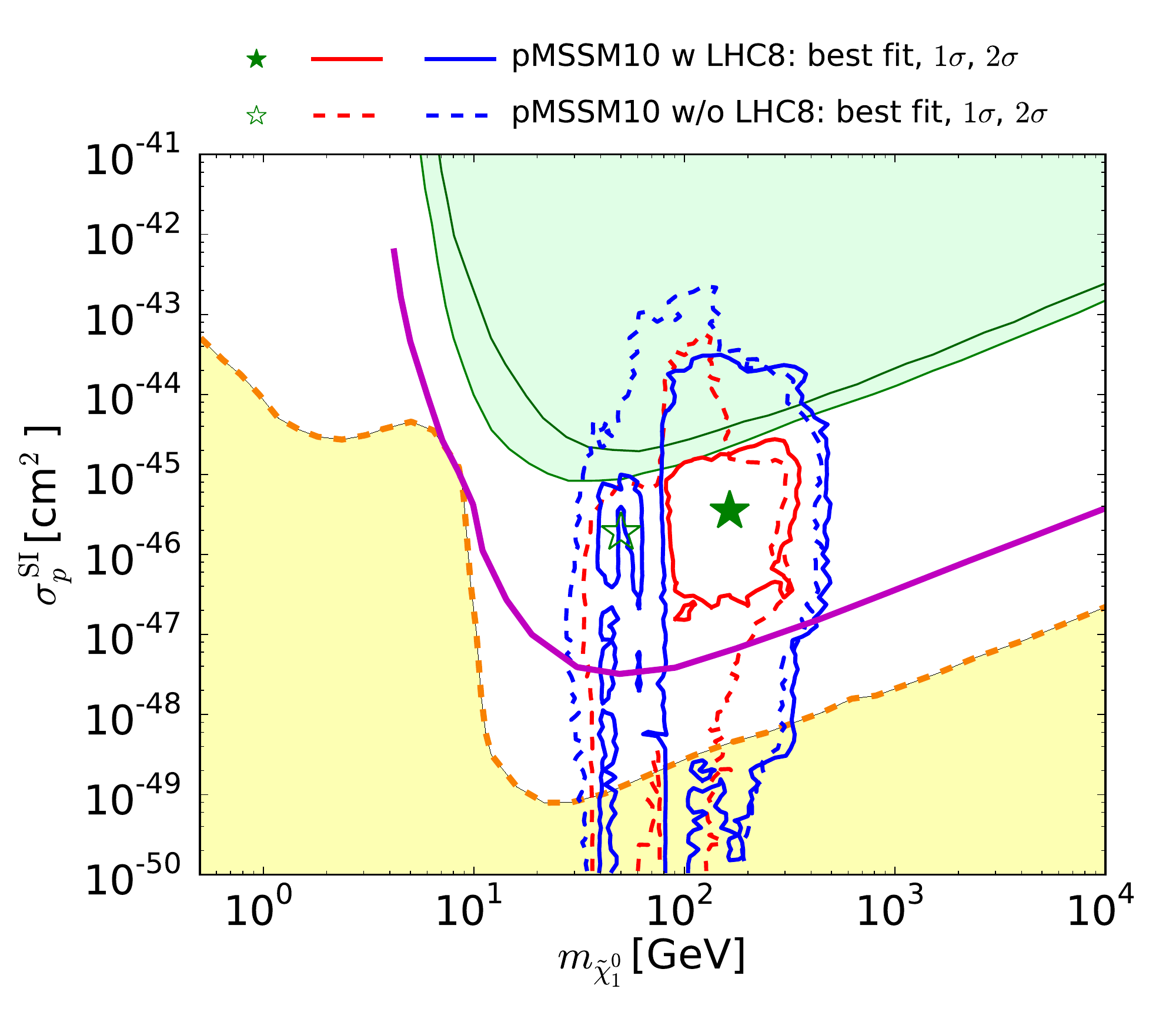}}
\end{center}
\vspace{-1cm}
\caption{\it Left panel: The one-dimensional profile likelihood in the pMSSM10
for $\mneu1$ (black line), compared with the NUHM2, the NUHM1 and the CMSSM
(solid, dashed and dotted blue lines, respectively).Right panel: The two-dimensional
profile likelihood function in the pMSSM in the $(\mneu1, \ssi)$-plane, showing the regions
excluded by the XENON100 and LUX experiments (shaded green), the neutrino `floor'
(shaded yellow), and the prospective sensitivity of the LZ
  experiment (purple)~\protect\cite{LZ}.
}
\label{fig:mneu-ssi}
\end{figure*}

The left panel of \reffi{fig:mneu-ssi} displays the one-dimensional profile likelihood in the pMSSM10
for $\mneu1$ with the same colour coding as in
\reffi{fig:onedimensional}.
We see that, in contrast to the other models, the
pMSSM10 favours a low mass for the $\neu1$, driven again by the \gmt\ constraint. 
The right panel of \reffi{fig:mneu-ssi} 
displays the two-dimensional profile
likelihood for the lightest neutralino mass versus the spin-independent
cross-section, where
the red and blue contours show the 68\% and 95\% CL levels respectively. 
The region that is excluded by LUX~\cite{lux} and XENON100~\cite{XENON100} is
shaded green, whereas the `floor' below which the background from
atmospheric neutrinos dominates is shaded yellow~\cite{nuback}.
The low-mass vertical 95\% CL
strips are due to points where the relic LSP density is brought into the
cosmological range by annihilations through direct-channel $Z$ and $h$ poles.

It is interesting to note that the pMSSM10 fit prefers rather high
values of the spin-independent cross section after application of the LHC8
constraints:
lower values could be reached for a Bino-like LSP, but the dark matter density constraint would then require stau
coannihilation and $t$-channel slepton exchange, which are, however,
disfavoured by the combined effects of the \lhcewk\ and \gmt\ constraints.
Our best-fit region is close to the present experimental upper limit on \ssi~\cite{lux}, and consequently within reach of future direct
detection experiments {such as LZ~\cite{LZ}, as indicated by the magenta line in the right panel of \reffi{fig:mneu-ssi}.}
On the other hand, we note that before applying these constraints, and even afterwards 
at the 95\% CL, there are values for \ssi\ that go far below this
neutrino `floor', highlighting the complementarity of direct detection
experiments and searches at the LHC. 
Since these very low values of \ssi\ are due to cancellations between
different contributions to the spin-independent scattering matrix
element, one may ask whether the spin-independent cross sections on
proton and neutron targets could be very different when this
cancellation occurs. More specifically, one may wonder whether, for
models in which \ssi\ is below the neutrino `floor', the cross-section \ssin\ for scattering on a
neutron target may be less suppressed, perhaps remaining above the neutrino
`floor'? As we see in \reffi{fig:ssipvsn}, 
the spin-independent
cross sections on proton and neutron targets are generally very 
similar when \ssi\ $> 10^{-47}$~cm$^2$, but may indeed be quite
different when \ssi\ $< 10^{-49}$~cm$^2$, which is approximately the
lowest level of the neutrino `floor', whose height varies as seen in the right
panel of \reffi{fig:mneu-ssi}. Points coloured black (green) [blue] \{{red}\} have both \ssi\ and \ssin\ above the neutrino `floor'
shown in the right panel of \reffi{fig:mneu-ssi} (\ssi\ below and \ssin\ above) [\ssi\ above and \ssin\ below]
\{{\ssi\ and \ssin\ both below}\}.
We see that there is
a significant population of models whose spin-independent scattering
cross sections on protons and neutrons are both below the `floor' (indicated in red), so
there is no `no-lose' theorem for dark matter scattering in the
pMSSM10~%
\footnote{We do not include in the right panel of \reffi{fig:mneu-ssi} 
and in \reffi{fig:ssipvsn} the contributions of loop-induced scattering off gluons~\cite{DN}. In general,
these contributions are relatively small~\cite{HIN}, but they would also shift slightly the parameters of the
models exhibiting strong cancellations in $\ssi$ and $\ssin$. We thank
N.~Nagata for discussions on these points.}%
.

\begin{figure*}[htb!]
\begin{center}
\resizebox{11cm}{!}{\includegraphics{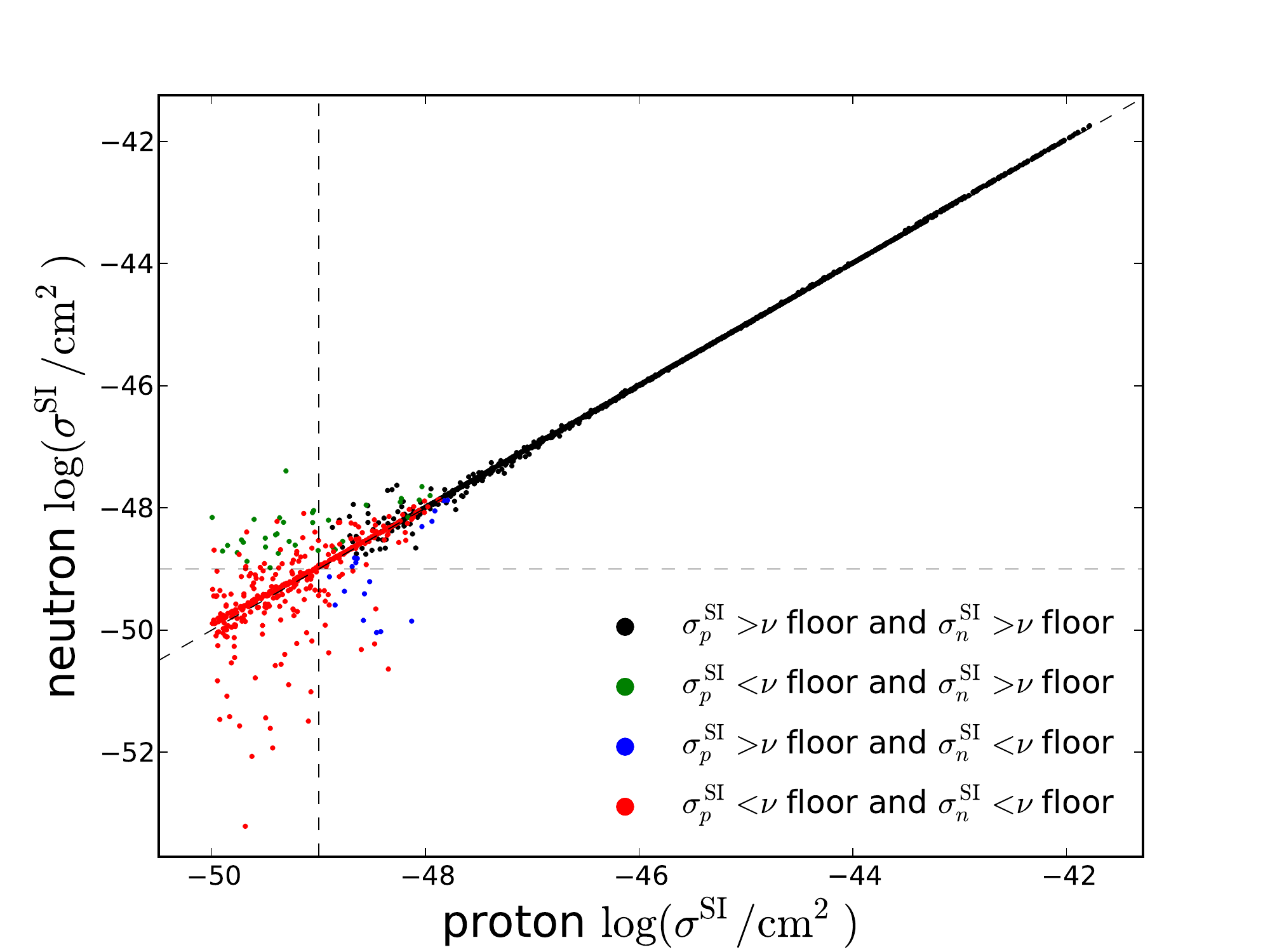}}
\end{center}
\vspace{-1cm}
\caption{\it Scatter plot of the cross sections for spin-independent
  scattering on a proton target (horizontal axis) and on a neutron
  target (vertical axis) {obtained from a sampling of pMSSM10 points within the
  95\% CL region.} The diagonal dashed line corresponds to equal
  spin-independent cross sections on proton and neutron targets.
  The colour-coding distinguishes between
points with either \ssi\ and/or \ssin\ above or below the neutino `floor' seen in \protect\ref{fig:mneu-ssi}.
 } 
\label{fig:ssipvsn}
\end{figure*}


\section{Extrapolation to High Scales}
\label{sec:tachyons}

In our analysis of the pMSSM10 we have not imposed any restriction
on the possible 
extrapolation of the (purely phenomenological) soft SUSY-breaking
parameters to high scales using the renormalization-group equations (RGEs).
In many cases, one could expect that renormalisation by the gaugino masses may drive
some soft supersymmetry-breaking {sfermion} masses-squared $m_0^2$
to negative values at high-energy scales~\cite{JKPR}. 
This raises cosmological issues that
have been studied, for example, in~\cite{EGLOS}, and such scenarios do not necessarily
lead to an unacceptable evolution of the Universe. However, it is interesting to
study the implications of requiring $m_0^2 >0$.
We emphasise that this cut reduces the data set significantly,
and one may anticipate that part of the parameter space would be recovered in a
dedicated scan. 
Nevertheless, we expect that the main features discussed here would be present also
in a more complete scan.

\reffi{fig:notachyons} displays the two-dimensional likelihood functions
in some relevant sparticle mass planes. In each panel, the red (blue) lines are the
68\% (95\%) CL contours, the solid (dashed) lines being after (before) a cut
requiring $m_0^2 > 0$ for all the sleptons and squarks at the GUT scale 
$\sim 2\cdot10^{16} \gev$.
The upper left panel shows the $(m_{\tilde q}, \mgl)$ plane, and can be compared
with \reffi{fig:up-down} (upper left plot). We see that the primary impact of the anti-tachyon cut is to remove all models
above a diagonal line where the negative renormalisation by $M_3$ drives the squark
masses-squared negative at the GUT scale. The upper right panel of \reffi{fig:notachyons} shows the impact
of the anti-tachyon cut on the $(\mst1, \mneu1)$ plane, which can be compared
with the middle left panel of \reffi{fig:planes}. Here the most obvious
impact is to remove the compressed stop region where 
$\mst1 - \mneu1<\mt$~\footnote{The 68\% CL region extends to much larger values of $\mst1$,
where larger values of $\mneu{1}$ are also found.}. The lower left panel of \reffi{fig:notachyons} shows the impact in the
$(\mcha1, \mneu1)$ plane, where we see that many models with small
$\mcha1 - \mneu1$ survive the anti-tachyon cut. 
{The anti-tachyon cut leads to a more pronounced
preference for small values of $\mneu1$ and in particular $\mcha1$.}
Finally, the lower right panel of
\reffi{fig:notachyons} displays the $(\msmu{R}, \mneu1)$ plane, where we
see that the anti-tachyon cut has very little effect, except to remove some points with small
$\msmu{R} - \mneu1$. In particular, the best-fit values of
$\msmu{R}$ and $\mneu1$ are little changed, and the mitigation of the \gmt\
anomaly in the pMSSM10 survives the anti-tachyon cut. However, we repeat that this
cut may even not be necessary~\cite{EGLOS}.

\begin{figure*}[htb!]
\vspace{-1cm}
\resizebox{7.5cm}{!}{\includegraphics{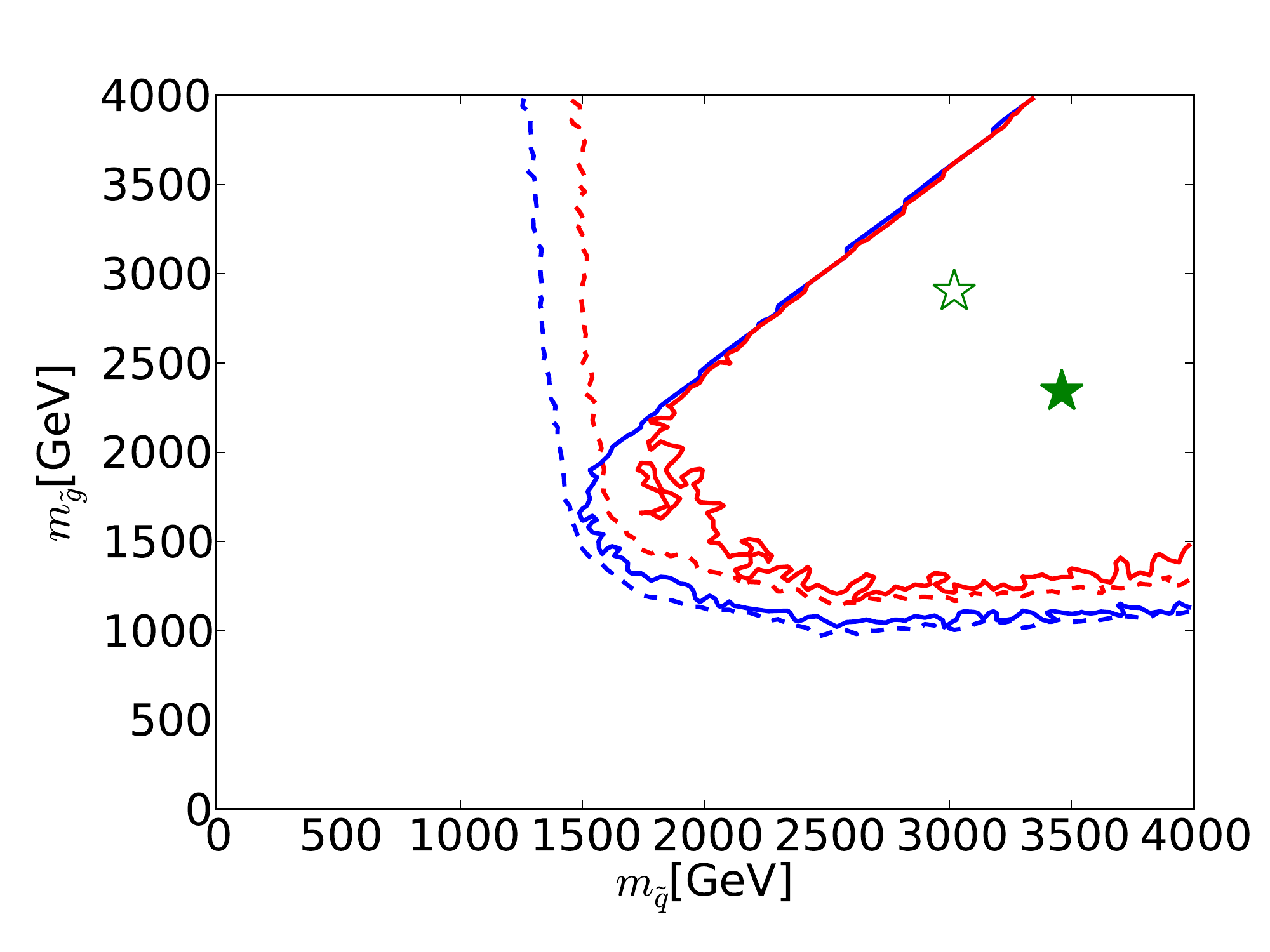}}
\resizebox{7.5cm}{!}{\includegraphics{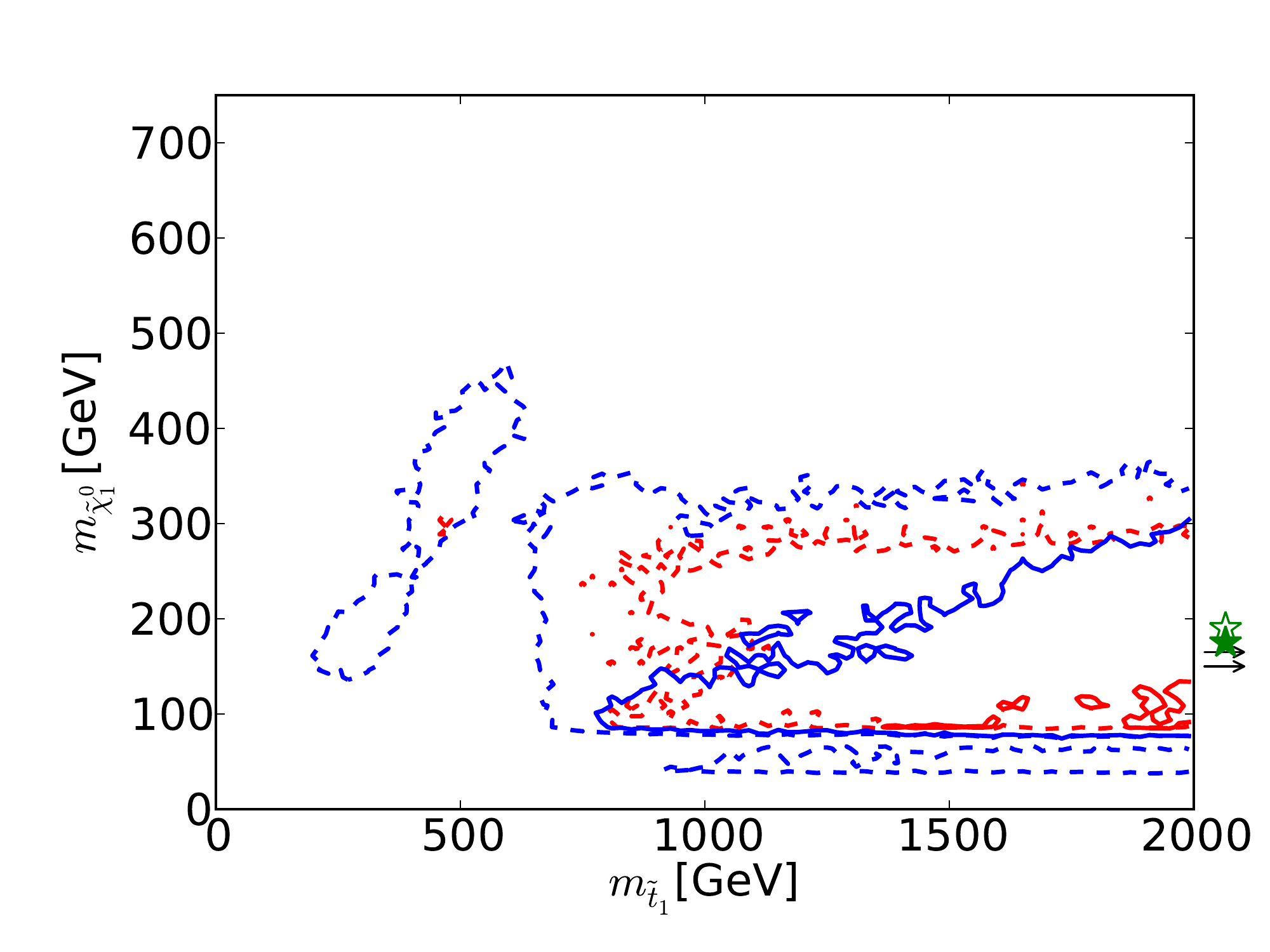}}\\
\resizebox{7.5cm}{!}{\includegraphics{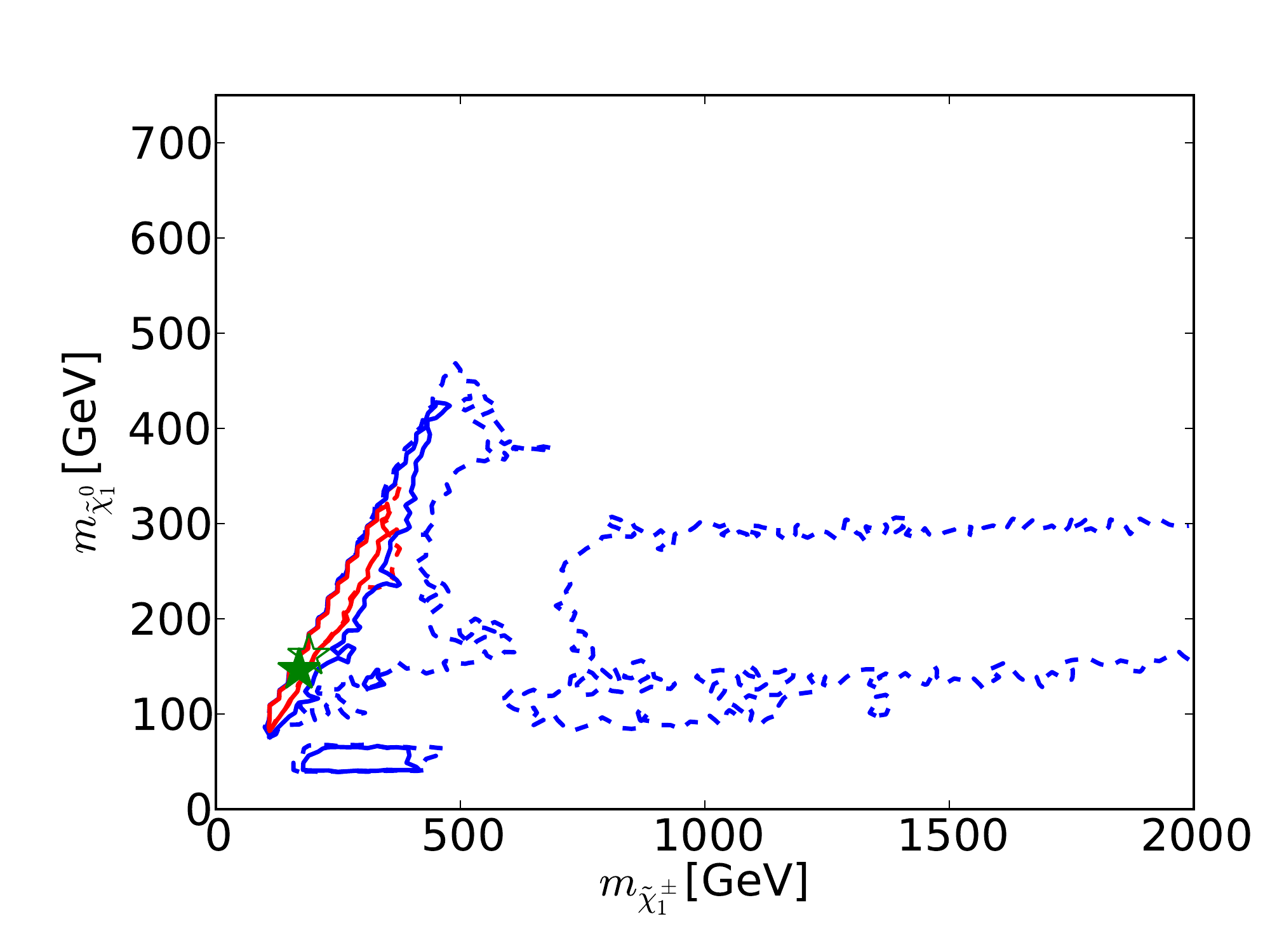}}
\resizebox{7.5cm}{!}{\includegraphics{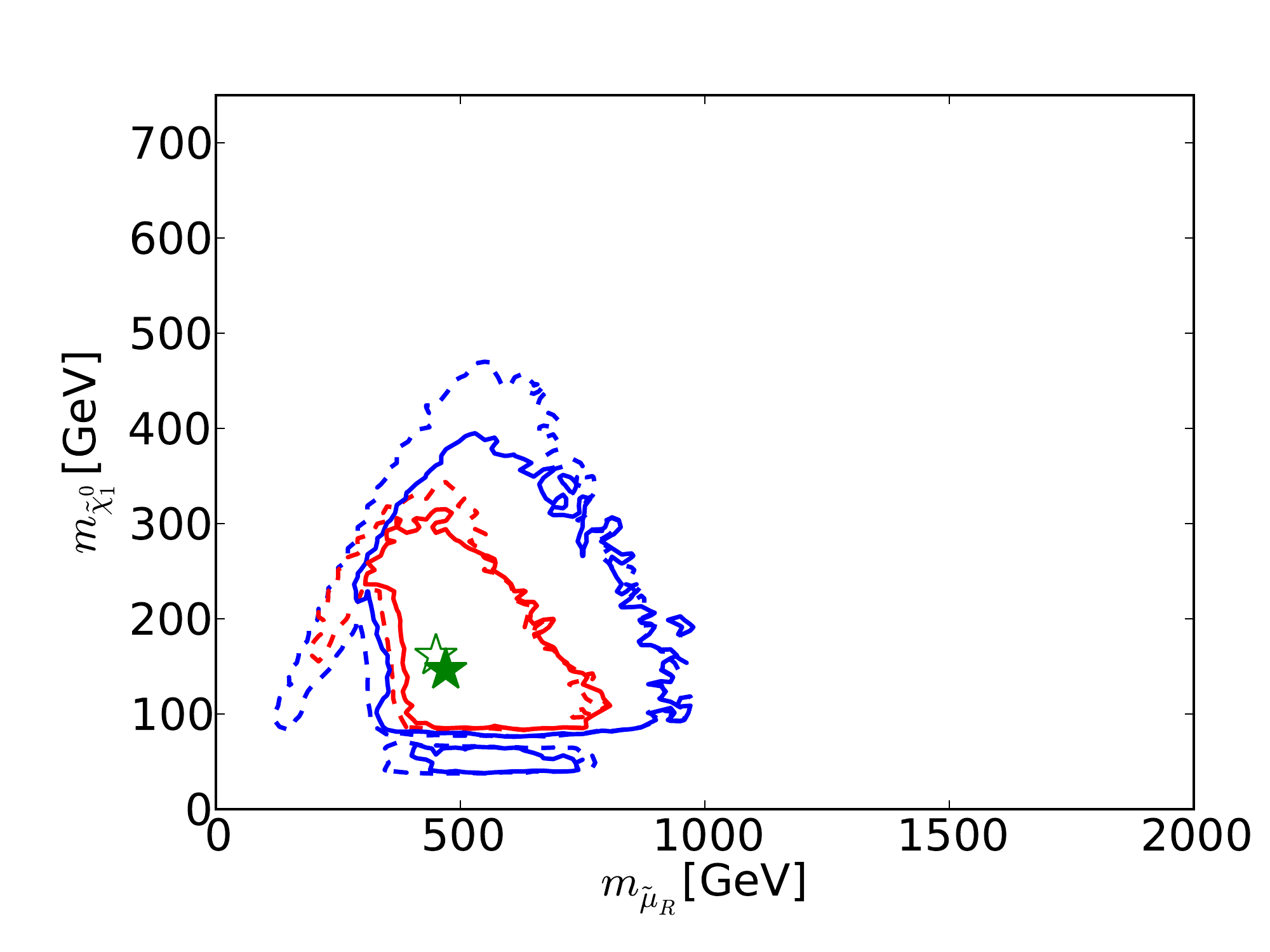}}\\
\vspace{-1cm}
\caption{\it The impacts of the optional anti-tachyon cut on the two-dimensional profile likelihood functions
in the $(m_{\tilde q}, \mgl)$, $(\mst1, \mneu1)$, $(\mcha1, \mneu1)$ and
$(\msmu{R}, \mneu1)$ planes. In each panel the solid (dashed) red/blue contours denote the 
$\Delta \chi^2 = 2.30/5.99$ level contours for the case where we
do (not) apply the anti-tachyon constraint, respectively.
The green filled and empty stars indicate the corresponding best-fit
points.
}
\label{fig:notachyons}
\end{figure*}

\reffi{fig:notachyons1D} shows the impacts of the optional anti-tachyon
cut on the one-dimensional profile likelihood functions for $\mgl$,
$m_{\tilde q}$, $\mst1$, $\mneu1$, $\mcha1$ and $\msmu{R}$ (from top
left to bottom right). We see that the $\chi^2$ function for the gluino
mass is little affected, whereas points with low $m_{\tilde q}$ are
systematically removed, as one might expect from enforcing $m_0^2 > 0$.
These effects can also be seen in the upper left panel of
\reffi{fig:notachyons}. As one would expect from the upper right panel
of \reffi{fig:notachyons}, points with low $\mst1$ are also removed by
the anti-tachyon cut, and the best-fit value of $\mst1$ is increased by
$\sim 1 \tev$. As seen in the middle right panel of
\reffi{fig:notachyons1D}, the one-dimensional likelihood function for
$\mneu1$ is little affected, whereas that for $\mcha1$ is squeezed
strongly. These effects reflect the behaviour in the $(\mcha1, \mneu1)$
plane seen in the lower left panel of \reffi{fig:notachyons}, where the
favoured points lie in a narrow $\cha1 - \neu1$ coannihilation
strip. These points have $M_1 \simeq M_2$ at the electroweak
scale, leading to the potential observability of
neutralino/chargino pair production at an $e^+e^-$ collider with a
centre-of-mass energy below $1000 \gev$, as we discuss later.
Finally, we see in the bottom right panel of
\reffi{fig:notachyons1D} that the likelihood function for $\msmu{R}$ is
little affected by the anti-tachyon cut, apart from the removal of some
low-mass points as seen already in the lower right panel of
\reffi{fig:notachyons}. However, as already commented, the removal of
these points does not prevent the pMSSM10 from addressing successfully
the \gmt\ problem. 
 
\begin{figure*}[htb!]
\resizebox{8.0cm}{!}{\includegraphics{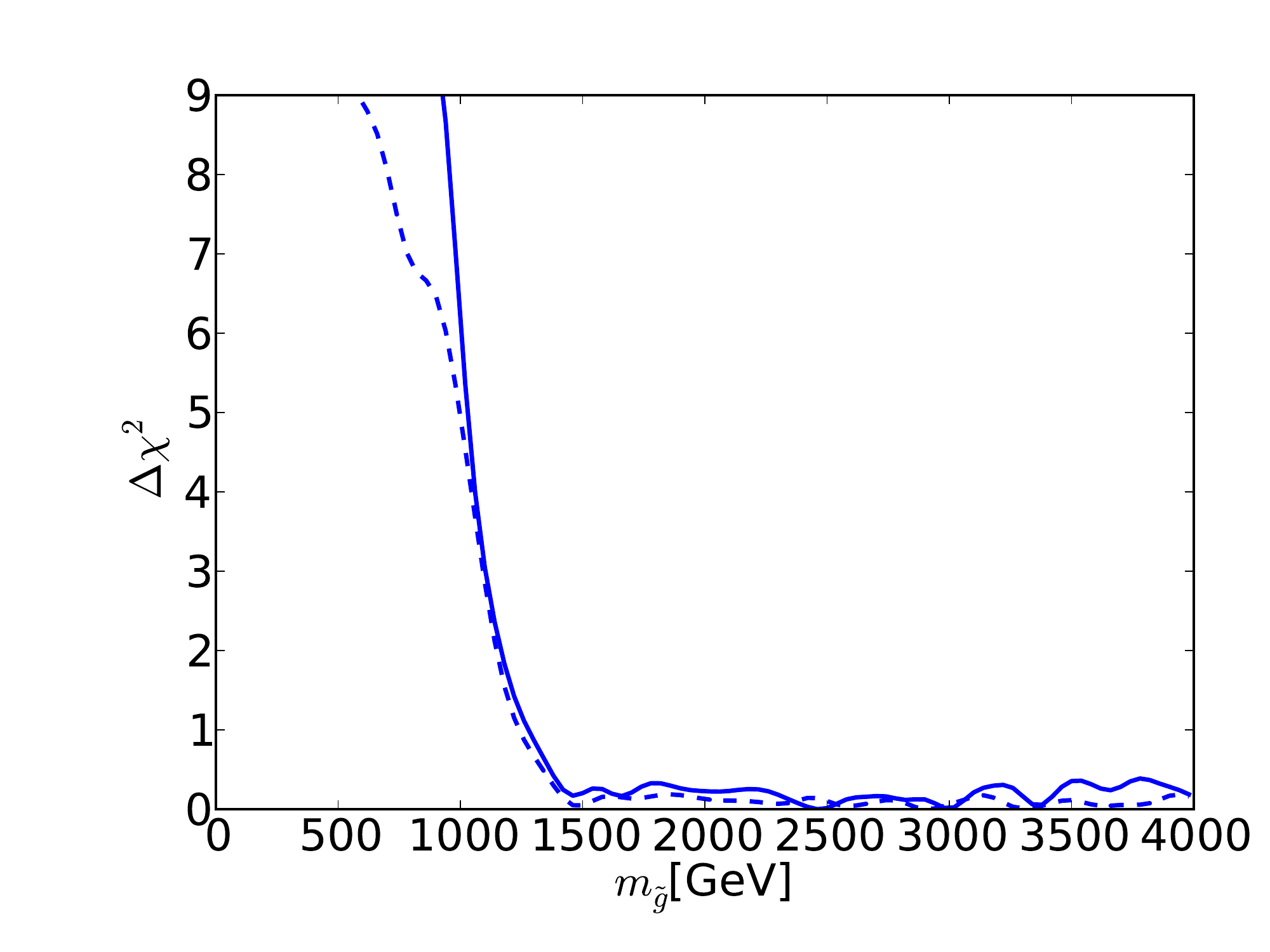}}
\resizebox{8.0cm}{!}{\includegraphics{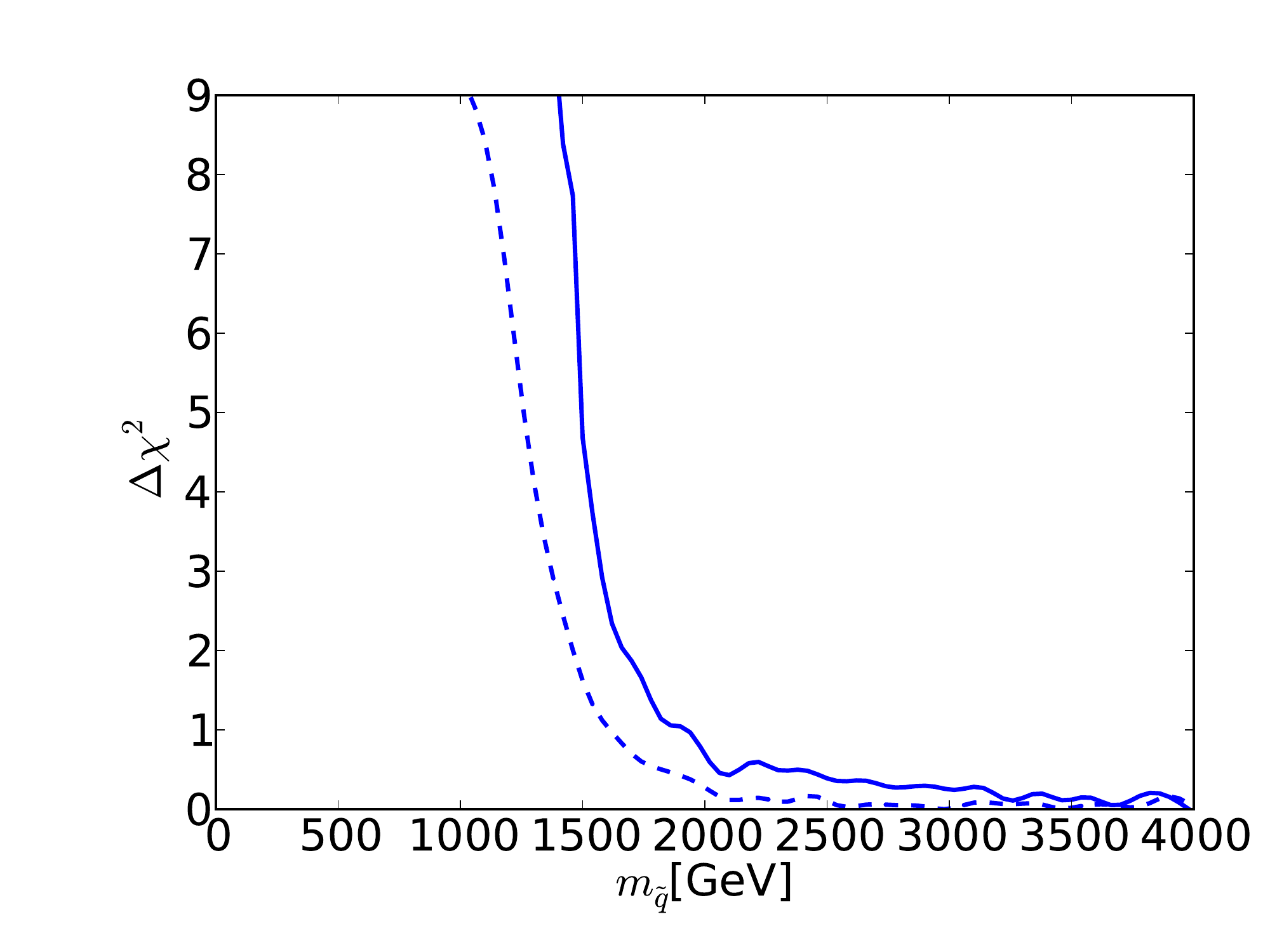}}\\
\hspace {0.5cm}
\resizebox{8.0cm}{!}{\includegraphics{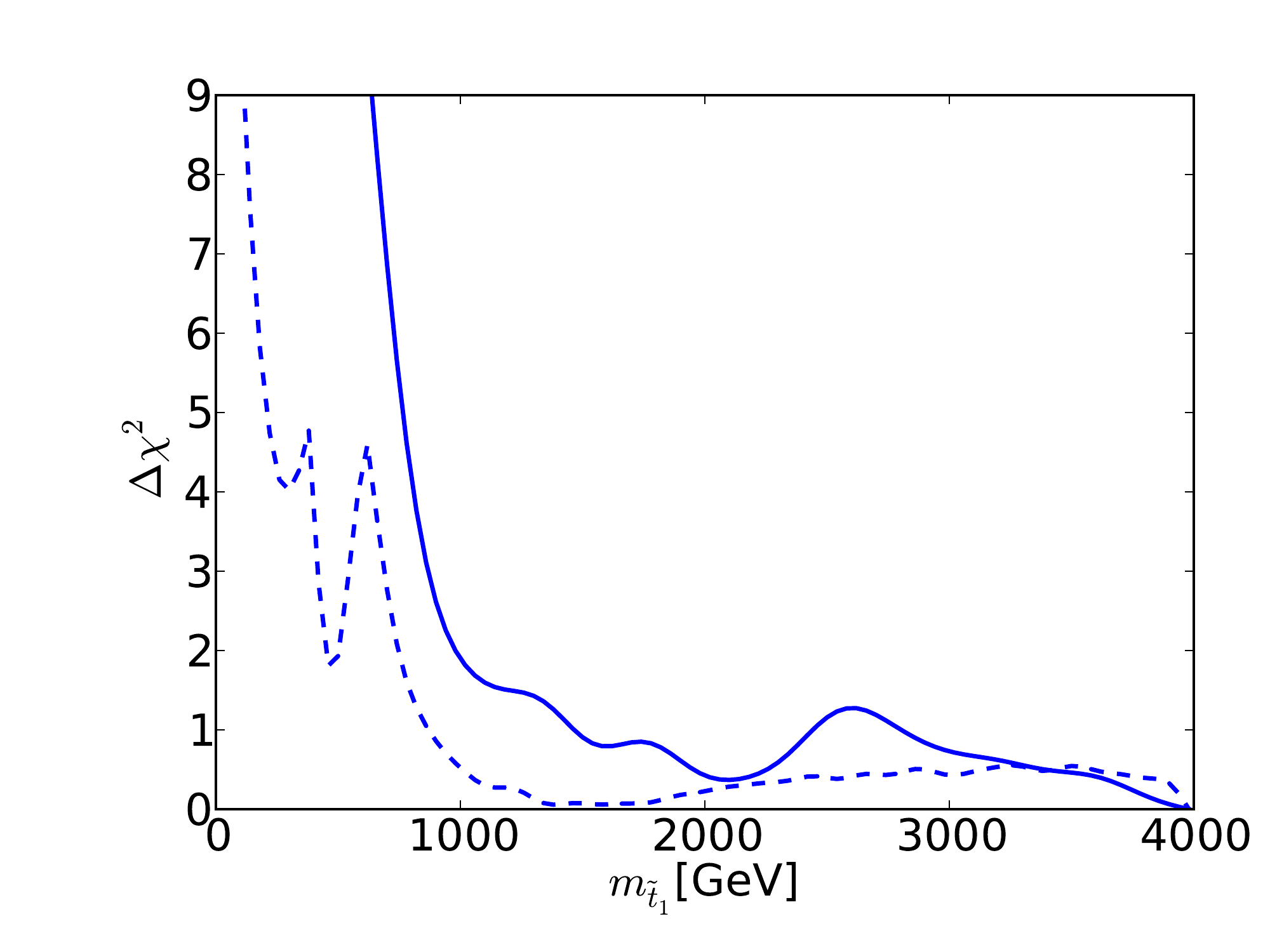}}
\resizebox{8.0cm}{!}{\includegraphics{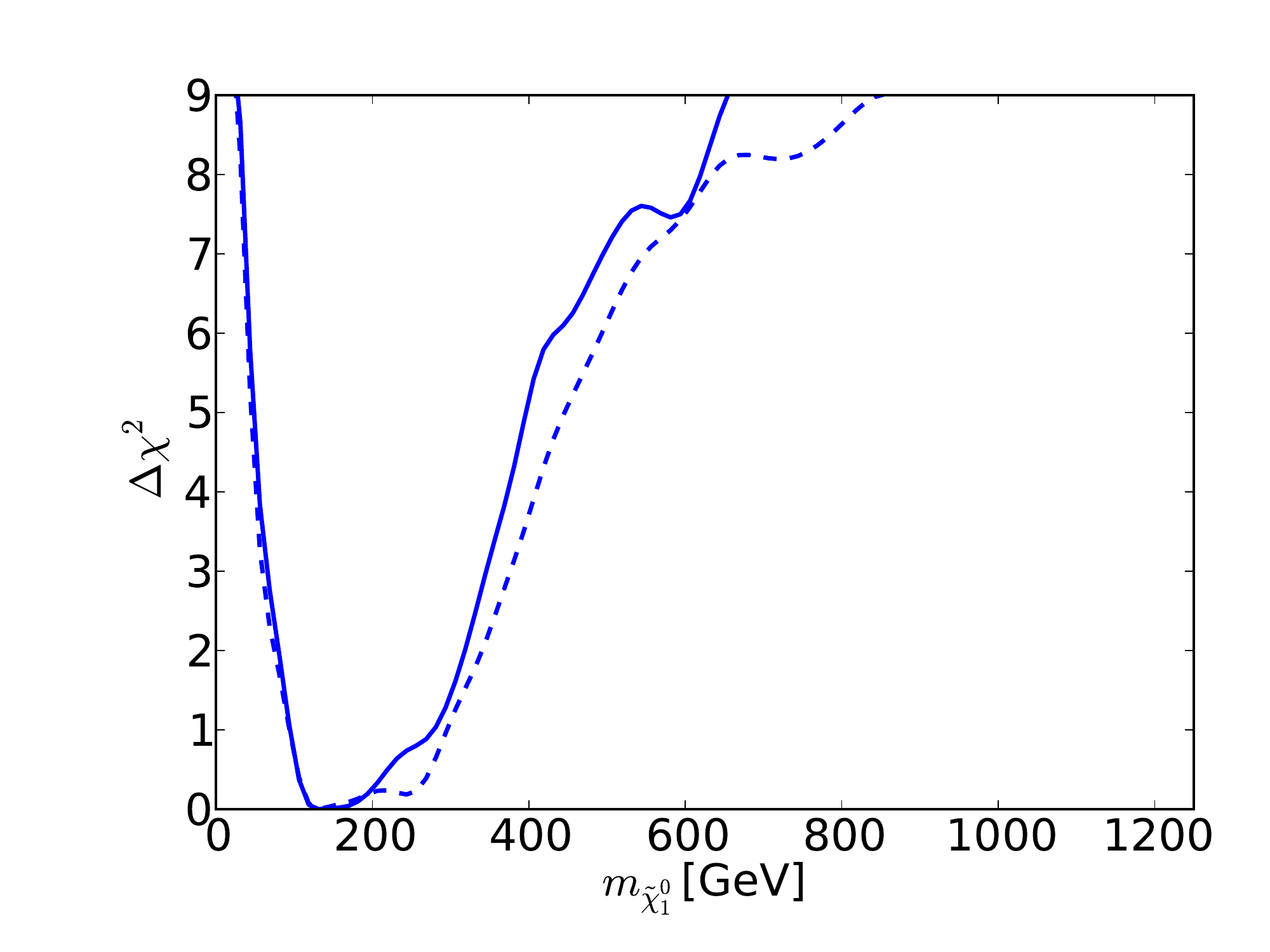}}\\
\hspace {0.5cm}
\resizebox{8.0cm}{!}{\includegraphics{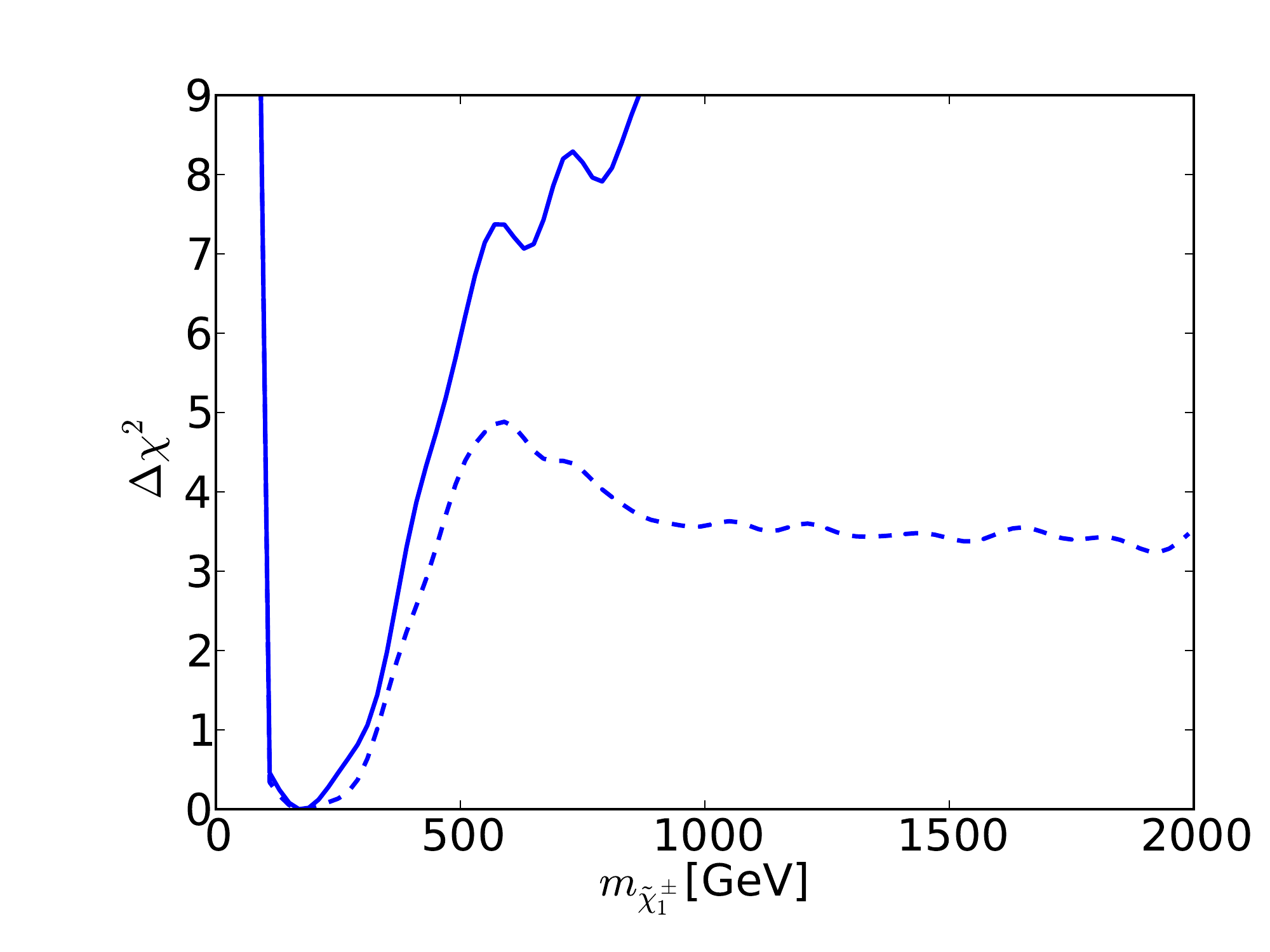}}
\resizebox{8.0cm}{!}{\includegraphics{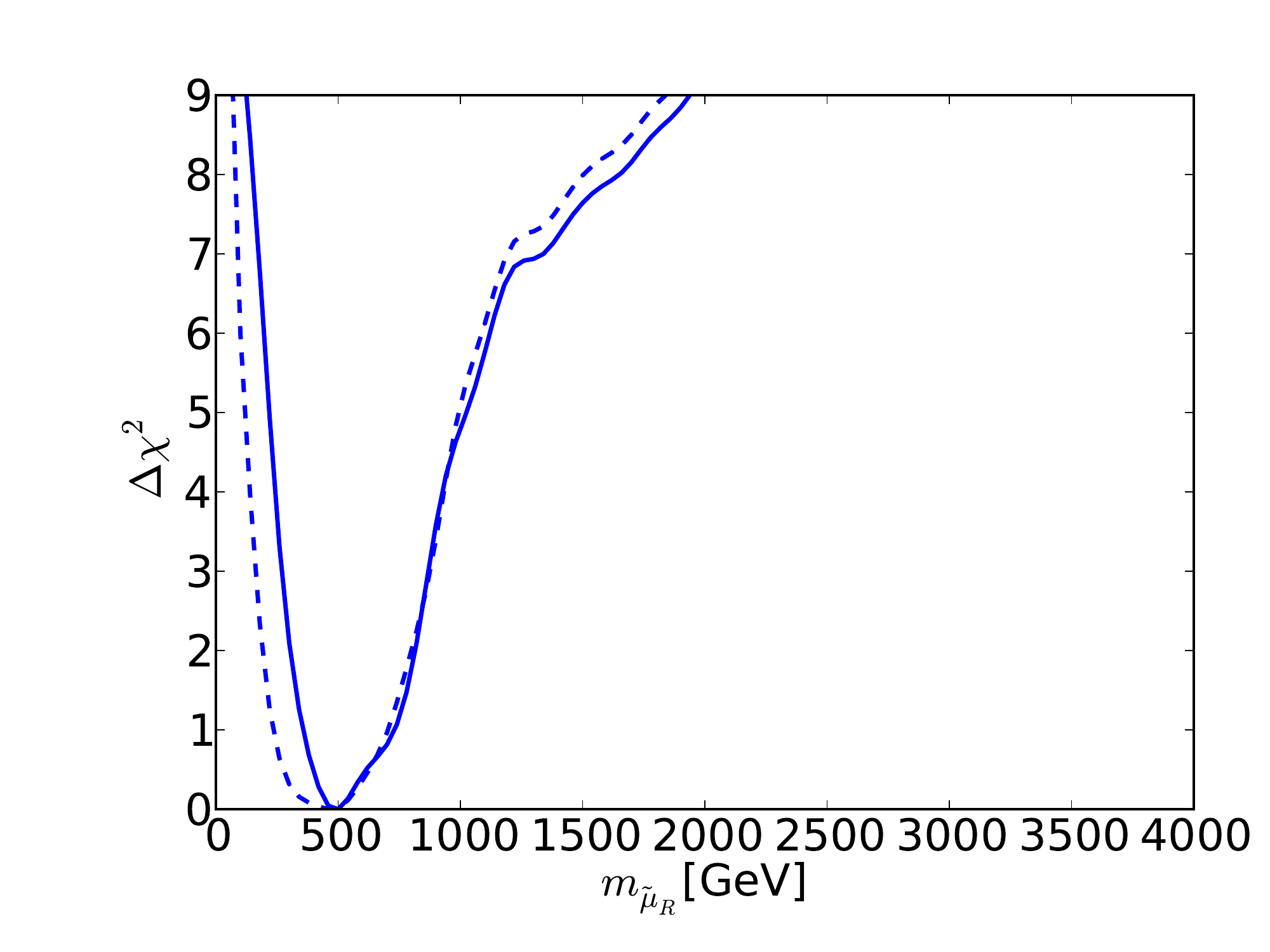}}\\
\vspace{-1cm}
\caption{\it The impacts of the optional anti-tachyon cut on the one-dimensional profile likelihood functions
for $\mgl$, $m_{\tilde q}$, $\mst1$, $\mneu1$, $\mcha1$ and
$\msmu{R}$. In each panel the solid (dashed) lines are for the cases where we
do (not) apply the anti-tachyon constraint, respectively.}
\label{fig:notachyons1D}
\end{figure*}

As a final topic in this Section, we discuss the departures from
universality of the soft supersymmetry-breaking parameters in the sample
that would survive the anti-tachyon cut. \reffi{fig:nonuniversality}
shows a plane of the root-mean-squared deviations from gaugino- and
sfermion-mass universality, defined by 
\begin{equation}
\sigma_{M, m} \equiv \sqrt{\sum_{i}^N (m_i - \bar{m})^2/N} \, ,
\end{equation}
where the $m_i$ denote, respectively, the various gaugino mass parameters and the square roots of the
(positive) squark and slepton $m^2_0$ parameters in the pMSSM10 at the GUT scale, and 
$\bar{m}$ denotes their respective averages. 
{Exact} unification of the gaugino (sfermion) masses is achieved when $\sigma_M$ ($\sigma_m$) vanishes.
We see that sfermion-mass universality is quite
strongly violated, and gaugino-mass universality is also disfavoured, though still possible at the 95\% CL.
As we have already commented, the favoured
points in the narrow $\cha1 - \neu1$ coannihilation strip must have
near-degenerate $\neu1$ and $\neu2$ and hence $M_2 \simeq M_1$ at the
SUSY-breaking scale, corresponding to a breakdown of universality by a
factor $\sim 2$ at the GUT scale, i.e.\ 
$M_1(M_{\rm GUT}) \sim 2 M_2(M_{\rm GUT})$.
As can also be inferred by comparing
the top left and middle right panels of \reffi{fig:notachyons1D}, a
violation of GUT-scale $M_3 - M_1$ universality is also suggested.
Thus, refined future fits based on more data might lead to a
preference for some different scenario for unification.

\begin{figure*}[htb!]
\begin{center}
\resizebox{11cm}{!}{\includegraphics{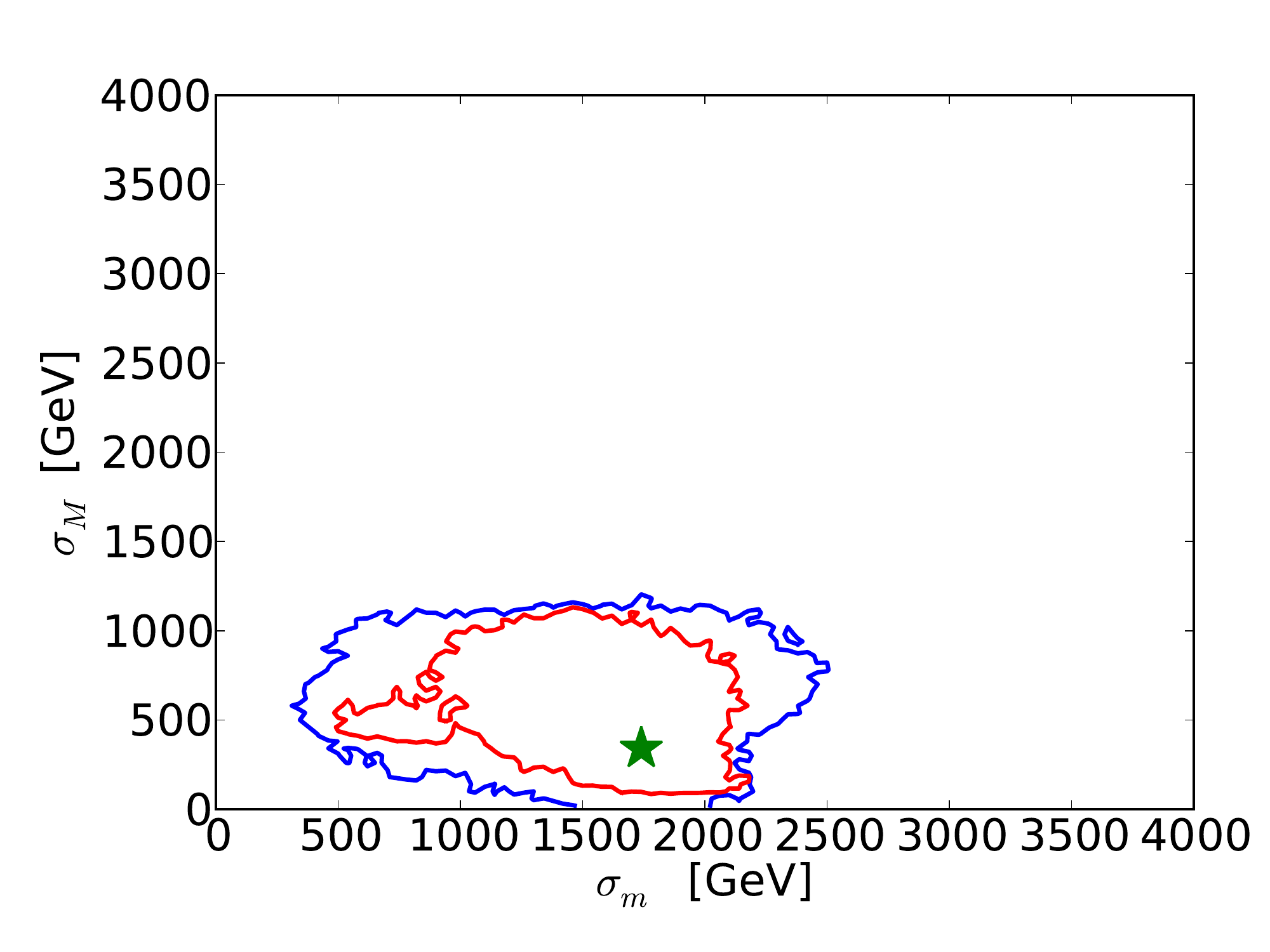}}
\end{center}
\vspace{-1cm}
\caption{\it Two-dimensional likelihood function in the plane of the
  root-mean-square deviations from 
sfermion- and gaugino-mass universality, $\sigma_m$ and $\sigma_M$,
  defined in the text.}
\label{fig:nonuniversality}
\end{figure*}


\section{Prospects for Sparticle Detection in Future LHC Runs}
\label{sec:run2prospects}

\begin{figure*}[htb!]
\centering
\resizebox{11cm}{!}{\includegraphics{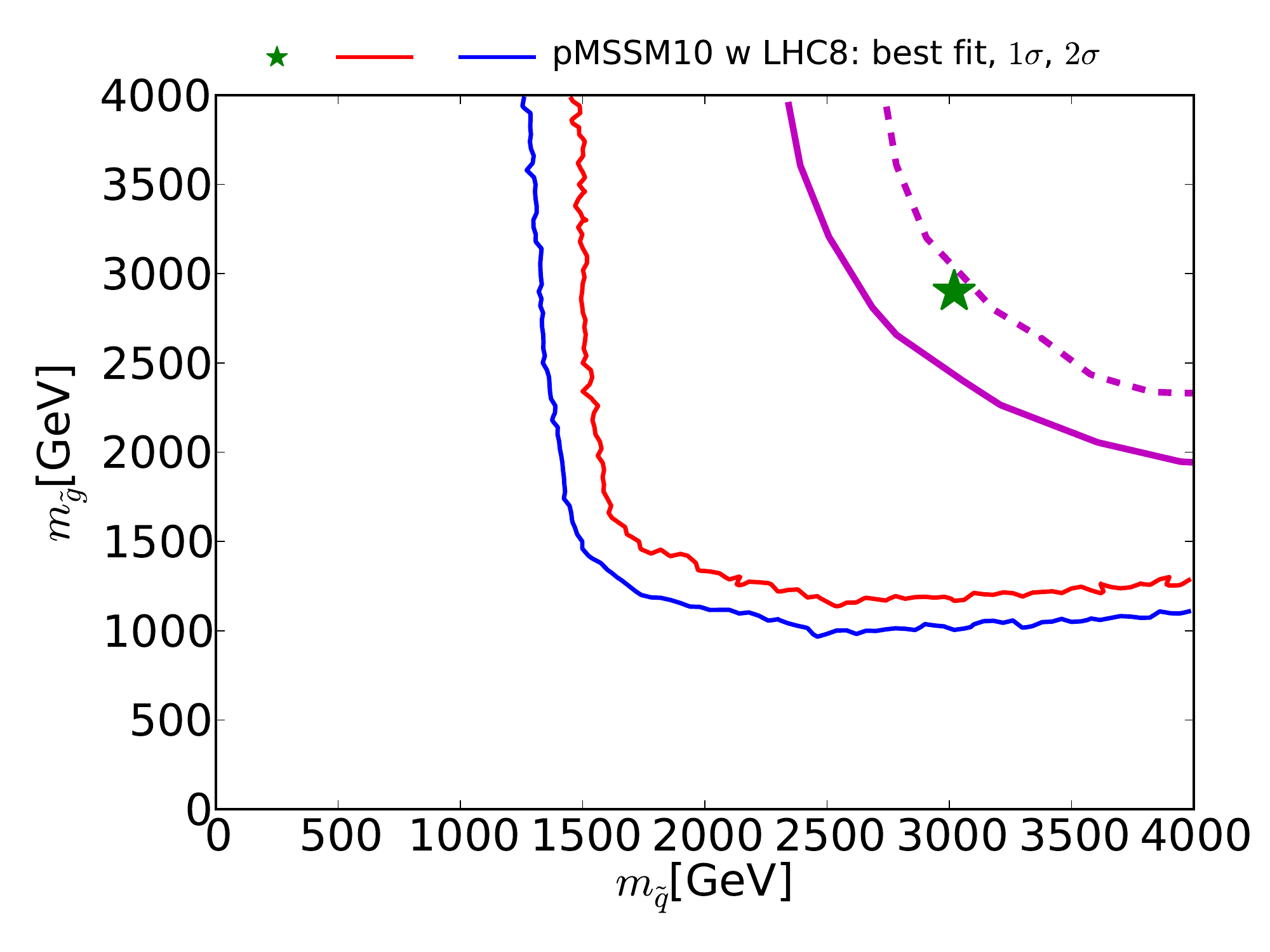}}
\vspace{-0.5cm}
\caption{\it The $(m_{\tilde q}, \mgl)$ plane with our 
68 and 95\% CL contours shown as solid red and blue lines, respectively,
and the best-fit point as a green star. Also shown as solid (dashed) magenta lines are the estimated
ATLAS sensitivities for 5-$\sigma$ discovery (95\% CL exclusion) of SUSY {via the generic
$\ETslash$ search} with 300~\ifb\ at 14\tev.}
\label{fig:colored-plane}
\end{figure*}

\begin{figure*}[htb!]
\resizebox{8.0cm}{!}{\includegraphics{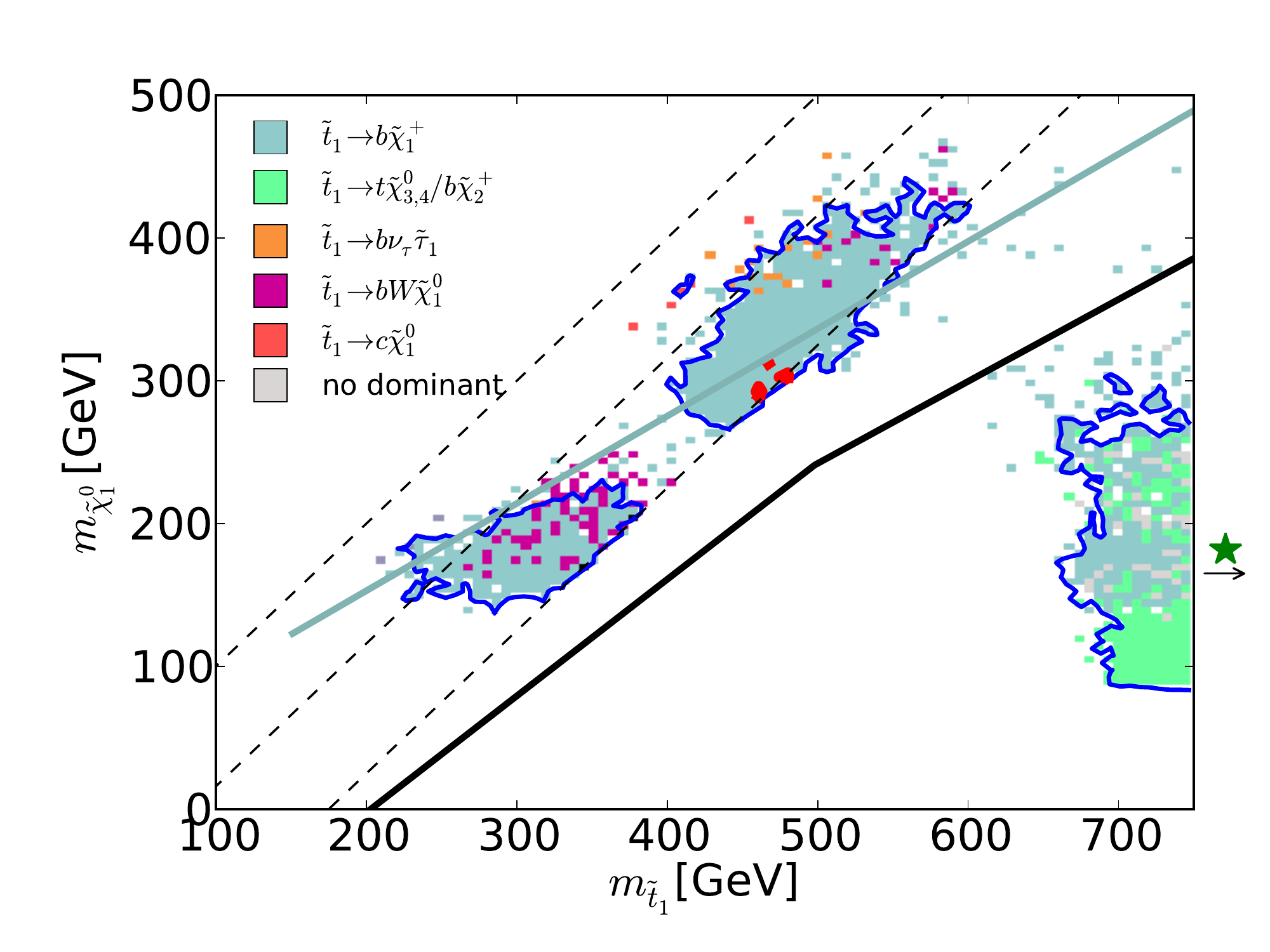}}
\resizebox{8.0cm}{!}{\includegraphics{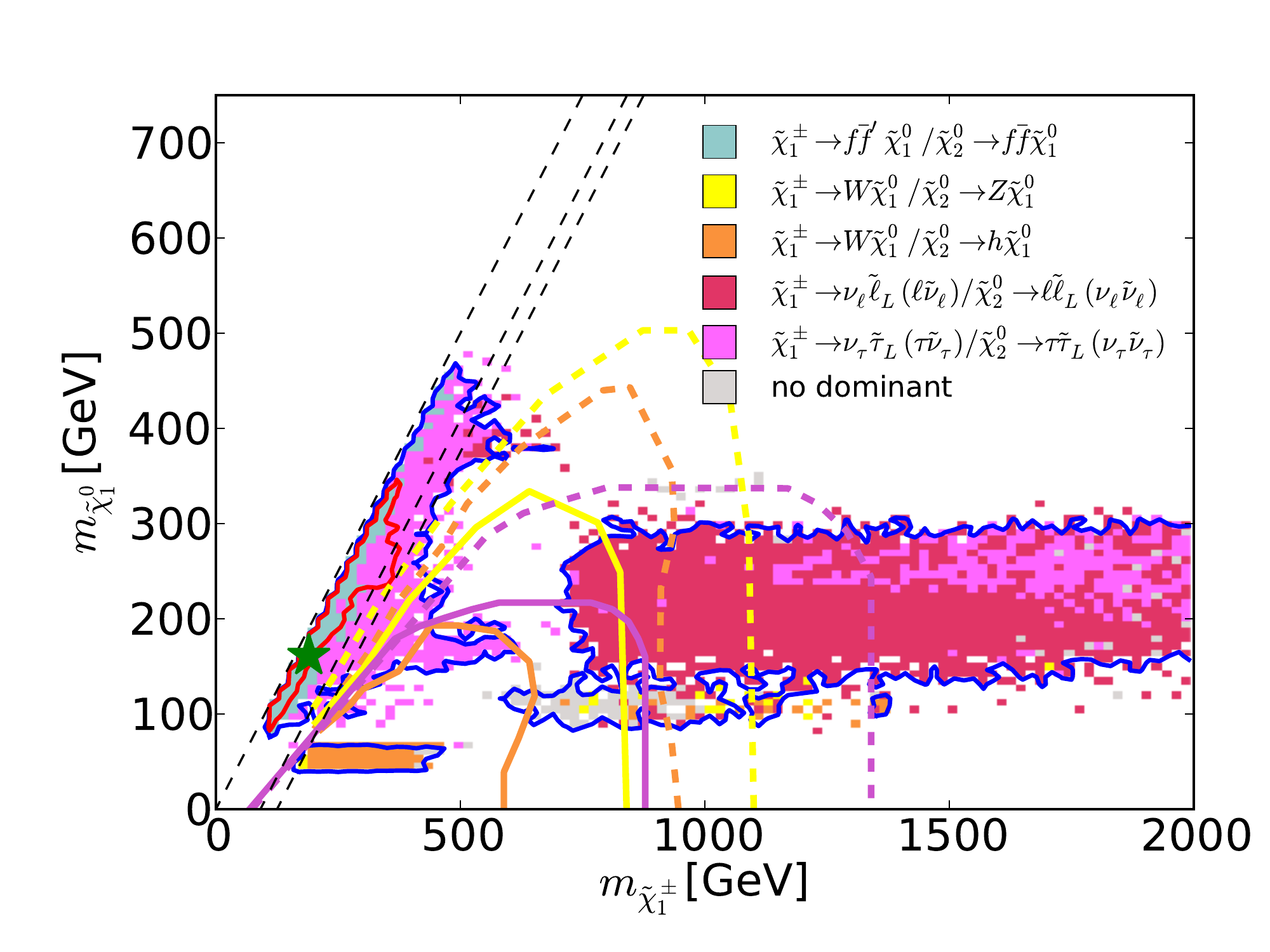}}  \\[1em]
\resizebox{8.0cm}{!}{\includegraphics{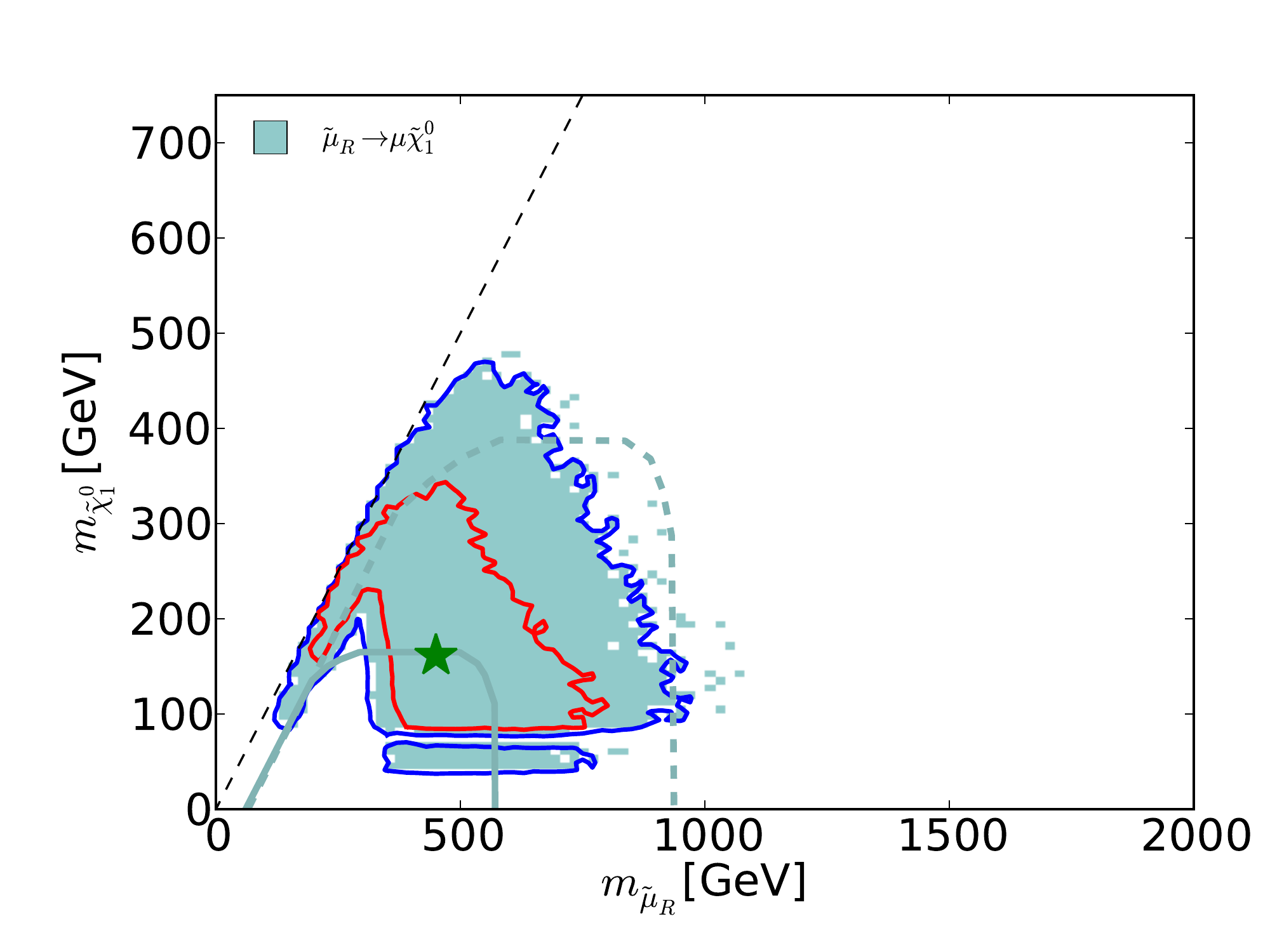}}
\resizebox{8.0cm}{!}{\includegraphics{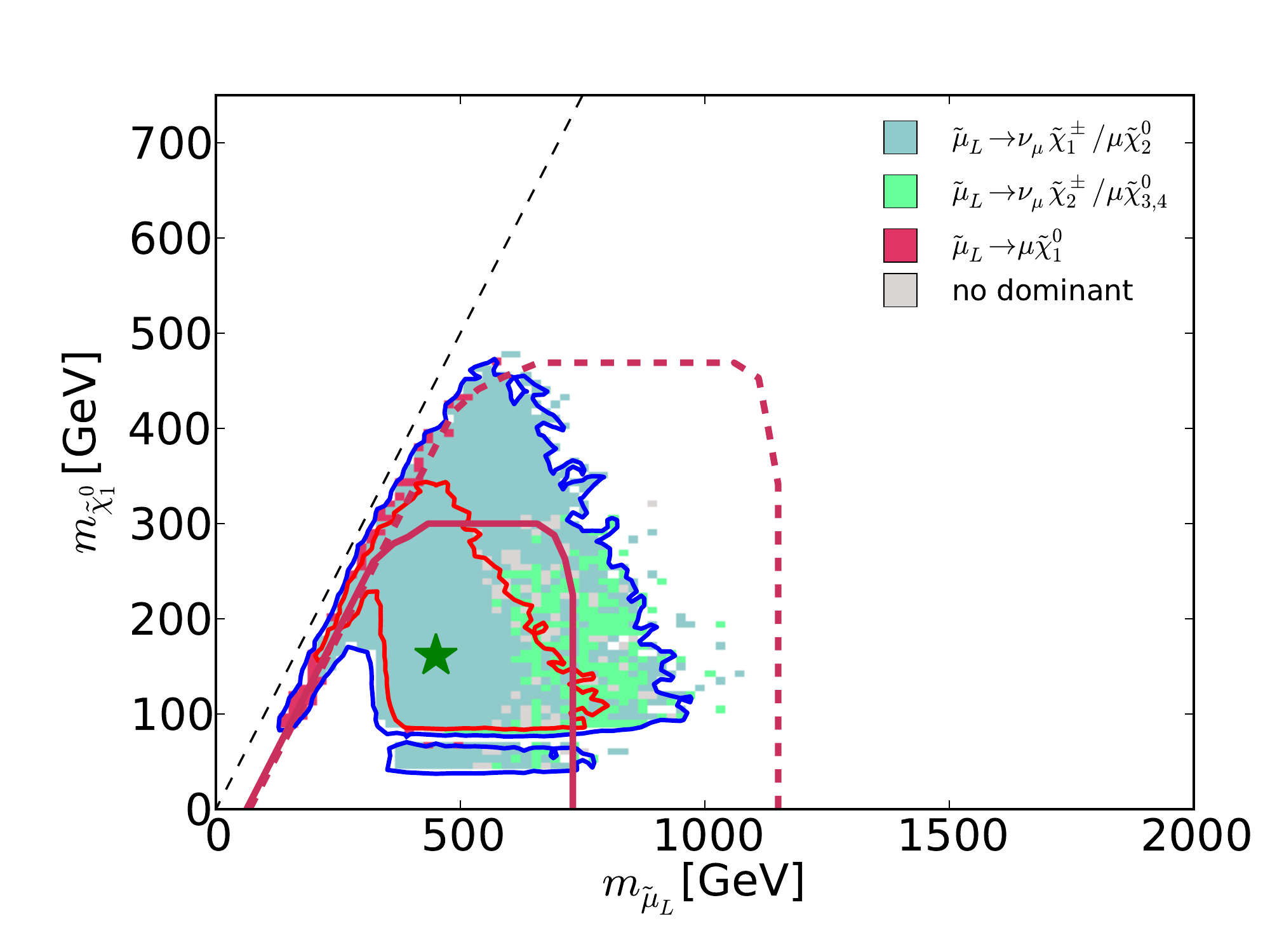}}
\vspace{-1.5cm}
\caption{\it Upper left panel: The $(\mstop1, \mneu1)$ plane with our 
68 and 95\% CL contours shown as solid red and blue lines, respectively, 
as well as coloured regions where the indicated branching ratios
exceed 50\%. The projected LHC sensitivity with 300~\ifb\ for $\sto1 \to \neu1 + t$ decays
is shown as a thick black line, 
and the corresponding
sensitivity for $\sto{1} \to \cha1 b$ decays is shown as a pale blue dashed line.
Upper right panel: The $(\mcha1, \mneu1)$ plane with our 
68 and 95\% CL contours shown as solid red and blue lines, respectively.
The shadings indicate where the branching ratios exceed 50\%.
Also shown as solid (dashed) yellow/orange/purple lines are the
projected LHC 95\% \cls\ exclusion reaches
for associated $\cha{1}$ and $\neu2$ production with decays via $W/Z$/$W/h$/${\tilde \ell_L}/{\tilde \nu_{\ell_L}}$/${\tilde
  \tau_L}/{\tilde \nu_{\tau_L}}$ with 300 (3000) \ifb\ of data if these
decays are dominant. 
Lower left panel: The $(\msmu{R}, \mneu1)$ plane with our 
68 and 95\% CL contours shown as solid red and blue lines, respectively, with pale blue shading showing also
where the  branching ratio for $\smu{R} \to \mu \neu1$ is dominant,
typically $\gtrsim 90\%$. The solid (dashed)
pale blue lines show our estimates of the LHC 95\% exclusion reach with 300 (3000)~\ifb.
Lower right panel: Similarly for the $(\msmu{R}, \mneu1)$ plane, displaying the regions where
the $\smu{L} \to \mu \neu1, \mu \neu2/\nu_\mu \cha1$ or $\mu \neu4/\nu_\mu \cha2$ decay modes
have branching ratios exceeding 50\%. The red lines indicate the
95\% exclusion reach with 300 (3000)~\ifb\
if $\smu{L} \to \mu \neu1$ were dominant, but are also indicative for the
decay into $\cha1/\neu2$ that have masses nearly degenerate with $\neu1$.} 
\vspace{1em}
\label{fig:other-planes}
\end{figure*}

\begin{figure*}[htb!]
\resizebox{8.0cm}{!}{\includegraphics{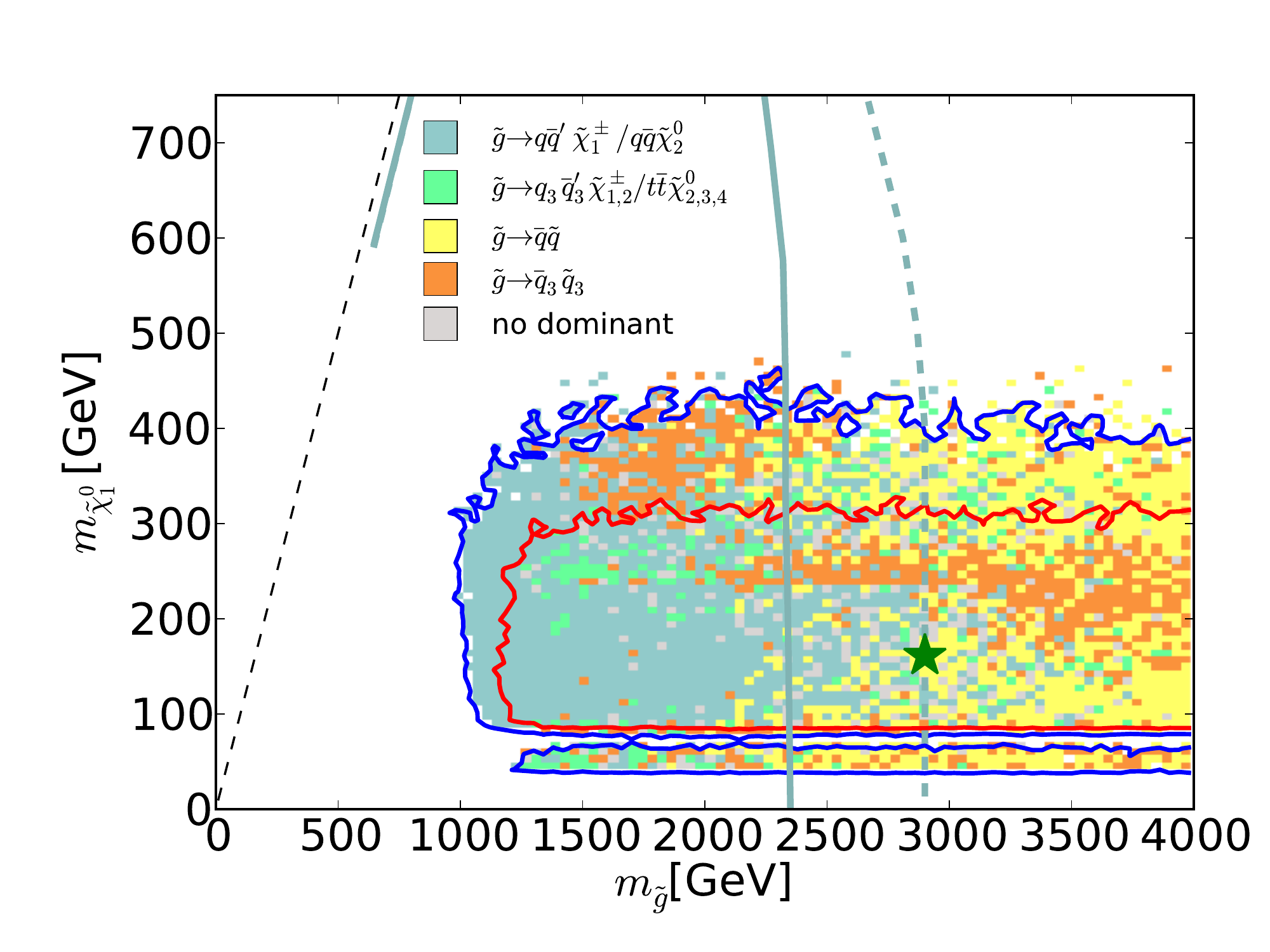}}
\resizebox{8.0cm}{!}{\includegraphics{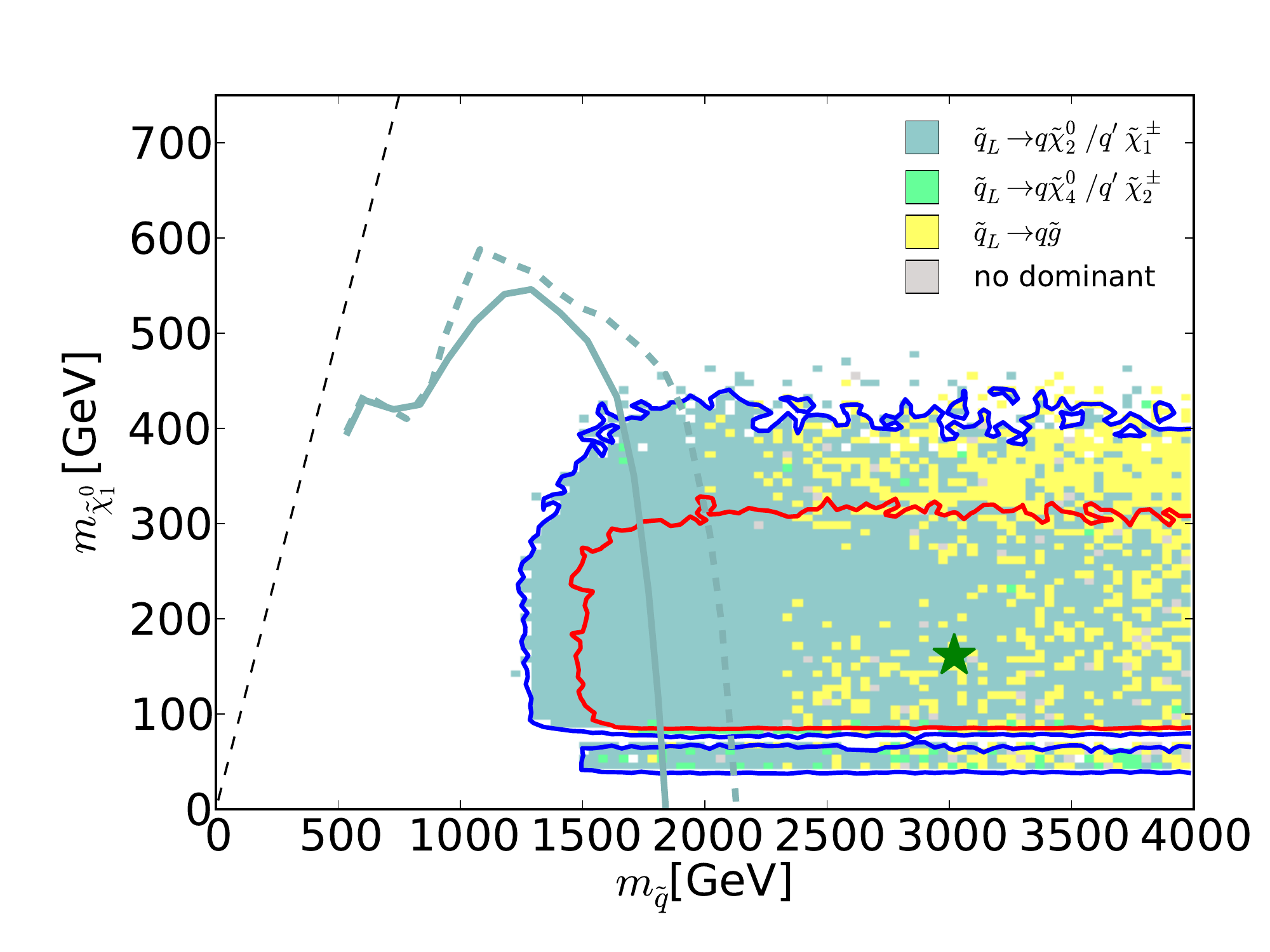}}  \\[1em]
\caption{\it As in Fig.~\protect\ref{fig:other-planes}, but for ${\tilde g}$ decays (left panel)
and ${\tilde q}$ decays (right panel). The pale blue solid (dashed) lines show the estimated sensitivities
with 300~\ifb (3000~\ifb) for
(left panel) ${\tilde g} \to q {\bar q}^\prime \cha{1}, q {\bar q} \neu{2}$ and (right panel)
${\tilde q_L} \to q \neu{2}, q^\prime \cha{1}$.}
\vspace{1em}
\label{fig:other-planes2}
\end{figure*}

At the time of writing, the LHC is starting Run~2, taking data at 
$13 \tev$, and
it is expected that an integrated luminosity of 300~\ifb\ will be collected by
the early 2020s. There are also plans for a subsequent high-luminosity upgrade
to accumulate 3000~\ifb.
In this Section we describe some prospects for future direct LHC searches for sparticles by
ATLAS and CMS that follow from our analysis of the pMSSM10.

With the increase of the LHC centre-of-mass energy from 8~TeV to 13~TeV for Run~2, there will be large
increases in the reaches for high-mass sparticle states.
As shown in Fig~\ref{fig:planes}, gluino masses $\sim 1.25$~TeV (top left panel) and first-
and second-generation squark masses $\sim 1.5$~TeV (top right panel) are
within our 68\% CL region. These masses will be probed by ATLAS and CMS
with just a few \ifb\ of data, demonstrating that already in an  early phase 
of Run~2 the
discovery of SUSY might well be possible. For third-generation squarks,
it is important to point out that besides masses of $\sim 800$~GeV for $\stopone$ (middle left panel) 
and $\sim 1$~TeV for sbottoms (middle left panel), we also find in our 95\% CL
region masses that are $\sim 200$ to 600~GeV in the compressed stop region and
$\sim 500$~GeV for sbottoms. These regions have not been excluded by the LHC searches so far,
but should become partly accessible in the first years of 13-TeV operation. 
As we comment later, in the cases of {compressed-spectrum} charginos (bottom left panel) and sleptons (bottom right panel)
comprehensive coverage of the preferred parameter space in the pMSSM10 by the LHC experiments
will be challenging. However, depending on the decay modes of the electroweakly produced sparticles,
early discovery at 13~TeV might also be possible.

Turning to the long-term prospects for the LHC, the ATLAS Collaboration has made physics studies that explore the discovery
and exclusion reach of ATLAS with 300 and 3000~\ifb\ at 14~TeV: 
see Fig.~13 of~\cite{atlasrun2}. 
In \reffi{fig:colored-plane} we display in the 
$(m_{\tilde q}, \mgl)$ plane our 68\% (95\%) {CL }contours in red
(blue) as well as the estimated 5-$\sigma$ discovery (95\% \cls\ exclusion) sensitivity with
300~\ifb\ as solid (dashed) magenta contours~%
\footnote{The 5-$\sigma$ discovery
contour for 3000~\ifb\ is almost coincident with the 95\% exclusion
contour for 300~\ifb.}%
. This shows that a substantial region of our preferred parameter space,
including our best-fit point, is within reach of
future LHC runs. However, we recall that the position of our best-fit
point in the $(m_{\tilde q}, \mgl)$ plane is rather poorly determined.

{In the following we revisit the mass planes of \reffi{fig:planes},
assessing carefully the decay modes of the respective SUSY particles. 
A recurring theme is that the $\cha1$ and $\neu2$ are nearly degenerate in
mass with $\neu1$ in the 68\%~CL region, so that squarks and sleptons decay via
$\cha1$ or $\neu2$ in large fractions of the preferred parameter space.
This general scenario is consistently indicated using pale blue shading.

With this in mind we turn
to \reffi{fig:other-planes}, where in the upper left panel we explore
the possible future LHC sensitivity to direct stop production in the
compressed-spectrum region. 
As previously, our present 68\% (95\%) CL contours are shown in red (blue).
The colour shadings code the regions where the corresponding
branching ratio, shown in the legend, exceeds 50\% for the point at each
location that minimises the $\chi^2$ function over the
remaining parameters, {and the thin diagonal dashed black lines correspond to
$\Delta m \equiv \mst{1} - \mneu{1} = 0, M_W + m_b$ and $m_t$.}
The solid dashed black lines show the projected LHC 95\%~\cls\ exclusion 
sensitivities for $\sto1 \to \neu1 t$
decays with 300~\ifb~\cite{ATL-PHYS-PUB-2013-011} (similar sensitivity is found in this region
with 3000~\ifb).
These do not cover the case of a compressed spectrum region, which
includes the 95\% CL region where the dominant $\sto1$ decays are to
$\cha1 b$. Here we rescale from the present 95\%~\cls\ limit from the dibottom
analysis, assuming that $\mcha1 - \mneu1 \sim 5 \gev$ and using the {\tt Collider Reach}
tool~\cite{CRtool} to rescale the production cross-section, 
and assume that future LHC searches maintain the same search
performance, i.e., 
the same signal yield after the event selection as present searches. We see that
a search with 300~\ifb\ of data (pale blue line) would already cover part of
the 95\% CL region in the compressed-spectrum region, and the estimate for 3000~\ifb\ is similar.

In the upper right panel of \reffi{fig:other-planes}, we explore
the possible future LHC sensitivity in
the $(\mcha1, \mneu1)$ plane~%
\footnote{We
recall that in the part of this region favoured at the 68\% CL
the $\neu1 - \neu2$ and $\neu2 - \cha1$ mass differences are
both small.}%
, where our current 68\% (95\%) {CL} contours are again shown in red (blue),
and the best-fit point is indicated by a green star, and the thin diagonal dashed black lines correspond to
$\Delta m \equiv \mcha{1} - \mneu{1} = 0, M_Z$ and $\Mh$.
We use colour-coding to display points in the $(\mcha1, \mneu1)$ plane
where the following $\cha1/ \neu2$ decay modes have branching ratios $> 50$\%: 
via virtual bosons $\tilde \chi^\pm_1\to f \bar f^\prime \neu{1}  / \tilde \chi^0_2 \to f \bar f \neu{1}$ (pale blue),
via on-shell bosons $\tilde \chi^\pm_1\to W \tilde \chi^0_1 / \tilde \chi^0_2 \to Z \tilde \chi^0_1 $ (yellow) or
$\tilde \chi^\pm_1\to W \tilde \chi^0_1 / \tilde \chi^0_2 \to h \tilde \chi^0_1 $ (orange),
via sleptons
$\tilde \chi^\pm_1\to \nu_\ell \tilde \ell_L (\ell \tilde \nu_\ell)/ \tilde \chi^0_2 \to \ell \tilde \ell_L ( \nu_\ell\tilde \nu_\ell)$ where $(\ell = e, \mu)$ (red)
and
$\tilde \chi^\pm_1\to \nu_\tau \tilde \tau_L (\tau \tilde \nu_\tau)/ \tilde \chi^0_2 \to \tau \tilde \tau_L ( \nu_\tau\tilde \nu_\tau)$ (purple), whereas
points with no
branching ratios exceeding 50\% are coloured grey. 
The ATLAS Collaboration has made available projections of its
sensitivities for some relevant searches for associated $\cha{1}$ and $\neu2$ production with 300 (3000) \ifb\ of
data~\cite{ATL-PHYS-PUB-2014-010}, which are also shown in the upper right plane of
\reffi{fig:other-planes} as solid (dashed) contours in the same colours as the
relevant decay modes: 
yellow for $\cha1\neu2$ via $WZ$,
orange for $\cha1\neu2$ via $Wh$, 
red for $\cha1\neu2$ via ${\tilde \ell_L}/\tilde \nu_\ell$ 
where $(\ell = e, \mu)$, 
and purple for $\cha1\neu2$ via ${\tilde \tau_1}/\tilde \nu_\tau$.
With 300 \ifb\ of data the $Wh$ search should already cover
essentially all of the 95\% CL island with $\mneu1 \lesssim 80 \gev$, where
these branching ratios exceed 50\%~\footnote{However, it should be kept in mind that the projected exclusion
regions always assume the relevant branching ratios to be equal to
one, and real exclusion bounds with ``mixed'' branching ratios would
in general be different.}.
However, even with 3000 \ifb\ of data these searches would have limited
impact on the other 95\% CL regions, since there the branching ratios
for these decays are typically small.
{More importantly,} they would have no impact on
the 68\% CL region. The most relevant searches in the region with
near-degenerate $\cha1, \neu2$ and $\neu1$ would be in hadronic final
states sensitive to 
$\tilde \chi^\pm_1\to f \bar f^\prime \neu{1}   / \tilde \chi^0_2 \to f \bar f \neu{1}$ and 
$\tilde \chi^\pm_1\to \nu_\tau \tilde \tau_L (\tau \tilde \nu_\tau)/ \tilde \chi^0_2 \to \tau \tilde \tau_L ( \nu_\tau\tilde \nu_\tau)$.}
Searches for compressed charginos/neutralinos have been explored in \cite{Schwaller:2013baa}, where some sensitivity was found
up to $\mcha1/\mneu2\lesssim 300\gev$ (after 3000~\ifb), 
although for very small mass differences $(\mcha1 \simeq \mneu2) - \mneu{1}\lesssim20\gev$ and with optimistic
assumptions about the possible systematic uncertainties.

The lower left panel of \reffi{fig:other-planes} provides
information about the dominant branching ratio for the $\smu{R}$ in 
the favoured region of the $(\msmu{R}, \mneu1)$ plane:
as in the previous panels, our current 68\% (95\%) CL contours are in red (blue).
We see that the branching ratio for $\smu{R} \to \mu \neu1$ exceeds 
{50\% in all of the 95\%~CL region}.
We also show projections of the possible future sensitivities of the LHC
with 300 (3000) \ifb\ of data to $\smu{R} \to \mu \neu1$ decay as solid
(dashed) pale blue lines. These projections were obtained via the
following steps: 1) the present LHC 95\%~\cls\ limit for large
$\msmu{R}/\mneu1$ was rescaled using the {\tt Collider Reach}
tool~\cite{CRtool} to estimate the $\smu{R}$ production cross-section, 
and assuming that future LHC searches maintain the same search
performance as present searches, and 2) we
assumed that the shapes of the future sensitivity curves for other
values of $\msmu{R}/\mneu1$ would be the same as for the current searches.
We see that with 300 \ifb\ the LHC would already explore a substantial
part of the current 68\% CL region in the $(\msmu{R}, \mneu1)$ plane,
and that most of the 95\% CL region could be explored with
3000~\ifb\ but missing a narrow band where $m_{\tilde \mu_r} - \mneu1$
is small. 

The favoured region of the $(\msmu{L}, \mneu1)$
plane, shown in the lower right panel of \reffi{fig:other-planes}, looks similar, but the dominant decay modes are more
varied: any of the decays 
$\smu{L} \to \mu \neu1,~\nu_\mu \cha1/\mu \neu2$ or $\nu_\mu \cha2/\mu \neu4$
have branching ratios exceeding 50\%. However, the $\smu{L} \to \mu \neu1$
decay mode dominates only when $\msmu{L} - \mneu1$ is very small.
We have used the same
approach as used above for projecting the $\smu{R}$ sensitivity to the $\mu \neu1$ decay mode also to
estimate the future $\smu{L}$ sensitivity, as shown by the solid (300~\ifb) and
dashed (3000~\ifb) deep red lines.
{Although this projection assumes decays directly into $\neu1$, it
may have similar sensitivity to the decay into $\cha1/\neu2$.

Fig.~\ref{fig:other-planes2} displays the corresponding $(m_{\tilde g}, \mneu{1})$ and
$(m_{\tilde q}, \mneu{1})$ planes. In the ${\tilde g}$ case (left panel), we see that a number
of different decay modes may have a branching ratio exceeding 50\%: 
via off-shell first- and second-generation squarks into charginos/neutralinos $\tilde g\to q   \bar q^\prime \tilde \chi^\pm_1 / q \bar q\tilde \chi^0_2 $ (pale blue),
through off-shell third-generation squarks into (heavier) charginos/neutralinos $\tilde g\to q_3 \bar q_3^\prime \tilde \chi^\pm_{1,2} / t \bar t \tilde \chi^0_{2,3,4} $ (green),
into first- and second-generation squarks $\tilde g\to \bar q \tilde q $ (yellow) 
or third-generation squarks $\tilde g\to \bar q_3 \tilde q_3$ (orange). However, 
none of these decay modes may have a branching ratio exceeding 50\% (grey).
In the ${\tilde q}$ case (right panel), the decays of ${\tilde q}\to q^\prime\cha{1}/q \neu{2}$
(pale blue) are usually dominant, particularly at lower masses, though in some cases decays into gluinos $\tilde q_L\to q\tilde g  $ and at low $\mneu{1}$ decays
into heavier neutralinos/charginos $\tilde q_L\to q\tilde \chi^0_4 / q^\prime\tilde \chi^\pm_2$ (pale green) are dominant.
The ATLAS Collaboration has presented projected exclusion limits for
$\tilde g \to q \bar q \neu1$ and $\tilde q \to q \neu1$ simplified models 
with 300 (3000) fb$^{-1}$ of data at 14 TeV
in \cite{ATL-PHYS-PUB-2014-010}.
Recalling that $m_{\cha1} (m_{\neu2}) \simeq m_{\neu1}$ in our 68\% CL region,
these simplified model limits may be applicable in the pale blue regions in Fig.~\ref{fig:other-planes2}
where $\tilde g \to q \bar q^\prime \cha1 / q \bar q \neu2$ ($\tilde q \to q' \cha1 / q \neu2$)
dominate the gluino (squark) decay modes.
We overlay these projected limits in Fig.~\ref{fig:other-planes2}.
The solid (dashed) curves corresponds to the 300 (3000) fb$^{-1}$ data. 
In the ($m_{\tilde g}$, $m_{\neu1}$) plane we can see that 
a large part of our 68\% and 95\% CL regions can be probed with 300 (3000) fb$^{-1}$ data.
Indeed, our best-fit point lies on the projected limit for 3000 fb$^{-1}$.
We also see in the ($m_{\tilde q}$, $m_{\neu1}$) plane that
some parts of our 68\% and 95\% CL regions can be explored with 300 (3000) fb$^{-1}$ data,
although the projected limit presented in \cite{ATL-PHYS-PUB-2014-010} with 3000 fb$^{-1}$ data
does not reach our best-fit point.

{Finally, we turn the prospects for discovery with 300~\ifb\ of our benchmark points, starting with 
our global best-fit point (\reffi{fig:bestfitspectrum}),
which is just inside the reach of generic $\ETslash$ searches (see \reffi{fig:colored-plane}),
well in reach for the slepton searches (lower panels of \reffi{fig:other-planes}), 
and even potentially within reach of the compressed-chargino/neutralino searches, 
as discussed in \cite{Schwaller:2013baa}, due to its small mass splitting $\mcha1-\mneu2\simeq20\gev$.
}

As for our local best-fit point in the low-$\mst1$ region (see the upper left panel of \reffi{fig:benchmarkspectra}),
it lies just within the reach of future searches in the compressed-stop region
(upper left panel of \reffi{fig:other-planes}), as well as
slepton searches (lower panels of \reffi{fig:other-planes}), but would be
difficult to access via chargino/neutralino searches, because of the low
mass splittings seen in \reffi{fig:benchmarkspectra} and the relatively high $\mneu1\simeq300\gev$. 
The relatively large contribution of the
\lhccol\ constraint for this point, seen in the third row of \refta{tab:breakdown},
indicates that this point may be accessible via jets + $X$ + \ETslash\ searches early in Run~2.

In the cases of the low-$\msq$ and/or -$\mgl$ points, by construction these points could
also be discovered early in Run~2 of the LHC, since they lie very close to the 
current 68\% CL boundary in the $(\msq, \mgl)$ plane shown in \reffi{fig:colored-plane}.
This feature is also indicated by the significant contributions to the global $\chi^2$
functions for these points that can also be seen the third row of \refta{tab:breakdown}.


\boldmath
\section{Prospects for Sparticle Detection at a Future \boldmath{$e^+ e^-$} Collider}
\label{sec:e+e-prospects}
\unboldmath

\begin{figure*}[htb!]
\vspace{-0.25cm}
\resizebox{8cm}{!}{\includegraphics{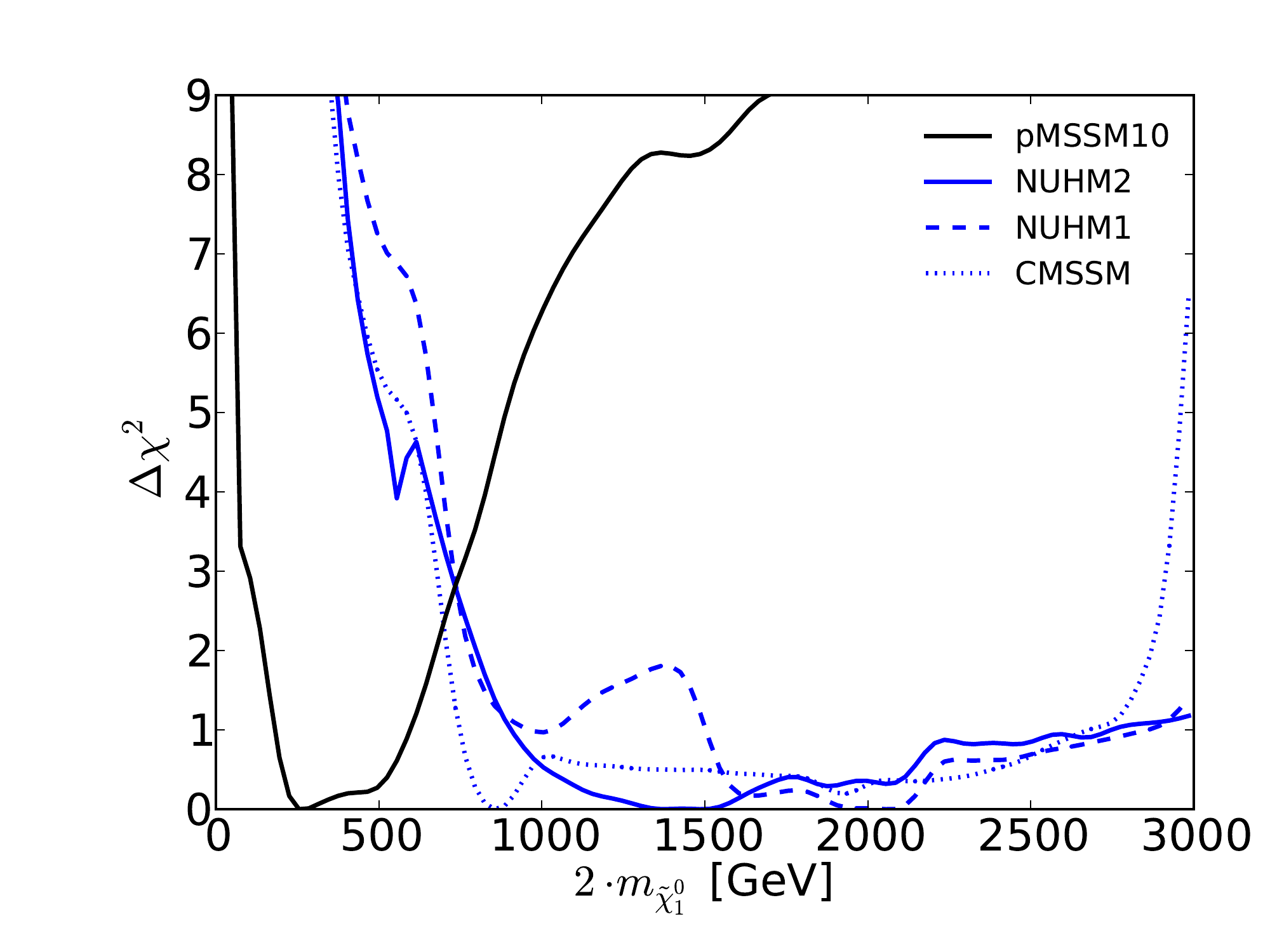}}
\resizebox{8cm}{!}{\includegraphics{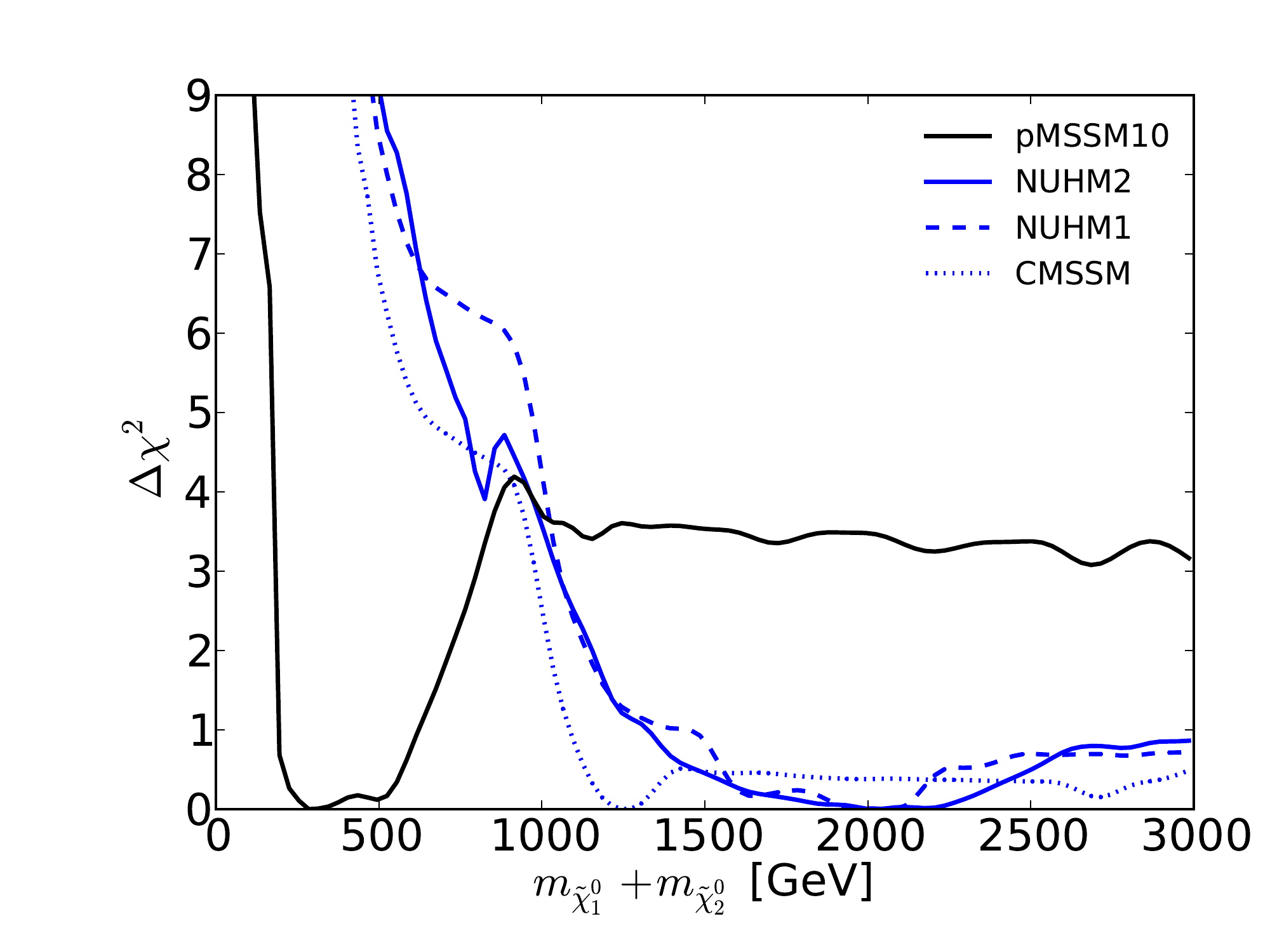}}  \\[1em]
\resizebox{8cm}{!}{\includegraphics{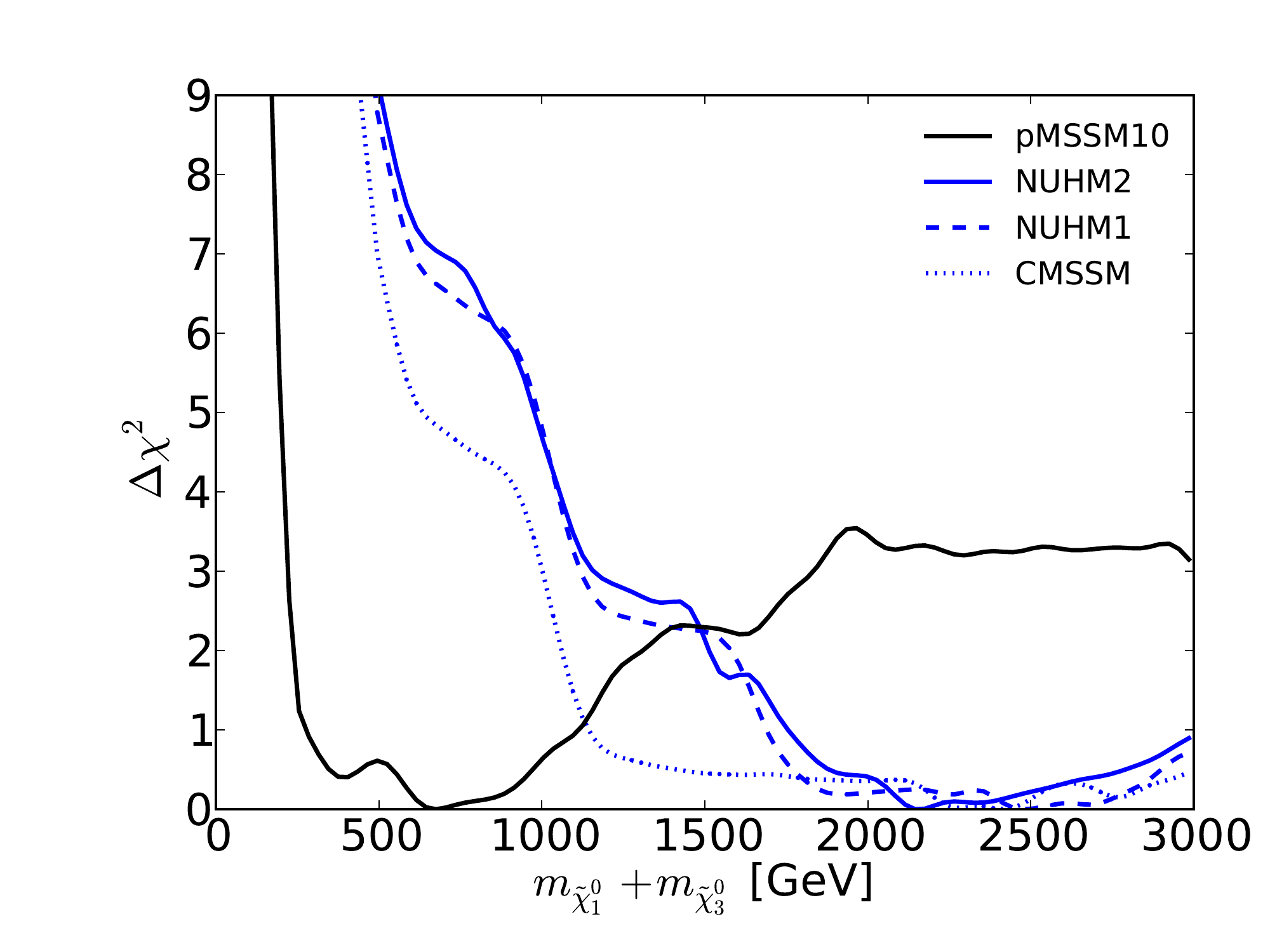}}
\resizebox{8cm}{!}{\includegraphics{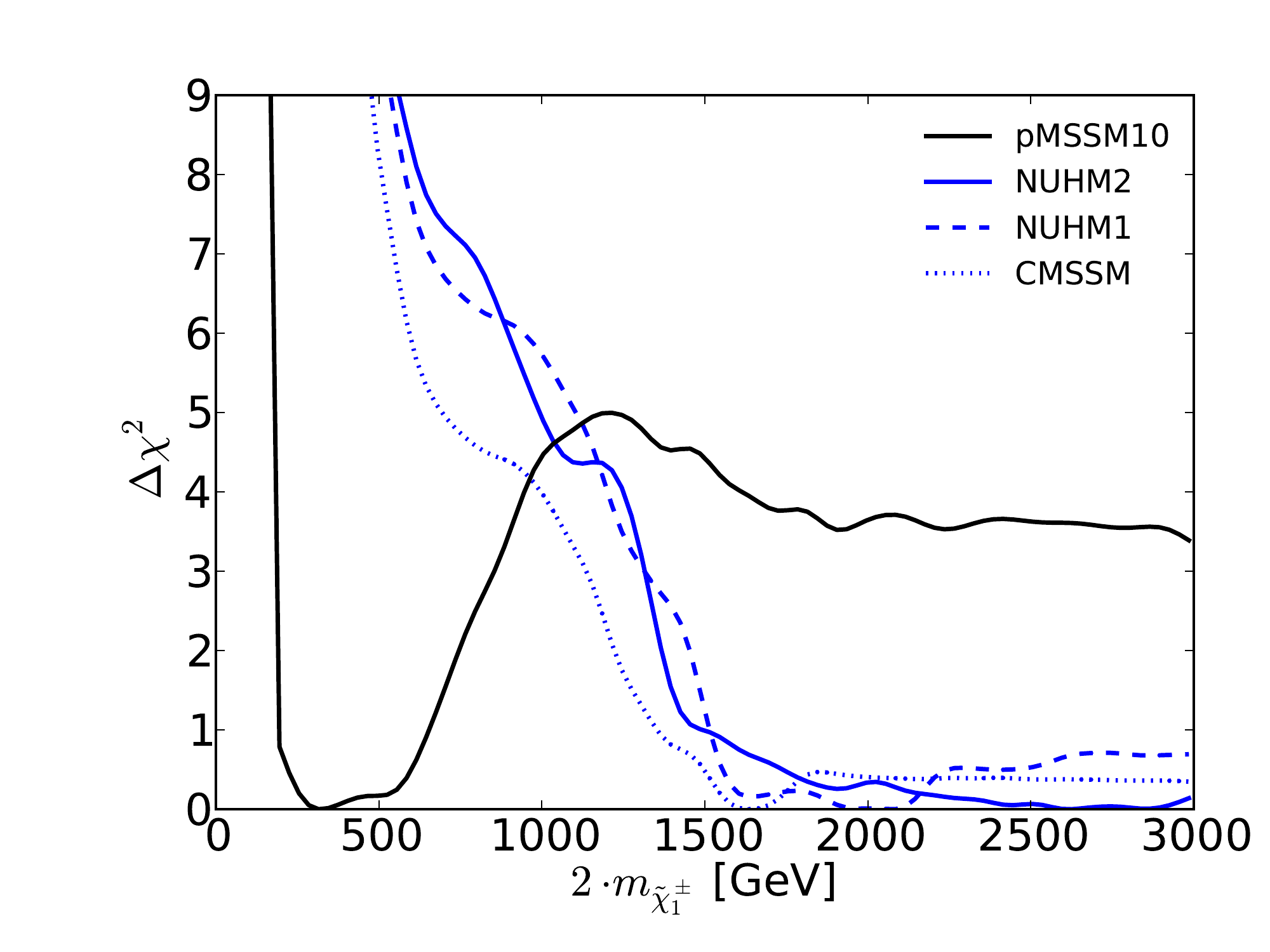}} \\
\vspace{-0.75cm}
\caption{\it The one-dimensional profile likelihood functions for
various thresholds in $e^+ e^-$ annihilation.
Upper left panel: The threshold for $\neu1 \neu1$ production.
Upper right panel: The threshold for associated $\neu1 \neu2$ production.
Lower left panel: The threshold for associated $\neu1 \neu3$ production.
Lower right panel: The threshold for $\cha1 \champ1$ production.}
\label{fig:e+e-chi2}
\end{figure*}

\reffi{fig:e+e-chi2} displays the one-dimensional $\chi^2$ functions for
the lowest particle pair- and associated {chargino and neutralino} production
thresholds in $e^+ e^-$ 
annihilation in the pMSSM10 (black), compared with their counterparts in the
CMSSM (dotted blue), NUHM1 (dashed blue) and NUHM2 (solid blue). In the cases
of $\neu1 \neu1$ (upper left panel), $\neu1 \neu2$ (upper right panel) and
$\cha1 \champ1$ (lower right panel) production, we see that the minima
of the 
$\chi^2$ functions in the pMSSM10 lie within reach of an $e^+ e^-$ collider
with centre-of-mass energy 500~GeV, and that threshold locations
favoured by $\Delta\chi^2 \le 3$
would be within reach of a 1000~GeV collider, whereas no upper limit can
be established at the 95\% CL. We also see
that, in the case of $\neu1 \neu3$ production (lower left panel)
(which is very similar to the cases of $\neu1 \neu4$, $\neu2 \neu3$ 
and $\cha1 \champ2$ production that we do not show) 
the minimum of the global $\chi^2$ function for the threshold
lies between 400~GeV and 
1000~GeV, again with no upper limit at the 95\% CL.
It should be noted, however, that the optional anti-tachyon cut would indeed yield
upper limits at the 95\% CL for those production modes.
Referring back to the bottom right panel of Fig.~\ref{fig:onedimensional}
and the right panel of Fig.~\ref{fig:smuon}, we see that slepton pair-production
thresholds may well also lie below 1000~GeV.
In all cases, the expected locations of the thresholds in the CMSSM,
NUHM1 and NUHM2 are at much higher centre-of-mass energies.

Thus, the accessibility of supersymmetric
particles at $e^+e^-$ colliders is vastly different in the pMSSM10 and
similar non-GUT models, as compared to the simplest GUT-based models.


\section{Conclusions}
\label{sec:summary}

We have performed in this paper the first global likelihood analysis of the
pMSSM using a frequentist approach that includes comprehensive treatments of the LHC8
constraints.
This analysis required many developments
and extensions of the {\tt MasterCode} framework that are described in earlier sections of the paper.
For example, in order to interpret the searches for coloured sparticles via jets + $X$ + $\ETslash$
signatures at LHC8, we combine searches sensitive to a variety of different cascade channels, 
whose relative probabilities depend on other model parameters. By combining a
sufficiently complete set of channels~\cite{Buchmueller:2013exa}, we capture essentially all the relevant decay channels,
and so achieve a reliable \lhccol\ constraint. In the cases of the \lhcewk\ constraints from searches for electroweak gauginos,
Higgsinos and leptons, we constructed computationally-efficient models for their contributions
to the global likelihood function that mimic closely the more computationally-intensive results
from the {\tt Atom} code. A similar procedure was used for the \lhcstop\ constraints from searches
for models with compressed stop spectra, with the addition that we constructed the likelihoods for
some simplified model searches using the {\tt Scorpion} code. These procedures have all been
validated extensively, as described in the text.

The results of our analysis of the pMSSM10 are described in Section~3, where we provide many details
of the global likelihood function. We give there the parameters of our best-fit pMSSM10 point, while
cautioning that its squark and gluino mass parameters are poorly constrained. On the other hand, some
of the pMSSM10 parameters in the electroweak sector are relatively tightly constrained. For example,
we find relatively narrow ranges of $\neu{1}$ and slepton masses,
which are quite light, and that $\mneu{1} \simeq \mneu{2} \simeq \mcha{1}$ in the region of parameter space that is
preferred at the 68\% CL. The light spectrum of electroweakly-interacting sparticles
is preferred by the \gmt\ constraint, and the neutralino and chargino mass degeneracies are then required to obtain a 
satisfactory cold dark matter density. In addition to the best-fit point, we have presented and analyzed
several alternative pMSSM10 points with low stop, squark and gluino masses that may serve as
benchmarks for LHC Run~2 analyses~\footnote{SLHA files~\cite{SLHA} for these points can be downloaded from
the MasterCode website~\cite{mcweb}.}.

One of the most striking features of our analysis is that the pMSSM10 can provide an excellent
fit to \gmt\ while respecting all the LHC8 constraints, something that is not possible in models with
universal soft supersymmetry-breaking terms at the GUT scale, such as the CMSSM, NUHM1 and NUHM2.
A corollary is that there are interesting prospects for exploring the preferred region of the pMSSM10
parameter space in future experiments. For example, LHC searches at 14~TeV have excellent
prospects for exploring the preferred regions of $m_{\tilde q}$ and $m_{\tilde g}$,
as well as light ${\tilde t_1}$, ${\tilde e}$ and ${\tilde \mu}$ masses.
Looking further ahead, the \gmt-friendly regions of the pMSSM10 could be explored in detail with
an $e^+ e^-$ collider operating at 500 to 1000~GeV in the centre of mass. 
{In particular, such a machine would have a significant discovery
potential in the preferred region for the lightest neutralino and chargino,
while those states would be difficult to access at the LHC with
the searches discussed in this paper.}
Also, we recall that
the region of the pMSSM10 parameter space that is favoured at the 68\% CL
after implementing the LHC8 constraints yields relatively large
values of \ssi\ that should be accessible to forthcoming experiments: see the right panel of Fig.~\ref{fig:mneu-ssi}.

It is a characteristic of the pMSSM that the possibility of extrapolation to high renormalisation scales is not
enforced, and indeed we find that most of our pMSSM10 parameter sets yield some tachyonic
sfermion masses at high renormalisation scales. It is not clear that such models should be rejected
out of hand~\cite{EGLOS}, but it is reassuring that many features of our pMSSM10 fit would, nevertheless, be
preserved if one required the absence of tachyons. On the other hand, the preferred region of
the pMSSM10 parameter space has non-universal gaugino and sfermion masses. The former
arise from the tension between \gmt\ (which favours small $M_{1, 2}$) and the \lhccol\ constraint
(which favours larger $M_3$) as well as the dark matter constraint (which favours $M_1 \simeq M_2$
at the electroweak scale, not at the GUT scale). In parallel, sfermion mass non-universality also arises
from the tension between \gmt\ (which favours small $m_{\tilde \mu}$) and the \lhccol\ constraint
(which favours large squark masses).

It would be desirable to extend our approach to more general variants of the pMSSM with fewer
restrictions on the parameters. For example, it would be
interesting to relax the assumption of a single 
slepton mass scale: this is unlikely to alter the preferred range
of the ${\tilde \mu_{L,R}}$, but would have important repercussions for dark matter density
calculations. It would also be desirable to revisit in more general pMSSM
scenarios the preferences we have found for neutralino and chargino mass degeneracies,
and the constraints we find in the $(\MA, \tb)$ plane, which are largely indirect (being due to
the interplay between constraints whose combination may have different implications in more
general pMSSM scenarios). However, we think that many features of our pMSSM10 analysis
would persist in more general scenarios.

Finally, when interpreting the impacts of experimental searches in our preferred pMSSM10 region, it is important to take
into account decay chains involving an intermediate chargino, which is required to be light
in order to fulfil the relic density constraint. In a large part of our preferred 
parameter space the chargino is almost mass degenerate with $\neu1$,
and there are also regions with a sizeable mass difference that exhibit distinctive decay chains. 
Therefore, the pMSSM10 motivates interpreting searches not only in terms of the
minimal decay chains of the simplified models presently being considered, but also with the $\cha1$
(and possibly also the $\neu2$) incorporated in the spectrum over a range of low masses.

We await with interest the verdict of future runs of the LHC.


\subsubsection*{Acknowledgements}

The work of K.J.dV., O.B., J.E., S.M., K.A.O. and K.S. is supported in part by
the London Centre for Terauniverse Studies (LCTS), using funding from
the European Research Council 
via the Advanced Investigator Grant 267352.
The work of R.C. is supported in part by the National Science Foundation under 
Grant No. PHY-1151640 at the University of Illinois Chicago and in part by Fermilab, 
operated by Fermi Research Alliance, LLC under Contract No. De-AC02-07CH11359 
with the United States Department of Energy.
This work of M.J.D.
is supported in part by the Australia Research Council. The work of J.E. is also supported in part by STFC
(UK) via the research grant ST/J002798/1. 
The work of S.H. is supported 
in part by CICYT (grant FPA 2013-40715-P) and by the
Spanish MICINN's Consolider-Ingenio 2010 Program under grant MultiDark
CSD2009-00064. The work of D.M.-S. is supported by FOM (NL) and by the European Research Council 
via Grant BSMFLEET 639068.
The work of K.A.O. is supported in part by DOE grant
DE-SC0011842 at the University of Minnesota. The work of G.W.\ is supported in 
part by the Collaborative Research Center SFB676 of the DFG, ``Particles, Strings and the early Universe", 
and by the European Commission through the ``HiggsTools" Initial Training Network PITN-GA-2012-316704. 



\begin{thebibliography}{99}


 \bibitem{ATLAS20}
G.~Aad {\it et al.}  [ATLAS Collaboration],
  arXiv:1405.7875 [hep-ex]; \\
{full ATLAS Run~1 results can be found at} \\ 
{\tt https://twiki.cern.ch/twiki/bin/view/} \\
{\tt AtlasPublic/SupersymmetryPublicResults}.   

\bibitem{CMS20}
S.~Chatrchyan {\it et al.}  [CMS Collaboration],
  JHEP {\bf 1406} (2014) 055
  [arXiv:1402.4770 [hep-ex]]; \\
{full CMS Run~1 results can be found at} \\
 {\tt https://twiki.cern.ch/twiki/bin/view/} \\
{\tt CMSPublic/PhysicsResultsSUS}.   

\bibitem{lhch}
G.~Aad {\it et al.}  [ATLAS Collaboration],
  Phys.\ Lett.\ B {\bf 716} (2012) 1
  [arXiv:1207.7214 [hep-ex]];
   S.~Chatrchyan {\it et al.}  [CMS Collaboration],
  Phys.\ Lett.\ B {\bf 716} (2012) 30 
  [arXiv:1207.7235 [hep-ex]].

 \bibitem{LHCbBsmm}
 R.Aaij {\it et al.}  [LHCb Collaboration],
 Phys.\ Rev.\ Lett.\  {\bf 111} (2013) 101805
 [arXiv:1307.5024 [hep-ex]].

 \bibitem{CMSBsmm}
  S.~Chatrchyan {\it et al.}  [CMS Collaboration],
 Phys.\ Rev.\ Lett.\  {\bf 111} (2013) 101804
 [arXiv:1307.5025 [hep-ex]].

 \bibitem{BsmmComb}
 R.Aaij {\it et al.}  [LHCb and CMS Collaborations],
LHCb-CONF-2013-012, CMS PAS BPH-13-007.

\bibitem{CMSLHCbBsmm}
  V.~Khachatryan {\it et al.}  [CMS and LHCb Collaborations],
  arXiv:1411.4413 [hep-ex], {\it accepted for publication in Nature}.

   \bibitem{HK}
  H.~P.~Nilles, Phys.\ Rept.\ {\bf 110} (1984) 1;
H.~E.~Haber and G.~L.~Kane,
  Phys.\ Rept.\  {\bf 117} (1985) 75.

  \bibitem{funnel}  
M.~Drees and M.~M.~Nojiri,
Phys.\ Rev.\ D {\bf 47} (1993) 376 [arXiv:hep-ph/9207234];
  H.~Baer and M.~Brhlik,
Phys.\ Rev.\ D {\bf 53} (1996) 597 [arXiv:hep-ph/9508321];
  Phys.\ Rev.\  D {\bf 57} (1998) 567
  [arXiv:hep-ph/9706509];
   H.~Baer, M.~Brhlik, M.~A.~Diaz, J.~Ferrandis, P.~Mercadante, P.~Quintana and X.~Tata,
    Phys.\ Rev.\  D {\bf 63} (2001) 015007
  [arXiv:hep-ph/0005027];
  J.~R.~Ellis, T.~Falk, G.~Ganis, K.~A.~Olive and M.~Srednicki,
  Phys.\ Lett.\ B {\bf 510} (2001) 236
  [hep-ph/0102098].

\bibitem{cmssm}
 G.~L.~Kane, C.~F.~Kolda, L.~Roszkowski and J.~D.~Wells,
  Phys.\ Rev.\  D {\bf 49} (1994) 6173
  [arXiv:hep-ph/9312272];
  J.~R.~Ellis, T.~Falk, K.~A.~Olive and M.~Schmitt,
Phys.\ Lett.\ B {\bf 388} (1996) 97
[arXiv:hep-ph/9607292];
Phys.\ Lett.\ B {\bf 413} (1997) 355
[arXiv:hep-ph/9705444];
J.~R.~Ellis, T.~Falk, G.~Ganis, K.~A.~Olive and M.~Schmitt,
Phys.\ Rev.\ D {\bf 58} (1998) 095002
[arXiv:hep-ph/9801445];
V.~D.~Barger and C.~Kao,
Phys.\ Rev.\ D {\bf 57} (1998) 3131
[arXiv:hep-ph/9704403];
J.~R.~Ellis, T.~Falk, G.~Ganis and K.~A.~Olive,
Phys.\ Rev.\ D {\bf 62} (2000) 075010
[arXiv:hep-ph/0004169];
L.~Roszkowski, R.~Ruiz de Austri and T.~Nihei,
JHEP {\bf 0108} (2001) 024
[arXiv:hep-ph/0106334];
  A.~Djouadi, M.~Drees and J.~L.~Kneur,
JHEP {\bf 0108} (2001) 055
[arXiv:hep-ph/0107316];
U.~Chattopadhyay, A.~Corsetti and P.~Nath,
Phys.\ Rev.\ D {\bf 66} (2002) 035003
[arXiv:hep-ph/0201001];
J.~R.~Ellis, K.~A.~Olive and Y.~Santoso,
New Jour.\ Phys.\  {\bf 4} (2002) 32
[arXiv:hep-ph/0202110];
H.~Baer, C.~Balazs, A.~Belyaev, J.~K.~Mizukoshi, X.~Tata and Y.~Wang,
JHEP {\bf 0207} (2002) 050
[arXiv:hep-ph/0205325];
R.~Arnowitt and B.~Dutta,
arXiv:hep-ph/0211417.

\bibitem{AbdusSalam:2011fc}
  S.~S.~AbdusSalam,  {\it et al.},
  Eur.\ Phys.\ J.\ C {\bf 71} (2011) 1835 
  [arXiv:1109.3859 [hep-ph]].

 \bibitem{nuhm1}
H.~Baer, A.~Mustafayev, S.~Profumo, A.~Belyaev and X.~Tata,
  Phys.\ Rev.\  D {\bf 71} (2005) 095008
  [arXiv:hep-ph/0412059];
            H.~Baer, A.~Mustafayev, S.~Profumo, A.~Belyaev and X.~Tata,
               {\em JHEP} {\bf 0507} (2005) 065, 
               hep-ph/0504001;
  J.~R.~Ellis, K.~A.~Olive and P.~Sandick,
  Phys.\ Rev.\  D {\bf 78} (2008) 075012
  [arXiv:0805.2343 [hep-ph]];
 J.~Ellis, F.~Luo, K.~A.~Olive and P.~Sandick,
  Eur.\ Phys.\ J.\ C {\bf 73} (2013) 2403
  [arXiv:1212.4476 [hep-ph]].

\bibitem{nuhm2}
J.~Ellis, K.~Olive and Y.~Santoso,
Phys.\ Lett.\  B~{\bf 539} (2002) 107
[arXiv:hep-ph/0204192];
J.~R.~Ellis, T.~Falk, K.~A.~Olive and Y.~Santoso,
Nucl.\ Phys.\ B {\bf 652} (2003) 259
[arXiv:hep-ph/0210205].

\bibitem{pMSSM}
See, for example,
C.~F.~Berger, J.~S.~Gainer, J.~L.~Hewett and T.~G.~Rizzo,
  JHEP {\bf 0902}, 023 (2009)
  [arXiv:0812.0980 [hep-ph]];
S.~S.~AbdusSalam, B.~C.~Allanach, F.~Quevedo, F.~Feroz and M.~Hobson,
  Phys.\ Rev.\ D {\bf 81}, 095012 (2010)
  [arXiv:0904.2548 [hep-ph]];
  J.~A.~Conley, J.~S.~Gainer, J.~L.~Hewett, M.~P.~Le and T.~G.~Rizzo,
  Eur.\ Phys.\ J.\ C {\bf 71}, 1697 (2011)
  [arXiv:1009.2539 [hep-ph]];
  J.~A.~Conley, J.~S.~Gainer, J.~L.~Hewett, M.~P.~Le and T.~G.~Rizzo,
  [arXiv:1103.1697 [hep-ph]];
  B.~C.~Allanach, A.~J.~Barr, A.~Dafinca and C.~Gwenlan,
  JHEP {\bf 1107}, 104 (2011)
  [arXiv:1105.1024 [hep-ph]];
  S.~Sekmen, S.~Kraml, J.~Lykken, F.~Moortgat, S.~Padhi, L.~Pape, M.~Pierini and H.~B.~Prosper {\it et al.},
  JHEP {\bf 1202} (2012) 075
  [arXiv:1109.5119 [hep-ph]];
  A.~Arbey, M.~Battaglia and F.~Mahmoudi,
  Eur.\ Phys.\ J.\ C {\bf 72} (2012) 1847
  [arXiv:1110.3726 [hep-ph]];
  A.~Arbey, M.~Battaglia, A.~Djouadi and F.~Mahmoudi,
  Phys.\ Lett.\ B {\bf 720} (2013) 153
  [arXiv:1211.4004 [hep-ph]];
  M.~W.~Cahill-Rowley, J.~L.~Hewett, A.~Ismail and T.~G.~Rizzo,
  Phys.\ Rev.\ D {\bf 88} (2013) 3,  035002
  [arXiv:1211.1981 [hep-ph]];
C.~Strege, G.~Bertone, G.~J.~Besjes, S.~Caron, R.~Ruiz de Austri, A.~Strubig and R.~Trotta,
  JHEP {\bf 1409} (2014) 081
  [arXiv:1405.0622 [hep-ph]];
  M.~Cahill-Rowley, J.~L.~Hewett, A.~Ismail and T.~G.~Rizzo,
  Phys.\ Rev.\ D {\bf 91} (2015) 5,  055002
  [arXiv:1407.4130 [hep-ph]];
  {L.~Roszkowski, E.~M.~Sessolo and A.~J.~Williams,
  JHEP {\bf 1502}, 014 (2015)
  [arXiv:1411.5214 [hep-ph]];
  {M.~E.~C.~Catalan, S.~Ando, C.~Weniger and F.~Zandanel,
  arXiv:1503.00599 [hep-ph];}
  J.~Chakrabortty, A.~Choudhury and S.~Mondal,
  arXiv:1503.08703 [hep-ph].}

   \bibitem{newBNL} 
 G.~Bennett et al.\ [The Muon g-2 Collaboration],
 {\it Phys. Rev. Lett.} {\bf 92} (2004) 161802, 
 [arXiv:hep-ex/0401008]; and
  {\em Phys.\ Rev.} {\bf D 73} (2006) 072003
  [arXiv:hep-ex/0602035].

\bibitem{mc10} O.~Buchmueller {\it et al.},
Eur.\ Phys.\ J.\ C {\bf 74} (2014) 12,  3212
 [arXiv:1408.4060 [hep-ph]].

\bibitem{multinest} F.~Feroz and M.P.~Hobson, 
  Mon.~Not.~Roy.~Astron.~Soc. {\bf 384} (2008) 449 
  [arXiv:0704.3704 [astro-ph]].
  F.~Feroz, M.P.~Hobson and M.~Bridges,
  Mon.~Not.~Roy.~Astron.~Soc. {\bf 398} (2009) 1601-1614
  [arXiv:0809.3437 [astro-ph]].
  F.~Feroz, M.P.~Hobson, E.~Cameron and A.N.~Pettitt,
  [arXiv:1306.2144 [astro-ph]].

\bibitem{Buchmueller:2013exa}
  O.~Buchmueller and J.~Marrouche,
  Int.\ J.\ Mod.\ Phys.\ A {\bf 29} (2014) 1450032
  [arXiv:1304.2185 [hep-ph]].

\bibitem{Atom}
M.~Papucci, K.~Sakurai, A.~Weiler and L.~Zeune, {\it {\tt Atom}: Automated Tests of Models}, {\em in preparation}.

\bibitem{Scorpion}
{\tt Scorpion} was first developed by J.~Marrouche, and developed further
by O.~Buchmueller, M.~Citron, S.~Malik and K.J.~de~Vries: details may be obtained
by contacting O.~Buchmueller.

\bibitem{mc9} O.~Buchmueller {\it et al.},
  Eur.\ Phys. J. C {\bf 74} (2014) 2922
  [arXiv:1312.5250 [hep-ph]].

\bibitem{mc7}
O.~Buchmueller {\it et al.},
Eur.\ Phys.\ J.\ C {\bf 72} (2012) 1878
[arXiv:1110.3568 [hep-ph]].

\bibitem{mc8} O.~Buchmueller {\it et al.},
  Eur.\ Phys.\ J.\ C {\bf 72} (2012) 2243
  [arXiv:1207.7315].

\bibitem{mc8.5} O.~Buchmueller {\it et al.},
  Eur.\ Phys.\ J.\ C {\bf 74} (2014) 2809
  [arXiv:1312.5233 [hep-ph]].

\bibitem{mcweb}
For more information and updates, please see {\tt http://cern.ch/mastercode/}.

\bibitem{Allanach:2001kg}
  B.~C.~Allanach,
  Comput.\ Phys.\ Commun.\  {\bf 143} (2002) 305
  [arXiv:hep-ph/0104145].

\bibitem{Svenetal}
  S.~Heinemeyer {\it et al.}, 
  JHEP {\bf 0608} (2006) 052
  [arXiv:hep-ph/0604147];
  S.~Heinemeyer, W.~Hollik, A.~M.~Weber and G.~Weiglein,
  JHEP {\bf 0804} (2008) 039
  [arXiv:0710.2972 [hep-ph]].

\bibitem{FeynHiggs}
 G.~Degrassi, S.~Heinemeyer, W.~Hollik, P.~Slavich and G.~Weiglein,
  Eur.\ Phys.\ J.\ C {\bf 28} (2003) 133
  [arXiv:hep-ph/0212020];
   S.~Heinemeyer, W.~Hollik and G.~Weiglein,
  Eur.\ Phys.\ J.\ C {\bf 9} (1999) 343
  [arXiv:hep-ph/9812472];
  S.~Heinemeyer, W.~Hollik and G.~Weiglein,
  Comput.\ Phys.\ Commun.\  {\bf 124} (2000) 76
  [arXiv:hep-ph/9812320];
   M.~Frank {\it et al.}, 
  JHEP {\bf 0702} (2007) 047
  [arXiv:hep-ph/0611326];
  See {\tt http://www.feynhiggs.de}~.

\bibitem{Mh-logresum} T.~Hahn, S.~Heinemeyer, W.~Hollik, H.~Rzehak and
  G.~Weiglein, 
  Phys.\ Rev.\ Lett.\  {\bf 112} (2014) 141801
  [arXiv:1312.4937 [hep-ph]].

\bibitem{SuFla}
 G.~Isidori and P.~Paradisi,
  Phys.\ Lett.\ B {\bf 639} (2006) 499
  [arXiv:hep-ph/0605012];
  G.~Isidori, F.~Mescia, P.~Paradisi and D.~Temes,
  Phys.\ Rev.\  D {\bf 75} (2007) 115019
  [arXiv:hep-ph/0703035], and references therein.

\bibitem{SuperIso}
F.~Mahmoudi,
  Comput.\ Phys.\ Commun.\  {\bf 178} (2008) 745
  [arXiv:0710.2067 [hep-ph]]; 
  Comput.\ Phys.\ Commun.\  {\bf 180} (2009) 1579
  [arXiv:0808.3144 [hep-ph]];
  D.~Eriksson, F.~Mahmoudi and O.~Stal,
  JHEP {\bf 0811} (2008) 035
  [arXiv:0808.3551 [hep-ph]].

\bibitem{MicroMegas}
  G.~Belanger, F.~Boudjema, A.~Pukhov and A.~Semenov,
  Comput.\ Phys.\ Commun.\  {\bf 176} (2007) 367
  [arXiv:hep-ph/0607059];
  Comput.\ Phys.\ Commun.\  {\bf 149} (2002) 103
  [arXiv:hep-ph/0112278];
  Comput.\ Phys.\ Commun.\  {\bf 174} (2006) 577
  [arXiv:hep-ph/0405253].

\bibitem{SSARD}  Information about this code is available from K.~A.~Olive: it contains important contributions 
from T.~Falk, A.~Ferstl, G.~Ganis, A.~Mustafayev, J.~McDonald, F. Luo, K.~A.~Olive, P.~Sandick, Y.~Santoso, V. Spanos, and M.~Srednicki. 

\bibitem{Muhlleitner:2003vg}
    M.~Muhlleitner, A.~Djouadi, Y.~Mambrini,
    Comput.\ Phys.\ Commun. {\bf 168} (2005) 46
    [arXiv:hep-ph/0311167].

\bibitem{HiggsSignals}
  P.~Bechtle, S.~Heinemeyer, O.~St{\aa}l, T.~Stefaniak and G.~Weiglein,
  Eur.\ Phys.\ J.\ C {\bf 74} (2014) 2,  2711
  [arXiv:1305.1933 [hep-ph]];
  JHEP {\bf 1411} (2014) 039
  [arXiv:1403.1582 [hep-ph]].

\bibitem{HiggsBounds}  
  P.~Bechtle, O.~Brein, S.~Heinemeyer, G.~Weiglein and K.~E.~Williams,
  Comput.\ Phys.\ Commun.\  {\bf 181} (2010) 138
  [arXiv:0811.4169 [hep-ph]],
  Comput.\ Phys.\ Commun.\  {\bf 182} (2011) 2605
  [arXiv:1102.1898 [hep-ph]];
  P.~Bechtle {\it et al.}, 
  Eur.\ Phys.\ J.\ C {\bf 74} (2014) 3,  2693
  [arXiv:1311.0055 [hep-ph]].

\bibitem{SLHA}
P.~Skands {\it et al.},
  JHEP {\bf 0407} (2004) 036
  [arXiv:hep-ph/0311123];
  B.~Allanach {\it et al.},
  Comput.\ Phys.\ Commun.\  {\bf 180} (2009) 8
  [arXiv:0801.0045 [hep-ph]].

  \bibitem{HFAG14}
Y.~Amhis {\it et al.}  [Heavy Flavor Averaging Group (HFAG) Collaboration],
  arXiv:1412.7515 [hep-ex].

  \bibitem{Misiak15}
M.~Misiak, H.~M.~Asatrian, R.~Boughezal, M.~Czakon, T.~Ewerth, A.~Ferroglia, P.~Fiedler and P.~Gambino {\it et al.},
  arXiv:1503.01789 [hep-ph].

\bibitem{Bellebtn}
B.~Kronenbitter {\it et al.}  [Belle Collaboration],
  arXiv:1503.05613 [hep-ex].

\bibitem{UTfit15}
A.~J.~Bevan {\it et al.}, UTfit Collaboration, {\tt http://utfit.org/UTfit.}

\bibitem{lux}
   D.~S.~Akerib {\it et al.}  [LUX Collaboration],
  Phys.\ Rev.\ Lett.\  {\bf 112}, 091303 (2014)
  [arXiv:1310.8214 [astro-ph.CO]].

  \bibitem{XENON100}
  E.~Aprile {\it et al.}  [XENON100 Collaboration],
  Phys.\ Rev.\ Lett.\  {\bf 107} (2011) 131302
  [arXiv:1104.2549 [astro-ph.CO]].

\bibitem{Aad:2015zhl} 
  G.~Aad {\it et al.}  [CMS Collaboration],
  arXiv:1503.07589 [hep-ex].

\bibitem{CMSHA}
  V.~Khachatryan {\it et al.}  [ CMS Collaboration],
  JHEP {\bf 1410} (2014) 160
  [arXiv:1408.3316 [hep-ex]].

  \bibitem{ATLASHA}
ATLAS Collaboration, \\{\tt https://cds.cern.ch/record/1744694/} {\tt files/ATLAS-CONF-2014-049.pdf}.

\bibitem{HBtautau}
  P.~Bechtle, S.~Heinemeyer, O.~St{\aa}l, T.~Stefaniak and G.~Weiglein,
SCIPP 15/05, 
{\em in preparation}.

\bibitem{ATLAS:2011tau}
    The~ATLAS~Collaboration, the~CMS~Collaboration and the~LHC~Higgs~Combination~Group
    (2011), \\
    {\tt https://cdsweb.cern.ch/record/1375842/
    files/ATL-PHYS-PUB-2011-011.pdf}.

\bibitem{rivet}
  A.~Buckley, J.~Butterworth, L.~Lonnblad, D.~Grellscheid, H.~Hoeth, J.~Monk, H.~Schulz and F.~Siegert,
  Comput.\ Phys.\ Commun.\  {\bf 184} (2013) 2803
  [arXiv:1003.0694 [hep-ph]].

\bibitem{PYTHIA6}
  T.~Sjostrand, S.~Mrenna and P.~Z.~Skands,
  JHEP {\bf 0605} (2006) 026
  [hep-ph/0603175].

\bibitem{atom_application}
  M.~Papucci, J.~T.~Ruderman and A.~Weiler,
  JHEP {\bf 1209} (2012) 035
  [arXiv:1110.6926 [hep-ph]];
  J.~S.~Kim, K.~Rolbiecki, K.~Sakurai and J.~Tattersall,
  JHEP {\bf 1412} (2014) 010
  [arXiv:1406.0858 [hep-ph]];
  P.~Grothaus, S.~P.~Liew and K.~Sakurai,
  arXiv:1502.05712 [hep-ph];
  T.~Jacques and K.~Nordstr{\" o}m,
  arXiv:1502.05721 [hep-ph].

\bibitem{atom_validation}
  M.~Papucci, K.~Sakurai, A.~Weiler and L.~Zeune,
  Eur.\ Phys.\ J.\ C {\bf 74} (2014) 11,  3163
  [arXiv:1402.0492 [hep-ph]];

\bibitem{DELPHES3}
  J.~de Favereau {\it et al.}  [DELPHES 3 Collaboration],
  JHEP {\bf 1402} (2014) 057
  [arXiv:1307.6346 [hep-ex]];
  see also S.~Ovyn, X.~Rouby and V.~Lemaitre,
  arXiv:0903.2225 [hep-ph] and
  {\tt http://www.fynu.ucl.ac.be/users/s.ovyn/} {\tt Delphes/index.html}.

\bibitem{LandS}
  {M.~S.~Chen and A.~Korytov, \\ {\tt http://mschen.web.cern.ch/mschen/}
    \\ LandS/}.

\bibitem{cms0l-aT} V.~Khachatryan {\it et al.}  [CMS Collaboration],
  Phys.\ Lett.\ B {\bf 698} (2011) 196
  [arXiv:1101.1628 [hep-ex]].

\bibitem{cms_mt2}
  V.~Khachatryan {\it et al.}  [CMS Collaboration],
  arXiv:1502.04358 [hep-ex].

\bibitem{CMSsearches}
    V.~Khachatryan {\it et al.}  [CMS Collaboration],
    arXiv:1408.3583 [hep-ex];
%
    S.~Chatrchyan {\it et al.}  [CMS Collaboration],
    Eur.\ Phys.\ J.\ C {\bf 73}, no. 12, 2677 (2013)
    [arXiv:1308.1586 [hep-ex]];
%
    S.~Chatrchyan {\it et al.}  [CMS Collaboration],
    Phys.\ Lett.\ B {\bf 718}, 815 (2013)
    [arXiv:1206.3949 [hep-ex]];
%
    S.~Chatrchyan {\it et al.}  [CMS Collaboration],
    JHEP {\bf 1401}, 163 (2014)
    [Erratum-ibid.\  {\bf 1501}, 014 (2015)]
    [arXiv:1311.6736, arXiv:1311.6736 [hep-ex]].
    S.~Chatrchyan {\it et al.}  [CMS Collaboration],
    Phys.\ Rev.\ D {\bf 90}, no. 3, 032006 (2014)
    [arXiv:1404.5801 [hep-ex]].

\bibitem{atlas_3L}
  G.~Aad {\it et al.}  [ATLAS Collaboration],
  JHEP {\bf 1404} (2014) 169
  [arXiv:1402.7029 [hep-ex]].

\bibitem{cms_EWK}
  V.~Khachatryan {\it et al.}  [CMS Collaboration],
  Eur.\ Phys.\ J.\ C {\bf 74} (2014) 9,  3036
  [arXiv:1405.7570 [hep-ex]].

\bibitem{atlas_2L}
  G.~Aad {\it et al.}  [ATLAS Collaboration],
  JHEP {\bf 1405} (2014) 071
  [arXiv:1403.5294 [hep-ex]].

\bibitem{Bharucha:2013epa}
  A.~Bharucha, S.~Heinemeyer and F.~von der Pahlen,
  Eur.\ Phys.\ J.\ C {\bf 73} (2013) 11,  2629
  [arXiv:1307.4237].

\bibitem{Han:2014nba}
  T.~Han, Z.~Liu and S.~Su,
  JHEP {\bf 1408} (2014) 093
  [arXiv:1406.1181 [hep-ph]].

\bibitem{arXiv:1308.2631}
G.~Aad {\it et al.}  [ATLAS Collaboration],
  JHEP {\bf 1310} (2013) 189
  [arXiv:1308.2631 [hep-ex]].

\bibitem{arXiv:1407.0583}
  G.~Aad {\it et al.}  [ATLAS Collaboration],
  JHEP {\bf 1411} (2014) 118
  [arXiv:1407.0583 [hep-ex]].

  \bibitem{atlas_stop-monojet}
  G.~Aad {\it et al.}  [ATLAS Collaboration],
  Phys.\ Rev.\ D {\bf 90} (2014) 5,  052008
  [arXiv:1407.0608 [hep-ex]].

\bibitem{atlas_stop-1L}
 G.~Aad {\it et al.}  [ATLAS Collaboration],
  JHEP {\bf 1411} (2014) 118
  [arXiv:1407.0583 [hep-ex]].

 \bibitem{LEPSUSY}
  {LEP SUSY Working Group, ALEPH, DELPHI, L3 and OPAL Collaborations,
  {\tt http://lepsusy.web.cern.ch/lepsusy/} {\tt Welcome.html}.} 

\bibitem{mc1} O.~Buchmueller {\it et al.},
   Phys.\ Lett.\ B {\bf 657} (2007) 87
  [arXiv:0707.3447 [hep-ph]];

\bibitem{James}
\gray{E.~L.~James, {\it Fifty Shades of Grey}, Vintage Books, a division of Random House, Inc., 2012 (New York).}

\bibitem{g-2}
  D.~Stockinger,
  J.\ Phys.\ G {\bf 34} (2007) R45
  [arXiv:hep-ph/0609168];
    J.~Miller, E.~de~Rafael and B.~Roberts,
   {\em Rept.\ Prog.\ Phys.} {\bf 70} (2007) 795
   [arXiv:hep-ph/0703049];
    J.~Prades, E.~de Rafael and A.~Vainshtein,
  arXiv:0901.0306 [hep-ph];
    F.~Jegerlehner and A.~Nyffeler,
  Phys.\ Rept.\  {\bf 477}, 1 (2009)
  [arXiv:0902.3360 [hep-ph]];
   M.~Davier, A.~Hoecker, B.~Malaescu, C.~Z.~Yuan and Z.~Zhang,
  Eur.\ Phys.\ J.\  C {\bf 66}, 1 (2010)
  [arXiv:0908.4300 [hep-ph]].
  J.~Prades,
  Acta Phys.\ Polon.\ Supp.\  {\bf 3}, 75 (2010)
  [arXiv:0909.2546 [hep-ph]];
    T.~Teubner, K.~Hagiwara, R.~Liao, A.~D.~Martin and D.~Nomura,
  arXiv:1001.5401 [hep-ph];
  M.~Davier, A.~Hoecker, B.~Malaescu and Z.~Zhang,
  Eur.\ Phys.\ J.\  C {\bf 71} (2011) 1515
  [arXiv:1010.4180 [hep-ph]].

\bibitem{Jegerlehner}
F.~Jegerlehner and R.~Szafron,
  Eur.\ Phys.\ J.\  C {\bf 71} (2011) 1632
  [arXiv:1101.2872 [hep-ph]];
  M.~Benayoun, P.~David, L.~DelBuono and F.~Jegerlehner,
  Eur.\ Phys.\ J.\ C {\bf 73} (2013) 2453
  [arXiv:1210.7184 [hep-ph]].

\bibitem{LZ}
 P.~Cushman, C.~Galbiati, D.~N.~McKinsey, H.~Robertson, T.~M.~P.~Tait, D.~Bauer, A.~Borgland and B.~Cabrera {\it et al.},
  arXiv:1310.8327 [hep-ex].

\bibitem{nuback}
P.~Cushman, C.~Galbiati, D.~N.~McKinsey, H.~Robertson, T.~M.~P.~Tait, D.~Bauer, A.~Borgland and B.~Cabrera {\it et al.},
  arXiv:1310.8327 [hep-ex].

\bibitem{DN}
M.~Drees and M.~M.~Nojiri,
  Phys.\ Rev.\ D {\bf 47} (1993) 4226
  [hep-ph/9210272].

\bibitem{HIN}
J.~Hisano, K.~Ishiwata and N.~Nagata,
  Phys.\ Rev.\ D {\bf 82} (2010) 115007
  [arXiv:1007.2601 [hep-ph]].

\bibitem{JKPR}
See, for example J.~Jaeckel, V.~V.~Khoze, T.~Plehn and P.~Richardson,
  Phys.\ Rev.\ D {\bf 85} (2012) 015015
  [arXiv:1109.2072 [hep-ph]].

\bibitem{EGLOS}
J.~R.~Ellis, J.~Giedt, O.~Lebedev, K.~Olive and M.~Srednicki,
  Phys.\ Rev.\ D {\bf 78} (2008) 075006
  [arXiv:0806.3648 [hep-ph]].

\bibitem{atlasrun2}
ATLAS Collaboration, \\
  arXiv:1307.7292 [hep-ex]

  \bibitem{ATL-PHYS-PUB-2013-011}
  ATLAS Collaboration,
  ATL-PHYS-PUB-2013-011.

 \bibitem{CRtool}
G.~Salam and A.~Weiler, \\
{\tt http://collider-reach.web.cern.ch/} \\ {\tt collider-reach/}.

\bibitem{ATL-PHYS-PUB-2014-010}
    ATLAS Collaboration, \\
    ATL-PHYS-PUB-2014-010.

\bibitem{Schwaller:2013baa}
  P.~Schwaller and J.~Zurita,
  JHEP {\bf 1403} (2014) 060
  [arXiv:1312.7350 [hep-ph]].

    
\end{thebibliography}
\end{document}